\documentclass{report} 
\usepackage{graphicx}
\usepackage{graphicx}
\usepackage{bm}
\usepackage{amsmath}
\usepackage{amssymb}
\usepackage{simplewick}
\newcommand{\ket}[1]{\ensuremath{\left|{#1}\right\rangle}}
\newcommand{\bra}[1]{\ensuremath{\left\langle{#1}\right|}}
\newcommand{\braket}[1]{\ensuremath{\left\langle{#1}\right\rangle}}
\newcommand{\opA}{\hat{A}}
\newcommand{\opB}{\hat{B}}
\newcommand{\opK}{\hat{K}}
\newcommand{\opH}{\hat{M}}

\newcommand{\opO}{\hat{O}}
\newcommand{\im}[1]{\ensuremath{\mathrm{Im}\,{#1}}}
\newcommand{\re}[1]{\ensuremath{\mathrm{Re}\,{#1}}}

\newcommand{\Hcal}{\mathcal{H}}
\newcommand{\HHcal}{\mathcal{H}^{(2)}}

\newcommand{\slpar}{\partial\!\!\!/}

\newcommand{\HRule}{\rule{\linewidth}{0.5mm}}
\begin{document}
\begin{titlepage}

\begin{center}

\includegraphics[width=0.2\textwidth]{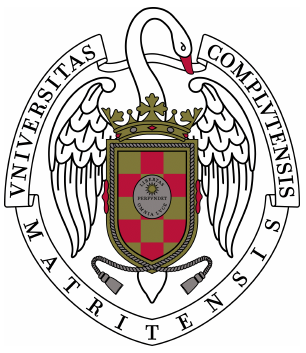}\\[1cm]    

\textsc{\Large Departamento de F{\'i}sica Te{\'o}rica I\\Universidad Complutense de Madrid}\\[1.5cm]

\textsc{\Large Doctoral Thesis}\\[0.5cm]

\HRule \\[0.4cm]
{ \huge \bfseries Bipartite entanglement of localized separated systems}\\[0.4cm]

\HRule \\[1.5cm]

\begin{minipage}{0.4\textwidth}
\begin{flushleft} \large
\emph{Author:}\\
Carlos \textsc{Sab{\'i}n Lestayo}
\end{flushleft}
\end{minipage}
\begin{minipage}{0.4\textwidth}
\begin{flushright} \large
\emph{Supervisor:} \\
Dr.~Juan \textsc{Le{\'o}n Garc{\'i}a}
\end{flushright}
\end{minipage}

\vfill
\textsc{\large Instituto de F{\'i}sica Fundamental\\Consejo Superior de Investigaciones Cient{\'i}ficas}\\[1.5cm]
\includegraphics[width=0.4\textwidth]{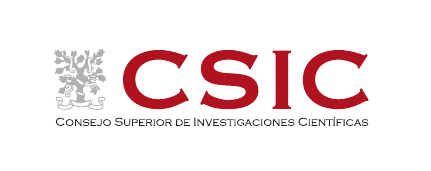}\\[1cm]    

{\large \date{}}

\end{center}
November 2011. Committee: M. A. Mart\'in-Delgado, A. Luis, I. Cirac, E. Solano, A. Cabello.
\end{titlepage}
\chapter*{}
\pagenumbering{roman} 
\begin{flushleft}
\textit{
Para Mencha.\\
\  Para Nicol{\'a}s.}
\end{flushleft}
\addcontentsline{toc}{chapter}{Agradecimientos}
\chapter*{Agradecimientos}
\begin{quote}
``People are afraid to face how great a part of life is dependent on luck'' (Woody Allen, \textit{Match Point})
\end{quote}

D\'ejenme que les cuente una historia sobre el azar.

El azar llev\'o a un estudiante sin ambiciones investigadoras de los \'ultimos cursos de F\'isica a una conferencia divulgativa de Guillermo Garc\'ia Alcaine sobre  Mec\'anica Cu\'antica, dirigida a los alumnos. Al final de la charla, Guillermo hizo publicidad sobre un Trabajo Acad\'emicamente Dirigido que le gustar\'ia dirigir, con un tema relacionado con el entrelazamiento. Puesto que mis compa\~neros estaban en clase (donde sin duda deber\'ia estar yo tambi\'en), pr\'acticamente nadie de los \'ultimos cursos recibi\'o esa propaganda, y fue la \'unica que se hizo. As\'i que no tuve ninguna competencia cuando al d\'ia siguiente aparec\'i por su despacho. \'Ese fue el comienzo de m\'as de dos a\~nos de inverstigaci\'on, que dieron lugar, adem\'as del mencionado trabajo, a una Tesis de M\'aster y un art\'iculo sobre clasificaci\'on y medida del entrelazamiento. El recuerdo de Guillermo y su sentido del humor me acompa\~na siempre, con su ejemplo de rigor, honestidad y perfeccionismo, cualidades que en m\'i se dan en tan peque\~nas dosis que son s\'olo torpes remedos de las suyas. S\'olo tengo para \'el admiraci\'on y gratitud.

Por motivos que no vienen al caso, Guillermo ya no dirig\'ia Tesis Doctorales. Por el camino yo me hab\'ia ido creyendo que pod\'ia dedicarme a eso de la investigaci\'on, pero el tiempo iba pasando sin que se me abriera ninguna puerta. El d\'ia que hice el \'ultimo examen de la carrera, mi costumbre de leer los carteles de las paredes (que tanto sol\'ia exasperar a mis acompa\~nantes) me llev\'o a leer un anuncio de Tesis Doctoral de un tal Juan Le\'on del Instituto de bla, bla, bla sobre...!`teor\'ia del entrelazamiento! ?`Ven lo que les dec\'ia sobre el azar?

La expresi\'on ``padre cient\'ifico'' se usa muchas veces con cierta alegr\'ia para designar la relaci\'on de cierto investigador con su director de Tesis. Pocas veces es tan cierta como en mi caso. Juan confi\'o en m\'i sin tener demasiados motivos para hacerlo y decidi\'o hacer frente, por pura bondad, a todas las dificultades que desde el principio se vio que yo le procurar\'ia. Tuvo varias ocasiones para deshacerse elegantemente de m\'i, pero no lo hizo. En lugar de eso se desvivi\'o por ayudarme. Me ense\~n\'o rigor, disciplina intelectual, ambici\'on investigadora, y despu\'es comprendi\'o que pod\'ia, que deb\'ia dejarme volar con libertad. Hoy yo le hablo ya como los hijos adolescentes hablan a sus padres, mezclando la indiferencia con el sarcasmo. Es pura m\'ascara: jam\'as olvidar\'e todo lo que Juan ha hecho por m\'i.

Guillermo, Juan... a hombros de gigantes tales, ?`qui\'en no acertar\'ia a ver con claridad? Pero recuerden que yo quer\'ia hablarles del azar. Yo ya llevaba un tiempo en el Consejo, aguantando cada ma\~nana la mirada entre ir\'onica (``?`a d\'onde va \'este?") y admonitoria (``!`haz algo, idiota!'') de D. Santiago Ram\'on y Cajal y D. Severo Ochoa, (cuyas estatuas se erigieron a los pocos d\'ias de llegar yo en sustituci\'on del busto de alg\'un olvidado pol\'itico de la dictadura), cuando Juanjo Garc\'ia Ripoll sac\'o plaza en nuestro grupo. Un d\'ia que comimos los dos juntos en el comedor, surgi\'o la idea de intentar ver qu\'e aplicaciones podr\'ian terner los resultados que hab\'iamos ido obteniendo Juan y yo en el campo emergente de los circuitos superconductores. Ah\'i empez\'o nuestra colaboraci\'on con Juanjo. Su asombrosa intuici\'on, cimentada sobre un abrumador conocimiento de toda la F\'isica experimental y te\'orica de los siglos XX y XXI, permitieron dar a esta Tesis otro salto de calidad. Gracias a \'el adem\'as pude entrar en contacto con Kike (as\'i quiere \'el que lo escribamos) Solano, cuyo entusiasmo y atrevimiento, su visi\'on po\'etica y rom\'antica del oficio, son un est\'imulo para todos los investigadores que nos cruzamos en su camino, quienes secretamente aspiramos a ser, cuando seamos j\'ovenes, como \'el.

Pero basta ya de hacer la pelota a los maestros. Durante la elaboraci\'on de esta Tesis, como es natural, conoc\'i a muchas personas. La mejor de todas ellas es mi buen amigo Borja Peropadre. Su bondad, sensatez, perspicacia y sabidur\'ia me hacen seguir confiando en la condici\'on humana y tratar de ser mejor. !`Qui\'en le iba a decir a un soci\'opata como yo que iba a disfrutar tanto compartiendo un despacho! Tambi\'en son buenos amigos el inimitable Marco del Rey, cuya exigencia intelectual y esp\'iritu cr\'itico elevaron considerablemente la calidad del cap\'itulo 5 de esta Tesis, el Dr. Edu Mart\'in Mart\'nez, el S\'uperManager Emilio Alba, Andrea Cadarso y el resto de gente de QUINFOG, as\'i como Alejandro Berm\'udez, Juan Manuel P\'erez Pardo, Fermando Lled\'o, Luis Garay, Lucas Lamata, Pol Forn, Jorge Casanova y qui\'en sabe cu\'anta gente que olvido, pero con la que habr\'e compartido al menos el caf\'e aguado de alg\'un congreso. Dir\'e tambi\'en que sin Paco y sin David, y el resto del Centro de Estudios Cuzco, no habr\'ia podido terminar esta Tesis.

Muchos padres intentan educar a sus hijos a su imagen y semejanza, confiando en que repitan sus doctrinas y consignas. Mis padres no me ense\~naron otra ideolog\'ia que el amor y los buenos sentimientos, y les estoy muy agradecido por ello. Si hubiera heredado la mitad de la originalidad y libertad de pensamiento de mi padre Jos\'e Ram\'on, de su inquietud y su curiosidad, de su insobornable integridad e individualidad, de su amor por la lectura y el conocimiento, de su profundidad para entender lo esencial, me dar\'ia por satisfecho. Si algo hubiera aprendido del amor incondicional de mi madre Rosal\'ia, de su entrega desinteresada por la familia, de su capacidad para seguir adelante por amargo que sea lo que le ofrezca la vida, para tirar de recursos y mantenerlo todo en pie y en orden con originalidad y creatividad, ya ser\'ia mucho. Espero que cuando tengan esta Tesis en sus manos sientan, aunque sea por un instante,  que en el fondo mereci\'o la pena.

Todo lo que soy, para bien o para mal, tiene que ver con mi hermano Mon, poeta, profesor, ingeniero y fil\'ologo (si mencionamos s\'olo lo m\'as evidente). El me ense\~n\'o, entre otras muchas cosas, a ser libre, a amar la cultura, a no tener miedo a pensar y a seguir tu propio camino, y aunque \'el no se acuerde, me convenci\'o sin pretenderlo para que estudiara F\'isica en lugar de Filosof\'ia. Su influencia sobre m\'i ha sido tan grande que lo he tenido que asesinar en sentido Freudiano muchas veces, para resucitarlo inmediatamente. Y mi hermana Mar\'ia, a la que llamaba cuando lloraba desconsolado de ni\~no, complet\'o mi formaci\'on como hombre, con su profundo sentido de la generosidad y la justicia. No puedo evitar acordarme ahora de mi t\'ia Carolita, que ya no podr\'a felicitarme desde el cielo.

Durante la elaboraci\'on de esta Tesis me cas\'e, mi mujer tuvo dos embarazos, naci\'o un hijo al que he visto crecer, aprender a andar y a construir sus primeras frases. Esta Tesis es para ellos. Para Menchita, que ha confiado en mi capacidad y en mis posibilidades mucho m\'as que yo mismo, que ha sacrificado tanto para que llegara este d\'ia, que con su sola presencia da sentido a la extra\~na comedia de estar vivo, de poner un pie delante de otro y caminar, uno, dos, inspirar, espirar, para Menchita que es el \'unico motivo por el que he seguido en pie tras esa noches en las que despertaba de pronto y sent\'ia tanto miedo. Sin ella, ``las estrellas, a pesar de su l\'ampara encendida perder\'ian el camino/ ?`qu\'e ser\'ia del Universo?'' Para Nicol\'as, que pronto cumplir\'a dos a\~nos, a quien deseo que no entienda jam\'as ni una palabra de lo que dice esta Tesis, que encuentre la felicidad m\'as f\'acilmente que su viejo, su pobre viejo que no podr\'a ense\~narle m\'as que alg\'un poema y aquello que dice esa vieja canci\'on de los 60, la misma que escuchaba Desmond Hume en el b\'unker:
\begin{quote}
``You have to make your own kind of music/ sing your own special song/ make your own kind of music/ even if nobody else sings along''
\end{quote}
Madrid, Septiembre del 2011.
\newpage{\ }
\thispagestyle{empty}

\addcontentsline{toc}{chapter}{List of publications}
\chapter*{List of publications}
\section*{}
\subsection*{Chapter 3}
\begin{itemize}
\item[*] (In this chapter, authors in alphabetical order)
\item ``Entanglement swapping between spacelike separated atoms'' , Juan Le\'on, Carlos Sab\'in, Phys. Rev. A \textbf{78}, 052314 (2008).
\item ``Photon exchange and correlation transfer in atom-atom entanglement dynamics'', Juan Le\'on, Carlos Sab\'in, Phys. Rev. A \textbf{79}, 012301 (2009).
\item ``Generation of atom-atom correlations inside and outside the mutual light cone'', Juan Le\'on, Carlos Sab\'in, Phys. Rev. A \textbf{79}, 012304 (2009).
\item ``A case of entanglement generation between causally disconnected atoms'', Juan Le\'on, Carlos Sab\'in, Int. J. Quant. Inf. \textbf{7}, 187-193 (2009).
\item ``Atom-atom entanglement generated at early times by two photon emission'', Juan Le\'on, Carlos Sab\'in, Phys. Scr. T\textbf{135} 014034 (2009). 
\end{itemize}
\subsection*{Chapter 4}
\begin{itemize}
\item ``Dynamics of entanglement via propagating microwave photons'', Carlos Sab\'in, Juan Jos\'e Garc\'ia-Ripoll, Enrique Solano, Juan Le\'on, Phys. Rev. B \textbf{81}, 184501 (2010).
\item ``Detecting ground-state qubit self-excitations in circuit QED: A slow quantum anti-Zeno effect'', Carlos Sab\'in, Juan Le\'on, Juan Jos\'e Garc\'ia-Ripoll, Phys. Rev B \textbf{84}, 024516 (2011).
\end{itemize}
\subsection*{Chapter 5}
\begin{itemize}
\item ``The Fermi problem with artificial atoms in circuit QED'', Carlos Sab\'in, Marco del Rey, Juan Jos\'e Garc\'ia-Ripoll, Juan Le\'on.  Phys. Rev. Lett. \textbf{107}, 150402 (2011).
\item ``Short-time quantum detection: probing quantum fluctuations'', Marco del Rey, Carlos Sab\'in, Juan Le\'on. Submitted to Phys. Rev. Lett. arXiv: 1108.0672.
\end{itemize}
\subsection*{Chapter 6}
\begin{itemize}
\item ``Quantum simulation of the Majorana equation and unphysical operations'', Jorge Casanova, Carlos Sab\'in, Juan Le\'on, I\~nigo L. Egusquiza, Rene Gerritsma, Christian F. Roos, Juan Jos\'e Garc\'ia-Ripoll, Enrique Solano. Accepted to Phys. Rev. X. arXiv:1102.1651.
\item \textbf{To be submitted} ``On Majorana Hamiltonians'', Carlos Sab\'in, Jorge Casanova, Juan Jos\'e Garc\'ia-Ripoll, Juan Le\'on, Enrique Solano, I\~nigo L. Egusquiza.
\item \textbf{To be submitted} ``Quantum simulation of relativistic potentials without potentials'', Carlos Sab\'in, Jorge Casanova, Juan Jos\'e Garc\'ia-Ripoll, Lucas Lamata , Enrique Solano, Juan Le\'on.
\end{itemize}
\section*{Other publications not included in this Thesis}
\begin{itemize}
\item ``A classification of entanglement in three-qubit systems'', Carlos Sab\' in, Guillermo Garc\'ia-Alcaine, European Physical Journal D, \textbf{48}, 435-442 (2008).
\end{itemize}

\tableofcontents
\newpage{\ }
\thispagestyle{empty}
\chapter{Introduction}
\pagenumbering{arabic}

In their famous paper of 1935, Einstein, Podolsky and Rosen (EPR) tried to show that Quantum Mechanics couldn't be a complete theory due to the prediction of strange correlations appearing among the parties in some states of composite systems \cite{epr}. Later that year, in a letter to Einstein, Schr\"odinger referred to the phenomenon with the German word  ``verschr\"ankung'', that was translated by himself to ``entanglement'' in \cite{schrodinger}.  Considered unphysical and paradoxical by EPR - ``no reasonable definition of reality could be expected to permit this"-, Bohr \cite{bohr} and Schr\"odinger \cite{schrodinger} immediately acknowledged entanglement as an essential feature of the theory. Today, after the series of experiments based on Bell inequalities \cite{bell} ruling out local hidden variable theories \cite{aspectexperiment}, Quantum Mechanics and its relativistic field version Quantum Field Theory are not considered just as  mere mathematical models that fit with outstanding precision a huge bulk of experimental data, but also a complete theory of Nature at the atomic and subatomic level. Being at the very heart of the theoretical building of Quantum Mechanics, entanglement is thus an essential feature of Nature. Besides its importance from this foundational viewpoint, entanglement is also a key resource for Quantum Information and Quantum Computation tasks.

In the early days after its discovery entanglement was considered as a consequence of the interaction. For instance, Schr\"odinger wrote \cite{schrodinger}:
\begin{quote}
``When two systems,[...] enter into temporary physical interaction due to known forces between them, and when after a time of
mutual influence the systems separate again, then they can no longer be described in the same way as before.[...] By the
interaction, the two representatives (or $\Psi$-functions) have become entangled''
\end{quote}
Therefore, the ``magic'' of  entanglement relies completely on the quantum state, which is generated by a direct interaction between the parties. Today, we know that this is not a complete picture. For instance, in the ``entanglement swapping'' protocol \cite{swapping} the parties that get finally entangled do not interact directly with each other. Instead if we want to entangle A and B with this protocol, A has to share an entangled state with some third party C and also B independently with the same C. Then, a measurement of the state of C leaves A and B in an entangled state. But in this case the origin of the A-B entanglement can be traced back to the
origin of the A-C and B-C entanglements which, following Schr\"odinger would be again the interaction. 

In some sense, the entangling interaction can be regarded as a``black box'' that generates the desired initial entangled state. In typical experiments involving entanglement, the black box may be a nonlinear crystal exhibiting Spontaneous Parametric Down-Conversion \cite{spdc}. A ``pump'' laser beam is directed to the crystal and some of the photons split into entangled pairs, which are then used in the experiment. But what it is inside the box? Where it is exactly the magic?  Can we have a deeper look onto the origin and the generation of entanglement? These are some questions that we will try to analyze in this Thesis. 

Actually, the above is deeply connected with modern investigations on the intriguing notion of ``vacuum entanglement''. That the vacuum of the field is an entangled state was discovered by Summers and Werner in the 1980's \cite{summerswerner}, but it was considered as a mere formal theoretical result until it attracted some attention from a different perspective \cite{reznik}. In \cite{reznik}, vacuum entanglement is exploited to generate entanglement between two spacelike separated detectors. After a finite time of interaction with a scalar field initially in the vacuum state,  the two-level detectors state is shown to evolve from an initial separable state to an entangled one, even if the detectors remain spacelike separated. The standard interpretation of this effect is that the entanglement initially contained in the vacuum state of the field is transferred to the detectors, like in the standard entanglement swapping described above but replacing C by two spacelike separated regions of the field and without measurements. An alternate view was given in \cite{franson}, where the entanglement between spacelike separated detectors is attributed to the exchange of virtual photons between them. Whatever the point of view, the mathematical structure underlying the phenomenon is the same, namely the Feynman propagator of the field. The fact that this object is not restricted to the lightcone was noticed by Feynman himself \cite{feynmanlightcone}. 

An important part of this Thesis is devoted to going a step further the previous results on entanglement generation between spacelike separated objects \cite{franson,reznik, reznikII} by moving to the physical framework of matter-radiation interaction and performing a thorough analysis of the phenomenon, keeping also in mind possible applications in realistic Quantum Information protocols. In a nutshell, one of the main goals was to propose the first realistic experimental proposal to test these effects in the lab. Unfortunately, as we will explain with more detail in the main text, a lot of experimental difficulties arise when dealing with real atoms and photons. This is the main reason to come to the different but related framework of circuit QED.

In circuit QED, superconducting qubits can play the role of artificial two-level atoms that can be coupled to transmission lines along which photons propagate. Thus, these systems mimic the standard matter-radiation interaction with the advantage of experimental amenability. Although they are interesting by themselves, circuit QED setups can be regarded as 1D Quantum Simulations of Quantum Optics setups. In this thesis, we will exploit the analogy to propose the desired experimental test of  ``outside the lightcone'' entanglement . Besides, we will explore some features of the novel physics showing up in this emerging field, as a consequence of the possibility of achieve very large values of the qubit-field coupling strength and the breakdown of the standard Rotating Wave Approximation. 

A natural question showing up when dealing with effects as those described above is: What about causality? Is it violated? Indeed the relationship between entanglement and causality has been present since the EPR paper. At first sight it may seem that the perfect correlations appearing for instance in the paradigmatic singlet state represent by themselves an instantaneous transmission of information. This seems to be also the viewpoint of Einstein himself when he write to Born the famous ``spukhafte Fernwirkung''-commonly translated as ``spooky action at a distance''.  Today, we know that Quantum Mechanics is actually a nonsignaling theory \cite{nonsignaling, gisin}, that is, the statistics of the measurements performed by B is completely independent of the measurements realized by A if A and B are spacelike separated. This entails that information between A and B cannot be transmitted at superluminal rates. In the singlet state, B can only get information on the state of A if both parties share some previous information on the shared entangled state and on the choice of the observable measured by A. In general, is well known in Quantum Information that quantum correlations have to be assisted with Classical Communication -which is of course subluminal- in order to transmit information. Therefore, the fact that a separable state can evolve to an entangled one ``faster than light'' does not represent a violation of Einstenian causality, since cannot be used by itself to transmit information. 

Although this general principle is clear, the translation to the particular setup consisting of  a pair of neutral atoms interacting with the electromagnetic field -the one that we extensively deal with in this Thesis-  has been an open theoretical problem -the so-called Fermi problem- since the 1930's. In particular, in 1932 Fermi \cite{fermi} proposed a ``gedanken'' experiment to check that the probability of excitation of one of the atoms behaves in a causal fashion (see Chapter 5 for more details). Although Fermi's conclusions were conceptually right about the causal nature of the model, a mathematical flaw of his computations opened a Pandora box that couldn't be closed neither by a theoretical consensus nor by an experimental test \cite{hegerfeldtfer, yngvasonfer}.  An important part of this Thesis is devoted to show that there are no causality problems in Fermi two-atom's system and to propose a feasible experimental test, taking advantage again of the possibilities offered by circuit QED. The results of such experiment could close an eight-decade old controversy on the foundations of Quantum Mechanics and Quantum Field Theory.

As we have explained above, circuit QED setups play a central role in this Thesis, as 1D Quantum Simulators of  matter-radiation interaction. In a nutshell, a Quantum Simulator, as Feynman envisioned \cite{Feynman82} is an experimentally amenable quantum device that mimics the dynamics of an unaccessible quantum model. The emerging field of Quantum Simulations is developing very fast a number of such devices in several branches of Physics. In particular, in this Thesis we pay attention to Quantum Simulations with trapped ions of Relativistic Quantum Mechanics systems. \cite{reviewrqm}. Relativistic Quantum Mechanics can be understood as a Quantum Field Theory with a fixed number of particles, thus it amounts to Quantum Field Theory below the energy threshold of pair creation. In this low-energy regime is the right model for relativistic particles. Surprisingly, its equations -such as Klein-Gordon or Dirac ones- come with a wealth of striking predictions, like ``Zitterbewegung'' or Klein paradox, which are very far from a direct detection in the lab \cite{thaller,greinerrqm}. In this sense, recent experiments with trapped ions simulators of the Dirac equation \cite{naturekike} have shed light on the nature and experimental implications of the mentioned phenomena. An important goal for the future would be to simulate the fully relativistic version of the two-atom model used in most part of this Thesis where the atoms would be replaced by relativistic particles following the Dirac equation. Steps in this direction are given at the end of this Thesis, where bipartite systems of  relativistic interacting particles are considered. In the meanwhile, we take a little diversion from the main course and propose the quantum simulation of a crucially different but related one- particle relativistic dynamics, namely the Majorana equation- see details in chapter 6-. The very interesting properties of this equation -which takes its roots in the prematurely lost overwhelming genius of Ettore Majorana \cite{donettore} but was actually introduced in \cite{majoranaeq1, majoranaeq2}- lead us to the introduction of a new class of mathematical objects with physical content, that we refer to in this Thesis as ``Majorana Hamiltonians". 

Besides the Majorana equation, this new class of generalized Hamiltonians is connected with a wider range of applications, going from the implementation of operations like time reversal and charge conjugation to the partial transpose- a crucial formal operation in Quantum Information due to its relationship with entanglement measures-. These operations are ``unphysical'', in the sense that they are not associated to operations that can be implemented directly in the laboratory. Instead, they entail a mathematical formal computation that has to be performed for instance in a classical computer and require the knowledge of the quantum state. With the techniques developed in this Thesis these operations can now be simulated with real physical operations, without stopping an ongoing experiment. From a fundamental viewpoint the similarities and differences between Majorana Hamiltonians and standard textbook Hamiltonians allows us to explore important questions lying at the foundations of Quantum Mechanics and Quantum Field Theory.

\section*{Structure of the Thesis. Brief summary of objectives and results.}

The structure of this thesis is the following. 
\begin{itemize}
\item In chapter 2 we will just provide the reader with some theoretical background on the formalism used in the main text.  

\item  In chapter 3 we will analyze different aspects of the entanglement dynamics in a system consisting of a pair of two-level neutral atoms with an electric dipole coupling to the electromagnetic field. We mainly focus in the relationship of entanglement with the existence or lack of a causal connection between the atoms.
\begin{itemize}
\item[*] In section 3.1 
we analyze whether a pair of neutral two level atoms can become entangled in a finite time while they remain causally
disconnected. The interaction with the electromagnetic field is treated perturbatively  in the electric dipole approximation. First, we start
from an initial state in which the field is in the vacuum and only one atom is excited. We obtain the final atomic correlations for the cases where $n = 0, 1,$ or 2 photons are produced
in a time $t$, and also when the final field state is unknown. Our results show that correlations are sizable inside and
outside the mutual light cone for $n= 1$ and 2, and also that quantum correlations become classical by tracing over the field
state. For $n = 0$ we obtain entanglement generation by photon propagation between the atoms, the correlations come from the
indistinguishability of the source for $n = 1$, and may give rise to entanglement swapping for $n = 2$. Finally, we consider a similar scenario but starting with both atoms excited as initial state. 
\item[*] In  section 3.2
we show a mechanism that projects a pair of neutral two-level atoms from an initially uncorrelated state to a maximally
entangled state while they remain spacelike separated. The atoms begin both excited in a common electromagnetic vacuum, and the
radiation is collected with a partial Bell-state analyzer. If the interaction time is short enough and a certain two-photon Bell
state is detected after the interaction, a high degree of entanglement, even maximal, can be generated while one atom is outside
the light cone of the other, for arbitrary large interatomic distances.
\item[*]In section 3.3
we analyze the entanglement dynamics of a system composed by a pair of neutral two-level atoms that are initially entangled, and
the electromagnetic field, initially in the vacuum state, within the formalism of perturbative quantum field theory up to the
second order. We show that entanglement sudden death and revival can occur while the atoms remain spacelike-separated and
therefore cannot be related with photon exchange between the atoms. We interpret these phenomena as the consequence of a
transfer of atom-atom entanglement to atom-field entanglement and viceversa. We also consider the different bi-partitions of the
system, finding similar relationships between their entanglement evolutions.
\end{itemize}

\item In chapter 4 we move to the framework of circuit QED with artificial atoms, considering these setups as Quantum Simulators of a 1D version of matter-radiation interaction.
\begin{itemize}
\item[*]
In section 4.1 we explore a consequence of the breakdown of the Rotating Wave Approximation in the ultrastrong coupling regime of circuit QED. In particular,
 we study an ultrastrong coupled qubit-cavity system subjected to slow repeated measurements. We demonstrate that even under a few imperfect measurements it is possible to detect transitions of the qubit from its free ground state to the excited state. The excitation probability grows exponentially fast in analogy with the quantum anti-Zeno effect. The dynamics and physics described in this section is accessible to current superconducting circuit technology.
\item[*] In section 4.2 we propose a simple circuit  QED experiment to test the generation of entanglement between two superconducting qubits that are initially in a separable state. Instead of the usual cavity QED picture, we study qubits which are coupled to an open transmission line and get entangled by the exchange of propagating photons. We compute their dynamics using a full quantum field theory beyond the rotating-wave approximation and explore a variety of regimes which go from a weak coupling to the recently introduced ultrastrong coupling regime. Due to the existence of single photons traveling along the line with finite speed, our theory shows a light cone dividing the spacetime in two different regions. In one region, entanglement may only arise due to correlated vacuum fluctuations, while in the other the contribution from exchanged photons shows up.

\end{itemize}
\item  In chapter 5 we show that the nonlocal quantum correlations phenomena considered in previous chapters  are perfectly compatible with Einstenian causality by showing explicitly that Fermi's two-atom system behaves in a causal way, and explore the theoretical consequences of going beyond RWA in a quantum detection.
\begin{itemize}
\item[*]
 In particular, in section 5.1 we propose  a feasible experimental test  of a 1-D version of the Fermi problem using superconducting qubits.  We give an explicit non-perturbative proof of strict causality in this model, showing that the probability of excitation of a two-level artificial atom with a dipolar coupling to a quantum field is completely independent of the other qubit until signals from it may arrive. We explain why this is in perfect agreement with the existence of nonlocal correlations and previous results which were used to claim apparent  causality problems for  Fermi's two-atom system. 
 \item[*] In section 5.2 we study the information provided by a detector click on the state of an initially excited two level system. By computing the time evolution of the corresponding conditioned probability beyond the rotating wave approximation, we show that a click in the detector is related with the decay of the source only for long times of interaction. For short times, non-rotating wave approximation effects like self-excitations of the detector, forbid a na\"{i}ve interpretation of the detector readings. These effects might appear in circuit QED experiments.
 \end{itemize}

\item Finally, in chapter 6 we consider different quantum simulations, namely of  relativistic quantum mechanics setups.
\begin{itemize}
\item[*] In section 6.1, we introduce the notion of Majorana Hamiltonians, objects that are in general neither linear nor antilinear but with an associated dynamics which conserves the norm. The Majorana equation, that is, the relativistic quantum mechanical equation of a fermion with Majorana mass term, is the archetypal example of this family. We analyze some amazing properties like the relevance of initial global phases in the dynamics that makes them detectable and, as a consequence, the inadequacy of the notion of mixed states. We provide an alternative description in terms of standard Hamiltonians, showing that a Majorana Hamiltonian can be mapped to a Hamiltonian in a real Hilbert space. In this way, Majorana Hamiltonians and antiunitary operators can be experimentally demonstrated.
\item[*] In section 6.2 we show how  to design a quantum simulator for the Majorana equation, a pseudohamiltonian - Majorana-hamiltonian- relativistic wave equation that might describe neutrinos and other exotic particles beyond the standard model. The simulation demands the implementation of charge conjugation, an unphysical operation that opens a new front in quantum simulations, including the discrete symmetries associated with complex conjugation and time reversal. Finally, we show how to implement this general method in trapped ions.
\item[*]Finally, in section 6.3 we show that a potential can be simulated with a free Dirac or Majorana Hamiltonian  since there is a local phase transformation between solutions of the equation with a potential and solutions of the free equation. The transformation depends on the potential, which is codified in the phase. In some cases, the probability density is unchanged, entailing that the particle behaves under the potential as if it were free. It is valid for a large class of potentials for the Majorana equation and a different class in the Dirac case, which explains why Dirac and Majorana particles exhibit different behavior under the same potentials. We extend the results to two-body equations with interaction potentials.

\end{itemize}
\end{itemize}
This thesis concludes with five appendices. The first three develop with more detail some involved computations related with chapters 3, 4 and 5. Finally, appendix D is devoted to a theoretical explanation on the 1-D Majorana equation and appendix E sheds light onto the implementation of the quantum simulation of the Majorana equation with trapped ions.
\chapter{Preliminaries}
\begin{quote}

\end{quote}
\section{Matter-radiation interaction. Hamiltonian}
\subsection{Introduction}
Throughout this work we will deal with a system consisting of one or more two-level systems (qubits in the language of Quantum Information) interacting with the electromagnetic quantum field. This section is devoted to provide a brief summary of the theoretical grounds of the employed model, with no pretension of completeness. Wider textbook treatments on these topics can be found in \cite{elcohen1}, \cite{elthiru}, \cite{elscully}, \cite{compagnobook}. 

We will start with the well known minimal-coupling Hamiltonian of an atom interacting with a quantum field:
\begin{equation}
H=\sum_i \frac{1}{2\,m_i}(\mathbf{p}_i-q_i\mathbf{A}(\mathbf{x}_i))^2+V (|\mathbf{x}-\mathbf{x}_i|)+\sum_{\lambda}\int d^3\mathbf{k}\hbar\,\omega_{\mathbf{k}}(a_{\mathbf{k}\lambda}^{\dagger}a_{\mathbf{k}\lambda}+\frac{1}{2}), \label{eq:2a}
\end{equation}
$\mathbf{p}_i$, $\mathbf{x}_i$, being the momentum and position, respectively of each particle of mass  $m_i$ and charge $q_{i}$ bound by a potential $V$ to a force center (nucleus) located at $\mathbf{x}$. The quantum electromagnetic field can be written down in terms of the creation and annihilation operators for each mode $\mathbf{k}$ (with frequency $\omega_k=c\,|\mathbf{k}|$) and polarization $\lambda$: $a_{\mathbf{k}\lambda}^{\dagger}$, $a_{\mathbf{k}\lambda}$ as:
\begin{equation}
\mathbf{A}(\mathbf{x}_{i})=\sqrt{\frac{\hbar}{2\,c\,\varepsilon_0\,(2\pi)^3}}\sum_{\lambda}\int \frac{d^3k}
{\sqrt{k}} (e^{i\mathbf{k}\,\mathbf{x_{i}}}\mathbf{\epsilon}(\mathbf{k},\lambda)\,a_{k\lambda} +
e^{-i\mathbf{k}\,\mathbf{x}_{i}}\mathbf{\epsilon}^*\,( \mathbf{k},\lambda)\,a^{\dag}_{k\lambda})\label{eq:2b}
\end{equation}
$\epsilon(\mathbf{k},\lambda)$ being the polarization vectors for each $\lambda$ and assuming the commutation relationships 
\begin{equation}
[a_{\mathbf{k}\lambda},a_{\mathbf{k'}\lambda'}^{\dagger}]=\delta^3(\mathbf{k}-\mathbf{k'})\delta_{\lambda\lambda'}.\label{eq:2c}
\end{equation}
(\ref{eq:2a}) would be the quantized version of the classical minimal-coupling Hamiltonian, which can be derived from a Lagrangian whose Euler-Lagrange equations give rise to the correct equations of motion of the system, namely the Maxwell equations for the field and the Lorentz equations for the particles. Notice that we can split (\ref{eq:2a}) into a ``free'' part $H_0$ and an interaction part $H_I$ :
\begin{equation}
H=H_0+H_I \label{eq:2d}
\end{equation}
with 
\begin{equation}
H_0= \sum_{i} \frac{\mathbf{p^2_{i}}}{2\,m_i}+\frac{q_i^2\,\mathbf{A}^2(x_{i})}{2\,m}+V (|\mathbf{x}-\mathbf{x}_i|)+\sum_{\lambda}\int d^3\mathbf{k}\hbar\,\omega_{\mathbf{k}}(a_{\mathbf{k}\lambda}^{\dagger}a_{\mathbf{k}\lambda}+\frac{1}{2}), \label{eq:2e}
\end{equation}
and
\begin{equation}
H_I=-\sum_{i}\frac{q_i\,\mathbf{p}_{i}\cdot\mathbf{A}(x_{i})}{m}. \label{eq:2f}
\end{equation}
But then:
\begin{equation}
H_0\neq H_{0R}+H_{0P},\label{eq:2g}
\end{equation}
that is, the ``free'' part of the Hamiltonian cannot be decomposed into the free Hamiltonian of the radiation,
\begin{equation}
H_{0R}=\sum_{\lambda}\int d^3\mathbf{k}\hbar\,\omega_{\mathbf{k}}(a_{\mathbf{k}\lambda}^{\dagger}a_{\mathbf{k}\lambda}+\frac{1}{2})\label{eq:2h}
\end{equation}
and the Hamiltonian of the particles without field:
\begin{equation}
H_{0P}=\sum_{i} \frac{\mathbf{p^2_{i}}}{2\,m_i}+V (|\mathbf{x}-\mathbf{x}_i|).\label{eq:2i}
\end{equation}
There is an extra, ``self-interaction'' term: $\frac{q_i^2\,\mathbf{A}^2(x_{i})}{2\,m_i}$. Thus the eigenstates of  $H_0$ (\ref{eq:2e}) are not product states of eigenstates of $H_{0R}$ and $H_{0P}$. To avoid this, instead of using the Hamiltonian in  (\ref{eq:2a}) we will move to a different equivalent description of the system. 

To this end we start by assuming the so-called ``long wavelength approximation'', which amounts to replacing $\mathbf{A} (\mathbf{x}_i)$ (\ref{eq:2b}) by $\mathbf{A} (\mathbf{x})$ in the Hamiltonian (\ref{eq:2a}). This approximation is only valid for those modes verifying $\mathbf{k}\cdot|\mathbf{x}-\mathbf{x_i}|<<1$, that is, for wavelengths much larger than the atomic size.  Then, we can apply the following unitary transformation:
\begin{equation}
U=e^{\frac{-i}{\hbar}\mathbf{d}(\mathbf{x})\cdot\mathbf{A}(\mathbf{x})} \label{eq:2j}
\end{equation} 
where we have introduced the electric dipole moment $\mathbf{d}$:
\begin{equation}
\mathbf{d} (\mathbf{x})=\,\sum_i q_i(\mathbf{x}_i-\mathbf{x}).\label{eq:2k} 
\end{equation}
This unitary transformation is an example of displacement operator, whose properties are well known in the framework of coherent states of the electromagnetic field. Using these properties and with a little algebra it is possible to arrive to the transformed Hamiltonian:
\begin{equation}
H'=U\,H\,U^{-1},\label{eq:2m}
\end{equation}
and assuming that we are dealing with a globally neutral atom,
\begin{equation}
\sum_i q_i =0, \label{eq:2n}
\end{equation}
$H'$ takes the form:
\begin{equation}
H'=H_{0R}+H_{0P}+V'+H'_I. \label{eq:2o}
\end{equation}
The first two terms are the free Hamiltonians of the particles and the field respectively (\ref{eq:2h})-(\ref{eq:2i}). The third term $V'$ is:
\begin{equation}
V'=\sum_{\lambda}\int\, d^3\mathbf{k} \frac{1}{2\,\varepsilon_0\,(2\,\pi)^3}(\mathbf{\epsilon}\cdot \mathbf{d}(\mathbf{x}))^2, \label{eq:2p}
\end{equation}
a divergent contribution which cancels out with one of the terms coming from the energy shift caused by emission and reabsorption of photons, as we will see in the next chapter. Thus, we can let this term drop. The last term is an interaction Hamiltonian:
\begin{equation}
H'_I=-\mathbf{d}(\mathbf{x})\cdot\mathbf{E} (\mathbf{x}) \label{eq:2q}
\end{equation}
where $\mathbf{E}(x)$ is the electric field vector operator:
\begin{equation}
\mathbf{E}(\mathbf{x})= i\sqrt{\frac{\hbar\,c}{2\,\varepsilon_0\,(2\pi)^3}}\sum_{\lambda}\int d^3k\,
\sqrt{k} (e^{i\mathbf{k}\,\mathbf{x}}\mathbf{\epsilon}(\mathbf{k},\lambda)\,a_{k\lambda} -
e^{-i\mathbf{k}\,\mathbf{x}}\mathbf{\epsilon}^*\,( \mathbf{k},\lambda)\,a^{\dag}_{k\lambda}).\label{eq:2r}
\end{equation}
So, finally we can split the new Hamiltonian $H'$ (\ref{eq:2o}) into: 
\begin{equation}
H'=H'_0+ H'_I \label{eq:2s}
\end{equation}
but now
\begin{equation}
H'_0=H_{0R}+H_{0P}, \label{eq:2t}
\end{equation}
in contrast with (\ref{eq:2g}). Obviously,  $H$ and $H'$ are equivalent descriptions of the system since they are related through a unitary transformation (\ref{eq:2j}), (\ref{eq:2m}), but states have to transform also according to the unitary. Throughout this work, we will the use the Hamiltonian given by (\ref{eq:2q}), (\ref{eq:2s}), (\ref{eq:2t}), whose `bare" eigenstates are product states of eigenstates of the free Hamiltonians for the particles and the field. Equivalent results would be obtained with the Hamiltonian in (\ref{eq:2a}), provided that the states are correctly transformed through (\ref{eq:2j}).
\subsection{Two level-atoms. Interaction picture}
Now we will particularize to the case of two-level atoms and move to the interaction picture. We consider that we have an atom with a ground state $\ket{g}$ and an excited state $\ket{e}$ such that:
\begin{equation}
H_0\ket{g}=E_g\ket{g}\, ,H_0\ket{e}=E_e\ket{e} \label{eq:21a}
\end{equation}
and
\begin{equation}
E_e-E_g=\hbar\,\Omega \label{eq:21b}.
\end{equation}
Under that conditions, we can use the completeness relation
\begin{equation}
\mathbf{1}=\ket{g}\bra{g}+\ket{e}\bra{e},\label{eq:21c}
\end{equation}
in order to write the electric dipole moment operator in a simple form. That is:
\begin{eqnarray}
\mathbf{d}(t)&=&(\ket{g}\bra{g}+\ket{e}\bra{e})e^{\frac{i\,H_0\,t}{\hbar}}\mathbf{d}\,e^{\frac{-i\,H_0\,t}{\hbar}}(\ket{g}\bra{g}+\ket{e}\bra{e})\nonumber\\&=& e^{\frac{-i\,\Omega\,t}{\hbar}}\,\mathbf{d}_{ge}\sigma^{-}+e^{\frac{i\,\Omega\,t}{\hbar}}\,\mathbf{d}_{eg}\sigma^{+}, \label{eq:21d}
\end{eqnarray}
with
\begin{equation}
\mathbf{d}_{ge}=\mathbf{d}^*_{eg}=\bra{g}\mathbf{d}\ket{e}\label{eq:21e}
\end{equation}
and 
\begin{equation}
\sigma^-=\ket{g}\bra{e}\,, \sigma^+=\ket{e}\bra{g}.\label{eq:21f}
\end{equation}
Thus the interaction Hamiltonian (\ref{eq:2q}) in the interaction picture takes the form:
\begin{equation}
H'_I (t) =-\mathbf{d} (\mathbf{x}, t)\cdot \mathbf{E}(\mathbf{x}, t), \label{eq:21g}
\end{equation}
with $\mathbf{d} (\mathbf{x}, t)$ given by (\ref{eq:21d}) and the standard expression for the electric field vector operator in the interaction picture:
\begin{eqnarray}
\mathbf{E}(\mathbf{x},t)= i\sqrt{\frac{\hbar\,c}{2\,\varepsilon_0\,(2\pi)^3}}\sum_{\lambda}\int d^3k
&\sqrt{k}& (e^{i(\mathbf{k}\,\mathbf{x}-\omega_k\,t)}\mathbf{\epsilon}(\mathbf{k},\lambda)\,a_{k\lambda} -\nonumber\\ & &e^{-i(\mathbf{k}\,\mathbf{x}-\omega_k\,t)}\mathbf{\epsilon}^*\,( \mathbf{k},\lambda)\,a^{\dag}_{k\lambda}).\label{eq:21h}
\end{eqnarray}
In the next chapter, we will extensively use the Hamiltonian (\ref{eq:21g}), and a 1-D version of it will be considered in chapters 4 and 5.
\subsection{Rotating wave approximation and beyond}
The Hamiltonian in (\ref{eq:21g}) can be written as a sum of four terms using (\ref{eq:21d}) and (\ref{eq:21h}), each of them containing two operators: an atomic ladder operator $\sigma^{\pm}$ and a creation or annihilation Fock operator $a_{k\lambda}$, $a_{k\lambda}^{\dagger}$. Thus we have two terms $\sigma^{-}a_{k\lambda}^{\dagger}$, $\sigma^{+}a_{k\lambda}$ associated to the frequency $\Omega-\omega$ and two terms $\sigma^{-}a_{k\lambda}$, $\sigma^{+}a_{k\lambda}^{\dagger}$ associated to the frequency $\Omega+\omega$. The rotating wave approximation (RWA) let the latter so-called counter-rotating terms drop, retaining only the first two ones. The usual argument is that the counter-rotating terms oscillate very quickly and average to zero in observable time scales. This is generally true in the Quantum Optics realm, due to the weak atom-field coupling. As we will see in this thesis (chapters 4 and 5), the argument fails for stronger couplings, as those appearing in circuit QED, where counter-rotating terms can be relevant at experimental times. Even for weak couplings, a theoretical analysis of the short-time behavior like the one in chapter 3 must take the non-RWA terms into account. Therefore, throughout this thesis the RWA will not be assumed.
\subsection{Rabi and Jaynes-Cummings models}
When the boundary conditions of the system are such that the electromagnetic field is restricted to a given region of space, engineered in such a way that the fundamental frequency of the field $\omega_0$ approximately matches with the frequency $\Omega$ of the qubit transition, then the multimode expansion of the electric field in the Hamiltonian (\ref{eq:2q}) (or (\ref{eq:21g}) in the interaction picture) can be restricted to only one mode $\omega_0$. In that case, the Rabi model is obtained. If the RWA is applied to the Rabi model, the resulting hamiltonian is the celebrated Jaynes- Cummings model, describing a two-level system in an optical cavity. We will deal with the Rabi model in circuit QED in chapter 4.
\section{Circuit QED}
Although the oscillations in electrical circuits are usually described by classical mechanical laws, at micro and nano scales and ultra-low temperatures, the thermal noise can be negligible compared to the spacing and width of the energy levels, unveiling the quantum nature of the system. Moreover, non-linear elements such as Josephson junctions do not obey the principle of correspondence, that is, the averages of the momentum and position operators do not follow the classical equations of motion, making apparent the necessity of a fully quantum-mechanical description of such elements. Thus non-trivial quantum effects can be observed in systems that can be considered as macroscopic, in the sense that they contain a huge number of elementary particles -the quantum numbers being collective degrees of freedom such as flux or charge-. Among all the possibilities opened up by this novel approach to Quantum Mechanics, in this thesis we are concerned with just one: circuit Quantum Electrodynamics.

A systematic quantization procedure of the usual elements of a circuit, such as inductances, capacitances and resistances is given in \cite{yurkedenker, devoret} within the approach of lumped-element circuits, valid when the relevant wavelengths are much larger than the element's size. This can also be extended to waveguides -also called transmission lines- which can be regarded as a chain of infinitesimally small LC lumped-element circuits, obtaining a quantized Hamiltonian which is the 1-D version of \ref{eq:2h}. With this lumped-element approach is also possible to obtain a quantum description of several types of superconducting qubits, using different combinations of superconductors and Josephson junctions \cite{reviewnature, johanssonthesis}. Engineering the length of the transmission line to match the fundamental wavelength of the qubit transition, a cavity QED system like the ones described in section 2.1.4 is obtained \cite{blais04, wallraff04}. Without this restriction the full hamiltonian of the matter-radiation interaction emerges \cite{guille, guillejuanjokike}, but in a range of experimental parameters different from the typical one in Quantum Optics. In this artificial 1D version of  atomic QED, the frequency of the qubit transition is in the order of Ghz, and thus the relevant wavelengths of the field are in the microwave regime.

In chapters 4 and 5 of this Thesis, we will exploit the possibilities of circuit QED as a quantum simulator of matter-radiation interaction, taking advantage of its experimental amenability, which includes the possibility of achieve very strong qubit-field couplings, entering into a regime in which non-RWA effects can be observed in the lab.

\section{Entanglement measures in two-qubit systems}
In this thesis, we will analyze the dynamics of the entanglement between qubits, like the two-level atoms described in the previous sections. In this section we give a brief description of the notion of entanglement and discuss different ways of quantify it. 

Let us start with pure states of two-qubit systems, which take the following general form: 
\begin{equation}
|{\Psi}\rangle=\sum_{i,j}\alpha_{ij}|{ij}\rangle, \label{eq:22a}
\end{equation}
$|{ij}\rangle$, $(i,j=e, g)$ being a basis of the Hilbert space $\mathcal{H}_{A}\otimes \mathcal{H}_{B}$ for qubits A and B. These pure two-qubit states are called separable if it can be written as:
\begin{equation}
\ket{\Psi}=\ket{\Psi}_A\ket{\Psi}_B,\label{eq:22b}
\end{equation}
$\ket{\Psi}_A$, $\ket{\Psi}_B$ being states of the Hilbert spaces $\mathcal{H}_A$, $\mathcal{H}_B$, respectively. Otherwise, the state is referred to as entangled, and exhibit correlations between the parties that cannot be achieved classically. 
For the non-pure case, a state $\rho$ is separable if they can be written as
\begin{equation}
\rho=\sum_{i}p_{i}\rho_{i}^{A}\otimes\rho_{i}^{B}, \label{eq:22c}
\end{equation}
where $\rho_{i}^{A}$ and $\rho_{i}^{B}$ are state operators of subsystems A
and B respectively, $p_{i}\geq0 $ $\forall i$, $\sum_{i}p_{i}=1$. Otherwise $\rho$ is entangled.

In general, an entanglement measure is some function of the state that is zero for separable states and non-zero for entangled ones. Besides, some technical properties must be fulfilled, the most standard one being the so- called monotonicity under Local Operations and Classical Communication (LOCC): these operations cannot create entanglement, thus any entanglement measure must non-increase under them. There are several measures of the entanglement of pure bipartite states; we shall cite only three of them: von Neumann's
entropy of reduced states, concurrence  \cite{wootters}, and negativity.  \cite{vidalwerner}.

The Von Neumann's entropy of a state $\rho$ is defined in Information Theory as 
\begin{equation}
S(\rho)=-\sum_{j}p_{j}\log_{2}p_{j}, \label{eq:22d}
\end{equation}
being
$p_{j}$ the eigenvalues of $\rho$. Reduced states $\rho_{A}$, $\rho^{B}$ are the result of tracing $\rho$ over qubits B and A respectively:
\begin{equation}
\rho_{A}= Tr_B\, \rho\, , \rho_{B}= Tr_A\, \rho. \label{eq:22e}
\end{equation}
Von Neumann's entropy of reduced states is the simplest measure of bipartite entanglement for pure states, but its extension to non-pure states does not correctly discriminate between separable and entangled states. Therefore, it is not a good measure of entanglement for general mixed states.

Concurrence is the more usual measure of entanglement for non-pure two-qubit states. It is defined as
\begin{equation}
\mathcal{C}(\rho)=max{\{0,\sqrt{\lambda_{1}}-\sqrt{\lambda_{2}}-\sqrt{\lambda_{3}}-\sqrt{\lambda_{4}}}\}, \label{eq:22f} 
\end{equation}
where ${\{\lambda_{i}}\}$, $(i=1...4)$ are the eigenvalues of 
\begin{equation}
R=\rho\widetilde{\rho}  \label{eq:22g}
\end{equation}
in decreasing order, with
\begin{equation}
\widetilde{\rho}=(\sigma_{y}\otimes\sigma_{y})\rho^{\star}(\sigma_{y}\otimes\sigma_{y}),  \label{eq:22h}
\end{equation}
and $\rho^{\star}$ being the complex conjugate of the state operator $\rho$. In arbitrary dimensions, operational generalizations of concurrence are known only for pure states. Therefore, it is a good measure of entanglement for general two-qubit states (pure or not) and for pure states in any dimension, but not for mixed bipartite states of arbitrary dimensions.

Finally, the negativity of a bipartite state ${\rho}$, defined in \cite{mirano} as:
\begin{equation}
N(\rho)=-2\sum_{i}\sigma_{i}(\rho^{TA}), \label{eq:22i}
\end{equation}
where ${\{\sigma_{i}(\rho^{TA})}\}$ are the negative eigenvalues of the partial transpose $\rho^{TA}$ of the total state $\rho$ respect to the subsystem A, defined as
\begin{equation}
\langle{i_{A},j_{B}}|{\rho^{TA}}|{k_{A},l_{B}}\rangle=\langle{k_{A},j_{B}}|{\rho}|{i_{A},l_{B}}\rangle, \label{eq:22j}
\end{equation}
A and B denoting the two subsystems: this is twice the value of the original definition of the negativity in   \cite{vidalwerner}. It can be proved that
for pure bipartite states of arbitrary dimensions the negativity (\ref{eq:22j}) is equal to the concurrence \cite{mirano}.  The main advantage of the negativity is that it can be evaluated in the same way for pure and non-pure states in
arbitrary dimensions, although there are entangled mixed states with zero negativity in every dimensions except $2\times2$ (two qubits) and $2\times 3$ (a qubit and a qutrit) \cite{horuno}, \cite{hordos}. Thus, negativity is not in general an ideal measure of bipartite entanglement of non-pure states: no measure discriminating separable from entangled states in the general non-pure
case is known. However, non-zero negativity is a sufficient condition for entanglement and for \textit{distillability} \cite{horuno} (a state is said to be distillable if a maximally entangled state can be obtained from it through (LOCC)), a very important property for Quantum Information tasks. 

\chapter{Dynamics of entanglement in matter-radiation interaction}
\begin{quote}
``A devil with merely local powers like a parish vestry would be too inconceivable a thing'' (Sir Arthur Conan Doyle,\textit{The
hound of the Baskervilles})
\end{quote}
In this chapter we will analyze exhaustively the entanglement dynamics in a system consisting of a pair of two-level atoms interacting through a quantum electromagnetic field. In Section 3.1 we start from different initial separable states and consider different projections onto final states of the field. In Section 3.2 a protocol for generating highly entangled states between spacelike separated atoms is considered. Finally, in section 3.3 we start from an initial entangled state and study the phenomena of entanglement sudden death and entanglement sudden birth. Throughout all the chapter, entanglement measures are used to compute the evolution of entanglement and relate its behavior to the spacetime region in which the atoms are placed.
\section{Generation of entanglement inside and outside the mutual light cone}
\subsection{Introduction}
Quantum superposition and entanglement are the cornerstones lying at the foundations of quantum information and the principal support of the new quantum technologies which are at different stages of conception and development at present. Putting entanglement to work, enabling its use as a resource, is the key to the success of these technologies. Therefore, a complete
understanding of entanglement, necessary  at the fundamental level, is  also important for these developments to occur.

As we have explained in Chapter 1, entanglement can be envisaged in very different forms; it  originally appeared in quantum mechanics~\cite{epr} as a direct connection between distant particles, a residue of past direct interaction between them \cite{schrodinger}. In quantum field theory entanglement can be traced back to the non-locality of the vacuum state ~\cite{summerswerner,summerswernerII} or, simply,
to field propagation. Similar arguments operate for a lattice of coupled oscillators~\cite{plenioeisert}.

In this section we analyze some features of entanglement generation closely related to the microscopic causality of quantum field theory. Put in simple words, this work attempts to ascertain whether a pair of spatially separated parties (say, a pair of neutral two-level atoms $A$ and $B$) can get entangled in a finite time while they remain causally disconnected
\cite{fermi,hegerfeldt,powerthiru}. Each party interacts locally with the electromagnetic field, the carrier of the interaction.
We stress here that, as shown in \cite{powerthiru}, perturbation theory produces non-signalling \cite{gisin} results for this system and that the apparent causality violations come from the nonlocal specification of some final states. At first sight, the question can be answered in the negative; if the parties remain causally separated from each other, they can not entangle.
However, the propagator $D(x,y)$ is finite even when $c (x-y)^0 < |\mathbf{x}-\mathbf{y}|$, and perhaps some correlations could be exchanged between both parties~\cite{franson}. Alternatively, the correlations could be blamed on the preexisting entanglement between different parts of the vacuum~\cite{reznik,reznikII}, which could be transferred to the atoms. Whatever the
point of view, correlations are exchanged through (time ordered) products, while only commutators are restricted to be causal \cite{fransonII}. Our analysis can not sidestep that the role of the field goes beyond that of a mere carrier, quanta could be absorbed from the field or escape in the form of photons \cite{cabrillo,lamata}. How does the entanglement between $A$ and $B$ depend on the state of the field? This question shapes our discussion below.

We will include in the final state all the perturbatively accessible field states, analyzing for each of them  the  correlations in the reduced atomic state. We compute the entanglement measures for different values of $(x-y)^0$ and $|\mathbf{x}-\mathbf{y}|$, that lie inside the atoms mutual light cone and beyond. The atomic state that results after tracing over the states of the field is separable, which means, in the scheme of \cite{reznik}, that there is no transference of vacuum
entanglement, only classical correlations. In \cite{reznik}, these correlations become entanglement when a suitable time dependent coupling with the scalar field is introduced. As pointed out in \cite{reznikII}, this would require an unrealistic control of the atom-field interaction in the electromagnetic case that we are dealing with here. As an alternative way to achieve entanglement between the atoms we consider a post-selection process of the field states with $n=0, 1, 2$ photons. This is a nonlocal operation and therefore entanglement generation is allowed. In \cite{franson}, only the vacuum case when $|\mathbf{x}-\mathbf{y}|\gg c (x-y)^0 $ was analyzed, and no entanglement measures were considered. We get quantum correlations for all the different field states. We also get useful hints on the nature of the correlations, whether they come from photon exchange, source indistinguishability, etc.

\subsection{The model}

We will consider the field initially  in the vacuum state, including the cases with 0, 1 and 2 final photons to analyze perturbatively the amplitudes and density matrices to order $\alpha$. We assume that the wavelengths relevant in the interaction with the atoms, and the separation between them, are much longer than the atomic dimensions.  As seen in the previous chapter, under these conditions the Hamiltonian can be split into two parts 
\begin{equation}
H = H_0 + H_I \label{eq:31a}
\end{equation}
that are separately gauge invariant. The first part is the Hamiltonian in the absence of interactions other than the potentials that keep $A$ and $B$ stable, 
\begin{equation}
H_0 = H_{0PA} + H_{0PB} + H_{0R}.\label{eq:31b} 
\end{equation}
$H_{0PA}$, $H_{0PB}$ being the corresponding ``free'' Hamiltonians (\ref{eq:2i}) of particles $A$ and $B$, respectively, and $H_{0R}$ is given by (\ref{eq:2h}). The second contains all  the interaction of the atoms with the field
\begin{equation}
H_I = - \sum_{n=A,B} \mathbf{d}_n(t)\,\mathbf{E}(\mathbf{x}_n,t) \label{a},
\end{equation}
where $\mathbf{E}$ is the electric field (\ref{eq:21h}), and 
\begin{equation}
\mathbf{d}_n \,=\,\sum_i\, q_i\,(\mathbf{x}_i-\mathbf{x}_n)\,\label{eq:31c}
\end{equation}
is the electric dipole moment (\ref{eq:2k}) of atom $n$, whose matrix elements (\ref{eq:21e}) we will take as real and of equal magnitude for both atoms 
\begin{eqnarray}
\mathbf{d}_A&=&\mathbf{d}_{Age}=\mathbf{d}_{Aeg}\nonumber\\
\mathbf{d}_B&=&\mathbf{d}_{Bge}=\mathbf{d}_{Beg}\nonumber\\
|\mathbf{d}_A|&=&|\mathbf{d}_B| \label{eq:31d}
\end{eqnarray}
$|\,e\,\rangle$ and $|\,g\,\rangle$ being the excited and ground states of the atoms, respectively.

In what follows we choose a system  given initially by the product state, 
\begin{equation}
|\,\psi\,(0)\rangle\,=\,
|\,e\,g\,\,0\,\rangle \label{eq:31e}
\end{equation}
 in which atom $A$ is in the excited state $|\,e\,\rangle$, atom $B$ in the ground state
$|\,g\,\rangle$, and the field in the vacuum state $|\,0\,\rangle$. The system then evolves under the effect of the interaction
during a lapse of time $t$ into a state:
\begin{equation}
|\,\psi\,(t)\rangle = T\,[e^{-i\, \int_0^t\,dt'\, H_I\,(t')/\hbar}\,]|\,\psi\,\rangle_0 ,\label{b}
\end{equation}
($T$ being the time ordering operator) that,  to  order $\alpha$, can be given in the interaction picture as
\begin{eqnarray}
|\mbox{atom}_1,\mbox{atom}_2,\mbox{field}\rangle_{t} =  ((1+A)\,|\,e\,g\rangle + X\,|\,g\,e\rangle)\,|\,0\rangle\nonumber\\
 +(U_A\,|\,g\,g\,\rangle+ V_B\,|\,e\,e\,\rangle)\,|\,1\,\rangle+
(F\,|\,e\,g\rangle+ G\,|\,g\,e\rangle)\,|\,2\rangle\  \label{c}
\end{eqnarray}
where
\begin{eqnarray}
A&=&\frac{1}{2}\langle0|T(\mathcal{S}_A^+ \mathcal{S}_A^- + \mathcal{S}_B^-\mathcal{S}_B^+)|0\rangle,\, X=
\langle0|T(\mathcal{S}^+_B
\mathcal{S}^-_A)|0\rangle\nonumber\\
U_A\,&=&\,\langle\,1\,|\, \mathcal{S}^-_A\,|\,0\,\rangle,\, V_B\,=\,\langle\,1\,|\,
\mathcal{S}^+_B\,|\,0\,\rangle \label{d}\\
F&=&\frac{1}{2}\langle2|T(\mathcal{S}_A^+ \mathcal{S}_A^- +\mathcal{S}_B^-\mathcal{S}_B^+)|0\rangle,\,
G=\langle2|T(\mathcal{S}^+_B \mathcal{S}^-_A)|0\rangle,\nonumber
\end{eqnarray}
being 
\begin{eqnarray}
\mathcal{S}_n\,=\,- \frac{i}{\hbar}  \int_0^t\, dt'\, H_{I}(t')=\mathcal{S}_n^{+}\, +\, \mathcal{S}_n^{-}\nonumber\\
\mathcal{S}^{\pm}_n= - \frac{i}{\hbar}  \int_0^t\, dt'e^{\mp\frac{i\,\Omega\,t}{\hbar}}\,\sigma^{\pm}\,\mathbf{d}_n\mathbf{E}(\mathbf{x},t),
 \label{eq:31f}
\end{eqnarray}
(the sign of the superscript in \ref{eq:31f} is associated to the energy difference between the initial and final atomic states of each emission) and  $|\,n\,\rangle,\,\, n=\,0,\,1,\,2$ is a shorthand for the state of $n$ photons with arbitrary  momenta and polarizations, i.e. 
\begin{equation}
|\,1\,\rangle\,\bra{1}=\sum_{\lambda}\int d^3\,\mathbf{k}\,|\mathbf{k},\, \mathbf{\epsilon}(\mathbf{k},\lambda)\,\rangle\bra{\mathbf{k},\, \mathbf{\epsilon}(\mathbf{k},\lambda)}, \label{eq:31g}
\end{equation}
etc. $\Omega$ is the frequency of the atomic transition, as defined in (\ref{eq:21b}).

Among all the terms that contribute to the final state (\ref{c}) only $X$ corresponds to interaction between both atoms, which is ``real'' interaction only
if $c\,t>r$ ($r$ being the interatomic distance). This would change at higher order in $\alpha$. Here, $A$ describes
intra-atomic radiative corrections, $U_A$ and $V_B$ single photon emission by one atom, and $G$ by both atoms, while $F$ corresponds
to two photon emission by a single atom. Details on the computations of these quantities would be given in Appendix A. 

In Quantum Optics terms like $V_B$, $F$ and $G$ are usually neglected by the introduction of a RWA. But here we are interested in the short time behavior, and therefore all the terms must be included, as in \cite{powerthiru,milonni,compagnoI}. Actually, only when all thesenon-RWA
effects are considered, it could be said properly that the probability of excitation of atom $B$ is completely independent of atom $A$ when $r>c\,t$ \cite{milonni, compagnoI} ($r$ being the distance between the atoms).

Note that the actions in   (\ref{eq:31f}) depends on the
atomic properties $\Omega$ and $\mathbf{d}_n$, and on the interaction time $t$. In our calculations we will take $(\Omega
|\mathbf{d}_n|/e c) = 5\,\cdot 10^{-3}$, which is of the same order as the 1s $\rightarrow$ 2p transition in the hydrogen atom,
consider $\Omega\,t \gtrsim 1$, and analyze the cases $(r/c\,t)\simeq 1$ near the mutual light cone, inside and outside.

Given a definite field state $|\,n\,\rangle$ the pair of atoms is in a pure two qubits state as shown in (\ref{c}). We will
denote these states by $|\,A\,B\,\rangle_n$, then
\begin{equation}
 \rho_{AB}^{(n)}\,=\,|A\,B\rangle_{nn}\langle A\,B\,|, \label{eq:31h}
\end{equation}
and the corresponding reduced states by
 \begin{equation}
  \rho_{A}^{(n)}\,=\,Tr_B\,\rho_{AB}^{(n)}\, ,\rho_{B}^{(n)}\,=\,Tr_A\,\rho_{AB}^{(n)}, \label{eq:31i}
\end{equation}
in the following, and will compute the entropy of entanglement $\mathbb{S}^{(n)}$ \cite{bennett}:
\begin{equation}
\mathbb{S}^{(n)}=Tr\, \rho_{A}^{(n)}\log{\rho_{A}^{(n)}}\label{f}
\end{equation}
and the concurrence $\mathbb{C}^{(n)}$  \cite{wootters}:
\begin{equation}
\mathbb{C}^{(n)}= max\{0, \sqrt{\lambda_i}-\sum_{j\neq i}\, \sqrt{\lambda_j}\} \label{g}
\end{equation}
 $\lambda_i$ being the largest of the eigenvalues $\lambda_j$ ($j=1,...,4$) of
\begin{equation}
R_{AB}=[(\sigma_y\otimes\sigma_y)\rho_{AB}*(\sigma_y\otimes\sigma_y)]\rho_{AB}\label{eq:31j}
\end{equation}
 for them.

\subsection{The case with $n=0$}
\begin{figure}[h!]
\includegraphics[width=\textwidth]{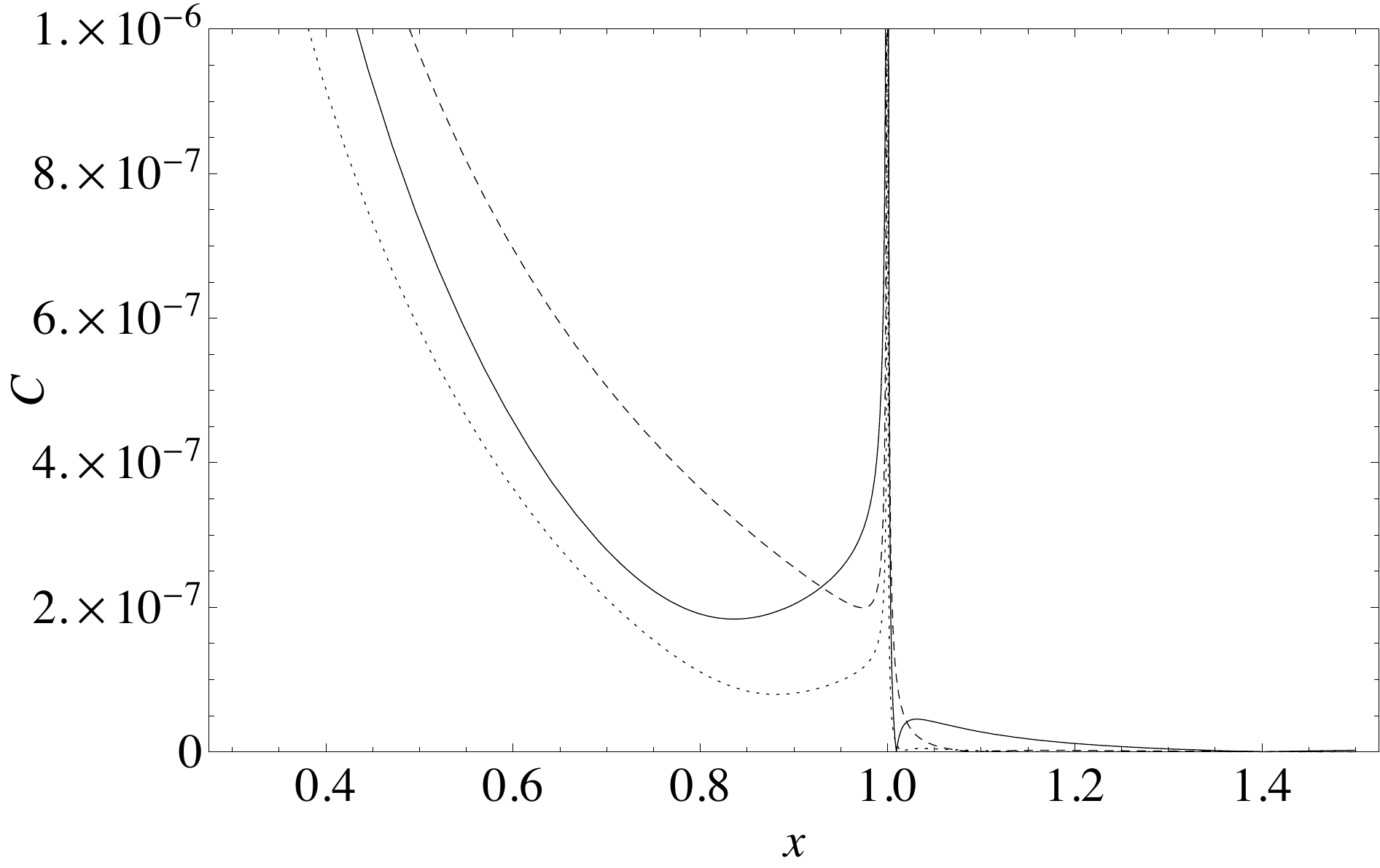}
\caption{Concurrence of the atomic state in the e.m. vacuum  $\rho_{AB}^{(0)}$ as a function of $x=(r/c\,t)$ for three values of
$z=(\Omega r/c)= 5$ (solid line), 10 (dashed line) and 15 (dotted line). The height of the peak is $\mathbb{C}^{(0)}=1$. }
\label{fig:1}
\end{figure}

We first consider the case $n=0$, where the field is in the vacuum state and, after (\ref{c}), the atoms are in the pure state
\begin{equation}
\ket{AB}_0=((1\,+\,A)\,|\,E\,G\,\rangle + X\,|\,G\,E\,\rangle) / c_0, \label{eq:31k}
\end{equation}
where $c_0\,=\, \sqrt{|1\,+\,A|^2 \,+\, |X|^2}$ is the
normalization, giving a concurrence
 \begin{equation} \mathbb{C}^{(0)} \,=\, 2\,
|X|\,|\,1\,+\,A\,|/c_0^2\ .\label{h}
 \end{equation}
 It is interesting to note that at lowest order the concurrence arises as an effect of the mutual interaction terms $X$ mediated
by photon exchange  or, in a different language, by the vacuum fluctuations. As expected,  at higher orders the radiative
corrections described by $A$ dress up these correlations. Analytic lowest order calculations (\cite{thiru}) showed that they can
persist beyond the mutual light cone, vanishing for $x = (r/c\,t) \rightarrow \infty$. We sketched in Fig. \ref{fig:1} the concurrence
$\mathbb{C}^{(0)}$ for $x$ around 1. Our computations were done for the illustrative case where both dipoles are parallel and
orthogonal to the line joining $A$ and $B$. We will adhere to this geometrical configuration for the rest in the following. It
would correspond to an experimental set up in which the dipoles are induced by suitable external fields.  $\mathbb{C}^{(0)}$
shows a strong peak (of height 1) inside a tiny neighborhood of $x=1$. The features outside the mutual light cone are
$\vartheta(|\mathbf{d}|/e r)^2 \simeq 10^{-6}$ here, and could be larger if $\Omega t < 1$ entering into the Zeno region
(incidentally, $|X|\,\propto t^4$ for very small $t$ \cite{thiru}). Notice the change of behavior between the region where the
atoms are spacelike separated ($x<1$) and the region where one atom is inside the light cone of the other ($x>1$). This
quantitative treatment complements the qualitative one given in \cite{franson}.

 The entropy of entanglement written in terms of the small quantity 
 \begin{equation}
 \eta_0\,=\,(|X|/c_0)^2 \in (0,1)\label{eq:31l} 
 \end{equation}
 is
\begin{equation}
\mathbb{S}^{(0)}\,=\, - (1-\eta_0)\, \log (1-\eta_0)\,-\,\eta_0\,\log \eta_0\ ,\label{i}
\end{equation}
this is a positive quantity in $(0,1)$, which attains its maximum possible value $\mathbb{S}^{(0)}\,=\,1$ when the state is
maximally entangled at $\eta_0=\,0.5$. This is well within the small neighborhood of $x=1$ mentioned above. Radiative
corrections would shift the maximum to $|X|\,=\,|1 + A|$, so the entropy is sensitive to the Lamb shift when this contributes to
the dipole radiative corrections.

\subsection{Photon emission}
\begin{figure}[h!]
\includegraphics[width=\textwidth]{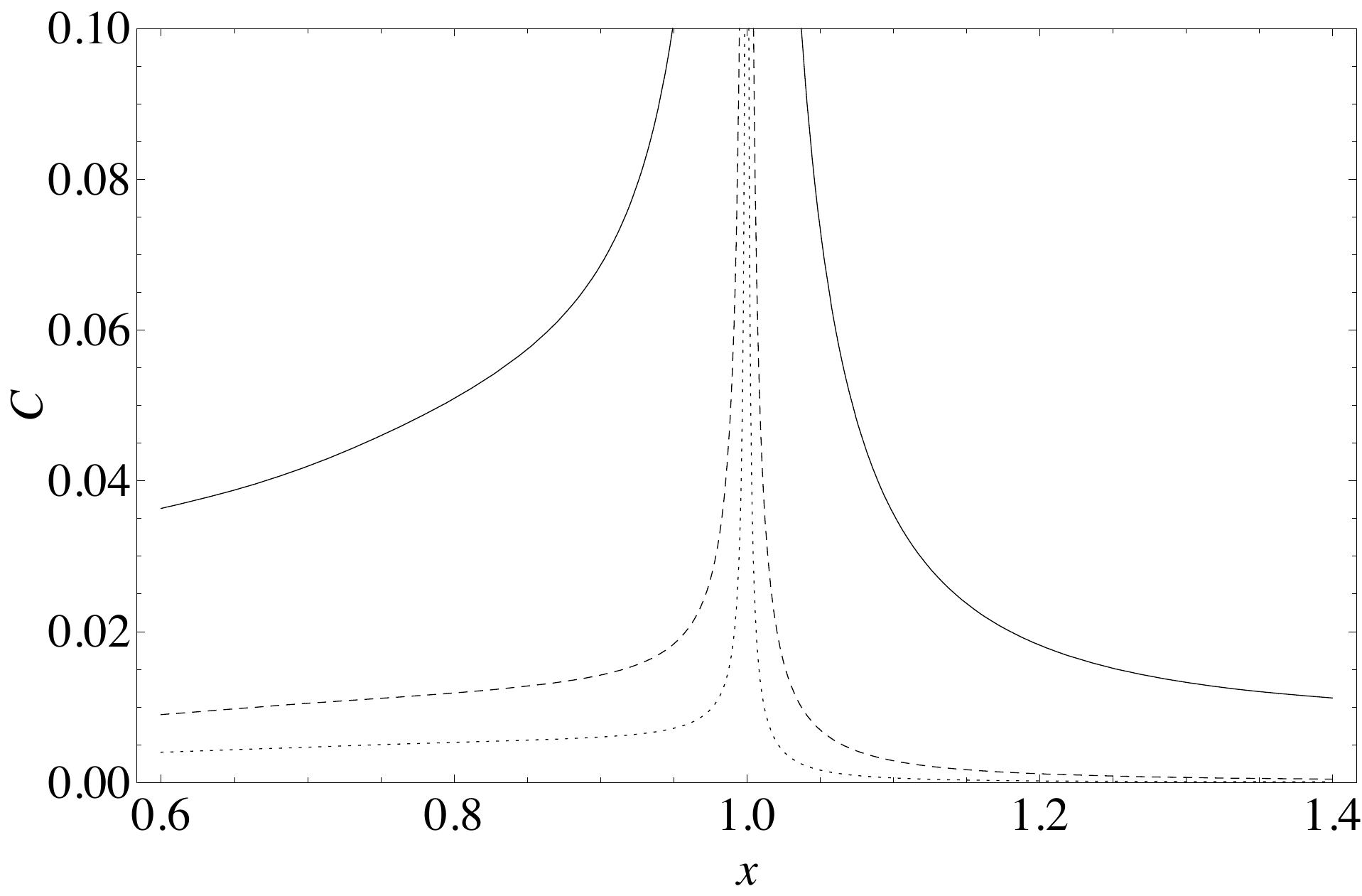}
\caption{Concurrence for one photon final state (\ref{31j}) as a function of $x={r/c\,t}$ for three values of $z=\Omega r/c= 5$
(solid line), 10 (dashed line) and 15 (dotted line). Entanglement vanishes as $t\rightarrow\infty$ ($x\rightarrow0$ for a given
$r$) and is sizeable for $x>1$.}
\label{fig:2}
\end{figure}

We now come to the case $n=1$, where the atoms excite one photon from the vacuum,  jumping to the state
\begin{equation}
\ket{A\,B}_1= (\,U_A\,|\,g\,g\,\rangle\,+\,V_B\,|\,e\,e\,\rangle)/c_1, \label{eq:31ll}
\end{equation}
(with $c_1 = \sqrt{|\,U_A\,|^2\,+\,|\,V_B\,|^2}$), during the time interval
$t$. The density matrix for this case contains the term 
\begin{equation}
l= V_B\,U_A^* = Tr_1\,\langle\,1\,|\,
\mathcal{S}^+_B\,|\,0\,\rangle\,\langle\,1\,|\, \mathcal{S}^-_A\,|\,0\,\rangle^*\,=\,\langle\,0\,|\,
\mathcal{S}^+_A\,\mathcal{S}^+_B\,|\,0\,\rangle ,\label{eq:31m}
\end{equation}
producing a concurrence
\begin{equation}
\mathbb{C}^{(1)} \,=\,2 |\,l\,|/c_1^2\ ,\label{31j}
 \end{equation}
so, even if this case only describes independent local phenomena attached to the emission of one photon by either atom $A$ or
$B$, the concurrence comes from the tangling between the amplitudes $u$ and $v$ which have different loci. The state of the
photon emitted by $A$ and the state of $A$ are correlated in the same way as the state of the photon emitted by $B$ with the
state of $B$ are. These independent field-atom correlations are transferred to atom-atom correlations when we trace out a photon
line with different ends, $A$ and $B$, when computing $V\,U^*$. In fact, while $|U|^2$ and $|V|^2$ are independent of the
distance $r$ between the atoms,
\begin{equation}
 l= - {c d_A^id_B^j \over \hbar \epsilon_0}\, \{(\delta_{ij}-\hat{r}_i \hat{r}_j)M''(r)+(\delta_{ij}+\hat{r}_i \hat{r}_j){ M'(r)\over r} \}\label{k}
 \end{equation}
where
\begin{equation}
 M(r) =
 \int_{0}^{\infty}dk \,{\sin{k r}\over r}\,\delta^t(\Omega+c k)\,\delta^t(\Omega-c k)\label{31l}
\end{equation}
which depends explicitly on $r$. Above we used $\delta^t(\omega)\,=\, \sin (\,\omega\,t/2)/(\pi \omega)$, which becomes
$\delta(\omega)$ in the limit $t\rightarrow \infty$. In Fig. \ref{fig:2} we represent $\mathbb{C}^{(1)}$ in front of $x={r/c\,t}$ for some
values of $z=\Omega r/c$. As the Figure shows, there may be  a significative amount of concurrence for all $x$, indicating that
$\rho^{(1)}$ is an entangled state inside and outside the mutual light cone. The peak at $x=1$ comes from the term with phase
$k(r - c t)$ that can be singled out from the linear combination of phasors in the integrand of (\ref{31l}).

Here we have a lone photon whose source we cannot tell. It might be $A$ or $B$, with the values of $l$ and $\mathbb{C}^{(1)}$
depending on their indistinguishability. Eventually, RWA will forbid the process $g \rightarrow e + \gamma$
for large interaction times. Therefore, $V_B$, $l$ and $\mathbb{C}^{(1)}$ will vanish as $t$ grows to infinity ($x\rightarrow0$
for each value of $z$ in Fig. 2), as can be deduced from the vanishing of $\delta^t(\Omega+c k)$ for $t\rightarrow \infty$.

The entropy of entanglement gives an alternative description of the situation. Its computation requires tracing over one of the
parts $A$ or $B$, so no information is left in $\mathbb{S}^{(1)}$ about $r$, but it still gives information about the relative
contribution of both participating states $|e\,e\,\rangle$ and $|g\,g\,\rangle$  to the final state.  In terms of  
\begin{equation}
\eta_1\,=\,
|\,V\,|^2/{c_1}^2 \in (0,1), \label{eq:31n}
\end{equation}
we have
 \begin{equation}
\mathbb{S}^{(1)}\,=\, - (1-\eta_1)\, \log (1-\eta_1)\,-\,\eta_1\,\log \eta_1 \label{m}
\end{equation}
Would not be for the difference between $\Omega + ck$ and  $\Omega -ck$, $V_B$ should be equal to $U_A$, $\eta_1\,= 0.5$, and
$\mathbb{S}^{(1)}$ would attain its maximum value. Not only this is not the case but, as said above, $V_B$ will vanish with time
and only $|g\,g\,\rangle$ will be in the final asymptotic state. Notice the result, indistinguishability was swept away because
for large $t$  we know which atom ($A$) emitted the photon. Therefore, the entropy will eventually vanish for large interaction
times.

There are two possibilities with $n=2$; one (with amplitude $F$) when both photons are emitted by the same atom, the other (with
amplitude $G$) when each atom emits a single photon. The final atomic state 
\begin{equation}
\ket{AB}_2=(F\,|\,e\,g\,\rangle\,+\,G\,|\,g\,e\,\rangle)/c_2,\label{eq:31o}
\end{equation}
with $c_2\,=\,\sqrt{|\,F\,|^2\,+\,|\,G\,|^2}$, is in the same subspace as for $n\,=\,0$. The normalization $c_2$ is
$\mathcal{O}(\alpha)$ like the expectation values $F$, $G$, so that all the coefficients in $\rho_{AB}^{(2)}$ may be large. The
concurrence is 
\begin{equation}
\mathbb{C}^{(2)}=2 |\,F\,G^*\,|/c_2^2. \label{eq:31p}
\end{equation}
Due to the tracing over photon quantum numbers,  $F\,G^*$ is a sum of
products containing not only  factors $U_A$ and $V_B$, but also $r$ dependent factors like $l$. The entropy $\mathbb{S}^{(2)}$ is
now given in terms of a parameter 
\begin{equation}
\eta_2\,=\, |\,G\,|^2/c_2^2. \label{eq:31q}
\end{equation}
Notice that 
\begin{equation}
|G|^2 = |U_A|^2 |V_B|^2+|l|^2. \label{eq:31r}
\end{equation}
Hence, both
$\mathbb{C}^{(2)}$ and $\mathbb{S}^{(2)}$, depend on $r$. This is different from the single photon case, where the only  $r$
dependence was in the coherences of $\rho^{(1)}_{AB}$, which did not feed into $\rho^{(1)}_{A}$. The correlations  came in that
case from the indistinguishability of the photon source. The case $n =2$ resembles that of the entanglement swapping paradigm
\cite{swapping}, where there are two independent pairs of down converted photons. Here we have two independent atom - photon
pairs. The swapping would arise in both cases from detecting one photon of each pair. But with the initial state we are
considering here, both $F$ and $G$ eventually vanish. More interesting would be the case with the initial atomic state
$|\,e\,e\,\rangle$, that will be considered in section 3.1.5.

\subsection{Tracing over the field}

We have seen that if the state of the field is defined, the atomic state is entangled inside and outside the light cone. But
what happens if the field state is ignored, that is, if we trace over the field degrees of freedom? Then the atomic state is
represented by the following density matrix (in the basis $\{|e\,e\rangle,|e\,g\rangle,|g\,e\rangle,|g\,g\rangle\}$):
\begin{eqnarray}
\rho_{AB}=\left( \begin{array}{c c c c}|V_B|^2&0&0&l\\0&|1+A|^2+|F|^2&(1+X)^*+FG^*&0\\0& X(1+A)^*+F^*G&|X|^2+|G|^2&0\\
l^*&0&0&|U_A|^2
\end{array}\right)N^{-1}\label{n}
\end{eqnarray}
where $l=\langle0|\,\mathcal{S}_{A}^+\,\mathcal{S}_{B}^+\,|0\rangle$ was used again, and $N=|1+A|^2
+|X|^2+|U|^2+|V|^2+|F|^2+|G|^2$. 
\begin{figure}[h!]
\includegraphics[width=\textwidth]{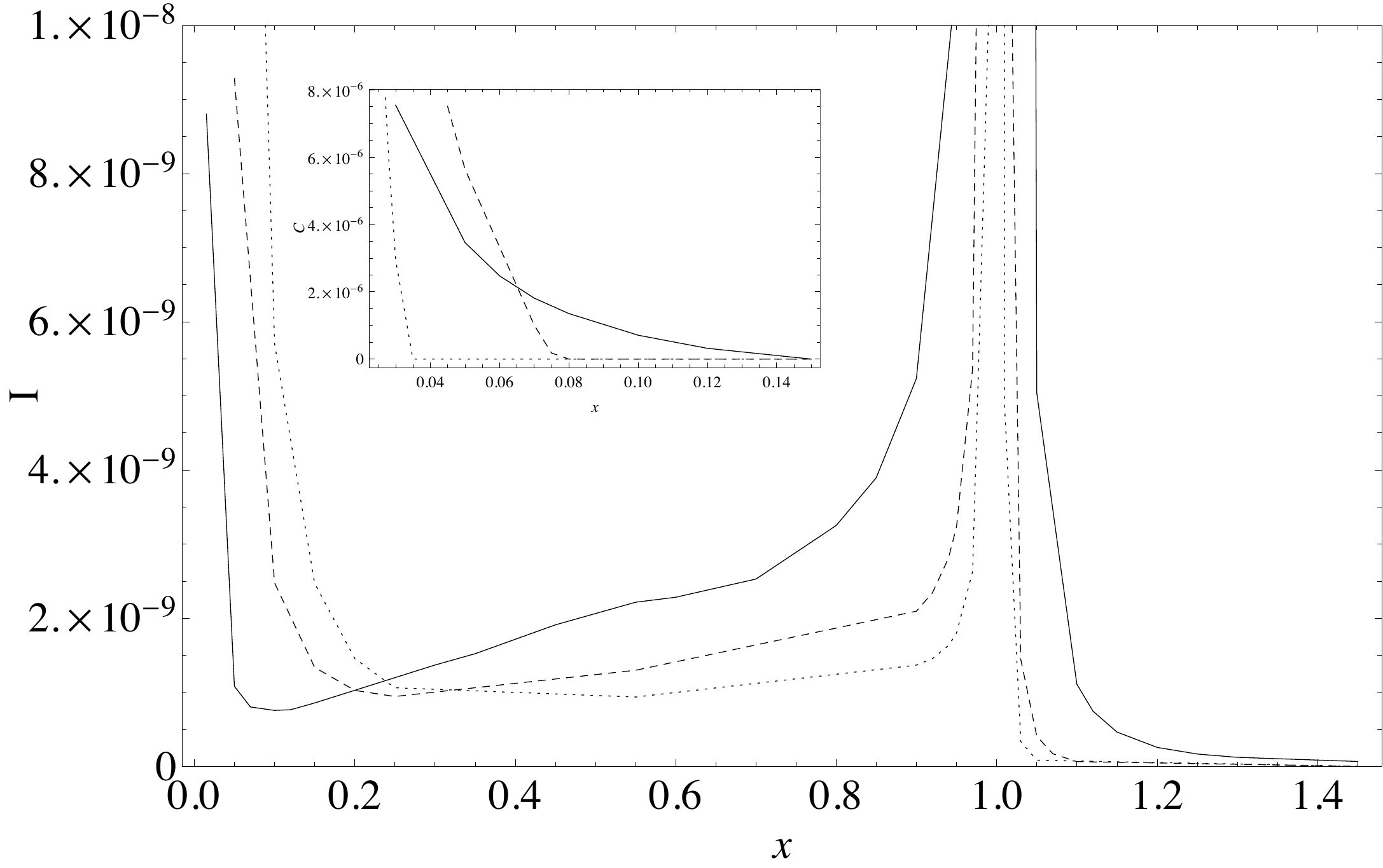}
\caption{Mutual information of $\rho_{AB}$ as a function of $x=r/c\,t$ for $z=\Omega r/c= 5$ (solid line), 10 (dashed line), 15
(dotted line). The inset shows the finite concurrences that are possible only for small values of $x$.}
\label{fig:3}
\end{figure}

The state in (\ref{n}) is an example of the so-called X-states, and the concurrence of such a state is given by:
\begin{equation}
C(\rho_{AB})=\frac{2}{c}\,\mbox{max}\left\{|\rho_{23}|-\sqrt{\rho_{11}\rho_{44}}\,,|\rho_{14}|-\sqrt{\rho_{22}\rho_{33}}\,,0\right\}\label{eq:31s}.
\end{equation}
Numerical computations show that the concurrence associated to this density matrix always vanishes except for a
bounded range of small values of $x$.
Beyond this range $\rho_{AB}$ is a separable state with no quantum correlations, either inside or outside the light cone. But
the atoms $A$ and $B$ are mutually dependent even for zero concurrence. Their mutual information 
\begin{equation}
\mathbb{I}(\rho_{AB})=
\mathbb{S}(\rho_{A})+\mathbb{S}(\rho_{B})-\mathbb{S}(\rho_{AB}), \label{eq:31t}
\end{equation}
which measures the total correlations between both parties, is
completely classical in this case, but may be finite. We show this quantity in Fig. \ref{fig:3} for different values of $z$ with an inset with the concurrence for small values of $x$.

\subsection{Different initial states}
\begin{figure}[h!]
\includegraphics[width=\textwidth]{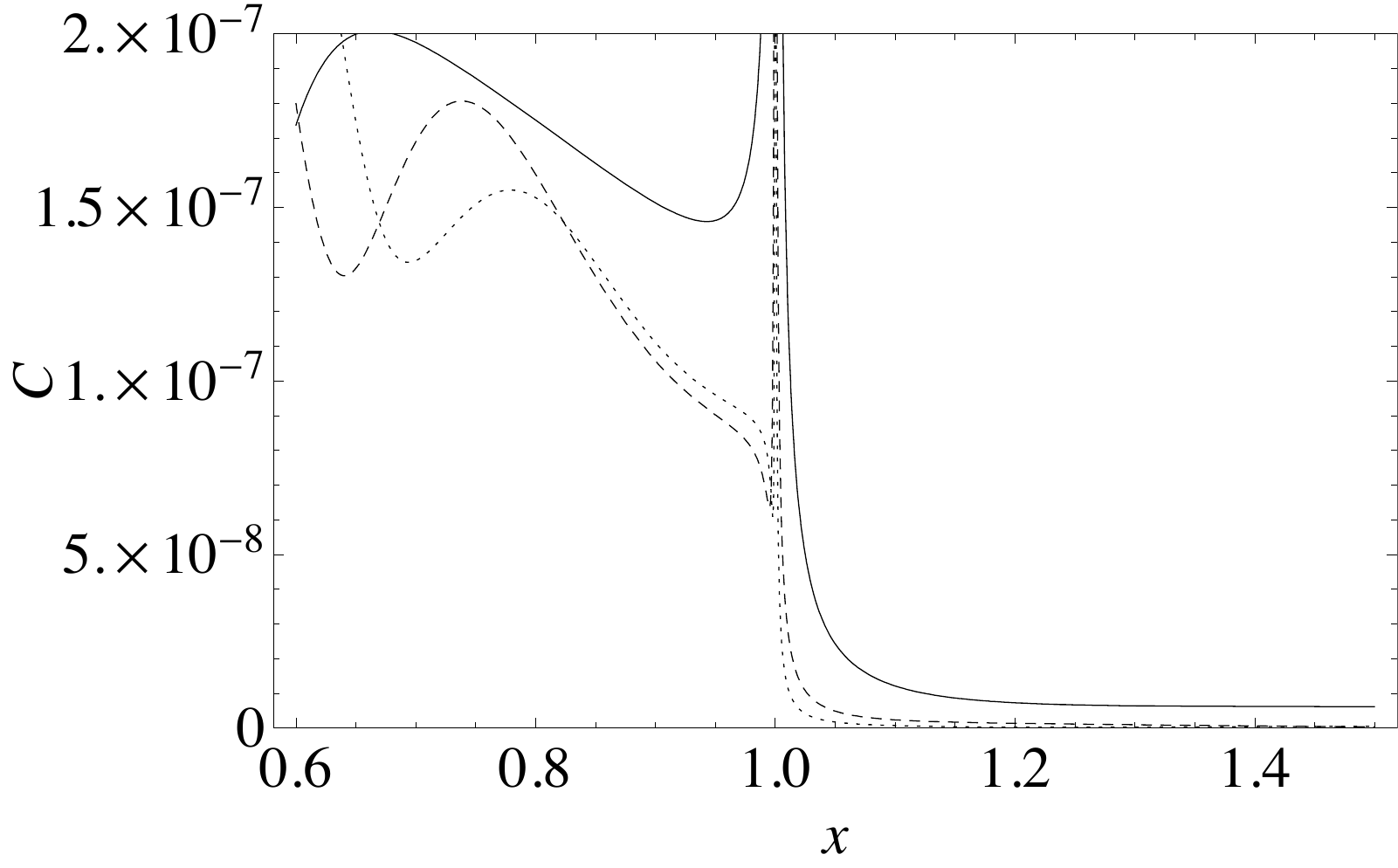}
\caption{Concurrence $\mathcal{C}^{(0)}$ of the atomic state in the e.m. vacuum  $\rho_{AB}^{(0)}$ as a function of $x=(r/c\,t)$
for $z=(\Omega r/c)=$ 5 (solid line), 10 (dashed line) and 15 (dotted line). The height of the peak is $\mathcal{C}^{(0)}=1$.
$x\rightarrow0$ ($t\rightarrow\infty$) is the region usually considered in Quantum Optics.}
\label{fig:4}
\end{figure}
\begin{figure}[h!]
\includegraphics[width=\textwidth]{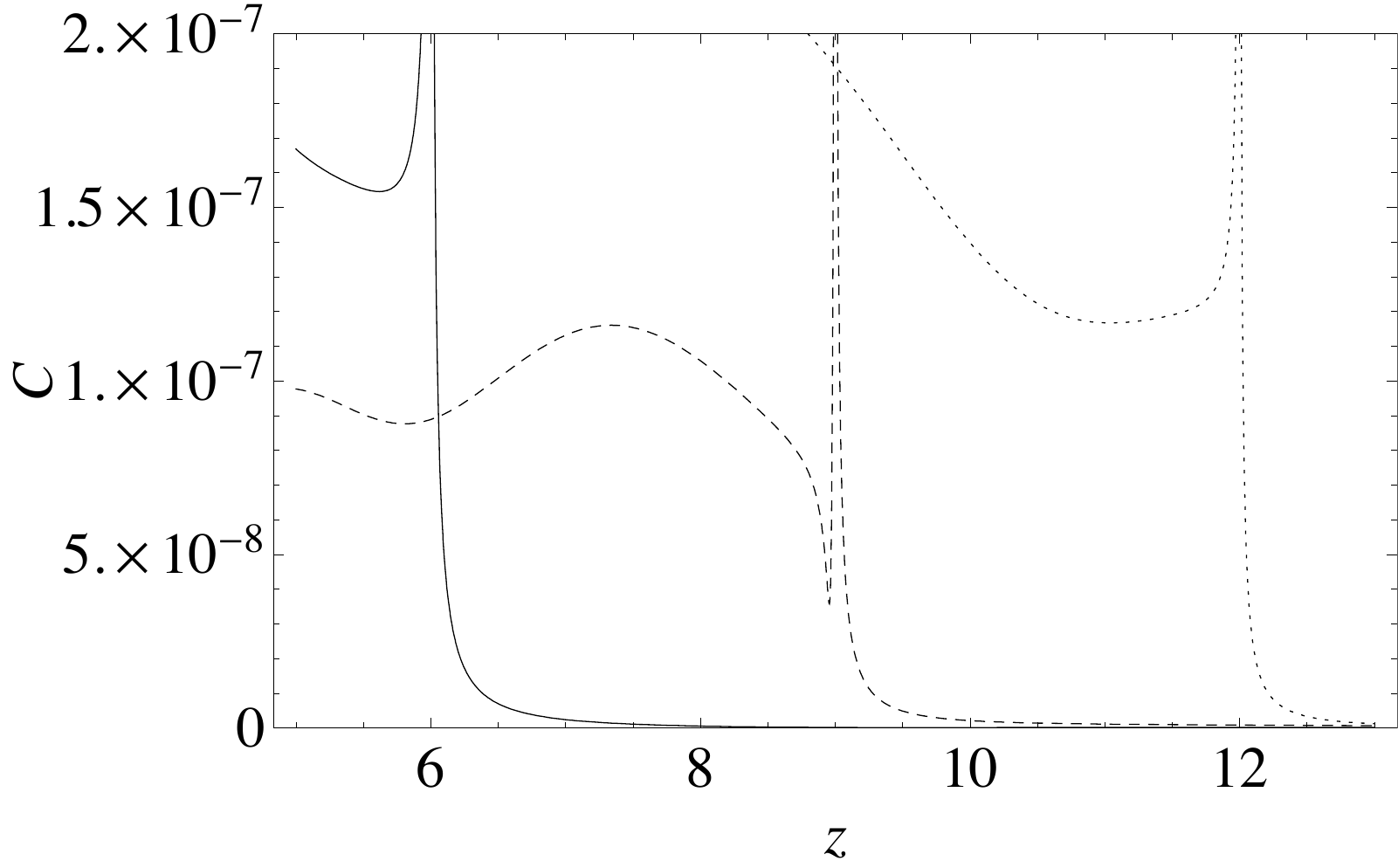}
\caption{Concurrence $\mathcal{C}^{(0)}$ of the atomic state in the e.m. vacuum  $\rho_{AB}^{(0)}$  if $|e\,e\,\rangle$ is the initial state as a function of
$z=(\Omega\,r/c)$ for $z/x=\Omega\,t=$ 6 (solid line), 9 (dashed line) and 12 (dotted line). $x>1$ amounts to $z>\Omega\,t$ in
each case. }
\label{fig:5}
\end{figure}
\begin{figure}[h!]
\includegraphics[width=\textwidth]{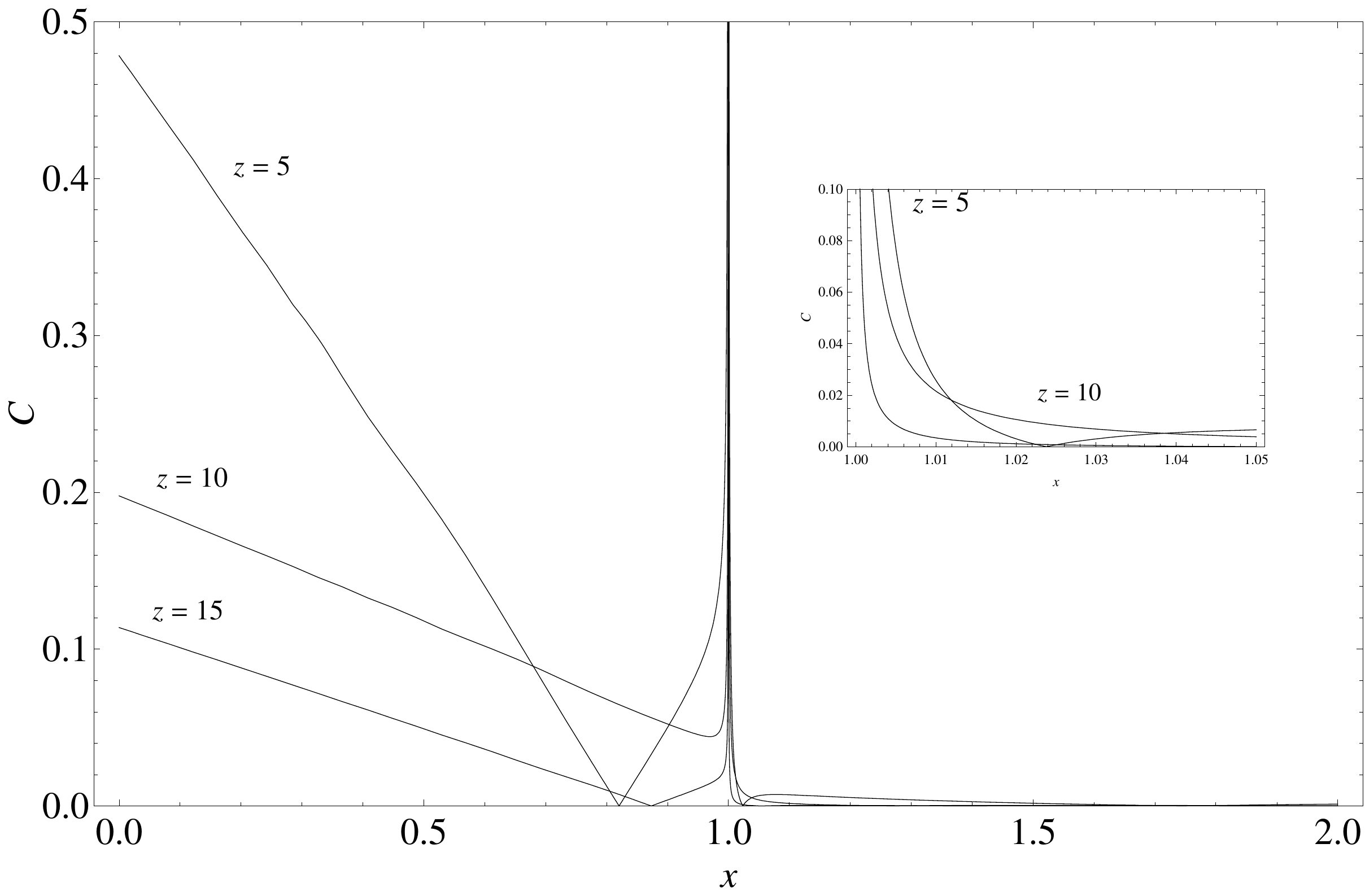}
\caption{Concurrence  $\mathcal{C}^{(1)}$ for one photon final state if $|e\,e\,\rangle$ is the initial state as a function of $x={r/ct}$ for three
values of $z=\Omega r/c$. The values of $\mathbb{C}$ for $x>1$ are of the same order as those displayed in Fig. .}
\label{fig:6}
\end{figure}

\begin{figure}[h!]
\includegraphics[width=\textwidth]{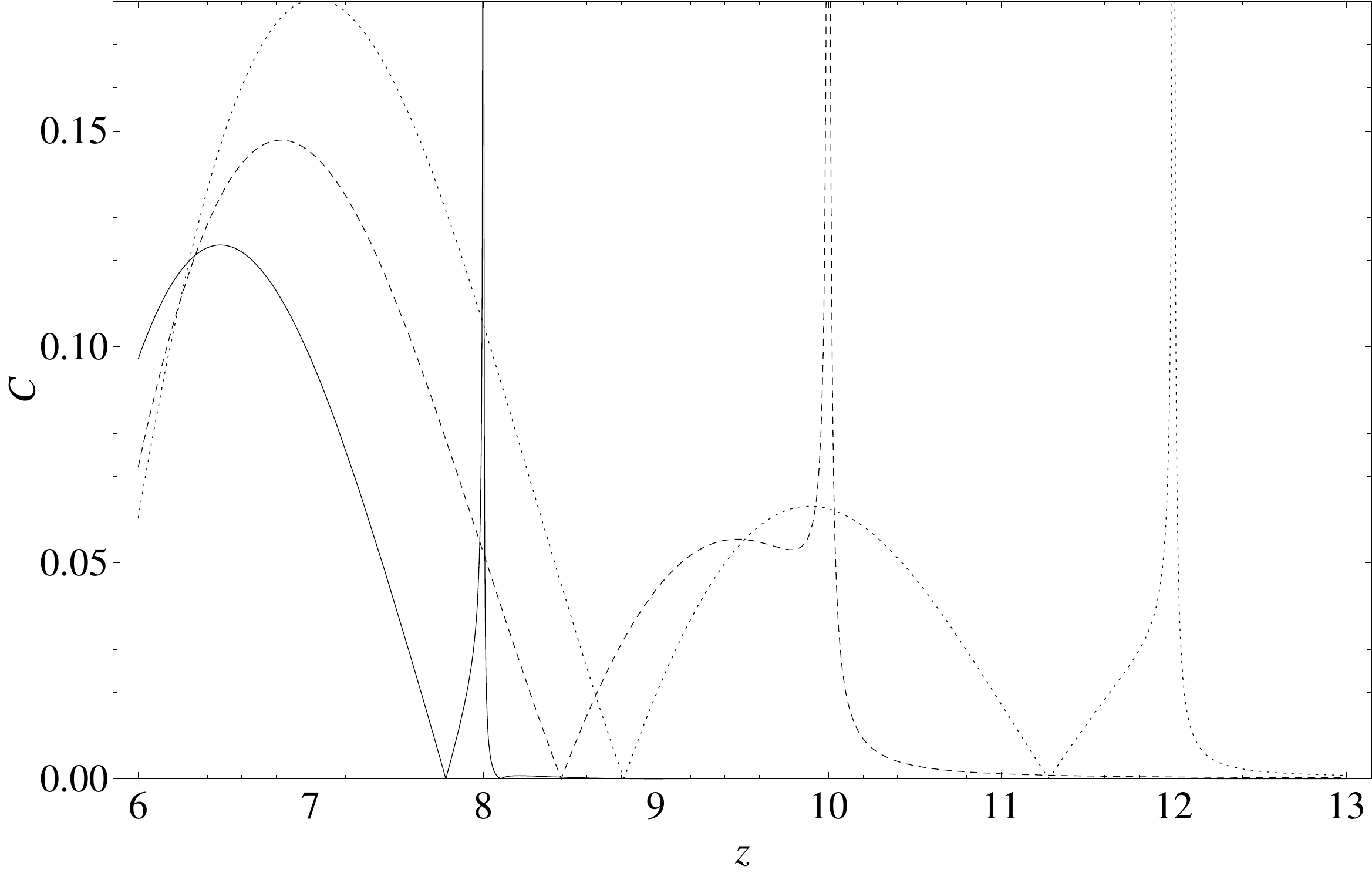}
\caption{Concurrence  $\mathcal{C}^{(1)}$ for one photon final state if $|e\,e\,\rangle$ is the initial state as a function of $z=\Omega r/c$ for
three representative values of the time $\Omega t= 8, 10, 12$ with peaks at $z=8, 10$ and $12$ respectively.}
\label{fig:7}
\end{figure}
\begin{figure}[h!]
\includegraphics[width=\textwidth]{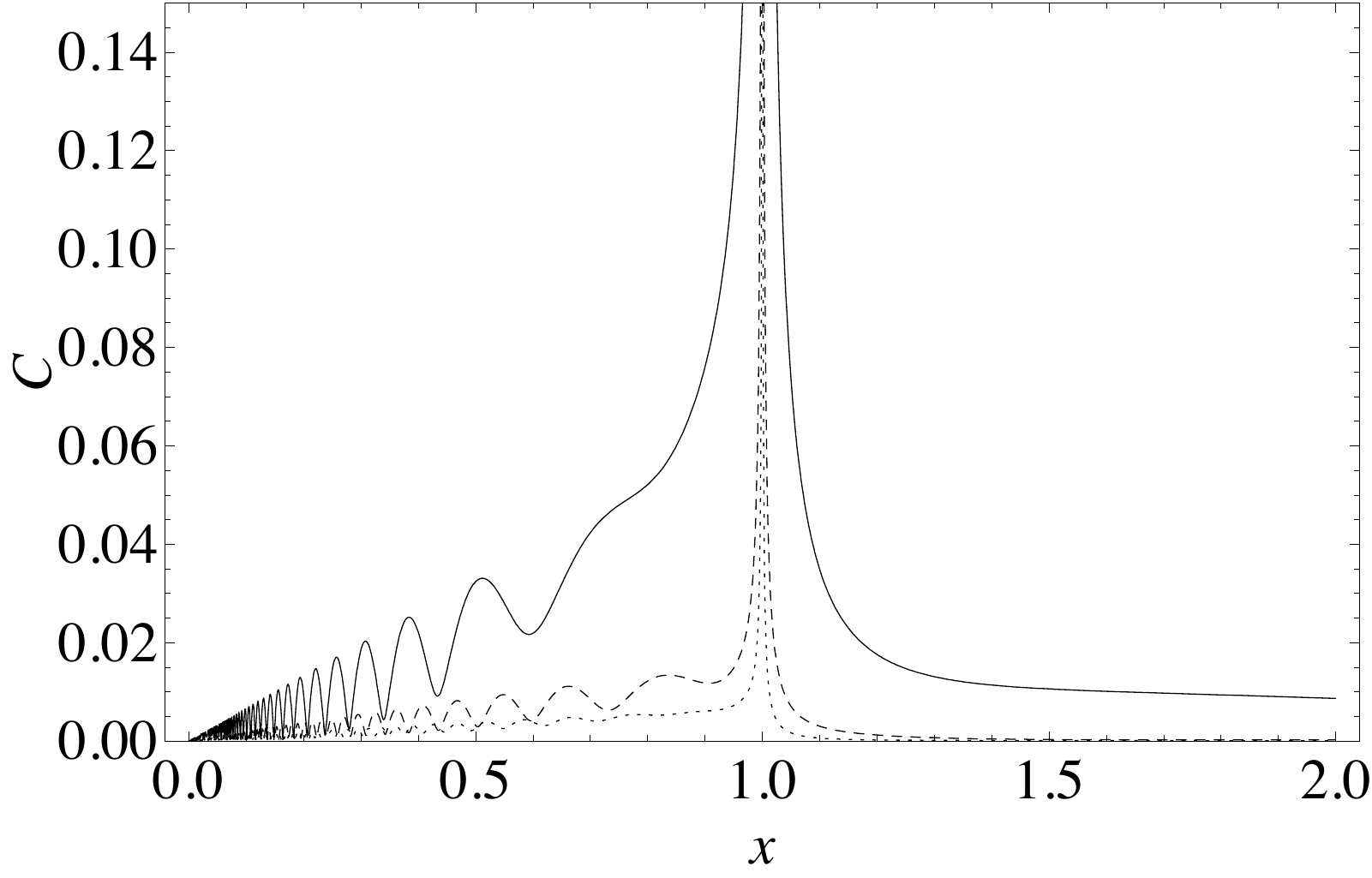}
\caption{Concurrence $\mathcal{C}^{(2)}$ of the atomic state with $n=2$ photons $\rho_{AB}^{(2)}$ in front of $x=r/(c\,t)$ for
$z=\Omega\,r/c=$ 5 (solid line), 10 (dashed line) and 15 (dotted line).}
\label{fig:8}
\end{figure}
\begin{figure}[h!]
\includegraphics[width=\textwidth]{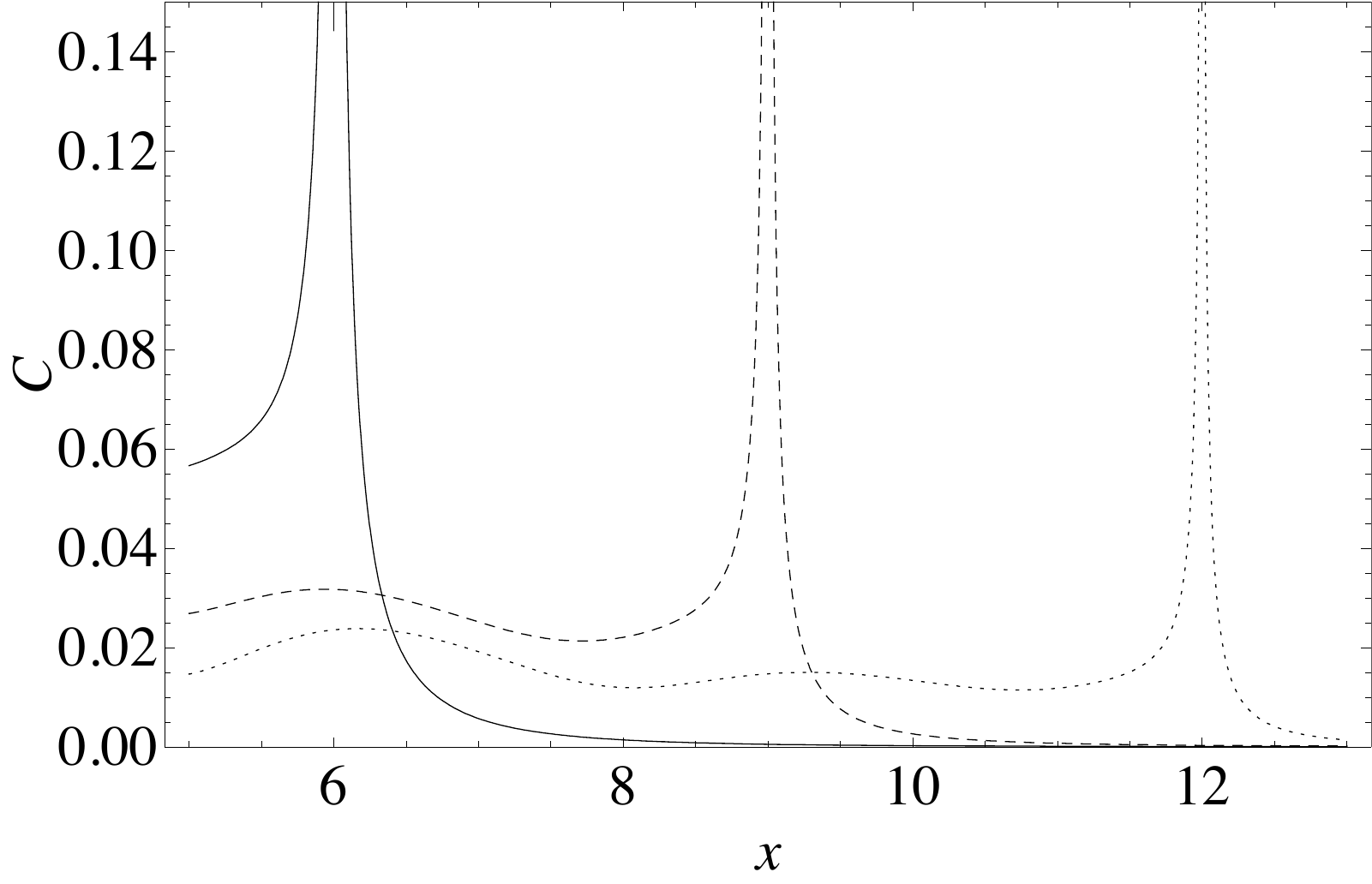}
\caption{Concurrence $\mathcal{C}^{(2)}$ of the atomic state with $n=2$ photons $\rho_{AB}^{(2)}$ in front of $z=\Omega\, r/c$
for $\Omega\,t=z/x=$ 6 (solid line), 9 (dashed line) and 12 (dotted line).}
\label{fig:9}
\end{figure}

In what follows we choose a system  given initially by the product state, 
\begin{equation}
|\,\psi\,(0)\rangle\,=\,
|\,e\,e\,0\,\rangle \label{eq:31u}
\end{equation}
 in which atoms $A$ and $B$ are in the excited state $|\,e\,\rangle$ and the field in the
vacuum state $|\,0\,\rangle$. The system then evolves under the effect of the interaction during a lapse of time $t$ into a state:
\begin{equation}
|\,\psi\,(t)\rangle = e^{-i\, \int_0^t\,dt'\, H_I\,(t')/\hbar}\,|\,\psi\,\rangle_0 \label{315b}
\end{equation}
that,  to  order $\alpha$, can be given in the interaction picture as
\begin{eqnarray}
|\mbox{atom}_1,\mbox{atom}_2,\mbox{field}\rangle_{t} =  ((1+A')\,|\,e\,e\rangle + X'\,|\,g\,g\rangle)\,|\,0\rangle\nonumber\\
 +(U_A\,|\,g\,e\,\rangle+ U_B\,|\,e\,g\,\rangle)\,|\,1\,\rangle+
(F'\,|\,e\,e\rangle+ G'\,|\,g\,g\rangle)\,|\,2\rangle\  \label{315c}
\end{eqnarray}
where
\begin{eqnarray}
A'&=&\frac{1}{2}\langle0|T(\mathcal{S}_A^+ \mathcal{S}_A^- + \mathcal{S}_B^+\mathcal{S}_B^-)|0\rangle,\, X'=
\langle0|T(\mathcal{S}^-_B\mathcal{S}^-_A)|0\rangle\nonumber\\
U_A\,&=&\,\langle\,1\,|\, \mathcal{S}^-_A\,|\,0\,\rangle,\, U_B\,=\,\langle\,1\,|\,
\mathcal{S}^-_B\,|\,0\,\rangle \label{315d}\\
F'&=&\frac{1}{2}\langle2|T(\mathcal{S}_A^+ \mathcal{S}_A^- +\mathcal{S}_B^+\mathcal{S}_B^-)|0\rangle,\,
G'=\langle2|T(\mathcal{S}^-_B \mathcal{S}^-_A)|0\rangle.\nonumber
\end{eqnarray}
The objects $\mathcal{S}^{\pm}_n$ has been defined in (\ref{eq:31f}). As in the previously analyzed case, in (\ref{315c}) only $X'$ corresponds to
interaction between both atoms, and now is a completely non-RWA term.  $A'$ describes intra-atomic radiative
corrections, $U_A$ and $U_B$ single photon emission by one atom, and $G'$ by both atoms, while $F'$ corresponds to two photon
emission by a single atom. 

We begin with the case $n=0$, where the field is in the vacuum state and, following (\ref{315c}), the atoms are in the projected
pure state 
\begin{equation}
|\,A\,B\,\rangle_0=((1\,+\,A')\,|\,e\,e\,\rangle + X'\,|\,G\,G\,\rangle) / c_0, \label{315e}
\end{equation}
where $c_0\,=\, \sqrt{|1\,+\,A'|^2
\,+\, |X'|^2}$ is the normalization, giving a concurrence
\begin{equation}
 \mathcal{C}^{(0)} \,=\, 2\,|X'|\,|\,1\,+\,A'\,|/c_0^2\ .\label{315h}
\end{equation}
The computation of $A'$ and $X'$ can be performed following the lines given in Appendix A where they were computed for the
case of a initial atomic state $|\,e\,g\,\rangle$. We will consider that the dipoles are parallel along the $z$ axis, while the
atoms remain along the $y$ axis. Under that conditions, using the dimensionless variables $x=r/c\,t$ and $z=\Omega\,r/c$:
\begin{eqnarray}
A'&=&\frac{4\,i\,K\,z^3}{3\,x}\,(\ln{|1-\frac{z_{max}}{z}|}\,+\,2\,i\,\pi),\nonumber\\
X'&=&\frac{\alpha\,d_i\,d_j}{\pi\,e^2}(-\mathbf{\nabla}^2\delta_{ij}+\nabla_i\nabla_j)\,I, \label{315i}
\end{eqnarray}
with $K=\alpha\,|\,\mathbf{d}\,|^2/(e^2\,r^2)$ and $I=I_+\,+\,I_-$, where:
\begin{eqnarray}
I_{\pm}&=&\frac{-i\,e^{-i\frac{z}{x}}}{2\,z}\,[\,\pm\,2\cos(\,\frac{z}{x}\,)\,e^{\pm\,i\,z}\,Ei(\mp\,i\,z)
+\,e^{-i\,z\,(1\pm\frac{1}{x})}\nonumber\\&\ &
Ei(i\,z\,(1\pm\frac{1}{x}))\,-\,e^{i\,z\,(1\pm\frac{1}{x})}\,Ei(-i\,z\,(1\pm\frac{1}{x}))\,]\label{315j}
\end{eqnarray}
for $x>1$ with the additional term $-2\,\pi\,i\,e^{i\,z\,(1-1/x)}$ for $x<1$. We use the conventions and tables of
\cite{bateman}.

We show in Fig. \ref{fig:4} the concurrence $\mathcal{C}^{(0)}$  (\ref{315h}) for $x$ near 1 for given values of $z$. Like the case where
$|\,e\,g\,\rangle$ is the initial atomic state, $\mathcal{C}^{(0)}$ jumps at $x=1$ and has different behaviors at both sides.
In Fig. \ref{fig:5} the concurrence is sketched as a function of $z$ for given values of $\Omega\,t=z/x$. The tiny values of the
concurrence for the region $z>\Omega\,t$ (which corresponds to $x>1$), diminish as $t$ grows and will eventually vanish, since
$X'$ is a non-RWA term.

In the case $n=1$  the final one photon atomic state would be 
\begin{equation}
|A\,B\,\rangle_1=(\,U_{A}\,|\,g\,e\,\rangle\,+\,U_{B}\,|\,e\,g\,\rangle)/c'_1, \label{315k}
\end{equation}
where $U_A$ is the same function of
$\mathbf{x}_A$ as $U_B$ is of $\mathbf{x}_B$, and  $c'_1 = \sqrt{2|\,U_A\,|^2}$ . Now, the indistinguishability of the photon source commented in Section 3.1.3 persists for large $t$
and so does entropy and concurrence. In particular, the $r$-dependent
concurrence is 
\begin{equation}
\mathbb{C}^{(1)} \,=\,2 |\,l'\,|/(c'_1)^2 \label{315l}, 
\end{equation}
where
\begin{equation}
 l'=\,\langle\,0\,|\,
\mathcal{S}^+_A\,\mathcal{S}^-_B\,|\,0\,\rangle. \label{315m}
\end{equation}
We represent it in Fig. \ref{fig:6}. In Fig. \ref{fig:7} we have represented the concurrence for the case where the
initial atomic state was $|e\,e\,\rangle$ in terms of the inter-atomic distance for three fixed values of time. What we obtain is
a shift of the concurrence features to longer $r$ as $t$ grows (so that they appear at the same $(r/ct)$), in such a way that,
even if  $t$ is just the duration of the interaction, it plays the role of propagation time for the generated correlations. They
are negligible small for large $r$, peak at the ``light cone" but, on the other hand grow, as we would expect, for larger
interaction times.

Now we will focus on the two photon case. The final atomic state 
\begin{equation}
|\,A\,B\rangle_2=(F'\,|\,e\,g\,\rangle\,+\,G'\,|\,g\,e\,\rangle)/c_2, \label{315n}
\end{equation}
with $c_2\,=\,\sqrt{|\,F'\,|^2\,+\,|\,G'\,|^2}$, is in the same subspace as for $n\,=\,0$. The normalization $c_2$ is
$\mathcal{O}(\alpha)$ like  $F'$, $G'$, so that all the coefficients in $\rho^{(2)}$ may be large.
Therefore, although the probability of attaining this state is small, the correlations are not. The concurrence is
\begin{equation}
\mathcal{C}^{(2)}=2 |\,F'\,G'^*\,|/c_2^2.\label{315o}
\end{equation}
We find that:
\begin{eqnarray}
F'&=&\theta(t_1-t_2)(\,V_A\,(t_1)\,U'_A\,(t_2)\,+\,U_A\,(t_1)\,V'_A\,(t_2)\nonumber\\
&+&\,V_B\,(t_1)\,U'_B\,(t_2)\,+\,U_B\,(t_1)\,V'_B\,(t_2)\,)\label{315l}\\
G'&=&U_B\,U'_A\,+\,U_A\,U'_B\nonumber
\end{eqnarray}
with $V_A\,=\,\langle\,1\,|\, \mathcal{S}^+_A\,|\,0\,\rangle$ and $V_B\,=\,\langle\,1\,|\, \mathcal{S}^+_B\,|\,0\,\rangle$. The
primes account for the two single photons, i.e.
\begin{equation}
|\,2\,\rangle=|\mathbf{k}\,\mathbf{\epsilon}_{\lambda},\,\mathbf{k'}\,\mathbf{\epsilon}_{\lambda'}\,\rangle. \label{315p}
\end{equation}
The quantities
$|\,U_A\,|^2\,=\,|\,U_B\,|^2\,=\,|\,U\,|^2$, $|\,V_A\,|^2\,=\,|\,V_B\,|^2\,=\,|\,V\,|^2$, $l\,=\,U_A\,V_B^*\,=\,U_B\,V_A^*$,
$U\,V^*\,=\,U_A^*\,V_A^*=\,U_B\,V_B^*$, $U_B\,U_A^*$ and $V_A\,V_B^*$ can be computed following the lines inAppendix A.

In Fig. \ref{fig:8}  we show $\mathcal{C}^{(2)}$ in front of $x$ for given values of $z$. When $x\rightarrow0$ ($t\rightarrow\infty$, i.e.
the Quantum Optics regime), $F'$ vanishes and the final atomic state would be the separable state $|\,g\,g\,\rangle$, with zero
concurrence. Entanglement is sizable for $x>1$, and could be maximized if a particular two photon state was detected, as we will explain in the next section.

In Fig. \ref{fig:9} ,  $\mathcal{C}^{(2)}$ is sketched as a function of $z$ for given values of $\Omega\,t=z/x$. Again, the concurrence for
the region $z>\Omega\,t$ ($x>1$), diminish as $t$ grows and will eventually vanish, since it is due to $f$, which is a non-RWA
term.
Interestingly, as we noted for the single photon emission, $x=1$ is a singular point that divides the
spacetime into two different regions. This occurs even if  $t$ is not the propagation
time of any physical signal between the atoms. This effect comes from the appearance of effective interaction terms like $l$, that would be missing if we could discriminate the source of emission of each photon.

\subsection{Conclusions}
 In this section we have studied the correlations between a pair of initially separable neutral two-level atoms that are allowed to interact with the
electromagnetic field, initially in the vacuum state. We have computed the concurrences that arise when the final state contains
$n=0,1,$ or 2 photons. They may be sizable for $x$ small ($t\rightarrow\infty$ for a given $r$) and also around $x=1$. Only
in the case $n=0$ there are interactions between both atoms, generating an entanglement that may persist asymptotically in the case that $\ket{eg}$ is the initial atomic state. We have
carefully taken into account all the terms contributing to the amplitude for finite time (they are $\propto t^4$, not $\propto
t^2$ as is sometimes assumed). A small amount of entanglement can be generated between spacelike separated parties due to the
finiteness of $X$ when $x>1$, but a change of behavior appears for $x<1$.  For $n=2$ the final atoms are in the same subspace
than for $n=0$. There are similar correlations that in this case can give rise to entanglement swapping, by measuring both
photons in a definite state for instance. Naturally, in this case entanglement may be sizable for spacelike separated parties,
as here this is not related to any kind of propagation. Entanglement in the case with only a final photon ($n=1$)  comes from
the indistinguishability of the photon source. It will vanish asymptotically in the case that the initial states is $\ket{eg}$ when eventually only one atom ($A$ in
the present case) may emit the photon, and it is also sizable when $x>1$. It is interesting how these correlations become
classical (except for small $x$) when the states of the field are traced over. We have shown through the mutual information the
residues of what were quantum correlations in the individual cases analyzed before.

\section{Entanglement swapping between spacelike separated atoms}
\subsection{Introduction}

There are mainly two different
known ways of generate entanglement: by interaction between the atoms (for instance, \cite{vanenk}) or by  detection of the emitted
photons \cite{cabrillo,feng,duan,simon,lamata}. Some of these proposals have been realized experimentally (for instance,
\cite{moehring}). For the latter cases, in principle, there is no reason to expect that the swapping \cite{swapping} of
atom-photon to atom-atom entanglement can only begin to occur when one atom enter into the light cone of the other.

The possibility of entanglement generation between spacelike separated atoms is of both theoretical and practical interest, and
was addressed from different points of view in \cite{franson,reznik,reznikII} and in the previous section of this Thesis. In section 3.1, we analyze this issue
perturbatively in a simple model of a pair of two-level atoms interacting locally with the electromagnetic field, initially in
the vacuum state. Tracing over the field states, the atoms are only classically correlated, but applying
$|\,n\rangle\langle\,n|$ ($n\,=\,0, 1, 2$ being the number of photons up to second order in perturbation theory), the atoms get
entangled. For $n\,=\,0$ the entanglement is generated by the interaction term and therefore is only relevant when one atom
enter into the light cone of the other, despite of the finiteness of the Feynman propagator beyond that region. But for $n\,=\,
1, 2$ entanglement may be sizeable, although small, if the interatomic distance is short enough. In \cite{reznik}, the trace
over the field states was considered in a model with a pair of two-level detectors coupled to a scalar field. The detectors may
get entangled if a suitable time dependent coupling is introduced, and this was applied to a linear ion trap in \cite{reznikII}.
In \cite{franson}, only the vacuum case when $t\rightarrow0$ was analyzed, and no entanglement measures were considered.

As commented in chapter 1, there are two possible interpretations for these effects: as a transfer of preexisting entanglement of the vacuum \cite{reznik,
reznikII} or as a consequence of the propagation of virtual quanta outside the light cone \cite{franson}. Both are compared and
discussed in \cite{franson}.

In this section we will go one step further and consider that the photons are detected with definite momenta and polarizations. We
show that, in principle, a high degree of entanglement, even maximal, can be generated between spacelike separated atoms if a
Bell state of the emitted photons is detected. We will consider a pair of neutral two-level atoms separated by a fixed and
arbitrary distance and study the evolution of an initially uncorrelated state under local interaction with the electromagnetic
field. We focus on the two-photon emission which, although has a smaller probability of success, shows a larger fidelity of the
projected state with the desired state and has a entanglement robust to atomic recoil \cite{morigi}. The photons pass through a
partial Bell-state analyzer \cite{zeilinger}, and we use entanglement measures to study the evolution of entanglement in the
projected atomic states after detection of the different photonic Bell states. The results show that interaction times must be
short, but interatomic distances can be as large as desired. The interaction time is independent of the photodetection time,
which is only related with the distance from the atoms to the detectors.  That distance can be such that the photodetection can
occur while the atoms remain spacelike separated.

The results can be interpreted as a transfer of part of the vacuum entanglement after a post-selection process. If no measurement were performed the atoms would have classical correlations transferred by the vacuum. In \cite{reznik} the classical correlations may become entanglement with a suitable time dependent coupling. The post-selection process can be seen as an
alternative way to achieve the entanglement transference. While the results in \cite{franson,reznik} are mainly theoretical, these could be probed experimentally, and would show for the first time the possibility of transfer entanglement from the vacuum state of the quantum field to spacelike separated atoms.

\subsection{Entanglement swapping between spacelike separated atoms}
\begin{center}
\begin{figure}[h!]
\includegraphics[width=\textwidth]{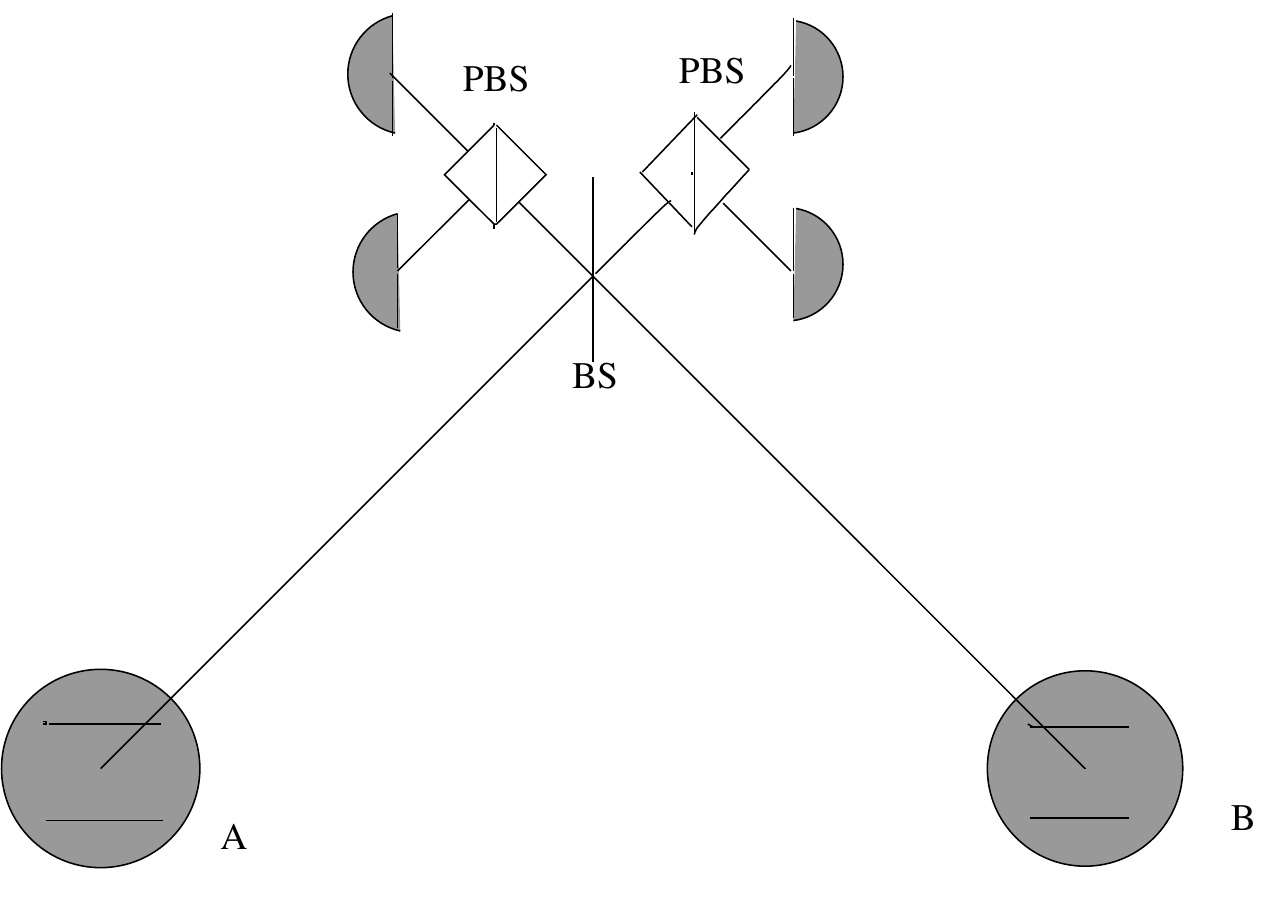}
\caption{Schematic setup for the entanglement swapping described in the text. The atoms $A$ and $B$  are at $(y,z)=(\mp r/2,0)$.
The emitted photons pass through a $50:50$ BS at $(0,r/2)$ and two PBS at $(\pm\,d/2\sqrt2,L/2+d/2\sqrt2)$, and there are four
single photon detectors at the outport ports of the two PBS, at $(\pm\,d/\sqrt2,r/2+d/\sqrt2)$ and $(\pm\,d/\sqrt2,r/2)$. Taking
into account that $|\Psi^-\rangle$ and $|\Phi^-\rangle$ are forbidden in our model, a $|\Psi^+\rangle$ is detected when there
are coincidence clicks in two detectors and $|\Phi^+\rangle$ when there is a double click in one detector. Then the atoms are
projected onto the atomic part of the state (\ref{32b}).}
\label{fig:10}
\end{figure}
\end{center}
In what follows we choose a system  given initially by the product state, 
\begin{equation}
|\,\psi\,(0)\rangle\,=\,
|\,e\,e\,0\,\rangle \label{eq:321}
\end{equation}
in which atoms $A$ and $B$ are in the excited state $|\,e\,\rangle$  and the field in the
vacuum state $|\,0\,\rangle$. As in the previous section,  the system then evolves under the effect of the interaction (\ref{a}) during a lapse of time $t$, and, up to
order $e^2$, 0, 1 or 2 photons may be emitted. If after that a two-photon state is detected,
\begin{equation}
|\Psi\rangle=|\mbox{photon}_1,\mbox{photon}_2\rangle=\sum_{\vec{k},\vec{k'},\lambda,\lambda'}\,c_{\vec{k}\,\vec{k'},\lambda,\lambda'}\,|\vec{k}\lambda,\,\vec{k'}\lambda'\rangle
\label{32a}
\end{equation}
(being $\hbar\,\vec{k}$, $\hbar\,\vec{k'}$ momenta and $\lambda$,$\lambda'$ polarizations), the projected state, up to order
$e^2$, can be given in the interaction picture as
\begin{equation}
|\mbox{photons},\mbox{atom}_1,\mbox{atom}_2\rangle_{t} = |\,\Psi\rangle(\frac{F_{\Psi}\,|\,e\,e\rangle+ G_{\Psi}\,|\,g\,g\rangle}{N})
\label{32b}
\end{equation}
where
\begin{equation}
F_{\Psi}=\frac{1}{2}\langle\Psi|T(\mathcal{S}_A^+ \mathcal{S}_A^-\,+\, \mathcal{S}_B^+\mathcal{S}_B^-)|0\rangle,\,
G_{\Psi}=\langle\Psi|T(\mathcal{S}^-_B \mathcal{S}^-_A)|0\rangle \label{32c}
\end{equation}
and $N=\sqrt{|\,F_{\Psi}\,|^2+|\,G_{\Psi}\,|^2}$. The objects $\mathcal{S}^{\pm}_n$ were defined in (\ref{eq:31f}).

Here, $G_{\Psi}$  describes single photon
emission by both atoms, while $F_{\Psi}$ corresponds to two photon emission by a single atom.  In Quantum Optics, $F_{\Psi}$ is usually
neglected by the introduction of a rotating wave approximation (RWA), but as we will see later, for very short interaction times
$F_{\Psi}$ and $G_{\Psi}$ may be of similar magnitude. Actually, a proper analysis of this model can be performed only beyond the RWA,
as we have commented in the previous section. Without RWA vacuum entanglement cannot be transferred to the atoms with this particular
post-selection process. In that case, a one photon post-selection process would entangle the atoms.

(\ref{32c}) can be written as:
\begin{eqnarray}
F_{\Psi}&=&\frac{1}{2}\,\theta(t_1-t_2)\langle\Psi|\mathcal{S}_A^+\,(t_1)
\mathcal{S}_A^-\,(t_2)+\mathcal{S}_B^+\,(t_1)\mathcal{S}_B^-\,(t_2))|0\rangle,\,\nonumber\\
G_{\Psi}&=&\langle\Psi|\mathcal{S}^-_B\,(t_1) \mathcal{S}^-_A\,(t_2))|0\rangle \label{32d}
\end{eqnarray}

The photons pass through a partial Bell-state analyzer \cite{zeilinger} consisting in a beam splitter (BS) and two polarization
beam splitters (PBS) with four single photon detectors at their output ports. If two detectors, one at one output port of one
PBS and one at an output port of the other, click at the same time, a state $|\Psi^-\rangle$ is detected, while if the two
clicks are in the two output ports of only one PBS, the state is $|\Psi^+\rangle$. If one of the four detectors emits a double
click, the state can be $|\Phi^+\rangle$ or $|\Phi^-\rangle$. Taking into account momenta and symmetrization, the Bell states
can be written as
\begin{eqnarray}
|\Psi^{\pm}\rangle&=&\frac{1}{\sqrt2}[|\vec{k}\downarrow,\,\vec{k'}\uparrow\rangle+|\vec{k'}\uparrow,\,\vec{k}\downarrow\rangle\nonumber\\
&\pm&(|\vec{k}\uparrow,\,\vec{k'}\downarrow\rangle+|\vec{k'}\downarrow,\,\vec{k}\uparrow\rangle)]\label{32f}\\
|\Phi^{\pm}\rangle&=&\frac{1}{\sqrt2}[|\vec{k}\downarrow,\,\vec{k'}\downarrow\rangle+|\vec{k'}\downarrow,\,\vec{k}\downarrow\rangle\nonumber\\
&\pm&(|\vec{k}\uparrow,\,\vec{k'}\uparrow\rangle +|\vec{k'}\uparrow,\,\vec{k}\uparrow\rangle)]\nonumber
\end{eqnarray}
where $\uparrow$ and $\downarrow$  are the photon polarizations, with polarization vectors
\begin{equation}
\epsilon\,(\vec{k},\uparrow)={-1\over\sqrt2}(\epsilon\,(\vec{k},1)+\epsilon\,(\vec{k},2)) \label{eq:32g}
\end{equation}
and
\begin{equation}
\epsilon\,(\vec{k},\downarrow)= {1\over\sqrt2}(\epsilon\,(\vec{k},1)-\epsilon\,(\vec{k},2)), \label{eq:32h}
\end{equation}
where 
\begin{equation}
\epsilon\,(\vec{k},1)=
(\cos{\theta_k}\cos{\phi_k},\cos{\theta_k}\sin{\phi_k},-\sin{\theta_k})\label{eq:32i}
\end{equation}
 and 
 \begin{equation}
 \epsilon\,(\vec{k},2)=
(-\sin{\phi_k},\cos{\phi_k},0). \label{eq:32j}
\end{equation}
Here 
\begin{equation}
|\vec{k}\lambda,\,\vec{k'}\lambda'\rangle=a^{\dag}_{k\lambda}\,
\,a^{\dag}_{k'\lambda'}|\,0\rangle.\label{eq;32k}
\end{equation}
We will use the concurrence (\ref{eq:22f}) $\mathbb{C}$ to compute the entanglement of the atomic states when the different
Bell states are detected. The concurrence of the atomic part of a state like (\ref{32b}) is just given by
\begin{equation}
\mathbb{C}=\frac{2 |\,F_{\Psi}\,G_{\Psi}^*\,|}{N^2} \label{32g}
\end{equation}
We assume that the atoms $A$, $B$ are along the $y$ axis, at $y=\,\mp L/2$ respectively, and the dipoles are parallel along the
$z$ axis, corresponding to an experimental set up in which the dipoles are induced by suitable external fields \cite{franson}.
We also take 
\begin{equation}
|\vec{k}|=|\vec{k'}|=\Omega/c. \label{eq:32l}
\end{equation}
Under that conditions, the first remarkable thing is that for $|\Psi^-\rangle$ and $|\Phi^-\rangle$,  we have $F_{\Psi}=\,G_{\Psi}=\,0$.
Therefore, at least while only E1 transitions are considered, in this model the Bell-sate analyzer is complete: if two different
detectors click the state is $|\Psi^+\rangle$, while if one detector clicks twice the state is $|\Phi^+\rangle$. First, we focus
on $|\Psi^+\rangle$.   Considering (\ref{32d}) and (\ref{32f}), with the mode expansion for the electric field and the
commutation relation for the creation and annihilation operators, a standard computation leads to:
\begin{eqnarray}
F_{\Psi}&=&\frac{K(\Omega,t,\theta)}{2}\,j(\Omega\,t)\cos{(\frac{z}{2}\,h_+(\theta,\phi))}\nonumber\\
G_{\Psi}&=& K(\Omega,t,\theta)\,\cos{(\frac{z}{2}\,h_-(\theta,\phi))},\label{32h}
\end{eqnarray}
with 
\begin{equation}
K(\Omega,t,\theta)= \frac{c\,\alpha|\mathbf{d}|^2\Omega t^2}{2\pi^2e^2}\sin{\theta_k}\,\sin{\theta_{k'}}, \label{eq:32m}
\end{equation}
($\alpha$ being the fine structure constant), 
\begin{equation}
j(\Omega\,t)= {|-1\,+\,e^{2i\,\Omega\,t}(1-2i\,\Omega\,t)|\over(\Omega\,t)^2},\label{eq:32n}
\end{equation}
and
\begin{equation}
h_\pm\,(\theta,\,\phi)=(\sin{\theta_k}\sin{\phi_k}\pm\sin{\theta_{k'}}\sin{\phi_{k'}}), \label{eq:32o}
\end{equation}
$\theta_k,\,\phi_k$ corresponding
to $\hat{k}$ and $\theta_{k'},\,\phi_{k'}$ to $\hat{k'}$, and $z=\,\Omega\,r/c$. Notice that $j(\Omega\,t)$ decreases as $t$
grows, and eventually vanish as $t\rightarrow\infty$ as required by energy conservation. The $r$ dependence is a result of the
individual dependence on the position of each atom, not on the relative distance between them.
\begin{figure}[h!]
\includegraphics[width=\textwidth]{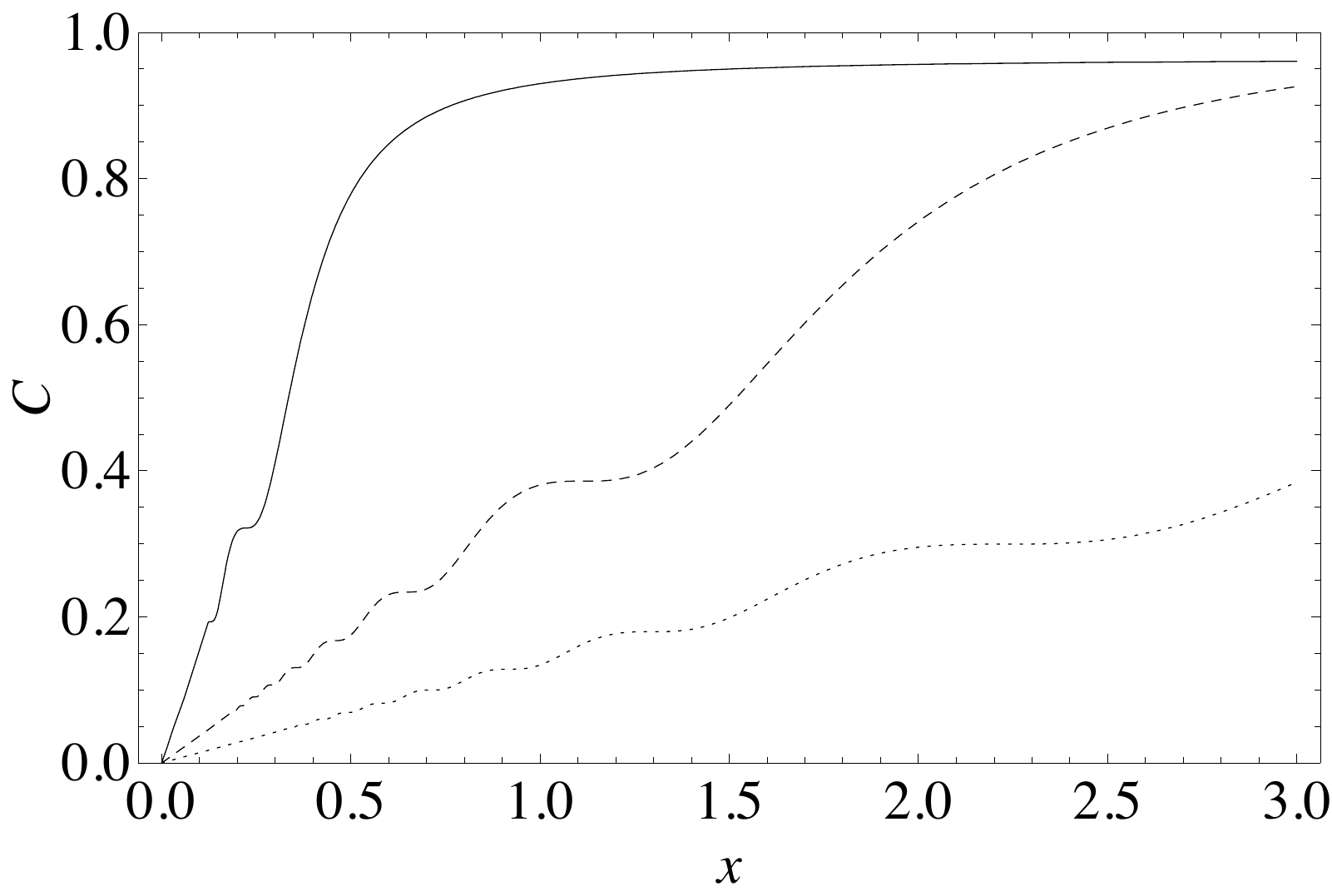}
\caption{Concurrence for the atomic state when a Bell state $|\Psi^+\rangle$ or $|\Phi^+\rangle$ of the photons is detected, as
a function of $x={r/ct}$ for $z=\Omega r/c=$1 (solid), 5 (dashed), 10 (dotted). The light cone is at $x<1$. For $x>1$ the
interaction time is short enough to have a significative amount of entanglement.}
\label{fig:11}
\end{figure}
\begin{figure}[h!]
\includegraphics[width=\textwidth]{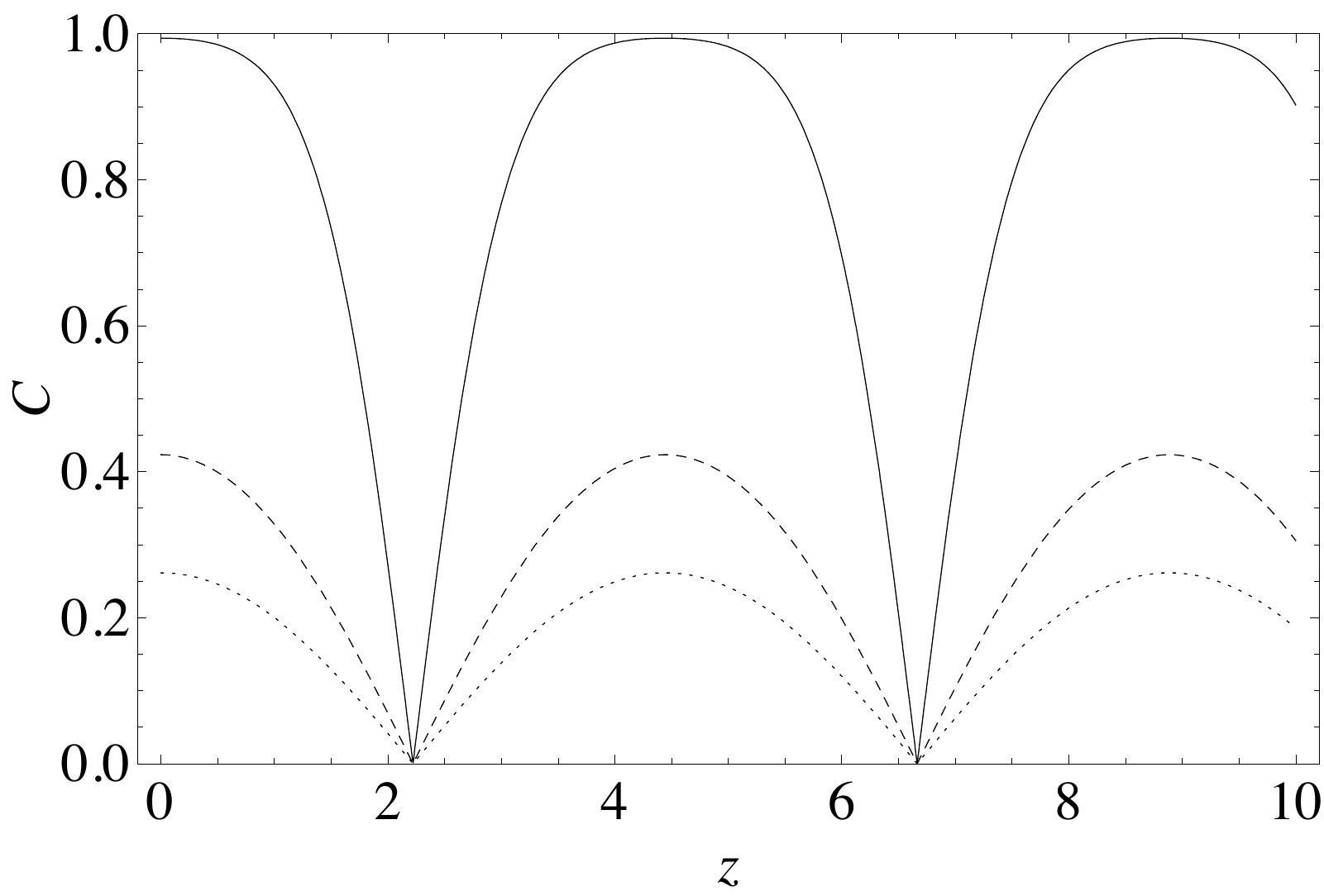}
\caption{Concurrence for the atomic state when a Bell state $|\Psi^+\rangle$  of the photons is detected, as a function of
$z=\Omega r/c$ for $\Omega\,t=$1 (solid), 4 (dashed), 7 (dotted). The light cone for each curve is at $z<\Omega\,t$ .}
\label{fig:12}
\end{figure}
\begin{figure}[h!]
\includegraphics[width=\textwidth]{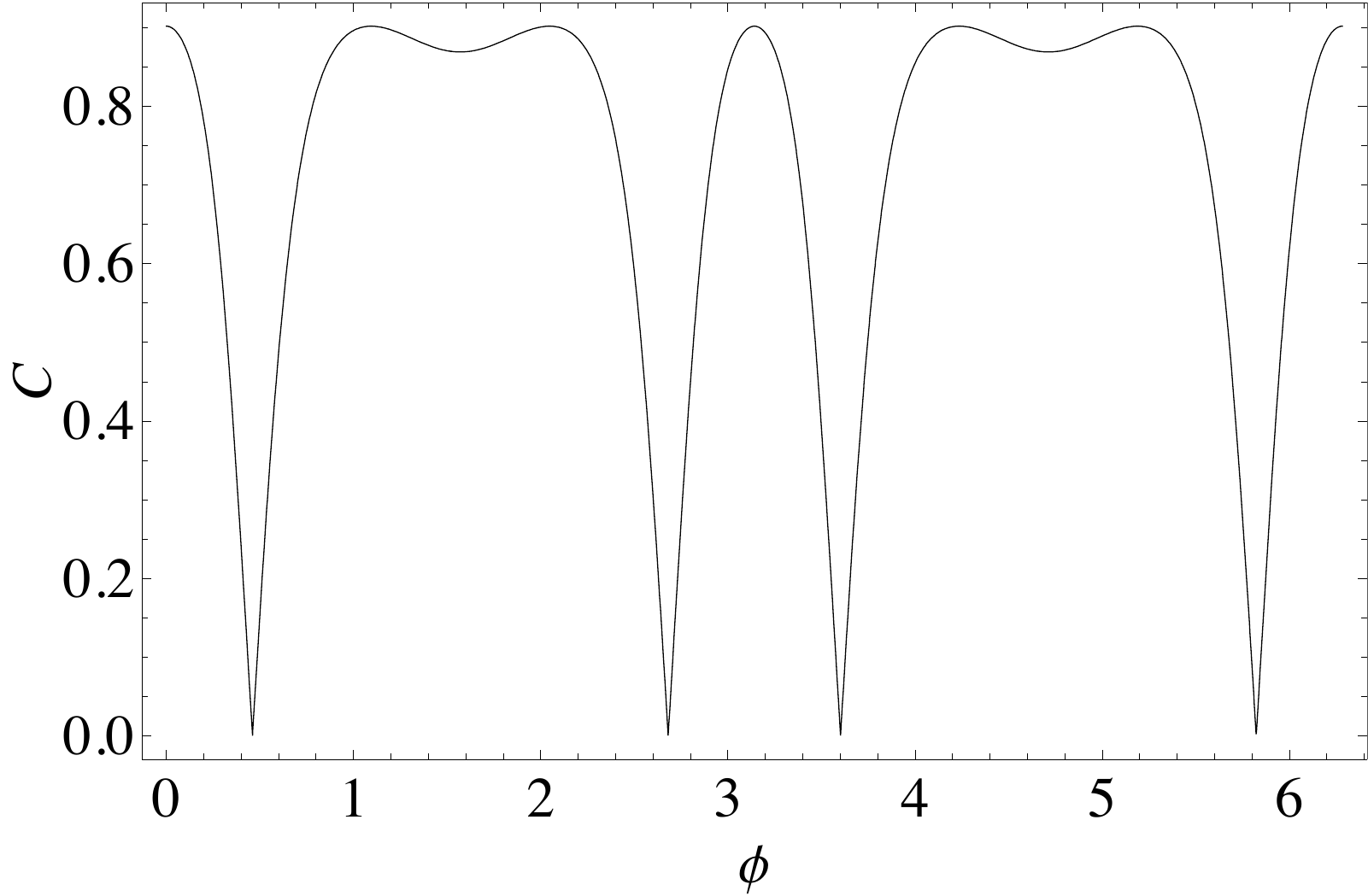}
\caption{Concurrence for the atomic state when a Bell state $|\Psi^+\rangle$ of the photons is detected, as a function of $\phi$
for $z=\Omega r/c=5$ and $x=r\,c/t=2.5$.}
\label{fig:13}
\end{figure}

Taking into account (\ref{32g}) and (\ref{32h}) the concurrence is given by:
\begin{equation}
\mathbb{C}={4|\cos{(\frac{z}{2}\,h_+(\theta,\,\phi))}\cos{(\frac{z}{2}\,h_-(\theta,\,\phi))}|\over\cos^2{(\frac{z}{2}\,h_+(\theta,\,\phi))}\,j(\Omega\,t)
+\cos^2{(\frac{z}{2}\,h_-(\theta,\,\phi))}\,{4\over j(\Omega\,t)}}, \label{32i}
\end{equation}

Now, we assume that the  50:50 BS is at $(y,z)=(0,r/2)$, the two PBS at $(\pm\,d/2\sqrt2,r/2+d/2\sqrt2)$ and the four detectors
at $(\pm\,d/\sqrt2,r/2+d/\sqrt2)$ and $(\pm\,r/\sqrt2,/2)$ (see Fig. \ref{fig:10}). (\ref{32h}) will not depend on the value of $d$, which is
the distance traveled by the photon to any detector after leaving the BS. Notice that, with this setup, $h_-=0$ and
$h_+=\sqrt2$.
In Fig. \ref{fig:11} we represent (\ref{i}) under that conditions as a function of $x=r/c\,t$ for three different values of $z$ (different
values of $r$). Notice that a high degree of entanglement, maximal for $x$ large enough (short enough interaction times $t$),
can be achieved in all cases when one atom is beyond the light cone of the other ($x>1$).
As $t\rightarrow\infty$ $(x\rightarrow0)$, the concurrence eventually vanish, in agreement with the fact that the only atomic
state allowed by energy conservation is just the separable state $|\,g,g\,\rangle$.

In Fig. \ref{fig:12} we represent (\ref{32i}) as a function of $z$ for three different values of $z/x=\Omega\,t$, to give an alternative
description. The mutual light cone corresponds to the region $z<\Omega\,t$ in each case. The concurrence oscillates with the
position of the atoms, and eventually vanish at $z=\sqrt2\,(n+1/2)\,\pi$ ($n=0,1,2...$), as a consequence of the vanishing of
$\cos{(z/\sqrt2)}$. For a given interaction time $t$, the maximum of the concurrence can be achieved for interatomic distances
as large as desired. In particular, a maximally entangled state is generated for $\Omega\, t=1$, which corresponds to
$t\simeq10^{-15}\, s$.
In Fig. \ref{fig:13} we sketch (\ref{32i}) as a function of $\phi=\phi_k=\phi_{k'}$ for given values of $x$ and $z$. Notice that the maximum
values for the entanglement are around $\phi=n\,\pi/2$ ($n=0,1,2...$), $\pi/2$ corresponding to the setup of Fig. \ref{fig:10}.

So far, we have focused on $|\Psi_+\rangle$, but, in principle, $|\Phi_+\rangle$ could be detected as well. The coefficients $F_{\Psi}$
and $G_{\Psi}$ would have opposite sign to those of $|\Psi_+\rangle$ and therefore the concurrence would be the same. But, due to the
interaction times considered here, the relaxation time of a single detector must be extremely short in order to emit a double
click.

\subsection{Conclusions}

In this section, we have shown that, in principle, two neutral two-level atoms can evolve from an initially uncorrelated state to
a highly entangled state in a time shorter than the time required for the light to travel between them. At the initial time,
both atoms are excited in a common electromagnetic vacuum. They are allowed to interact with the field due to an induced dipole
during a time $t$  and, up to second order in perturbation theory, $n=0,1,2$ photons may be emitted. After that, the emitted
radiation pass through a partial Bell-state analyzer. For interaction times $t\simeq10^{-15} s$ and if a two-photon Bell state
$|\Psi^+\rangle$ or $|\Phi^+\rangle$ (the other two are forbidden in this model) is detected after that, the atoms are projected
into an entangled state, which may be maximally entangled for short enough $t$. For a given $t$, the degree of entanglement
oscillates periodically with the distance and the maximum degree available can be achieved for interatomic distances $r$ as
large as desired. Notice that the interaction time $t$, which must be $t\simeq10^{-15}\, s$, is absolutely independent of the
time $t'$ at which the photodetection takes place. Since the distance traveled by the photons from the atoms to the detector is
$r/\sqrt2 + d$, $d$ being arbitrary, the photodetection can occur after a time $t'\lesseqgtr r/c$. A suitable choice of $d$ is
necessary in order to ensure that the atoms may remain spacelike separated. The degree of entanglement is independent of $d$.

\section{Entanglement Sudden Death and Sudden Birth. Photon exchange and correlations transfer in atom-atom entanglement dynamics}
\subsection{Introduction}

Entanglement between qubits may disappear in a finite time when the qubits interact with a reservoir. This is commonly known as
``entanglement sudden death'' (ESD). After its discovery \cite{zyczk,diosi,yueberly}, the phenomenon has attracted great
attention (for instance, \cite{yueberlyII,Yonac,jamroz,ficektanasI,compagnoII,paz,lastrasolano,ficek,cole}) and has been
observed experimentally \cite{mafalda}.

ESD shows up in a variety of systems that can be roughly divided in two sets: those in which the qubits interact individually
with different reservoirs and those in which they interact with a common environment. In particular, in
\cite{jamroz,ficektanasI,ficektanasII} a system of a pair of two-level atoms interacting with a common electromagnetic vacuum is
considered. The dynamics of the system is given in all the cases by the Lehmberg-Agarwal master equation \cite{lehmberg,
agarwal} which is derived with the rotating wave approximation (RWA) and the Born-Markov approximation. Recently, non-Markovian
\cite{compagnoII} and non-RWA \cite{treschinos} effects have been considered in systems of qubits coupled individually to
different reservoirs. There are good reasons for going beyond the Markovian and RWA scenario in the case of a pair of two-level
atoms in the electromagnetic vacuum. For short enough times non-RWA contributions are relevant (see Chapter 3) and a proper
analysis of causality issues can only be performed if they are taken into account \cite{powerthiru,milonni,compagnoI}. Besides,
as we shall show in this section the death of the entanglement between the atoms is related with the birth of entanglement between
the atoms and the field, and therefore the field is actually a non- Markovian reservoir. This was also the case in
\cite{Yonac,lastrasolano,ficek} with different reservoirs.

In the previous sections, we have applied the formalism of perturbative quantum electrodynamics (QED) to the system
of a pair of neutral two-level atoms interacting locally with the electromagnetic field, and for initially separable states
analyzed the generation of entanglement. This is a non-Markovian, non-RWA approach. The use of the Lehmberg-Agarwal master
equation can be seen as a coarse-grained in time approximation to the perturbative treatment \cite{cohentannoudji}. The first
goal of this section is to apply also the QED formalism to analyze the ESD in these systems for initially entangled atomic states,
comparing the results with the previously obtained \cite{jamroz,ficektanasI} with master equations. We will focus mainly on the
range $r/(c\,t)\approx1$, $r$ being the interatomic distance and $t$ the interaction time, in order to investigate the role of
locality. We will also consider for the first time in these systems the rest of pairwise concurrences, namely the entanglement
of each atom with the field, and multipartite entanglement, following the spirit of \cite{Yonac,lastrasolano,ficek,cole}. While
the mentioned papers deal with a four qubit model, our model here consists in two qubits (the atoms) and a qutrit (the
electromagnetic field, which may have 0, 1 or 2 photons). We shall show that the phenomenon of revival of entanglement after the
ESD \cite{ficektanasI} can occur for $r>c\,t$, and therefore is not related with photon exchange as is usually believed. We will
see that atom-atom disentaglement is connected with the growth of atom-field entanglement and viceversa. Similar relationship
will be obtained among the ``atom-(atom+field)'' and ``field-(atom+atom)'' entanglements.

The reminder of this section is organized as follows. In  subsection 3.3.2 we will describe the Hamiltonian and the time evolution from
the initial state of the system. In section 3.3.3 we will obtain the reduced state of the atoms and analyze the behavior of its
entanglement. In section 3.3.4 the same will be performed with the reduced state of each atom and the field, comparing the
entanglement cycle with the one obtained in the previous section. Tripartite entanglement will be considered in section 3.3.5 in
terms of the entanglement of all the different bi-partitions of the system, and we conclude in section 3.3.6 with a summary of our
results.

\subsection{Hamiltonian and state evolution}

In what follows we choose a system  given initially by an atomic entangled state, with the field in the vacuum state
$|\,0\rangle$:
\begin{equation}
|\,\psi\,(0)\rangle\,=(\,\alpha\, |\,e\,e\,\rangle +\,\beta\,|\,g\,g\,\rangle) |\,0\,\rangle.\label{33b}
\end{equation}
The system then evolves under the effect of the interaction (\ref{a}) during a lapse of time $t$ into a state:
\begin{equation}
|\,\psi\,(t)\rangle = T (e^{-i\, \int_0^t\,dt'\, H_I\,(t')/\hbar})\,|\,\psi\,\rangle_0, \label{33c}
\end{equation}
Up to second order in perturbation theory, (\ref{33c}) can be given in the interaction
picture as
\begin{eqnarray}
|\mbox{atom} 1,\mbox{atom} 2,\mbox{field}(t)\rangle = \,\alpha\,  |\,e\,e\,0\,(t)\rangle
 +\,\beta\,|\,g\,g\,0\,(t)\rangle \label{33d}
\end{eqnarray}
where
\begin{eqnarray}
|\,e\,e\,0\,(t)\rangle=\,&&((1+A')\,|\,e\,e\rangle + X'\,|\,g\,g\rangle)\,|\,0\rangle
 +(U_A\,|\,g\,e\,\rangle+ U_B\,|\,e\,g\,\rangle)\,|\,1\,\rangle+\nonumber\\
&&(F'\,|\,e\,e\rangle+ G'\,|\,g\,g\rangle)\,|\,2\rangle \label{33e}
\end{eqnarray}
and
\begin{eqnarray}
|\,g\,g\,0\,(t)\rangle=\,&&((1+A'')\,|\,g\,g\rangle + X''\,|\,e\,e\rangle)\,|\,0\rangle
 +(V_A\,|\,e\,g\,\rangle+ V_B\,|\,g\,e\,\rangle)\,|\,1\,\rangle+\nonumber\\
&&(F''\,|\,g\,g\rangle+ G''\,|\,e\,e\rangle)\,|\,2\rangle \label{33f}
\end{eqnarray}
where
\begin{eqnarray}
A'&=&\frac{1}{2}\,\theta(t_1-t_2)\langle\,0|\mathcal{S}_A^+\,(t_1)
\mathcal{S}_A^-\,(t_2)+\mathcal{S}_B^+\,(t_1)\mathcal{S}_B^-\,(t_2)|0\rangle\nonumber\\
A''&=&\frac{1}{2}\,\theta(t_1-t_2)\langle\,0|\mathcal{S}_A^-\,(t_1)
\mathcal{S}_A^+\,(t_2)+\mathcal{S}_B^-\,(t_1)\mathcal{S}_B^+\,(t_2))|0\rangle\nonumber\\
X'&=&\langle\,0|T(\mathcal{S}^-_B\, \mathcal{S}^-_A)\,|0\rangle, \,X''=\langle\,0|T(\mathcal{S}^+_B\,
\mathcal{S}^+_A)\,|0\rangle,\nonumber\\
U_A\,&=&\,\langle\,1\,|\, \mathcal{S}^-_A\,|\,0\,\rangle,\, V_A\,=\,\langle\,1\,|\,
\mathcal{S}^+_A\,|\,0\,\rangle \label{33g}\\
U_B\,&=&\,\langle\,1\,|\, \mathcal{S}^-_B\,|\,0\,\rangle,\, V_B\,=\,\langle\,1\,|\, \mathcal{S}^+_B\,|\,0\,\rangle\nonumber\\
F'&=&\frac{1}{2}\,\theta(t_1-t_2)\langle\,2|\mathcal{S}_A^+\,(t_1)
\mathcal{S}_A^-\,(t_2)+\mathcal{S}_B^+\,(t_1)\mathcal{S}_B^-\,(t_2)|0\rangle\nonumber\\
F''&=&\frac{1}{2}\,\theta(t_1-t_2)\langle\,2|\mathcal{S}_A^-\,(t_1)
\mathcal{S}_A^+\,(t_2)+\mathcal{S}_B^-\,(t_1)\mathcal{S}_B^+\,(t_2))|0\rangle,\nonumber\\
G'&=&\langle\,2|T(\mathcal{S}^-_B\, \mathcal{S}^-_A\,)|0\rangle,\,G''=\langle\,2|T(\mathcal{S}^+_B\,
\mathcal{S}^+_A\,)|0\rangle\nonumber
\end{eqnarray}
with the definitions and conventions of Section 3.1. Here, $A'$ and $A''$ describe intra-atomic radiative
corrections, $U_A\, (U_B)$ and $V_A\, (V_B)$ single photon emission by  atom $A$ ($B$), and $G'$ and $G''$ by both atoms, while
$F'$ and $F''$ correspond to two photon emission by a single atom.  Only $X'$ and $X''$ correspond to interaction between both
atoms.  $A'$, $X'$, $X''$, $V_A$, $V_B$, $F'$, $F''$ and $G'$ are  non-RWA terms. As in the previous sections of this chapter,  in our calculations we will take $(\Omega
|\mathbf{d}|/e c) = 5\,\cdot 10^{-3}$, which is of the same order as the 1s $\rightarrow$ 2p transition in the hydrogen atom,
consider $\Omega\,t \gtrsim 1$, and focus mainly on the cases $(r/c\,t)\simeq 1$. In this case, we will take into account  $|\,e\,\rangle$ is actually a
triply degenerate state $|\,e\,,m\rangle$ with $m=0,\pm1$ and we will average over two different independent possibilities for
dipole orientations: $\mathbf{d}_A=\mathbf{d}_B=\mathbf{d}=d\,\mathbf{u}_z$ for transitions with $\Delta m=0$ \cite{milonniII}
and $\mathbf{d}=d\,(\mathbf{u}_x \pm i \mathbf{u}_y)/\sqrt{2}$ \cite{milonniII} for transitions with $\Delta m=\pm1$.

\subsection{Sudden death and revival of atom-atom entanglement}
\begin{figure}[h!]
\includegraphics[width=\textwidth]{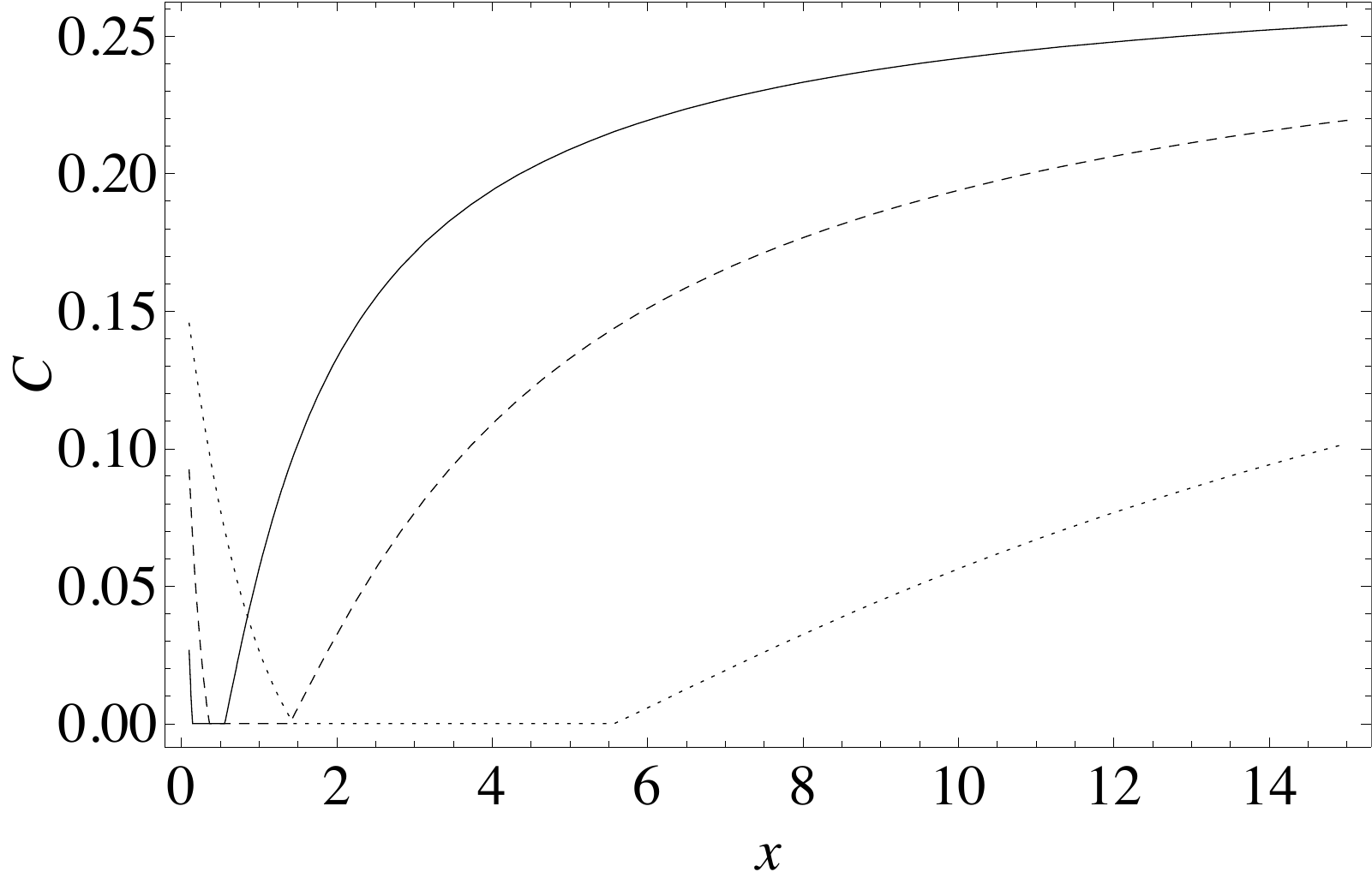}
\caption{Concurrence $\mathbb{C}(\rho_{AB})$ in front of $x=r/c\,t$ for $p=0.98$ and $z=\Omega r/c= 2\cdot10^6$ (solid line),
$5\cdot10^6$ (dashed line) and $2\cdot10^7$ (dotted line). In the latter case sudden death and revival of entanglement occur for
$x>1$.}
\label{fig:14}
\end{figure}
\begin{figure}[h!]
\includegraphics[width=\textwidth]{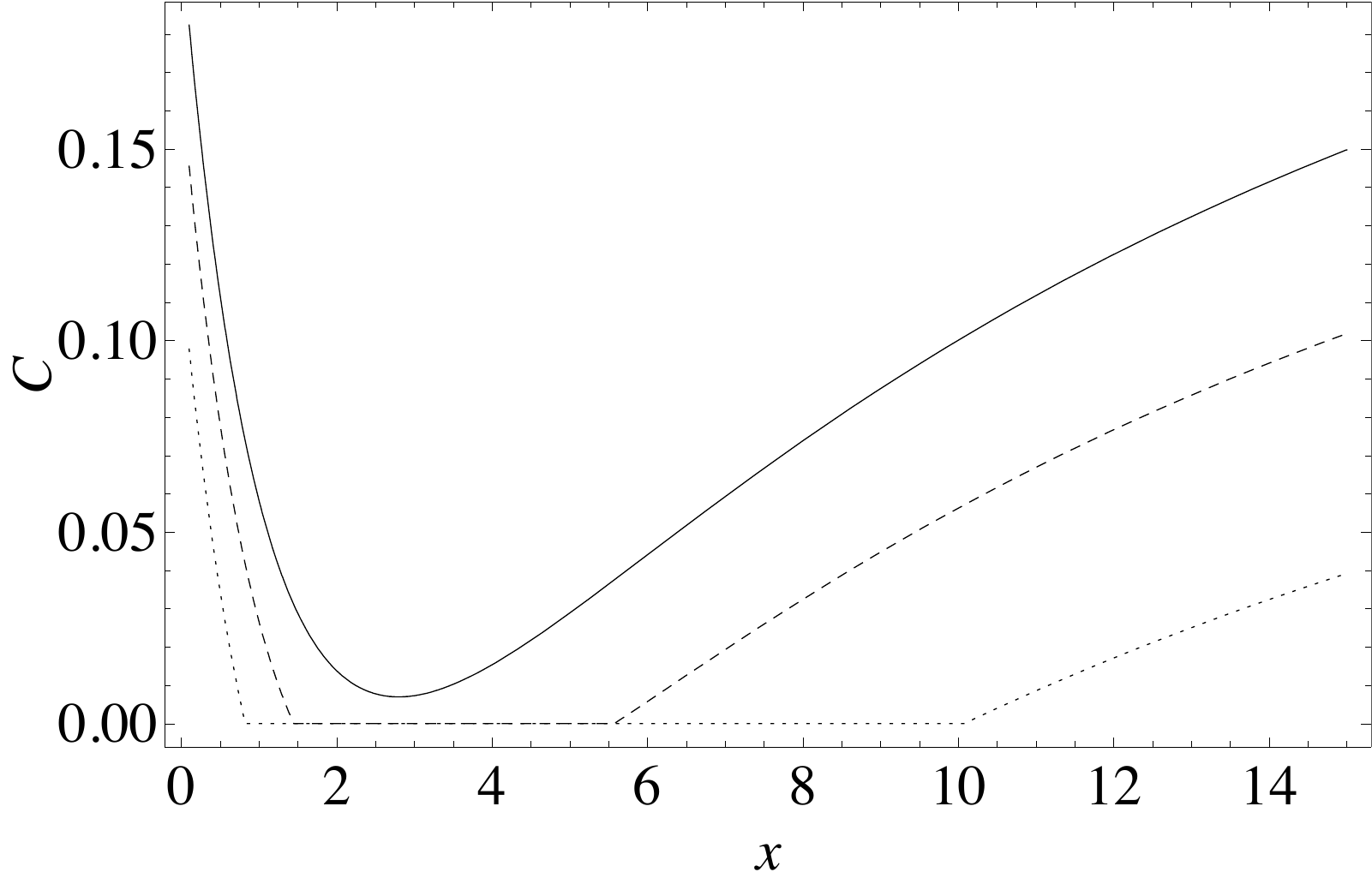}
\caption{Concurrence  $\mathbb{C}(\rho_{AB})$ in front of $x=r/c\,t$ for $z=\Omega r/c= 2\cdot10^7$  and $p=0.97$  (solid line),
$p=0.98$ (dashed line) and $p=0.99$ (dotted line). In the first case, entanglement decreases as $t$ grows up to a minimum value
and begin to grow since then. This behavior becomes entanglement sudden death and revival when the minimum value is 0 for higher
values of $p$. $\mathbb{C}(\rho_{AB})$ tends to 0 as $x\rightarrow\infty$ and $p\rightarrow1$.}
\label{fig:15}
\end{figure}

After tracing over all the states of the field, the density matrix of the atomic state $\rho_{AB}$ takes the form (in the basis
$\{|e\,e\rangle,|e\,g\rangle,|g\,e\rangle,|g\,g\rangle\}$):
\begin{eqnarray}
\rho_{AB}=\frac{1}{N}\left( \begin{array}{c c c c}\rho_{11}&0&0&\rho_{14} \\0&\rho_{22}&\rho_{23}&0\\0&\rho_{23}^*&\rho_{33}&0\\
\rho_{14}^*&0&0&\rho_{44}\end{array}\right)\label{33i}
\end{eqnarray}
where
\begin{eqnarray}
\rho_{11}&=&|\,\alpha\,(1\,+\,A')\,+\,\beta\,X''\,|^2\,+\,|\,\alpha\,F'\,+\,\beta\,G''\,|^2,\nonumber\\
\rho_{22}&=&\rho_{33}=|\,\alpha\,|^2\,|\,U\,|^2\,+\,|\,\beta\,|^2\,|\,V\,|^2+2\,Re\,(\alpha\,\beta^*\,l^*)\nonumber\\
\rho_{44} &=&|\,\alpha\,b\,+\,\beta\,(1\,+\,A'')\,|^2\,+\,|\,\alpha\,G'\,+\,\beta\,F''\,|^2,\label{33j}\\
\rho_{14} &=& |\,\alpha\,|^2\,((1\,+\,A')\,X'*\,+\,F'\,G'^*)+|\,\beta\,|^2\,((1\,+\,A'')^*\,X''\nonumber\\
&+&\,G''\,F''^*)+\alpha\,\beta^*((1+A')\,(1+A'')\,+\,F'\,F''^*)\nonumber\\
&+& \beta\,\alpha^*(X''\,X^{'*}+\,G''\,G^{'*})\nonumber\\
\rho_{23} &=& |\,\alpha\,|^2\,U_B\,U_A^*+|\,\beta\,|^2\,V_A\,V_B^*\,+\,2\,Re\,(\alpha\,\beta^*\,U\,V^*)\nonumber\\
N&=&\,\rho_{11}\,+\,\rho_{22}\,+\,\rho_{33}+\,\rho_{44}\nonumber
\end{eqnarray}
where
\begin{eqnarray}
 |\,U\,|^2\,=\,|\,U_A\,|^2\,=\,|\,U_B\,|^2,\nonumber\\
  |\,V\,|^2\,=\,|\,V_A\,|^2\,=\,|\,v_B\,|^2,\nonumber\\
l\,=\,U_A\,V_B^*\,=\,U_B\,V_A^* \label{eq:33a}
\end{eqnarray}
and 
\begin{equation}
U\,V^*\,=\,U_A^*\,V_A^*=\,U_B\,V_B^*. \label{eq:33b}
\end{equation}
The computation of $A'$, $X'$, etc. can be performed following the lines given in the Appendix A, where they
are computed for the initial state $|\,e\,g\,\rangle$ and only for $\Delta m=0$. In terms of $z\,=\,\Omega\,r/c$ and
$x\,=\,r/c\,t$, being $r$ the interatomic distance, we find:
\begin{eqnarray}
A'&=&\frac{4\,i\,K\,z^3}{3\,x}\,(\ln{(1-\frac{z_{max}}{z})}\,+\,i\,\pi),\nonumber\\
A''&=&\frac{-4\,i\,K\,z^3}{3\,x}\,\ln{(1+\frac{z_{max}}{z})}\nonumber\\
X'&=&=X^{''*}=\frac{\alpha\,d_i\,d_j}{\pi\,e^2}(-\mathbf{\nabla}^2\delta_{ij}+\nabla_i\nabla_j)\,I, \label{33k}
\end{eqnarray}
with $K=\alpha\,|\,\mathbf{d}\,|^2/(e^2\,r^2)$ and $I=I_+\,+\,I_-$, where:
\begin{eqnarray}
I_{\pm}&=&\frac{-i\,e^{-i\frac{z}{x}}}{2\,z}\,[\,\pm\,2\cos(\,\frac{z}{x}\,)\,e^{\pm\,i\,z}\,Ei(\mp\,i\,z)
+\,e^{-i\,z\,(1\pm\frac{1}{x})}\nonumber\\&\ &
Ei(i\,z\,(1\pm\frac{1}{x}))\,-\,e^{i\,z\,(1\pm\frac{1}{x})}\,Ei(-i\,z\,(1\pm\frac{1}{x}))\,]\label{33l}
\end{eqnarray}
for $x>1$, having the additional term $-2\,\pi\,i\,e^{i\,z\,(1-1/x)}$ otherwise.

$|\,U\,|^2$, $|\,V\,|^2$, $l$, $U_B\,U_A^*$, $V_A\,V_B^*$ and $U\,V^*$ have been computed in Section 3.1. Besides:
\begin{eqnarray}
G'&=&U_B\,U'_A\,+\,U_A\,U'_B\, , G''=V_A\,V'_B\,+\,V_B\,V'_A\nonumber\\
F'&=&\theta(t_1-t_2)(\,V_A\,(t_1)\,U'_A\,(t_2)\,+\,U_A\,(t_1)\,V'_A\,(t_2)\nonumber\\
&+&\,V_B\,(t_1)\,U'_B\,(t_2)\,+\,U_B\,(t_1)\,V'_B\,(t_2)\,)\label{33m}\\
F''&=&\theta(t_1-t_2)(\,U_A\,(t_1)\,V'_A\,(t_2)\,+\,U'_A\,(t_1)\,V_A\,(t_2)\nonumber\\
&+&\,U_B\,(t_1)\,V'_B\,(t_2)\,+\,U'_B\,(t_1)\,V_B\,(t_2)\,)\nonumber
\end{eqnarray}
where the primes in the $U$'s and $V$'s are introduced to discriminate between the two single photons.
\begin{figure}[h!]
\includegraphics[width=\textwidth]{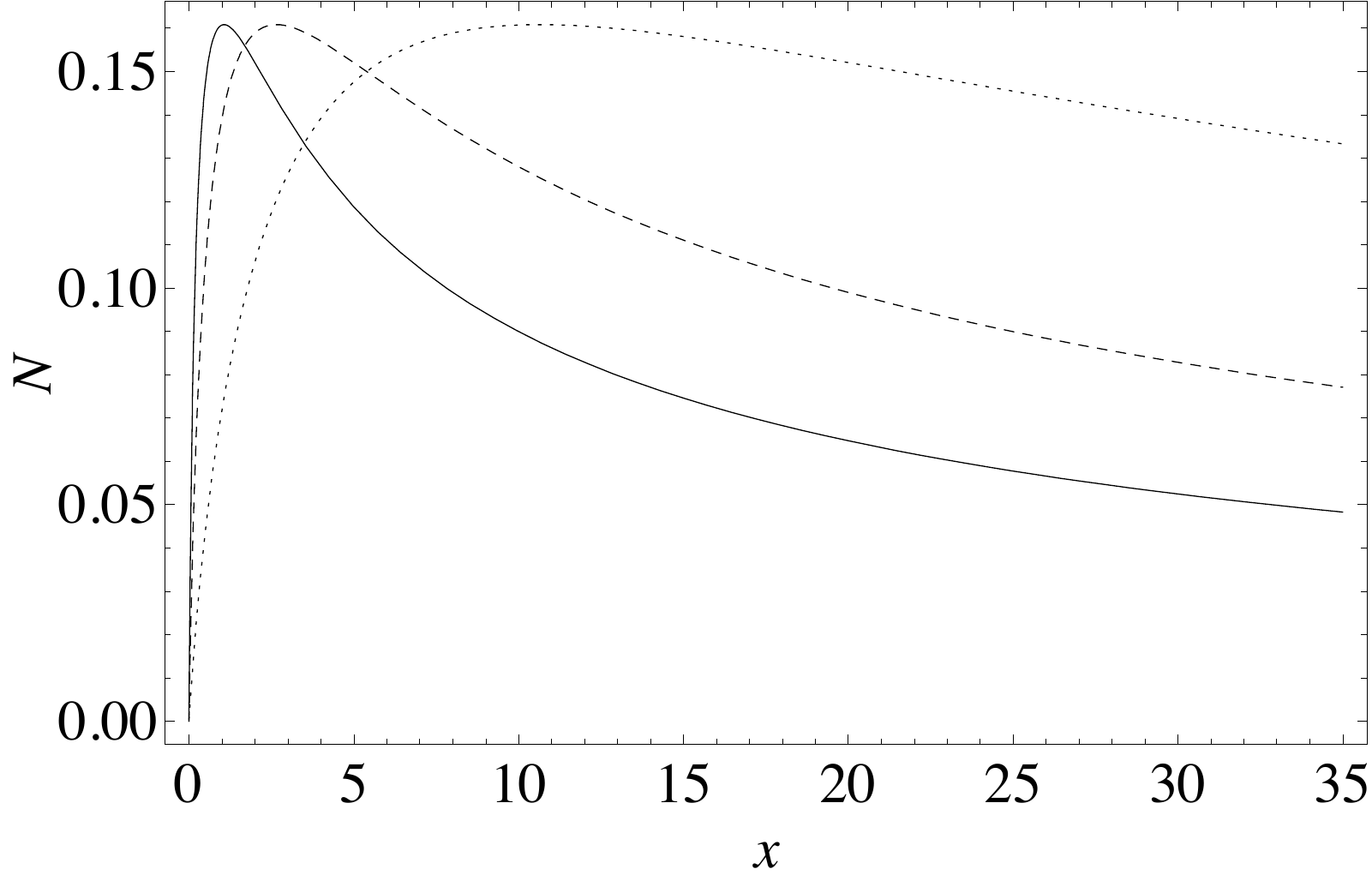}
\caption{Negativity  $\mathbb{N}(\rho_{BF})=\mathbb{N}(\rho_{AF})$ in front of $x=r/c\,t$ for $p=0.98$ and $z=\Omega r/c=
2\cdot10^6$ (solid line), $5\cdot10^6$ (dashed line) and $2\cdot10^7$ (dotted line). Entanglement increases from 0 at
$x\rightarrow\infty$ up to a maximum value and then decreases and vanishes eventually.}
\label{fig:16}
\end{figure}
\begin{figure}[h!]
\includegraphics[width=\textwidth]{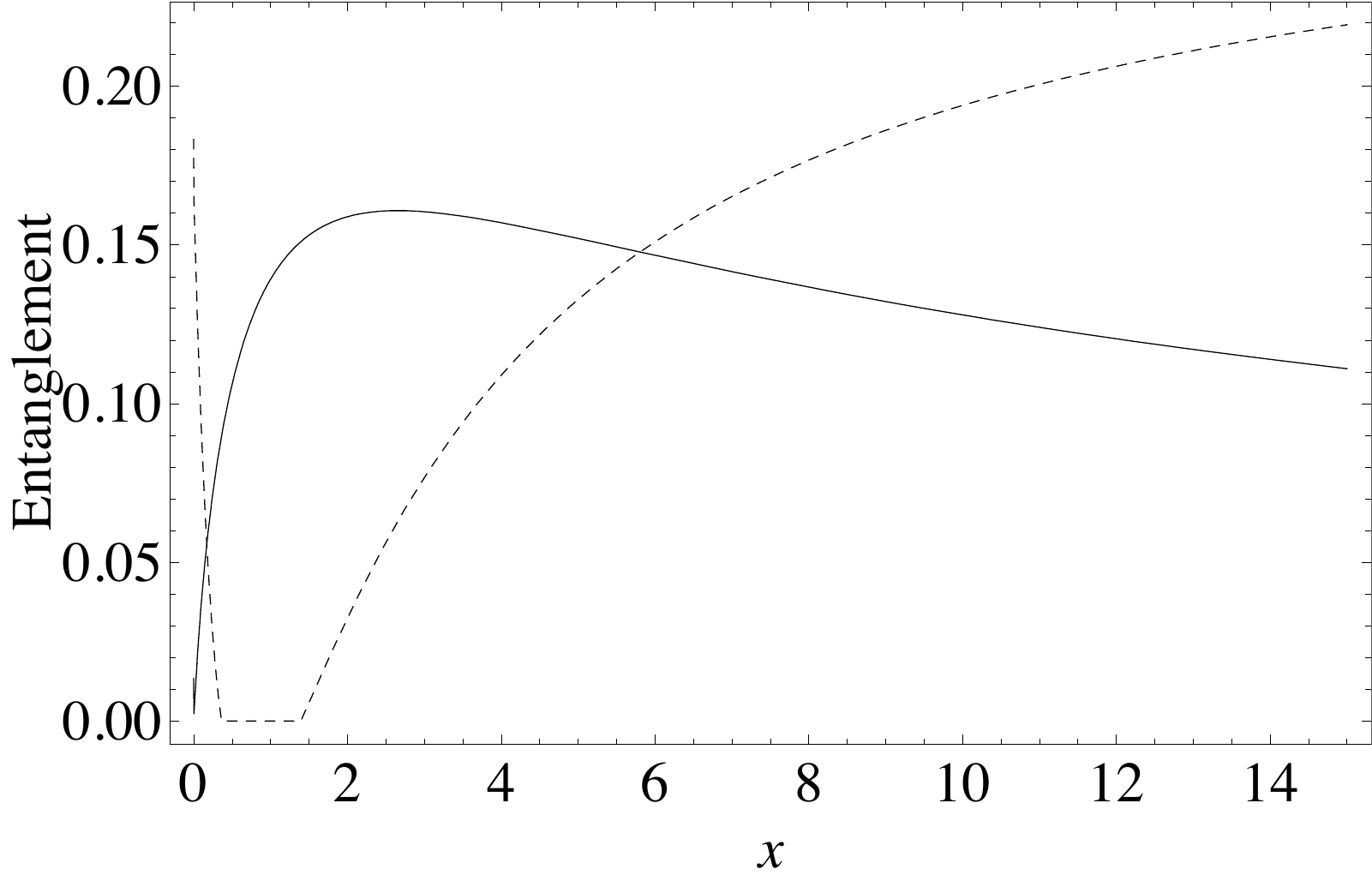}
\caption{Negativity $\mathbb{N}(\rho_{BF})=\mathbb{N}(\rho_{AF})$ (solid line) and concurrence $\mathbb{C}(\rho_{AB})$ (dashed
line) in front of $x=r/c\,t$ for $p=0.98$ and $z=\Omega r/c= 5\cdot10^6$. Entanglement atom-atom cycle is clearly correlated
with the atom-field cycle, although the sum is not a conserved quantity. Although atom-field entanglement may change while the
other remains zero, both entanglements cannot increase or decrease at the same time.}
\label{fig:17}
\end{figure}
\begin{figure}[h!]
\includegraphics[width=\textwidth]{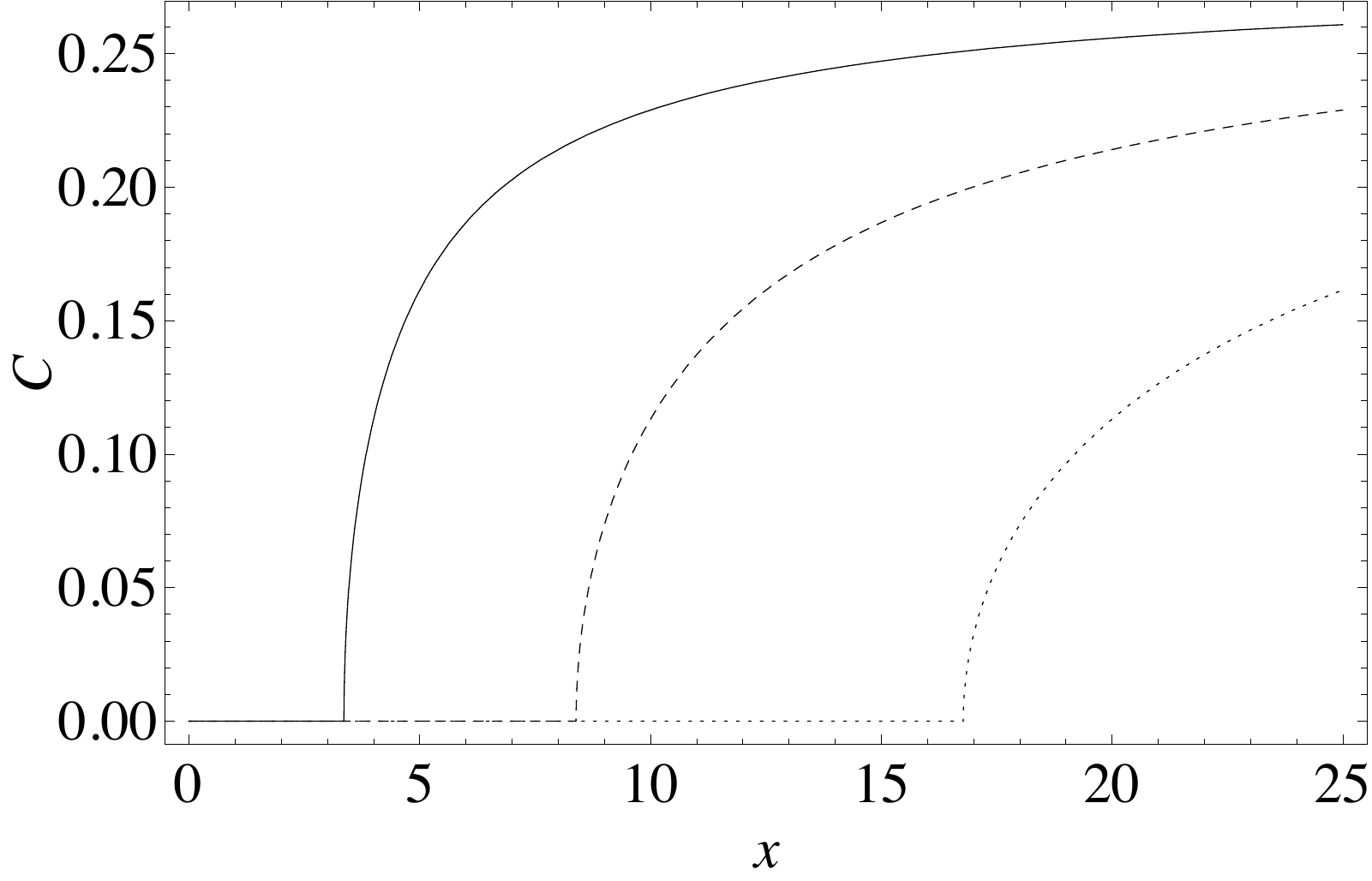}
\caption{I concurrence $\mathbb{C}_{A-BF}=\mathbb{C}_{B-AF}$  in front of $x=r/c\,t$ for $p=0.98$ and $z=\Omega r/c=
2\cdot10^5$(solid line), $5\cdot10^5$ (dashed line) and $1\cdot10^6$ (dotted line). Entanglement disappears faster than the
entanglement between the atoms (Fig. \ref{fig:14}) and remains 0 since then.}
\label{fig:18}
\end{figure}
\begin{figure}[h!]
\includegraphics[width=\textwidth]{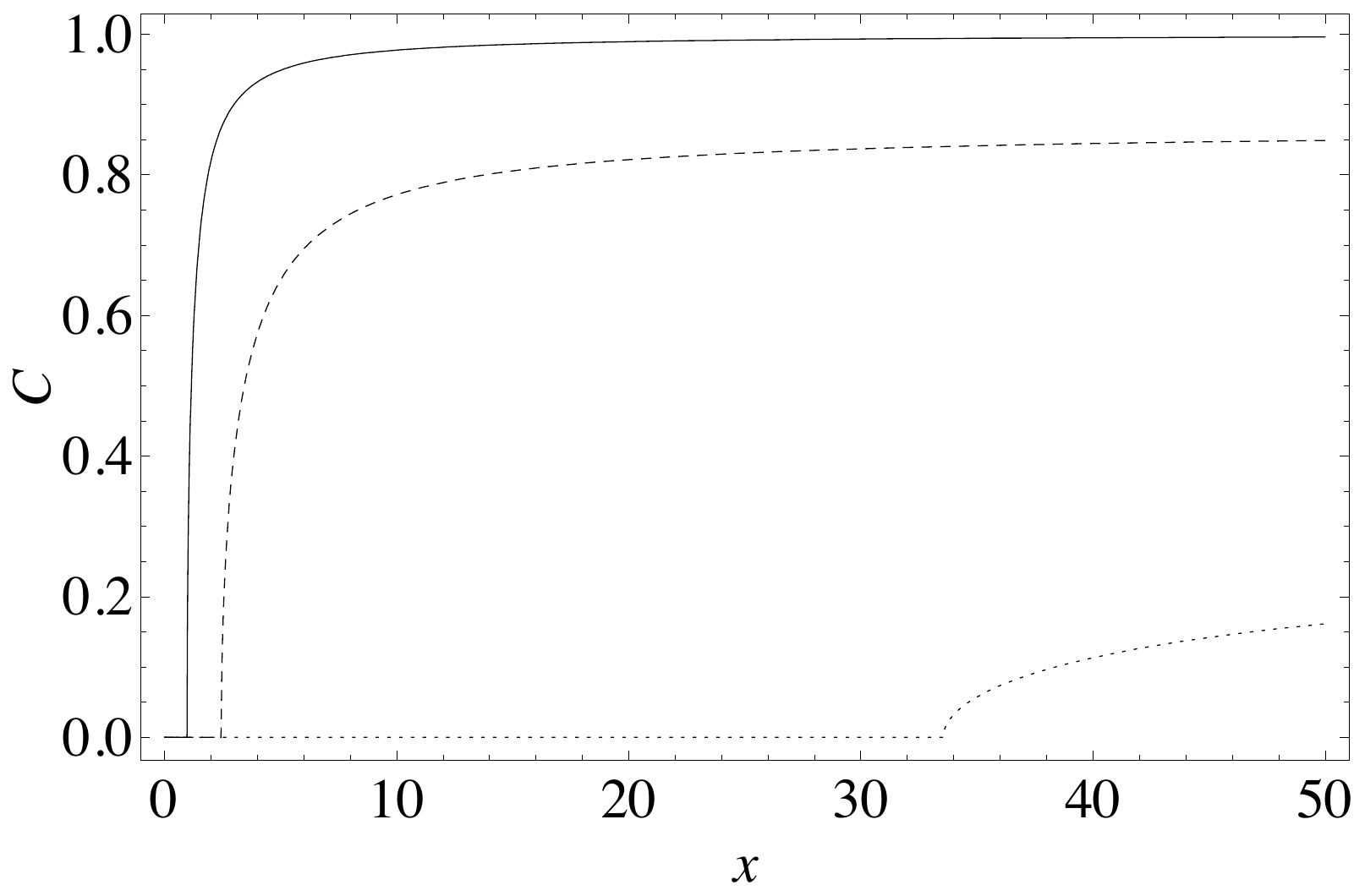}
\caption{I concurrence $\mathbb{C}_{A-BF}=\mathbb{C}_{B-AF}$  in front of $x=r/c\,t$ for $z=\Omega r/c= 2\cdot10^6$ and $p=0.50$
(solid line), $p=0.75$ (dashed line) and $p=0.98$ (dotted line). Entanglement sudden death occurs for a wider range than the
entanglement between the atoms (Fig. \ref{fig:15}).}
\label{fig:19}
\end{figure}
\begin{figure}[h!]
\includegraphics[width=\textwidth]{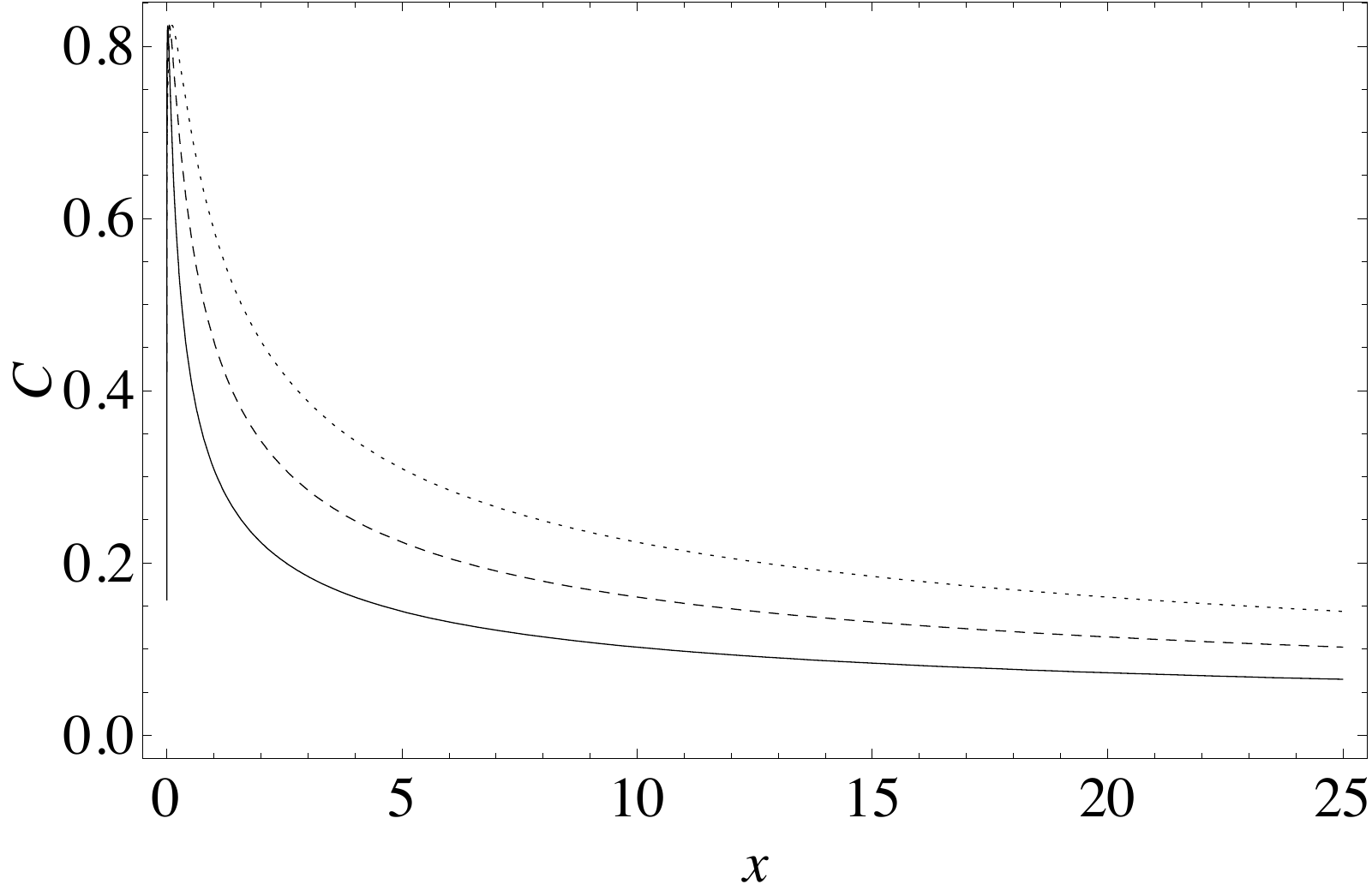}
\caption{I concurrence $\mathbb{C}_{F-AB}$  in front of $x=r/c\,t$ for $p=0.98$ and $z=\Omega r/c= 2\cdot10^5$ (solid line),
$5\cdot10^5$ (dashed line) and $1\cdot10^6$ (dotted line). Entanglement grows from 0 to its maximum value at $x\approx0.1$ and
then decreases.}
\label{fig:20}
\end{figure}
\begin{figure}[h!]
\includegraphics[width=\textwidth]{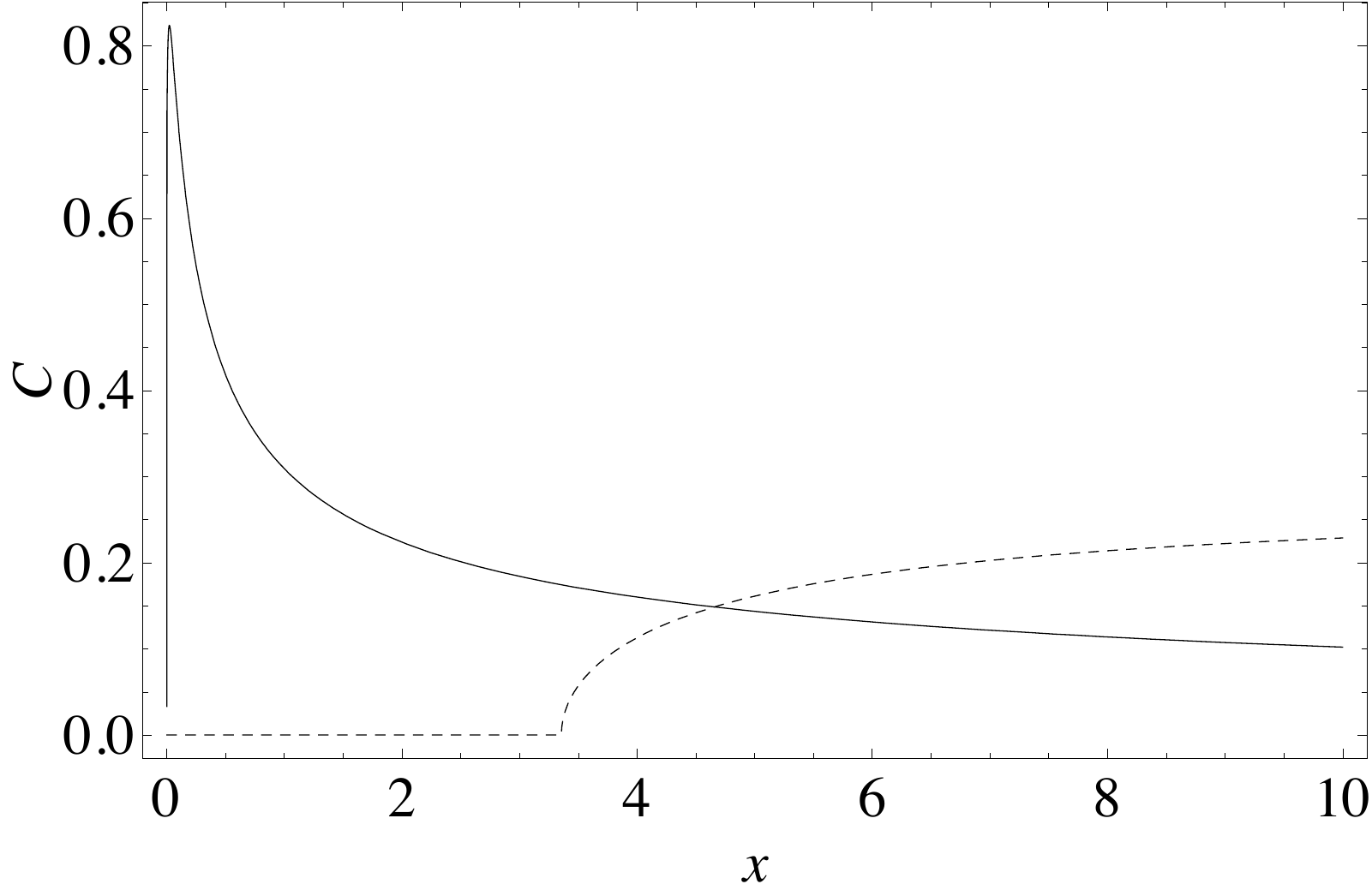}
\caption{I concurrence $\mathbb{C}_{F-AB}$ (solid line) and $\mathbb{C}_{A-BF}= \mathbb{C}_{B-AF}$ (dashed line) in front of
$x=r/c\,t$ for $z=\Omega r/c= 2\cdot10^5$ and $p=0.98$. Both magnitudes cannot increase or decrease at the same time.}
\label{fig:21}
\end{figure}

We will use the concurrence $\mathbb{C}(\rho)$ \cite{wootters} to compute the entanglement, which for a X-state like (\ref{33i}) is given by (\ref{eq:31s}).
If we take 
\begin{equation}
\alpha=\sqrt{p}\,, \beta=\sqrt{1-p},  \label{eq:33c}
\end{equation}
we find that ESD appears at a range of values of $p$ that decreases with
increasing $r$, in agreement with \cite{ficektanasI}. Although this would suggest that ESD disappear for $r$ large enough, we
find that there are high values of $p$ for which ESD exists for arbitrary large $r$. In Fig. \ref{fig:14}, we represent
$\mathbb{C}(\rho_{AB})$ in front of $x$ for different values of $z$ and $p=0.98$. ESD occurs at $z/x=\Omega\,t$ of the order of
$10^7$. Thus, as $z$ (that is $r$) grows, ESD is shifted to higher values of $x$. It is also interesting to analyze the
phenomenon of entanglement revival, discovered in these systems in \cite{ficektanasI}. We find that the dark periods
\cite{ficektanasI}  between death and revival has larger time durations for increasing $z$. Besides, although in
\cite{ficektanasI} the revival is described as a consequence of the photon exchange, for $r$ sufficiently large both the ESD and
the revival can occur for $x>1$, where photon exchange is not allowed. We think that the explanation for entanglement revival is
closer to the spirit of \cite{Yonac,ficek} where entanglement revival between noninteracting atoms is interpreted as coming from
entanglement transfer between different parts of the system. We shall discuss this point in the following sections.
In Fig. \ref{fig:15} we sketch the dependence with $p$. Although sudden death and revivals appear in a very restricted range of the
parameter, they are only a particular case of the generic behavior of entanglement observed in a wider range, which can be
described as disentanglement up to a minimum value and growth of quantum correlations since then.
\subsection{Atom-field entanglement}

Tracing (\ref{33d}) over states of atom A (B) the reduced atom-field density matrix $\rho_{BF}$ ($\rho_{AF}$) is obtained. Taking
the basis $\{|\,e\,0\,\rangle,|\,e\,1\,\rangle,|\,e\,2\,\rangle,|\,g\,0\,\rangle,|\,g\,1\,\rangle,|\,g\,2\,\rangle\}$, we have:
\begin{eqnarray}
\rho_{BF}=\rho_{AF}=\frac{1}{N'}\left( \begin{array}{c c c c c c}\rho'_{11}&0&\rho'_{13}&0&\rho'_{15}&0
\\0&\rho'_{22}&0&\rho'_{24}&0&\rho'_{26}
\\ \rho'^{*}_{13}&0&\rho'_{33}&0&\rho'_{35}&0\\
0&\rho'^{*}_{24}&0&\rho'_{44}&0&\rho'_{46}\\
\rho'^{*}_{15}&0&\rho'^{*}_{35}&0&\rho'_{55}&0\\0&\rho'^{*}_{26}&0&\rho'^{*}_{46}&0&\rho'_{66}
\end{array}\right)\label{33p}
\end{eqnarray}
with
\begin{eqnarray}
\rho'_{11}&=&|\,\alpha\,(1\,+\,A')\,+\,\beta\,X''\,|^2,\,\rho'_{22}=\rho'_{55}=\rho_{22}\nonumber\\
\rho'_{33}&=&|\,\alpha\,F'\,+\,\beta\,G''\,|^2,\,\rho'_{44}=|\,\alpha\,X'\,+\,\beta\,(1\,+\,A'')\,|^2\nonumber\\
\rho'_{66}&=&|\,\alpha\,G'\,+\,\beta\,F''\,|^2,\,\rho'_{13}=(\alpha\,(1\,+\,A')+\beta\,X'')\,(\alpha\,F'\,+\,\beta\,G'')\nonumber\\
\rho'_{15}&=&(\alpha\,(1\,+\,A')+\beta\,X'')\,(\alpha\,U_{B}\,+\,\beta\,V_A)^* \label{33q}\\
\rho'_{24}&=&(\alpha\,U_{A}\,+\,\beta\,V_B)\,(\beta\,(1+A'')\,+\,\alpha\,X')^*\nonumber\\
\rho'_{26}&=&(\alpha\,U_{A}\,+\,\beta\,V_B)\,(\alpha\,G'\,+\,\beta\,F'')^*\nonumber\\
\rho'_{35}&=& (\alpha\,F'\,+\,\beta\,G'')\,(\alpha\,U_{B}\,+\,\beta\,V_{A})^*\nonumber\\
\rho'_{46}&=& (\alpha\,X'\,+\,\beta\,(1\,+A''))\,(\alpha\,G'\,+\,\beta\,F'')^*\nonumber\\
N'&=&\,\rho'_{11}\,+\,\rho'_{22}\,+\,\rho'_{33}+\,\rho'_{44}+\,\rho'_{55}+\,\rho'_{66}\nonumber
\end{eqnarray}
There are no operational generalizations of concurrence for mixed states in $2\times3$ dimensions like the ones in Eq.
(\ref{33p}). We will use the negativity \cite{vidalwerner} $\mathbb{N}(\rho)$, which is the absolute value of the sum of the
negative eigenvalues of the partial transposes of a state $\rho$. For the $2\times2$ and $2\times3$ cases $\mathbb{N} (\rho)>0$
is a necessary and sufficient condition for $\rho$ to be entangled.

Up to second order in perturbation theory, we have that $N'=N$ and that the nonzero eigenvalues of the partial transposes of
both $\rho_{BF}$ and $\rho_{AF}$ are
\begin{equation}
\lambda_{\pm}=\frac{\rho'_{11}+\rho'_{55}\pm\sqrt{(\rho'_{11}-\rho'_{55})^2+4|\rho'_{24}|^2}}{2\, N'}\label{33r}
\end{equation}
and
\begin{equation}
\lambda'_{\pm}=\frac{\rho'_{44}+\rho'_{55}\pm\sqrt{(\rho'_{44}-\rho'_{22})^2+4|\rho'_{15}|^2}}{2\, N'}\label{33s}
\end{equation}
being zero the other two. In Eqs. (\ref{33r}) and (\ref{33s}) only the terms up to second order are retained.  Therefore, if
$|\rho'_{24}|^2>\rho'_{11}\,\rho'_{55}$ then $\lambda_{-}<0$ and if $|\rho'_{15}|^2>\rho'_{22}\,\rho'_{44}$ then
$\lambda'_{-}<0$.
In Fig. \ref{fig:16} we represent $\mathbb{N}(\rho_{BF})=\mathbb{N}(\rho_{AF})$ in front of $x$ for same values of $p$ and $z$ of Fig. 1.
We see that the negativity grows from 0 at $x\rightarrow\infty$ ($t=0$) to its maximum value and then starts to decrease and
eventually vanishes, following the opposite cycle to the entanglement of $\rho_{AB}$.

Although it would be interesting to look for conservation rules of entanglement like the ones in \cite{Yonac,ficek,cole}, this
search is beyond the focus of this paper since in our study we are using different entanglement measures in Hilbert spaces of
different dimensions. Besides, except for the concurrence between atoms $A$ and $B$, the rest of the concurrences in the
mentioned papers have not obvious counterparts in our case. But it is clear that in general the entanglement cycle between atoms
is correlated with the entanglement cycle between each atom and the field, as can be seen in Fig. \ref{fig:17} in a particular case.
Although atom-field entanglement may change while the other remains zero, both entanglements cannot increase or decrease at the
same time.

\subsection{Tripartite entanglement}

Tripartite entanglement has been widely studied in terms of the entanglement of the different bipartitions $A-BC$, $B-AC$,
$C-AB$ in the system \cite{shan,conguillermo,pascazio}, where $A$, $B$ and $C$ stand for the three parties. Here, we will
compute the I concurrences \cite{rungta} $\mathbb{C}_{A-BF}$, $\mathbb{C}_{B-AF}$, $\mathbb{C}_{F-AB}$, where
\begin{equation}
\mathbb{C}_{J-KL}=\sqrt{2(1-Tr\, \rho_{J}^2)}, \label{eq:Iconcurrencedef}
\end{equation}
where $J$ runs form $A$ to $F$ and $KL$ from $BF$ to $AB$ respectively, being
$\rho_{J}$ the reduced density matrix of $J$. $A$ and $B$ stand for the atoms, and $F$ for the field.

Tracing (\ref{d}) over $BF$ $(AF)$, we find the following density matrices $\rho_{A}$ $(\rho_{B})$:
\begin{eqnarray}
\rho_{A}=\rho_{B}=\frac{1}{N_{A}}\left( \begin{array}{c c }\rho_{A11}&0
\\0&\rho_{A22}\end{array}\right)\label{33t}
\end{eqnarray}
where 
\begin{equation}
\rho_{A11}=\rho'_{11}+\rho'_{33}+\rho_{22} \label{eq:roa11}
\end{equation}
and 
\begin{equation}
\rho_{A22}=\rho'_{44}+\rho'_{66}+\rho_{22}\label{eq:roa22}
\end{equation}
 and
$N_{A}=\rho_{A11}+\rho_{A22}$. In Fig. \ref{fig:18} we sketch the behavior of $\mathbb{C}_{A-BF}$ and $\mathbb{C}_{B-AF}$ in front of $x$
for different values of $z$. Entanglement vanishes before the death of the entanglement between $A$ and $B$, and does not have a
revival. Besides, ESD appears in a wider range of $p$, as can be seen in Fig. \ref{fig:19}.
Now, tracing (\ref{33d}) over $AB$ we obtain the reduced density matrix of the field $\rho_{F}$:
\begin{eqnarray}
\rho_{F}=\frac{1}{N_{F}}\left( \begin{array}{c c c}\rho_{F11}&0&\rho_{F13}
\\0&\rho_{F22}&0\\\rho^*_{F13}&0&\rho_{F33}\end{array}\right)\label{33u}
\end{eqnarray}
where 
\begin{eqnarray}
\rho_{F11}=\rho'_{11}+\rho'_{44}, \, \rho_{F22}=2\rho_{22}, \,\nonumber\\ \rho_{F33}=\rho'_{33}+\rho'_{66},
\rho_{F02}=\rho'_{13}+\rho'_{46}, \label{eq:33d}
\end{eqnarray}
and $N_{F}=\rho_{F11}+\rho_{F22}+\rho_{F33}$. In Fig. \ref{fig:20} we represent $\mathbb{C}_{F-AB}$ in
front of $x$ for the same values of $z$ and $p$ as in Fig. \ref{fig:18}. Entanglement grows from 0 to a maximum value at $x\approx0.1$ and
then decreases. The growth of $\mathbb{C}_{F-AB}$ is correlated with the decrease of $\mathbb{C}_{A-BF}$ and $\mathbb{C}_{B-AF}$
in the same way as the magnitudes analyzed in the previous section, as can be seen in Fig. \ref{fig:21} for a particular case.

\subsection{Conclusions}
We have analyzed in a previously unexplored spacetime region the entanglement dynamics of a system consisting in a pair of
neutral two-level atoms $A$ and $B$ interacting with a common electromagnetic field $F$. At $t=0$ atoms are in the Bell state
$\sqrt{p}\,|\,e\,e\,\rangle\,+\,\sqrt{1-p}\,|\,g\,g\,\rangle$ and the field in the vacuum state. The evolution of this state has
been considered within the non-Markovian, non-RWA approach of quantum electrodynamics up to second order in perturbation theory.
We find ESD and revival of entanglement in the reduced state of the atoms, in a range of $p$ that decreases with the interatomic
distance $r$, in agreement with the results obtained with master equations \cite{ficektanasI}. For $r$ large enough, we find
that the revival of entanglement can occur with $r>c\,t$ and therefore is not a consequence of photon exchange between the
atoms. We find that this phenomenon is strongly related to the transfer of entanglement between the different subsystems of two
parties that coexist in the entire system: we obtain sort of entanglement cycle for the atom-field reduced states opposite to
the atom-atom one. We have considered also the different bi-partitions of the system, namely $A-BF$, $B-AF$ and $F-AB$, finding
similar relationships between their entanglement cycles.

\chapter{Circuit QED as a quantum simulation of matter-radiation interaction beyond RWA}
\begin{quote}
``GALILEO: I improved it.
  LUDOVICO: Yes, sir. I am beginning to understand science. [...]  Yes, a pretty red. When I saw it first it was covered in green."
  (Bertold Brecht, \textit{Galileo}.)
\end{quote}
The experimental implementation of the theoretical results introduced in Chapter 3 is extremely challenging, mostly because of the impossibility of turning the interaction on and off at will, among other reasons. In this chapter we will move to the framework of circuit QED, which can be understood as 1-D version of matter-radiation interaction with artificial atoms and photons, with the advantage of experimental amenability. An important feature of circuit QED is that stronger values of the coupling strength can be achieved, entering into the so-called ``ultrastrong'' coupling regime in which effects beyond RWA become accessible to experiment. An important consequence of this will be analyzed in section 4.1, in which a possible experimental protocol to detect with certainty ground state qubit self-excitations is introduced. In section 4.2 we present an experiment proposal to test the entanglement generation between superconducting qubits coupled to a quantum field initially in the vacuum state, that is the circuit QED version of some of the results presented in chapter 3. 
\section{Detecting ground state qubit self-excitations in circuit QED:\\ slow quantum anti-Zeno effect}
\subsection{Introduction}
The model of a two-level system interacting with one or more harmonic oscillators can be implemented in circuit QED combining a superconducting qubit with a microwave resonator or a transmission line~\cite{blais04,wallraff04,chiorescu04}. Compared to experiments in Quantum Optics with microwave cavities~\cite{raimond01,walther06} or with trapped ions~\cite{leibfried03}, the superconducting circuit experiments have one important advantage: the strength of the qubit-photon coupling. The fact that superconducting resonators and superconducting qubits follow essentially the same physical laws makes it possible not only to reach the strong coupling regime~\cite{wallraff04,chiorescu04}, in which multiple Rabi oscillations are possible within the decoherence of the cavity or the qubit, $g \gg \kappa, \gamma,$ but also entering the ultrastrong coupling regime, $g \sim \omega,$ in which the internal and interaction energies become similar~\cite{niemczyk10,forn-diaz10}. In this new regime the dynamics is very fast and the usual approximations such as the Rotating Wave Approximation (RWA) in the Jaynes-Cummings model break down~\cite{irish05,irish07,ashhab10}.

One of the most astounding predictions of the ultrastrong coupling regime is that a single qubit can distort its electromagnetic environment, giving rise to a ground state in which the qubit is dressed with photons. As we will show in the following sections, in the case of a qubit and a single harmonic oscillator, this translates into a state which is a superposition of a desexcited qubit and a vacuum, with other states in which the qubit, the oscillator or both are populated with excitations and photons, respectively~\cite{irish07,lizuain10,hausinger10}. This is a completely non-RWA effect which requires large values of the coupling to be observed. More precisely, the excitation probability grows approximately as $p_e \propto (g/\omega)^2$ and $g$ has to become comparable to the energies of a photon, $\hbar \omega,$ or of a qubit, $\hbar\omega_0,$ making the interaction dynamics both very strong and very fast. From the experimental point of view it would thus seem unfeasible to probe a physics that takes place at speeds of $\omega\sim1-10$ GHz, while the typical measurement apparatus in circuit-QED have response times which are much slower, of about 50 ns. There are four routes to escape this problem: making the ultrastrong coupling switchable by design~\cite{peropadre10}, dynamically turning it off by external drivings~\cite{hausinger10b}, engineering faster measurement apparatus or looking for new ways to extract information out of slow measurement devices.

In this section we take the slow route,  showing that is possible to extract valuable information from the fast dynamics of the system with current measurement technologies. We will study what happens to an ultra-strongly coupled qubit-cavity system when the qubit is subject to repeated measurements by a detector with a slow repetition rate that is only capable of performing weak measurements of the state of the qubit. The main goal is to detect the qubit in its excited state starting from the ground state of the system. The first measurement has already a small probability of success, as commented in the previous paragraph. In case of failure the system is projected to a non-equilibrium state which rapidly exhibits a dynamics with an oscillatory probability of excitation, mainly due to non-RWA transitions from the ground state of the qubit $\ket{g}$ to the excited one $\ket{e}$. By means of performing repeated measurements, we will show that the detector is able to probe these usually considered as ``virtual''  excitations of the qubit and the cavity and at the same time reveal information of the interaction model. More precisely, the repeated measurements accumulate information exponentially fast and behave like an anti-Zeno effect \cite{saveriozeno} in which the qubit is projected onto its excited state, revealing those ground-state excitations that we were looking for. We show that  this anti-Zeno ``decay'' $\ket g \rightarrow \ket e,$ is very efficient and does only require a \textit{short number of repeated measurements} with a repetition rate which is much slower than in the standard anti-Zeno effect.

Like other proposals for probing the ultrastrong coupling limit~\cite{lizuain10}, (see also section 4.2), the anti-Zeno dynamics in this work is supported by the counter-rotating terms in the qubit-resonator interaction, using as seed the ground state excitations of these systems. The phenomenon is absent in the limit of RWA in Jaynes-Cummings models. Let us remark that the non-RWA effects are being extensively studied not only in the ultrastrong coupling regime of circuit-QED but also in other fields like Quantum Optics \cite{almutbeige}. Models of repeated measurements on superconducting qubits were considered for instance in Ref. \cite{calarcoonofrio} and  have been implemented in the lab \cite{measurementsqubits1, measurementqubits2}.

The structure of this section is as follows. In  4. 1. 2 we will show that the eigenstates of the hamiltonian, and in particular the ground state of a qubit-cavity system in the ultrastrong coupling regime are not separable, $\ket{g,2n}$ or $\ket{e,2n+1},$  but linear combinations of these vacua and excitations. More precisely, the qubit-resonator ground state contains a contribution of $\ket{e,1}$ which grows with the coupling strength and becomes relevant in the ultrastrong coupling regime, $g\sim \omega.$ We will see that after a few ideal periodic projective measurements of the qubit state, the probability of finding that it is in the state $\ket g$ tends quickly to 0, even if an uncertainty in the time taken by the measurement is considered. In section 4.1.2.5 we will consider a realistic model of  measurement  in which large  amounts of errors are allowed, showing the robustness of our method. Section 4.1.2.6 is devoted to the analysis of the role of relaxation and dephasing. We conclude in section 4.1.3 with a summary of our results.

\subsection{Detecting ground state qubit self-excitations}

\subsubsection{The Rabi model}

We will consider the following Hamiltonian, corresponding to a qubit-cavity system
\begin{equation}
\label{eq:hamiltonian}
H = H_0 + g H_I =
\hbar \omega a^\dag a+ \frac{\hbar\omega_0}{2}\sigma^z + \hbar g \sigma^x (a + a^\dagger),
\end{equation}
where $\hbar \omega_0$ is the energy splitting between the two levels of the qubit $\ket e$ and $\ket g,$ $\omega$ the frequency of the photons in the cavity or resonator field and $g$ the coupling strength.

In the weak and strong coupling regimes, in which the coupling $g \ll \omega,\omega_0$ is only compared to the decay rates of the cavity and the qubit, one may treat $H_I$ as a small perturbation on top of the bare qubit and resonator states. In this limit the counter-rotating terms $a^\dagger\sigma^+, a\sigma^-$ average out, and the total Hamiltonian becomes equivalent to the Jaynes-Cummings model, whose ground state is a separable combination of the qubit ground state and a cavity vacuum, $\ket{g,0}.$

\begin{figure}[h]
  \centering
  \includegraphics[width=\linewidth]{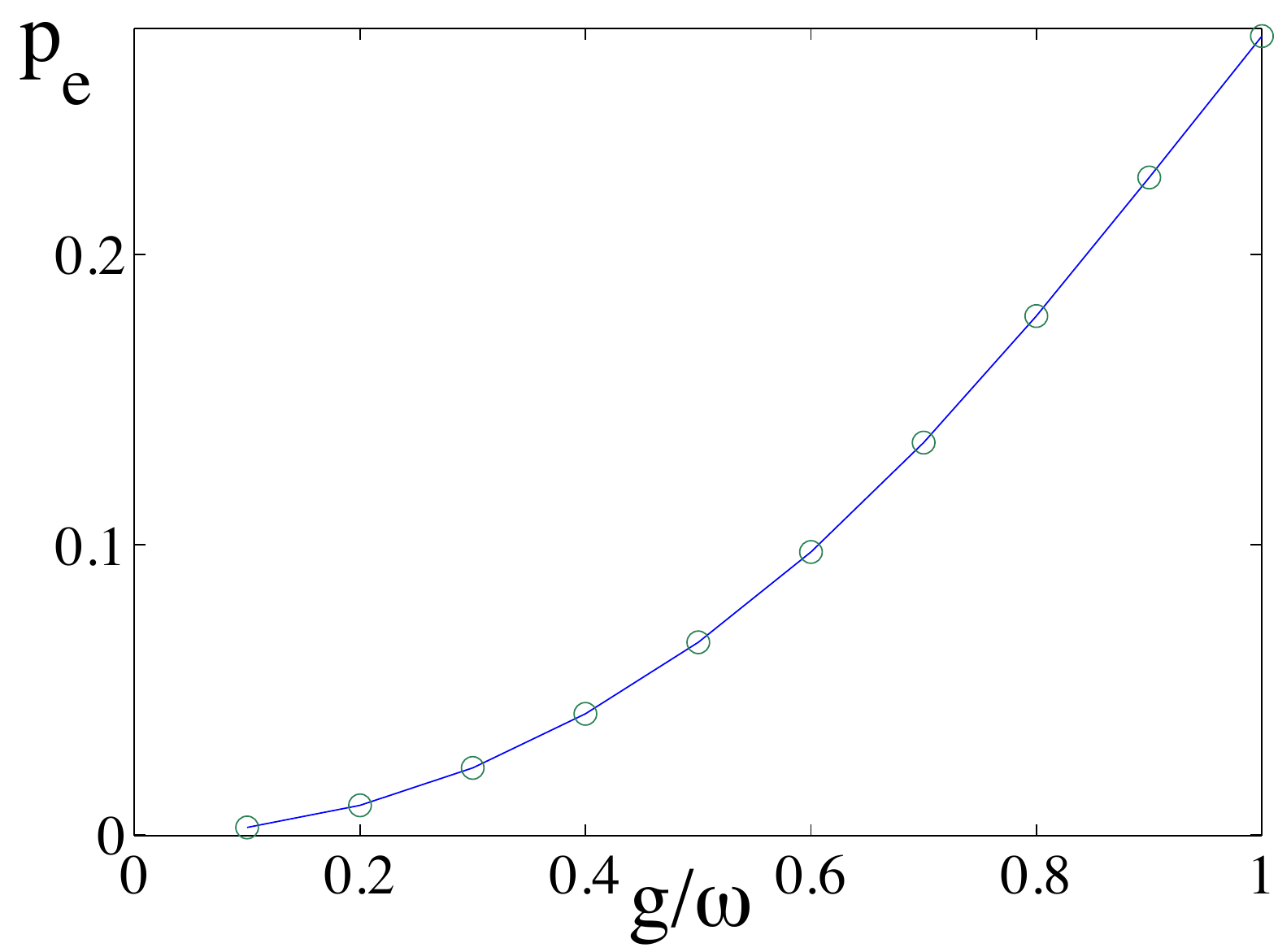}
  \caption{(Color online) Probability of excitation for of a qubit $p_e$ (blue,solid) vs. the dimensionless ratio $g/\omega,$ for a qubit-resonator system [Eq.~(\ref{eq:hamiltonian})] in the ground state of a ultrastrong coupled limit $\omega=\omega_0=1$ GHz. This line is undistinguishable from a quadratic fit (green,dashed). }
  \label{fig:groundstate}
\end{figure}

In this work we are interested  however in the ultrastrong coupling regime, in which $g$ approaches the qubit and photon frequencies, $\omega$ and $\omega_0.$ In this case it is more convenient to look at the state space in the language of parity subspaces~\cite{lizuain10}, and treat $H_0$ and $H_I$ on equal footing. Within this picture, the Hilbert space splits up in two different chains of states coupled by $H_I,$ and in particular the ground state of the system becomes a linear combination of states in the even parity sector
\begin{equation}
\label{eq:groundstate}
\ket{G}= c_0\ket{g0}+c_1\ket{e1}+c_2\ket{g2}+c_3\ket{e3}+ \ldots
\end{equation}
where the coefficients $c_i$ depend on $g,\omega,\omega_0.$

\subsubsection{Detecting excitations with one measurement}

One of the goals of this paper is design a protocol for measuring the tiny excitations in the ground states ---$|c_1|^2+|c^3|^2+|c_5|^2+\ldots$ in Eq.~(\ref{eq:groundstate})---. Let us assume for now that we have a good measurement apparatus and that we perform a single measurement of the qubit in the ground $\ket{G}$ of the system. In Fig.~\ref{fig:groundstate} we plot the probability of finding the qubit excited after just one measurement
\begin{equation}
p_{e}=\braket{\hat{P}_e}_G =: \braket{\ket{e}\bra{e}}_G \label{eq:pe}
\end{equation}
against different values of the coupling strength, assuming always $\omega=\omega_0$ and $g/\omega \le1$. For the strongest couplings the values of $p_e$ are sizable. Moreover, we have:
\begin{equation}
p_{e}=\lambda \frac{g^2}{\omega^2}, \,\, (\omega=\omega_0,\,\frac{g}{\omega}\le1) \label{eq:pequadratic}
\end{equation}
This quadratic behavior comes as no surprise. The main contribution to $p_{e}$ is $|c_1|^2.$ If we think of $\ket G$ as the free vacuum $\ket{g,0}$ dressed by the interaction $H_I,$ then $|c_1|$ may be computed from perturbation theory in interaction picture, the leading term  being proportional to $\left|\braket{e,1|H_I|g,0}\right|^2.$ It is interesting to see how these contributions quickly grow as $g$ approaches $\omega,$ but that at the same time the signal in current experiments with $10\%$ coupling strengths, might have a too small excitation signal to be accurately detected.

This work is born from the idea that perfect projective measurements in c-QED might be too difficult, as existing measurement apparatus may be too slow or not have enough sensitivity to capture those excitations. The constraint of time is found, for instance, in flux qubit measurement devices based on SQUIDs, which roughly work as follows: A very short current pulse is sent to the SQUID, instantaneously changing its potential from a periodic function to a washboard potential. In this brief period of time, one of the flux qubit states which is sitting inside the SQUID may provide, through its intrinsic current and flux, enough additional energy for the SQUID to tunnel into a voltage state. This stochastic process is random in time and does no have a 100$\%$ success rate. Moreover, it requires an additional sustained current that keeps the SQUID in that voltage state during an integration time large enough for the electronics to realize that the measurement succeeded. Adding the excitation and integration phases, the best experimental setups bring the detection time down to tens of nanoseconds, which is still slower than the qubit-resonator dynamics --$1.6$ ns for a $600$ MHz coupling, and much shorter for the qubit and resonator periods, $1/\omega.$

An additional complication of the ultrastrong coupling limit is that an arbitrary measurement device might not have enough good coupling to either the qubit or the resonator in the ultrastrong coupling regime. If we assume that both quantum systems interact so strongly that their eigenstates are highly entangled states with large energy gaps, $\sim g,\omega,\omega_0,$ the detector could have problems coupling to those states and breaking their energy level structure. In other words, the measurement device couples through an operator, $\sigma^z,$ which typically represents a perturbation of the qubit-resonator model, and if that perturbation, which aims at breaking the linear combinations~(\ref{eq:groundstate}), is not strong enough, it might not extract any information from the system, or the amount of information might be reduced, becoming an off-resonant, weak dispersive measurement.

\begin{figure}[h]
  \centering
  \includegraphics[width=\linewidth]{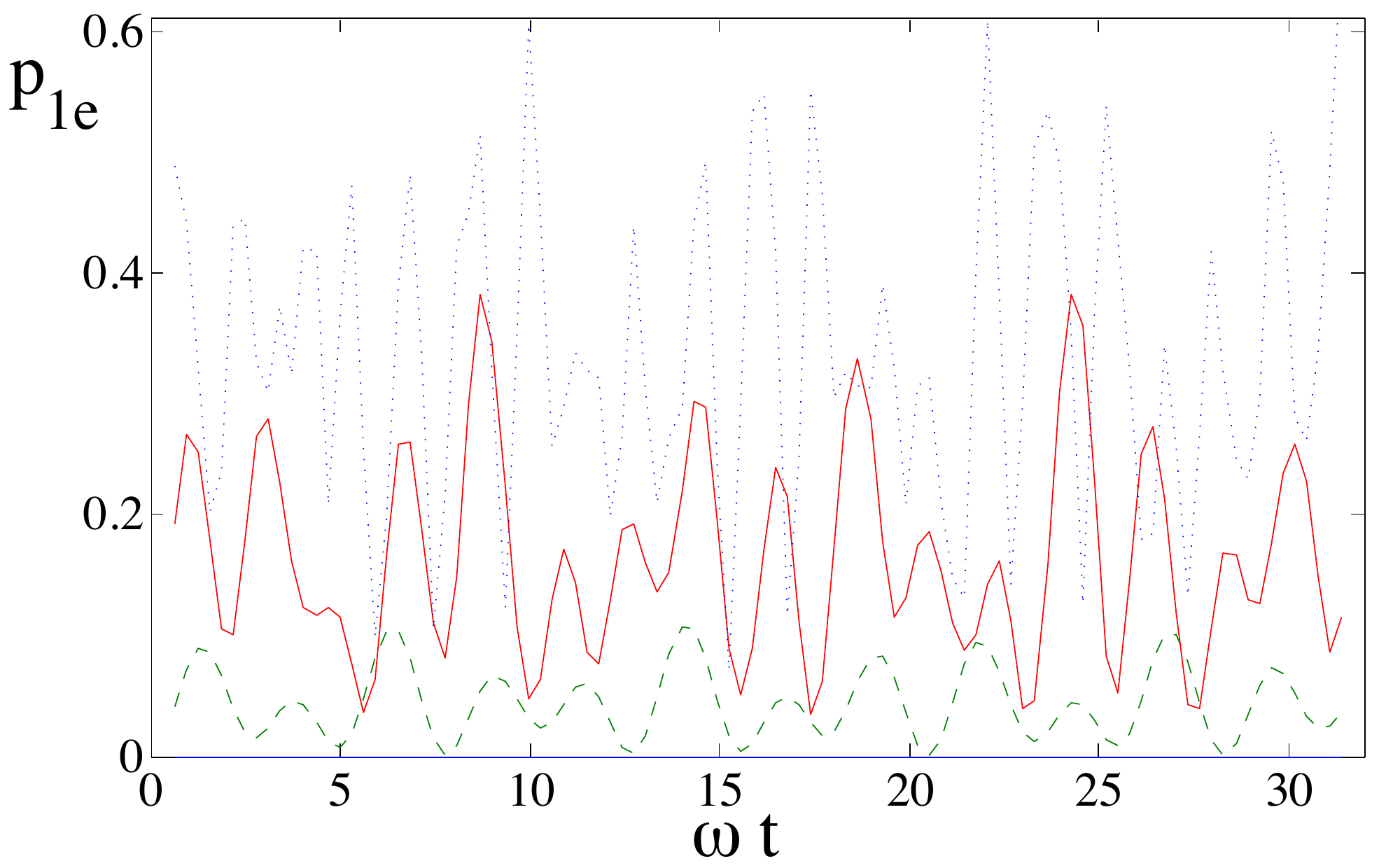}
  \caption{(Color online) After the qubit has been measured once, the qubit in the qubit-resonator system is left in a non-equilibrium state, $\ket{\Psi}.$ Here we plot the probability of excitation for the qubit $p_{1e}$ as a function of the dimensionless time $\omega t,$ shortly after that measurement. We show three situations, $\omega=\omega_0= 1$ GHz and $g/\omega=1/3 $ (dashed), $g/\omega=2/3$ (solid) and $g/\omega=1$ (dotted), which exhibit fast dynamics.} \label{fig:onemeasurement}
\end{figure}

All these considerations brought us to the idea of using more than one measurement steps in the same experiment, with the aim of increasing the amount of information that it is extracted from the same state. This can be done because the kind of measurements done in experiments are non-destructive: the same qubit can be continued to be measured at another time. It is true, however, that the interval between measurements might carry a strong, fast and almost chaotic dynamics~[Fig.~\ref{fig:onemeasurement}], arising from the fact that the measurement brings the system into a non-equilibrium state, even if it did not produce any information. We will show that this is not a limitation, but a plus, and that the repeated measurements may characterize the intermediate dynamics.

\subsubsection{Repeated measurements: survival probability}
\label{sec:repeated}

If we measure the qubit once, the measurement apparatus does not click and we are working with a perfect projective measurement, we conclude that the qubit-resonator system has been projected onto the state
\begin{equation}\label{eq:projection}
  \ket{\Psi} = \sum_{n}\hat{c}_{2n} \ket{g,2n},
\end{equation}
which is a (normalized) linear combination of deexcited qubits and some photons in the cavity. By measuring the ground state in an improper basis, we have created a non-stationary state that will evolve very quickly, with frequencies that are close to $g,\omega$ and $\omega_0.$ Lacking any other relaxation mechanism than the cavity and qubit decoherence times, these oscillations will be sustained for a large period of time, causing the qubit to get reexcited multiple times. The excitation probability
\begin{equation}
  p_{1e}(t,0)=\braket{\Psi(t)\left|\hat{P}_e\right|\Psi(t)}, \label{eq:p1e}
\end{equation}
may be computed from the initially measured state as
\begin{equation}
  \ket{\Psi(t)} \propto e^{-iHt} (1-\hat{P}_e) \ket{G}.
\end{equation}
As Fig.~\ref{fig:onemeasurement} shows, $p_{1e}$  exhibits very fast oscillations, but  also  average to a nonzero value, which is always close to the ground state excitation probability of the qubit, $p_e = \sum_n |c_{2n+1}|^2.$ Consequently, if we perform a second measurement at a later time $t_1$ we will have again a certain probability of success $p_{1e}(t_1,0)$ of detecting the state $\ket{e},$ and a certain probability of failure $p_{1g}(t_1,0)=1-p_{1e}(t_1,0).$ In the latter case the system is projected to a new state with a new time dependent probability $p_{2e}(t_2,t_1),$ and so on. After a few measurements we can define the survival probability as the probability that we have never detected a state $\ket{e}$ in the qubit
\begin{equation}
P_g^N=p_g p_{1g} (t_1,0) p_{2g}(t_2,t_1)...p_{Ng}(t_N,t_{N-1}).\label{eq:survival}
\end{equation}
A key idea in the interpretation of this formula is the fact that the intermediate probabilities $p_{ng}$ are on average very similar, and almost independent of the timespan among measurements. For the range of couplings that are within intermediate reach in experiments, $g/\omega \sim 0.1 - 1,$ we have verified numerically and perturbatively that this probability is well approximated by a quadratic law
\begin{equation}
p_{ng} \sim 1- \chi_n \frac{g^2}{\omega^2} 
\end{equation}
with minor differences among realizations, $\chi_n.$  The accumulation of products in Eq.~(\ref{eq:survival}) leads to an approximately exponential decrease of the survival probability
\begin{equation}
  P_g^N \sim \prod_{n=1}^N \left(1- \chi_n \frac{g^2}{\omega^2}\right) \sim \exp \left(-N\bar\chi\frac{g^2}{\omega^2}\right),
  \label{eq:exponential}
\end{equation}
as long as $\bar\chi\frac{g^2}{\omega^2}<<1$. This exponential behavior is typical of the so called anti-Zeno effect, in which repeated measurements of a quantum system accelerate the transition of a quantum system between two states. In our case the repeated measurements are rather creating a non-unitary evolution that excites the qubit from $\ket{g}$ to $\ket{e}$ using as seed the nonzero excitation probability $p_{1e}=\sum_n |c_{2n+1}|^2$ which is present in the equilibrium state of the qubit-resonator system. This last point is particularly important because this anti-Zeno evolution is impossible when the ground state of the qubit and the resonator is the vacuum $\ket{g,0}.$ In this case $g/\omega$ is so small, and $p_{1e}$ so close to zero, that all measurements will give no signal at all and the qubit will remain in the state $\ket{g}$ for the duration of the experiment. As we will see in the following, there is a key difference between the effects described in this section and the standard anti-Zeno effect: we need only a few measurements and they can be widely spaced in time.

In the following sections we will summarize extensive numerical studies of the anti-Zeno dynamics. We have contrasted these with various semi-analytical methods, one of which, the use of truncated Hilbert spaces, helps us in understanding the reason for this behavior. For the range of couplings of current interest, $g/\omega \sim 0.1 - 1,$ it suffices to take two photons, and the ground, $\ket{G},$ plus the two excited states $\ket{E'}$, $\ket{E''}$ within the same parity subspace.  All states can be expanded as in Eq.~(\ref{eq:groundstate}) with coefficients $c_i,\,c_i',\, c_i''.$, as linear combinations of  $\ket{g0}$, $\ket{e1}$, $\ket{g2}$. After the first measurement, the qubit will end up in an excited state with probability $|c_1|^2$ and it will remain in the unexcited state with $|c_0|^2\simeq 1 - |c_1|^2,$ ending up in a combination
\begin{equation}
 \hat{P}_g\ket{G} = c_0 \ket{G} + c_0'\ket{E'} + c_0''\ket{E''}+\ldots
\end{equation}
The crudest approximation would be to neglect all excited state contributions and assume that after each measurement, either the state $\ket{e}$ is detected, or the system ends up in $\ket{G}.$ In this case the survival probability would be exactly exponential
\begin{eqnarray}
  P_g^N = (1 - |c_1|^{2} )\prod_{i=i}^N|c_0|^{2} = |c_0|^{2N+2}.
  \label{eq:truncated-exp}
\end{eqnarray}
In practice, however, the combined system does not end up only on the ground state, but gets excited state contributions from $\ket{E'},\ket{E''}.$ When we average the contributions over the period in which the measurement takes place, we find that already after the first measurement step, the excited states add up to the total probability, $\braket{p_{1e}}_T=|c_1|^2\,|c_0|^2+|c_1'|^2\,|c_0'|^2+|c_1^{''}|^2\,|c_0^{''}|^2+\ldots,$ enhancing the original behavior.

\begin{figure}[h]
  \centering
  \includegraphics[width=\linewidth]{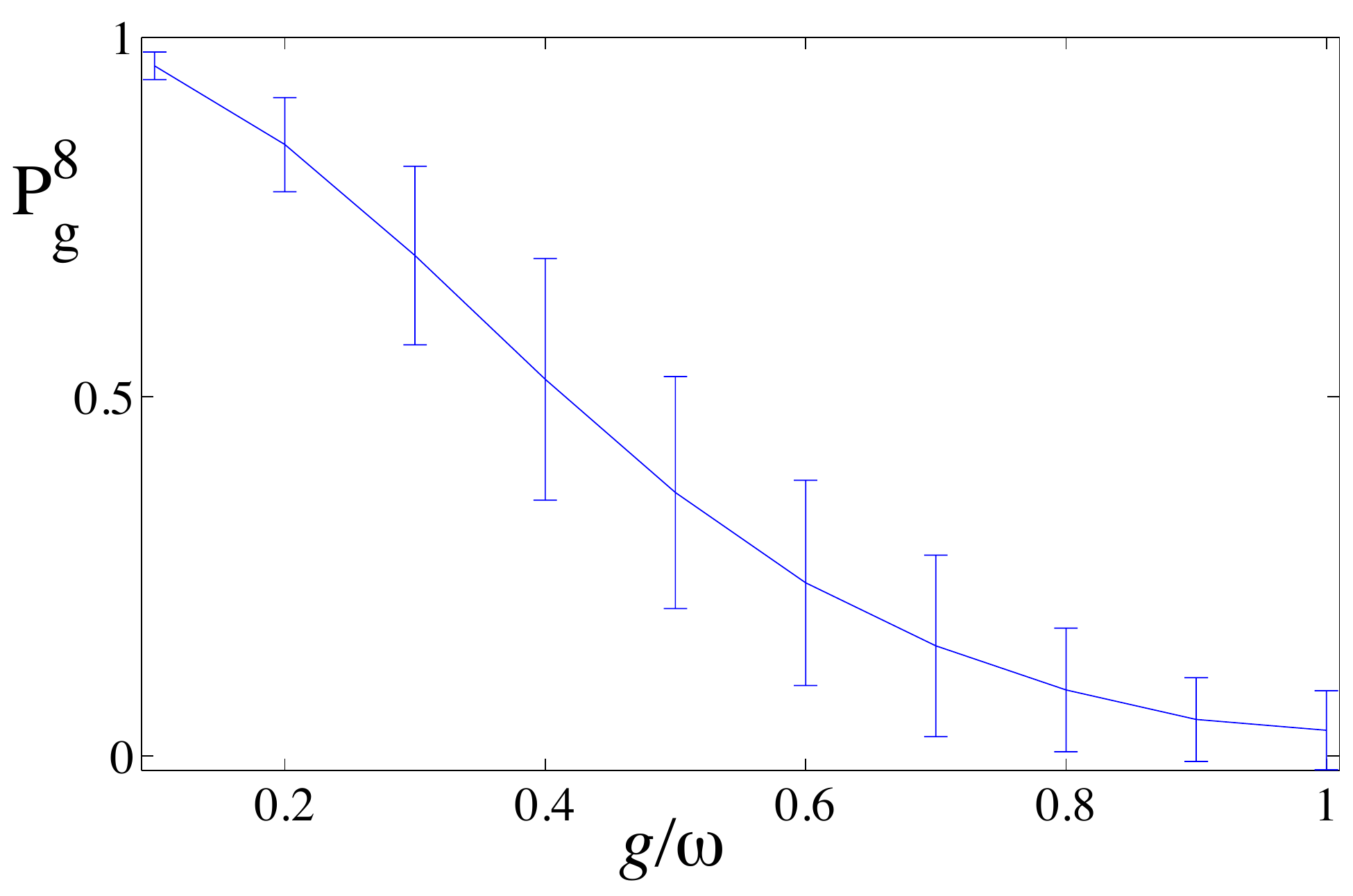}
  \caption{Survival probability after eight measurements $P_g^8$ vs. $g/\omega,$ with $\omega=\omega_0.$ The measurements are performed with periods $T_1,$ $T_2=\sqrt{2} T_1$ and averaged over 100 values of $T_1$ within the interval $2\pi [0.1,5].$ Note how the law approximates the Gaussian behavior in Eq.~(\ref{eq:exponential}).} \label{fig:8M}
\end{figure}

\subsubsection{Numerical experiments}

We have verified the anti-Zeno dynamics and the exponential law~(\ref{eq:exponential}) by means of exact numerical simulations in which we compute the outcome of repeated measurements on a qubit-resonator Dicke model~(\ref{eq:hamiltonian}). We will now explain the main results of this study.

From an experimental point of view it might be interesting to maximize the exponent $\bar\chi,$ optimizing the measurement repetition rate to hit all the maxima in the evolution of the excitation probability~[See Fig.~\ref{fig:onemeasurement}]. However we found that this is very difficult and demands a lot of precision on the measurement apparatus; for small errors or some measurement randomization this procedure drives the apparatus into exactly the opposite regime: always hitting the minima of excitation. Seeking a more robust, less demanding approach we opted for using two incommensurate periods, $T_1$ and $T_2 \simeq \sqrt{2}T_1,$ simulating measurement at times
\begin{equation}
  \label{eq:times}
  t_n \in \{T_1, T_1+T_2, 2T_1+T_2, 2T_1+2T_2,\ldots\},
\end{equation}
and at most optimizing the value of $T_1.$

\begin{figure}[h!]
  \centering
  \includegraphics[width=\linewidth]{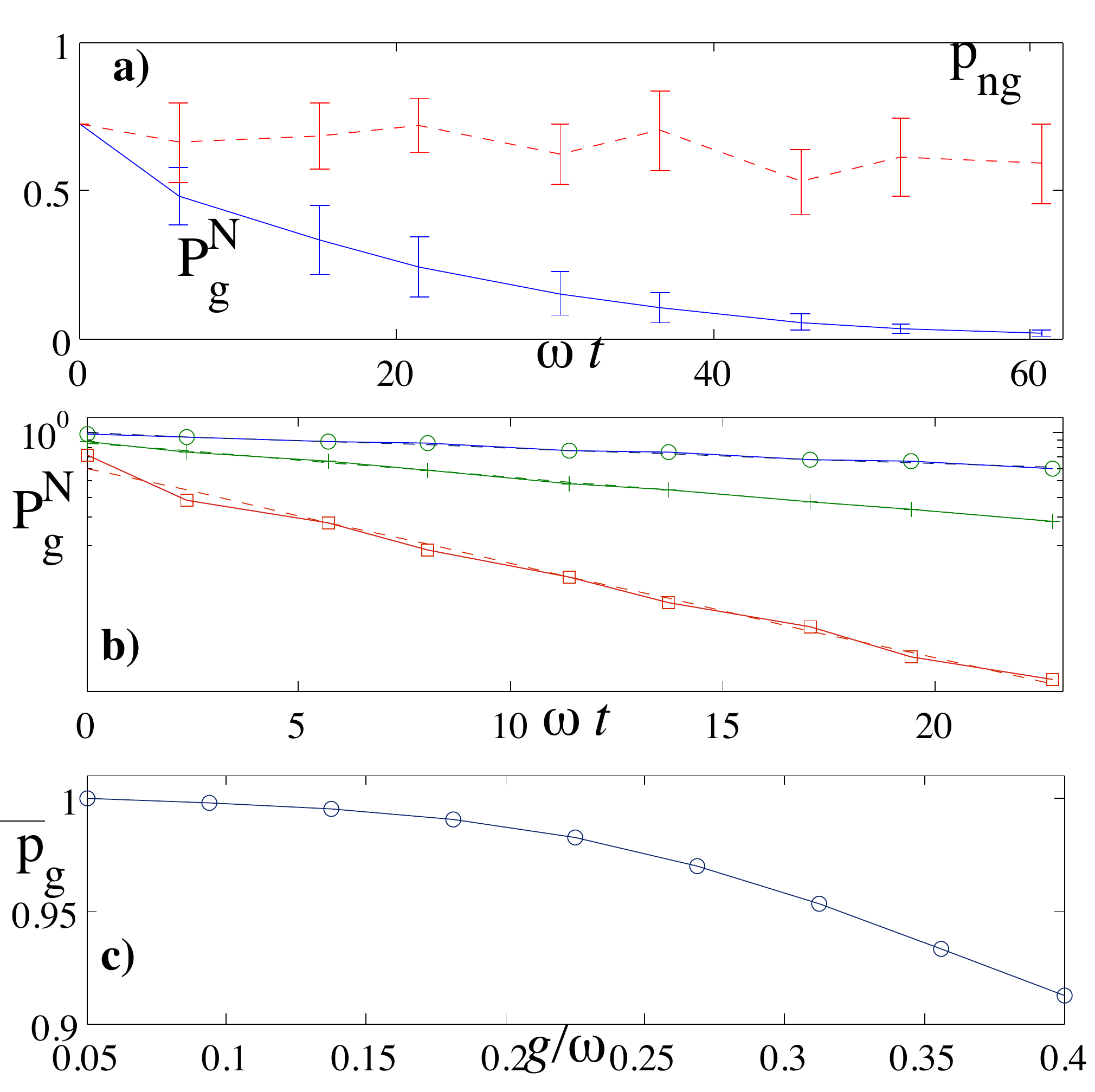}
  \caption{(a) Probability after the $n-$th single measurement, $p_{ng},$ (dashed) and accumulated survival probability, $P_g^N = \prod_{n=1}^N p_{ng}$ (solid) vs. dimensionless time $\omega t.$ We use $g/\omega=1$ and perform measurements with approximate periods $\omega T_1=\omega T_2/\sqrt{2}=2\pi,$ averaging over random perturbations of the actual measurement time, $t_n,$ within the interval $\omega t_n+[-0.2\pi,+0.2\pi].$ (b) Survival probability $P_g^N$ (solid lines) and the corresponding exponential fits (dashed lines) for $g/\omega=1/3$ (circles), $2/3$ (crosses) and $1$ (squares). (c) Mean value $\bar{p_g}$ (solid)  and the corresponding quadratic fit (dashed) vs. dimensionless coupling strength $\frac{g}{\omega}.$ All plots assume $\omega=\omega_0$ and (b,c) use $\omega T_1=3\pi/4$}
  \label{fig:8Mtime} 
\end{figure}

With this  approach, and exploring different values of $T_1,$ we have studied the survival probability and concluded that the exponential laws are really accurate. As shown in Fig.~\ref{fig:8M}, if we fix the total number of measurements to be $N=8$ and sample various periods, $T_1,$ we recover on average the Gaussian behavior $\exp(-N\bar\chi g^2/\omega^2)$ deduced in Eq.~(\ref{eq:exponential}). Instead of fixing the number of measurements, we can also study the same law and verify the exponential decay with respect to $N.$ This is shown in Figs.~\ref{fig:8Mtime}a-b, where we plot the accumulated survival probability, $P^N_g,$ as a function of time, and fit it against the same exponential~(\ref{eq:exponential}).

It is important to remark that the exponential decay is a robust signature that survives even when the measurement does not take place at  precise times, $t_n,$ from the list given before~(\ref{eq:times}). This has been verified by simulating multiple runs in which $t_n$ is randomly perturbed around its average value, and computing the survival probability. We want to remind the reader the importance of this robustness, because some measurement apparatus such as SQUIDs behave stochastically and produce a signal at a random time that can not be determined a-priori. The fact that the measurement protocol and the resulting physical behavior are independent of a precise control is encouraging.

The accuracy of the exponential law~(\ref{eq:exponential}) suggests that the survival probability of a single measurement remains constant throughout a single experiment, $p_g \simeq p_{ng}\;\forall n.$ This is qualitatively confirmed by Fig.~\ref{fig:8Mtime}a, where we show that these values oscillate around a mean one that is close to the average population of $\ket{g}$ in the ground state, i.~e. to $p_{g}.$ This suggests us to consider average values and approximate
\begin{equation}
  \bar{p}_{g}=\sum_n \frac{p_{ng}}{N}\simeq 1-\bar\chi\frac{g^2}{\omega^2}\label{eq:pgmean} 
\end{equation}
which has the expected quadratic behavior. This estimate is confirmed by Fig.~\ref{fig:8Mtime}c, where the quadratic fit is almost undistinguishable from the actual behavior.
\begin{figure}[h!]
  \centering
  \includegraphics[width=\linewidth]{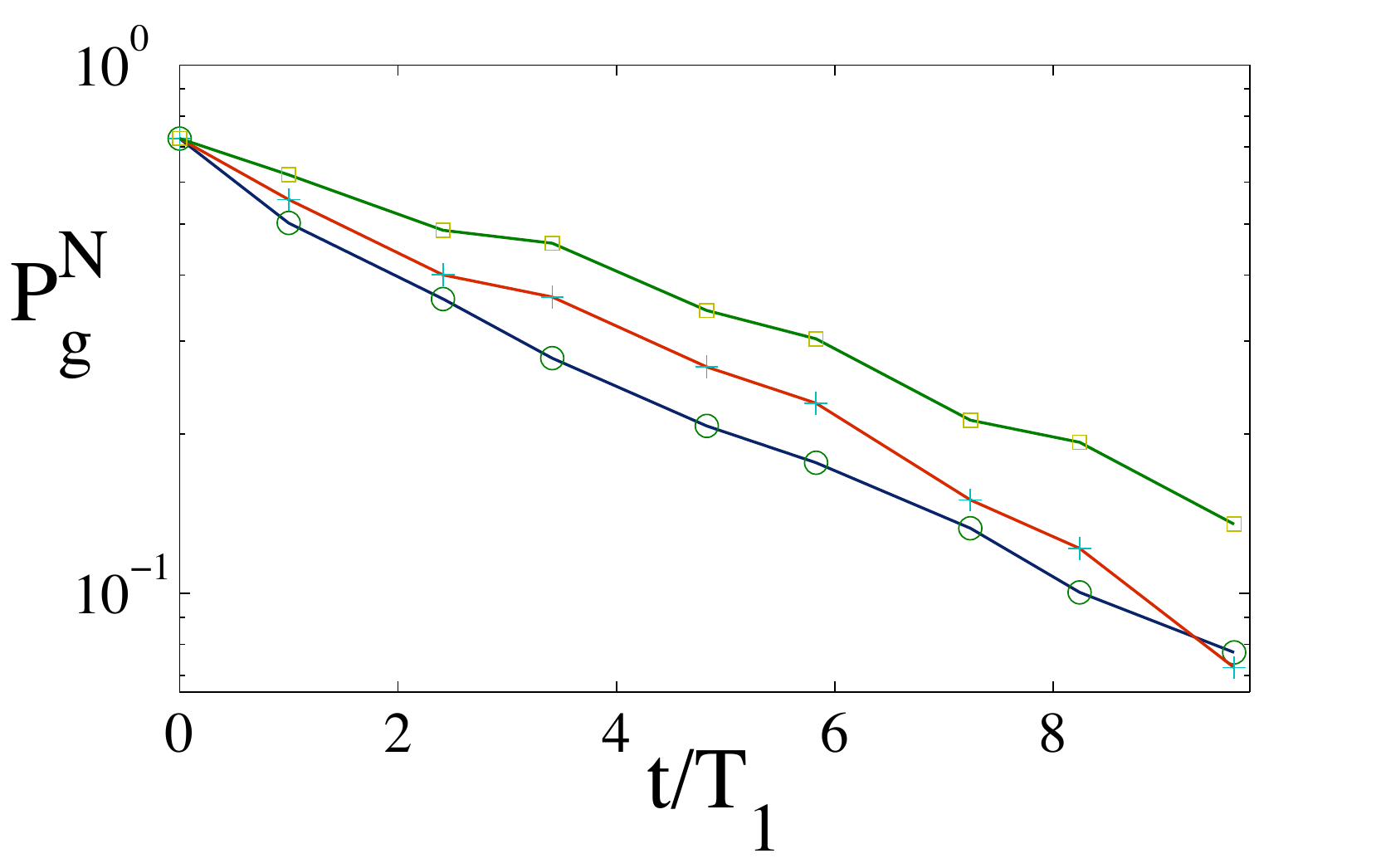}
  \caption{Survival probability $P_g^N$ vs. $ t/T_1$, with $g=\omega=\omega_0$ GHz and $\omega T_1=\pi $ (blue, circles), $2\pi $ (red, crosses) and $3\pi$(green, squares). Each marker corresponds to a measurement.} \label{fig:collapsed}
\end{figure}
The final question which remains to be answered is whether the exponent $\bar\chi$ depends on the frequency of the measurements or not. For that we have fixed the coupling strength and explored three values of the period, $T_1,$ studying the average exponential behavior. The result is shown in Fig.~\ref{fig:collapsed}, collapsing all numerical simulations in the dimensionless quantity $t/T_1,$ and finding that they have very similar slopes.

\subsubsection{Weak measurements}

So far we have considered ideal projective measurements, introducing only some stochasticity in the time at which the measurement event is produced. We will now add another ingredient to our measurement model, which is the possibility that the detector only performs a partial measurement, leaving the state ``untouched'' with a nonzero probability, $\epsilon.$

We can easily model an imperfect detector using the formalism of completely positive maps, operations that transform density matrices into density matrices. If $\rho$ and $\rho'$ are the states of the qubit-resonator system before and after the measurement, we will write, up to normalization
\begin{equation}
\rho'= (1-\epsilon) (\mathbf{1} - \hat{P}_e) \rho  (\mathbf{1} - \hat{P}_e) + \epsilon \rho. \label{eq:errors}
\end{equation}
This is read as follows. With probability $\epsilon$ the measurement device will do nothing, leaving the state untouched. With probability $(1-\epsilon)$ the measurement device will detect the state of the qubit. In this case it will either give us a positive signal, moment at which we will stop the experiment, or it will not produce anything at all, and we will continue with the projected state $(\mathbf{1}-\hat{P}_e)\rho(\mathbf{1}-\hat{P}_e),$ that has the qubit deexcited, $\ket{g}.$

This qualitative model describes measurements from a SQUID \cite{measurementsqubits1,measurementqubits2}, where we place ourselves on the verge of metastability and assume that if the qubit is in the excited state, $\ket{e},$ the SQUID will tunnel to the voltage state with probability $(1-\epsilon),$ giving no signal for $\ket{g}.$ Note that with probability $\epsilon$ the SQUID may not tunnel and then we will gain no information about the qubit or the resonator.

\begin{figure}[h!]
  \centering
  \includegraphics[width=\linewidth]{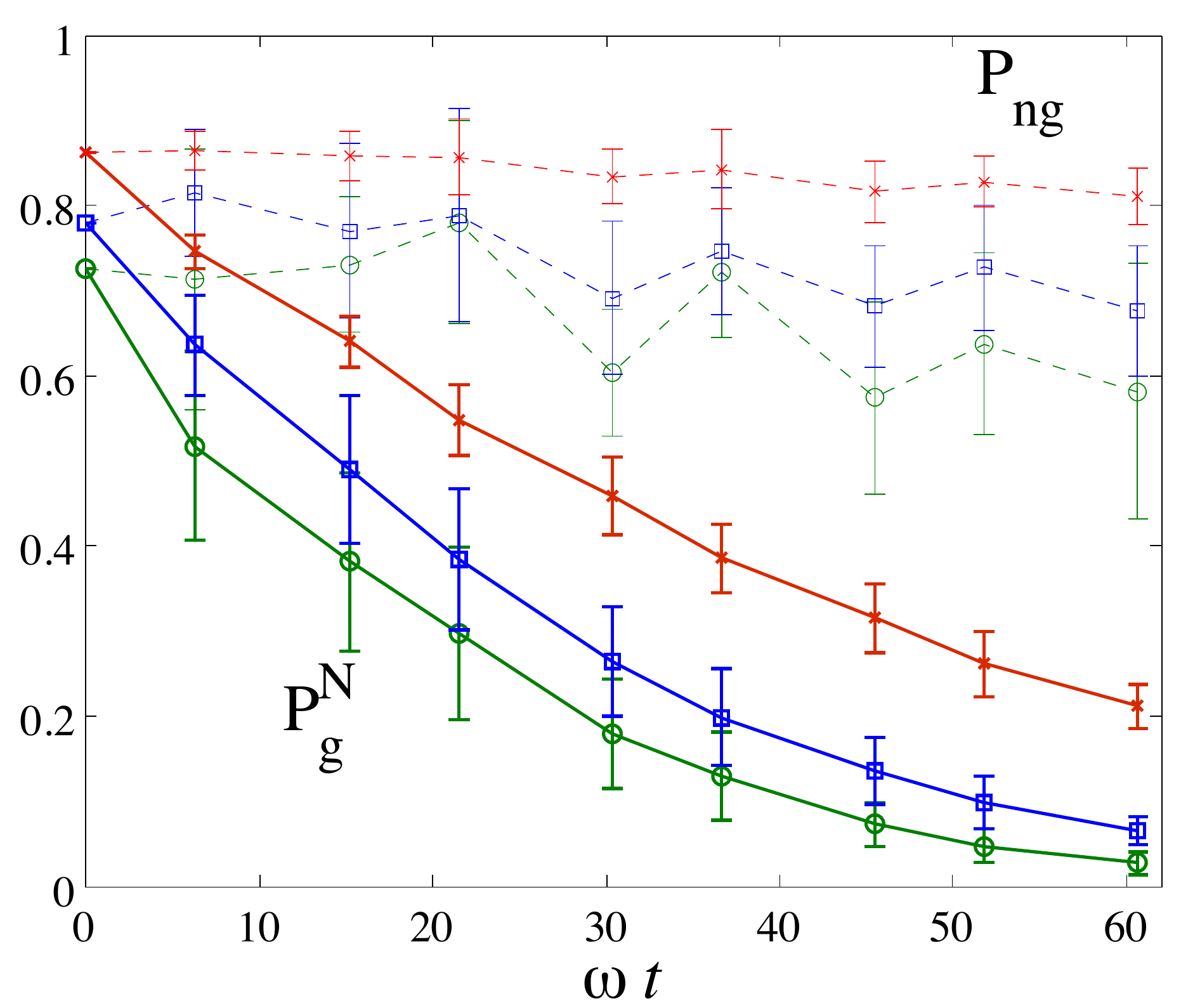}
  \caption{(Color online) Survival probability $P_g^N$ (solid lines) and $p_{ng}$ (dashed lines) vs. $\omega t,$ with $g=\omega=\omega_0= 1$ GHz , $\omega T_1=2\pi$ and $\epsilon=0.2, 0.1$ and $0$ (crosses, squares and circles, respectively). The probability of a measurement at a time $\omega t=\omega t_0$ is averaged over 20 random values  within the interval $[\omega t_0-0.2\pi,\omega t_0+0.2\pi]$} \label{fig:8Mtimeepsilon} 
\end{figure}

In Fig. \ref{fig:8Mtimeepsilon} we analyze the impact of $\epsilon$ in our previous results. Even for large errors $\epsilon=0.2$ we retain the exponential behavior observed in Fig. \ref{fig:8Mtime}a, with acceptable error bars that decrease with increasing number of measurements --- in other words, the qubit is still efficiently projected to the excited state.
  
\subsubsection{Relaxation and dephasing}

Throughout this work we have considered in the numerical simulations the model given by the Hamiltonian in Eq. (\ref{eq:hamiltonian}) which do not include effects like relaxation or qubit dephasing, usually included in master equation approaches.

We want to remark that it is still an open question, both experimentally and theoretically, to understand and model the dissipation and decoherence processes of quantum circuits in the presence of ultrastrong qubit-cavity coupling. One popular approach~\cite{werlang08,dodonov10} is to combine the usual photon leakage mechanism from quantum optics models, $\mathcal{L}(\rho)\sim 2 a\rho a^\dagger - a^\dagger a \rho - \rho a^\dagger a,$ with the qubit-cavity Hamiltonian. Note that in such a combination, the asymptotic states of the dissipation (the vacuum) and of the interaction (populated cavity) are incompatible, and one may find excitations induced by the dissipative terms, an infinite stream of photons leaking out of the cavity and other controversial phenomena.

These effects disappear when one rederives the master equation from first principles, using the qubit-cavity eigenstates of the ultrastrong coupling model and the usual zero temperature baths. In the resulting models the main relaxation mechanisms are found to be the decay to the ground state $\ket{G}$ and a dephasing of the joint cavity-qubit states ---in other words, dissipation and decoherence in the proper basis---. If we assume this reasonable model, then we can conclude that the exponential laws derived in this manuscript are not significantly distorted. To begin with, relaxation to the ground state $\ket{G}$ just makes the experiment closer to the truncated Hilbert space model considered in Sect.~\ref{sec:repeated}, and in particular to the exponential law from Eq.~(\ref{eq:truncated-exp}). For strong couplings, decoherence amounts to random modulations of the qubit-cavity energy levels, without significantly affecting the populations, $|c_i|^2.$ Since this is the most relevant quantity in all the previous discussions, we can also expect that, up to minor changes in the rates, the anti-Zeno effect will also survive.

\subsection{Conclusions}

We have considered a system consisting in a superconducting qubit coupled to a closed transmission line, operating in the ultrastrong coupling regime. The ground state in such scheme is not just a product of the ground states of the qubit and the cavity, as is the case for weaker couplings. On the contrary, the vacuum of the system is dressed by the interaction and so it contains a relevant probability of finding the qubit excited. This probability is proportional to the square of the coupling strength. We have introduced a protocol for detecting that excitations with certainty, maximizing the small probabilities that are obtained with only one measurement.

Our main result is that, after a number of periodic measurements of the qubit, the probability of finding it in the ground state in all the measurements goes exponentially to zero, even if the measurements are weak and are performed with a slow repetition rate in comparison with the fast dynamics of the interaction. We refer to this as slow quantum anti-Zeno effect. Like the well known quantum anti-Zeno effect, the result is the acceleration of a transition, in this case the exotic transition $\ket g\rightarrow \ket e,$ which becomes relevant in this regime due to the breakdown of the RWA. But this procedure is less experimentally demanding, since it requires a smaller number of measurements and a shorter duration of the period at which they are performed. We have shown that the protocol is robust to large errors in the measurement process, when a realistic SQUID readout is considered.

This is one of the first experimentally accessible consequences of the new ultrastrong coupling regime and can only be derived beyond the RWA. The physical nature of the ground state qubit self-excitations, commonly considered as  a virtual process without possible experimental record, seems now to be clear. Moreover, although the ultrastrong coupling entails a very fast dynamics, we have shown that valuable information of the interaction can be extracted efficiently with the current slow and imperfect measurement technologies.

Finally, we want to remark that strong qubit excitations have also been found theoretically in models that combine the full Rabi coupling with traditional dissipative contributions~\cite{werlang08,dodonov10}. However, the form of those dissipative terms is questionable in non-RWA setups, and furthermore, there is no justification to equate the sparse measurement setup in this work to a particular dissipative model. This lack of equivalence between models manifests in the fact that, as we have seen numerically, the sparsely repeated measurements can hit certain resonances that invalidate the anti-Zeno dynamics.

\section{Entanglement dynamics via propagating microwave photons}
\subsection{Introduction}

Quantum mechanics does not allow us in general to consider two arbitrary distant systems as separate~\cite{schrodinger}. In some cases there exist quantum correlations that cannot be generated by local operations and classical communication between remote systems. Time enters this picture through two different questions. The first one is related to the speed bound of a hypothetical superluminal influence which could explain all quantum correlations, estimated to be $10^4 c$ in a recent experiment~\cite{salart08}. The second question is of a more practical nature inside the quantum theory~\cite{reznik,reznikII,franson} (and Chapter 3 of this Thesis) : what is the speed at which two distant systems become entangled?

Quantum field theory (QFT) fulfills the principle of microscopic causality by which two space-like separated events cannot influence each other~\cite{microcausalidad} and thus cannot be used to transfer information~\cite{powerthiru,gisin}. We may then ask whether microcausality also sets a limit on the speed at which entanglement can be created between two separate systems. More precisely, can two subsystems, supported at regions $(\mathbf{x},t)$ and $(\mathbf{x}',t')$, become entangled while they are still space-like separated? Or in simple terms, can finite quantum correlations develop before signals arrive? As we have analyzed in Chapter 3, the answer to this far reaching question is yes, it is possible. After all, Feynman propagators are finite beyond the light cone and even before photon arrival there exist correlations between the vacuum fluctuations at any two space-like separated events.

In this section we demonstrate that circuit QED is arguably one of the most suitable fields to study the dynamics of entanglement between distant systems. One reason is the existence of various choices of high quality superconducting qubits, the so-called artificial atoms~\cite{martinis85,bouchiat98,mooij99,transmon}. Another reason is the possibility of coupling those qubits strongly with traveling photons using microwave guides and cavities~\cite{blais04,wallraff04,chiorescu04}. Furthermore, those coupling strengths can reach the ultrastrong coupling regime~\cite{abdumalikov08,bourassa09,guenter09,niemczyk10,forn-diaz10}, where the qubit-photon interaction approaches the energies of the qubit and photons. In this case, the rotating-wave approximation (RWA) breaks down and a different physical structure emerges. Such regimes can be activated and deactivated~\cite{peropadre10}, facilitating the creation of a fairly large amount of entanglement in a time-dependent way, as we will see in this work.

\begin{figure}[h!]
  \centering
  \includegraphics[width=\linewidth]{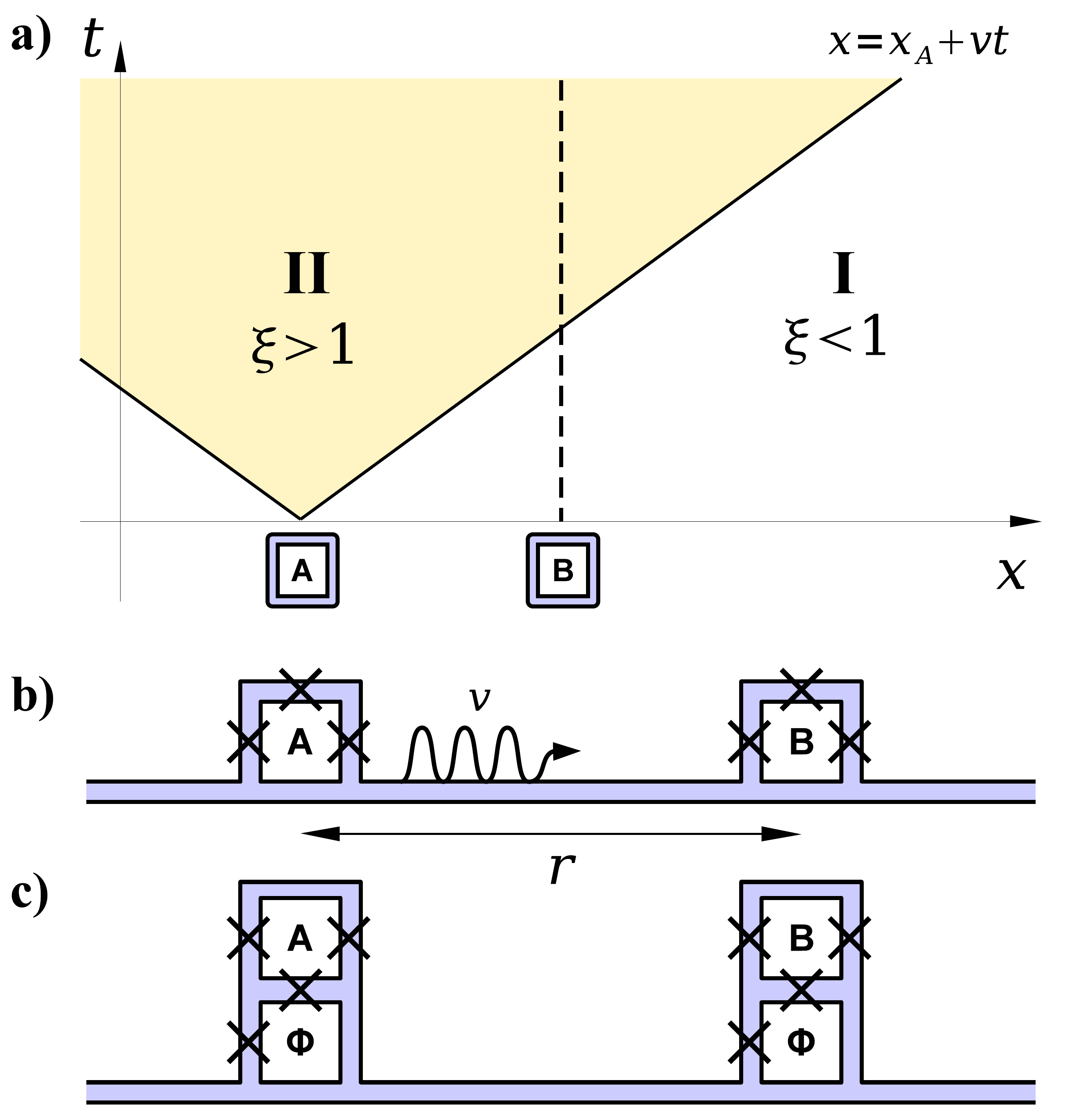}
  \caption{(a) Qubits that interact via traveling photons with finite velocity $v$ can be space-like (I, white) or time-like (II, shaded) separated, depending on the value of $\xi=vt/r.$ While only in II they are causally connected, entanglement may appear already in region I. (b) A possible implementation of these ideas consists of flux qubits ultra-strongly coupled to a common transmission line. (c) With a slight modification, the coupling of the qubits to the line can be dynamically tuned via fast magnetic fluxes, $\Phi$ (Color online). }
  \label{fig:setup}
\end{figure}

We will discuss some of the preceding questions in the framework of a precise circuit QED setup, see Fig.~\ref{fig:setup}, consisting on two well separated superconducting qubits coupled ultra-strongly to an open transmission line [Fig.~\ref{fig:setup}b]. The waveguide provides a continuum of microwave photons propagating with uniform velocity, $v,$ mediating an interaction between the qubits. Given an intitial separable state in which only qubit A is excited, we have studied the evolution of correlations and related it to the propagation of photons between qubits. The main results are: (i) Outside the light cone, that is in region I of Fig.~\ref{fig:setup}a where $\xi=vt/r < 1$, the excitation probability of qubit B is independent of the distance $r$ to qubit A. (ii) Still in region I, entanglement between the qubits always takes a finite value and grows with time. (iii) Once the qubits are time-like separated, that is as soon as we cross into region II, entanglement grows faster than the excitation probability of qubit B and takes sizeable values. Result (i) is a manifestation of the fact that our Quantum Field Theory (QFT) model satisfies microscopic causality, which formally translates into the vanishing of commutators associated with observables at space-like separations, $[{\mathcal{Q}}(\mathbf{x},t),{\mathcal{Q'}}(\mathbf{x}',t')]= 0 $ for $|\mathbf{x}-\mathbf{x}'|^2-c^2(t-t')^2 >0$. Furthermore, it shows that two qubits which are space-like separated cannot be used to communicate superluminal information. Result (ii), on the other hand, reveals the fact that correlations between vacuum fluctuations at separate points can be established at arbitrarily short times, even though they are non-signalling and cannot transmit information.

It is important to remark that the previous questions have been posed theoretically using model detectors \cite{reznik}, two-level atoms \cite{franson} (and Chapter 3 of this Thesis), scalar fields \cite{reznik,reznikII} and photons \cite{franson} (and Chapter 3 of this Thesis), yet no experimental test has been accomplished. However in this work, we show that the access to the ultrastrong couplings in circuit QED allows us to explore these ideas with very advantageous parameter ranges.

\subsection{Superconducting qubits coupled to a quantum field}

Our setup consists of two qubits, $A$ and $B,$ interacting via a quantum electromagnetic field. The qubits have two stationary states $\ket{e}$ and $\ket{g}$ separated by an energy $\hbar\Omega$ and interact with a one-dimensional field, which propagates along the line connecting them,
\begin{eqnarray}
  V(x)=i\,\int dk \sqrt{N\omega_k}\left[e^{ikx}a_k +\mathrm{H.c.}\right].\label{field}
\end{eqnarray}
This field is described by a continuum of Fock operators 
\begin{equation}
[a_k,a^{\dag}_{k'}]=\delta_{kk'}, \label{commutation}
\end{equation}
 and a linear spectrum, 
\begin{equation} 
 \omega_k = v|k|, \label{omega}
\end{equation} 
where $v$ is the propagation velocity of the field and plays the role of the speed of light. The normalization and the speed of
photons depend on the microscopic details. In particular 
\begin{equation}
v=1/\sqrt{cl}, \label{propagationvelocity}
\end{equation}
where $c$ and $l$ are the capacitance and inductance per unit length.

We consider qubits that are much smaller than the relevant wavelengths, $\lambda=v/\Omega,$ and lay well separated. Under these conditions we can split the Hamiltonian, $H = H_0 + H_I,$ into a free part for the qubits and the field
\begin{equation}
  H_0 = \frac{1}{2}\hbar\Omega(\sigma^z_A + \sigma^z_B) + \int dk \hbar\omega(k)
  a^{\dagger}_ka_k, \label{free}
\end{equation}
and a point-like interaction between them
\begin{equation}
  H_I = -\sum_{J=A,B} d_J\,V(x_J). \label{interaction}
\end{equation}
Here $x_A$ and $x_B$ are the fixed positions of the atoms, and 
\begin{equation}
d_J=d\times \sigma^x_{J} \label{dipole} 
\end{equation}
is equivalent to the dipole moment in the case of atoms interacting with the electromagnetic field.

In what follows we choose the initial state 
\begin{equation}
|\psi(0)\rangle = |e\,g\,0\rangle, \label{initial}
\end{equation}
where only qubit $A$ has been excited, while both $B$ and the field remain in their ground and vacuum states, respectively. In the interaction picture given by the ``free'' Hamiltonian $H_0,$ the system evolves during a lapse of time $t$ into the state
\begin{equation}
  \ket{\psi(t)} = {\cal T}[e^{-i \int_0^tdt' H_I(t')/\hbar}]\ket{eg}\otimes\ket{0},\label{42c}
\end{equation}
${\cal T}$ being the time ordering operator. Up to second order in perturbation theory the final state can be written as
\begin{eqnarray}
  \ket{\psi(t)}  = && \!\!\!\!\!  \left[(1+A)\ket{eg} + X\ket{ge}\right]\otimes\ket{0} +
  \label{wavefunction} \nonumber \\
 &&  (U_A\ket{gg} + V_B\ket{ee})\otimes\ket{1} \nonumber \\ && + (F\ket{eg} +  G \ket{ge})\otimes\ket{2} + {\cal O}(d^3).
\end{eqnarray}
The coefficients for the vacuum, single-photon, and two-photon states,
are computed using the action $(J=A,B)$
\begin{eqnarray}
  \mathcal{S}^+_J \! = \! - \frac{i}{\hbar}
  \int_0^t
  e^{i\Omega t'}\braket{e_J|d\sigma^x_J|g_J} V(x_J,t') dt'
  = -(\mathcal{S}^{-}_J)^\dagger\label{42f}
\end{eqnarray}
among different photon number states $\ket{n}, n=0,1,2\ldots$, being 
\begin{equation}
\ket{n}\bra{n}=\frac{1}{n!}\int dk_1....\int dk_n\ket{k_1...k_n}\bra{k_1...k_n} \label{numberstates}
\end{equation}
and 
\begin{equation}
\ket{k}=a_k^{\dagger}\ket{0}. \label{fock}
\end{equation} 
Only one term corresponds to interaction
\begin{equation}
  X = \langle0|T(\mathcal{S}^+_B \mathcal{S}^-_A)|0\rangle . \label{exchange}
\end{equation}
This includes photon exchange only inside the light cone, $vt>r,$ and vacuum fluctuations for all values of $t$ and $r$, being 
\begin{equation}
r=x_B-x_A \label{distance}
\end{equation}
the distance between the qubits.  The remaining terms are
\begin{eqnarray}
A & \!\! = \!\! & \frac{1}{2}\bra{0}T(\mathcal{S}_A^+ \mathcal{S}_A^- +
\mathcal{S}_B^-\mathcal{S}_B^+)\ket{0}\label{42e}\\
U_A & \!\! = \!\! & \bra{1}\mathcal{S}^-_A\ket{0},
V_B = \bra{1}\mathcal{S}^+_B\ket{0} , \nonumber\\
F & \!\! = \!\! & \frac{1}{2}\bra{2}T(\mathcal{S}_A^+ \mathcal{S}_A^-
+\mathcal{S}_B^-\mathcal{S}_B^+)\ket{0} \! , \, G = \bra{2}T(\mathcal{S}^+_B \mathcal{S}^-_A)\ket{0}.\nonumber
\end{eqnarray}
Here, $A$ describes intra-qubit radiative corrections, while $U_A, V_B, F$ and $G$ correspond to single-photon emission events by one or more qubits.

All the above can be understood as the 1-D c-QED version of the formalism in chapter 3
The coefficients in Eq.~(\ref{wavefunction}) can be computed analytically (Appendix B) as a function of two dimensionless parameters, $\xi$ and $K.$ The first one, 
\begin{equation}
\xi=vt/r, \label{chi}
\end{equation}
 was introduced before and it distinguishes the two different spacetime regions [Fig.~\ref{fig:setup}a], before and after photons can be exchanged. The second parameter is a dimensionless coupling strength
\begin{equation}
  K=\frac{4d^2N}{\hbar^2 v} = 2\left(\frac{g}{\Omega}\right)^2\label{42g}.
\end{equation}
Note that the qubit-line coupling 
\begin{equation}
g=\frac{d\sqrt{N\Omega}}{\hbar} \label{couplingstrength}
\end{equation}
corresponds to the qubit-cavity coupling that appears by taking the same transmission line and cutting to have a length $L=\lambda$ thus creating a resonator Refs.~\cite{wallraff04,blais04}. This formulation has the advantage of being valid both for inductive and capacitive coupling, the details being hidden in the actual expressions for $d$ and $N.$

Tracing over the states of the field, we arrive at the following
reduced density matrix
\begin{eqnarray}
\rho_{AB}=\frac{1}{c}\left( \begin{array}{c c c c}
\rho_{11}&0&0&\rho_{14} \\
0&\rho_{22}&\rho_{23}&0\\
0&\rho_{23}^*&\rho_{33}&0\\
\rho_{14}^*&0&0&\rho_{44}
\end{array}\right) , \label{state}
\end{eqnarray}
representing the two-qubit state in the basis formed by $\ket{ee},$
$\ket{eg},$ $\ket{ge},$ and $\ket{gg}.$ The coefficients with the
leading order of neglected contributions are
\begin{eqnarray}
\rho_{11}&=&|V|_B^2+\mathcal{O}(d^4),~
\rho_{22}=1+2\mathrm{Re}(A)+\mathcal{O}(d^4)\nonumber\\
\rho_{33}&=&|X|^2+|G|^2+\mathcal{O}(d^6),~
\rho_{44}=|U|_A^2+\mathcal{O}(d^4) \nonumber\\
\rho_{14}&=&U_A^*V_B+\mathcal{O}(d^4)=\langle0|\mathcal{S}_A^+ \mathcal{S}_B^+|0\rangle +\mathcal{O}(d^4)\label{42j}\\
\rho_{23} &=& X^*+\mathcal{O}(d^4) , \nonumber
\end{eqnarray}
and the state is normalized $c=\sum_i \rho_{ii}.$

Let us now remark the validity of the perturbative methods applied in this work. The leading corrections to $\mathcal{C}(\rho_{AB})$ (see (\ref{eq:31s})) come from the leading order corrections to $\rho_{23}, \rho_{11}, \rho_{44}$ (Eq.~(\ref{42j})). In the case of $\rho_{23}$ we have 
\begin{equation}
\rho_{23}(d^4)=(1+A) X^*+FG^* \label{corrections1}
\end{equation}
and 
\begin{equation}
\rho_{23}(d^6)=\rho_{23}(d^4)+X_1+X_2,\label{corrections2} 
\end{equation}
where $X_1$ comes from the interference of one and two photon exchange amplitudes and $X_2$ comes from  the probability amplitude of three photon exchange. A rough upper bound for these two terms is given by $2\,|X|^3$.  For $\rho_{11}$ and $\rho_{44}$ they involve a number of photon emissions and re-absorptions by the same atom or by the other, giving a term 
\begin{equation}
\rho_{11}\rho_{44} (d^6)=|U_A|^2|V_B|^2+A_1+A_2, \label{corrections3}
\end{equation}
where rough upper bounds to $A_1$ and $A_2$ are  $2 |A||U_A|^2|V_B|^2$ and  $2|X||U_A|^2|V_B|^2 $ respectively . All these products are shown to be small for the regions of interest discussed here, $\xi < 2.$ The same techniques can be extended to all orders in perturbation theory since the bounds to the different contributions can be grouped and treated as power series, giving rise to corrections that remain negligible as long as $|A|$, $|X|$, $|U_A|^2$ and $|V_B|^2$ are small enough, like in the parameter range explored in this work. Finally, note that similar calculations and results can be obtained in the case in which the qubits have close but different frequencies.

\subsection{Entanglement dynamics and single photons}

We will use the concurrence $\mathcal{C}$ to compute the entanglement of the X-state in (\ref{state}) which is given by (\ref{eq:31s}).

Since all quantities depend only on two dimensionless numbers, $\xi$ and $K,$ we can perform a rather exhaustive study of the dynamics of entanglement between both qubits. To cover the widest possible spectrum of experiments, we have chosen coupling strengths (\ref{42g}) over three orders of magnitude, 
\begin{equation}
\frac{K}{K_0}= 1, 10, 100, 1000. \label{couplings}
\end{equation}
 The smallest value 
 \begin{equation}
 K_0= 1.5\cdot10^{-4},\label{smallest} 
 \end{equation}
 which corresponds to $g/\pi\simeq175$ MHz and $\Omega/2\pi\simeq10$ GHz, that is for instance a charge qubit in the strong coupling limit with a transmission line~\cite{blais04}. The largest value, $K=1000\,K_0$ corresponds to $g\simeq2\pi\times 500$ MHz and $\Omega \simeq 2\pi\times 2$ GHz, and typically corresponds to a flux qubit directly coupled to a transmission line~\cite{bourassa09,peropadre10}, as shown in Fig.~\ref{fig:setup}b-c. 
 \begin{figure}[h!]
\includegraphics[width=\linewidth]{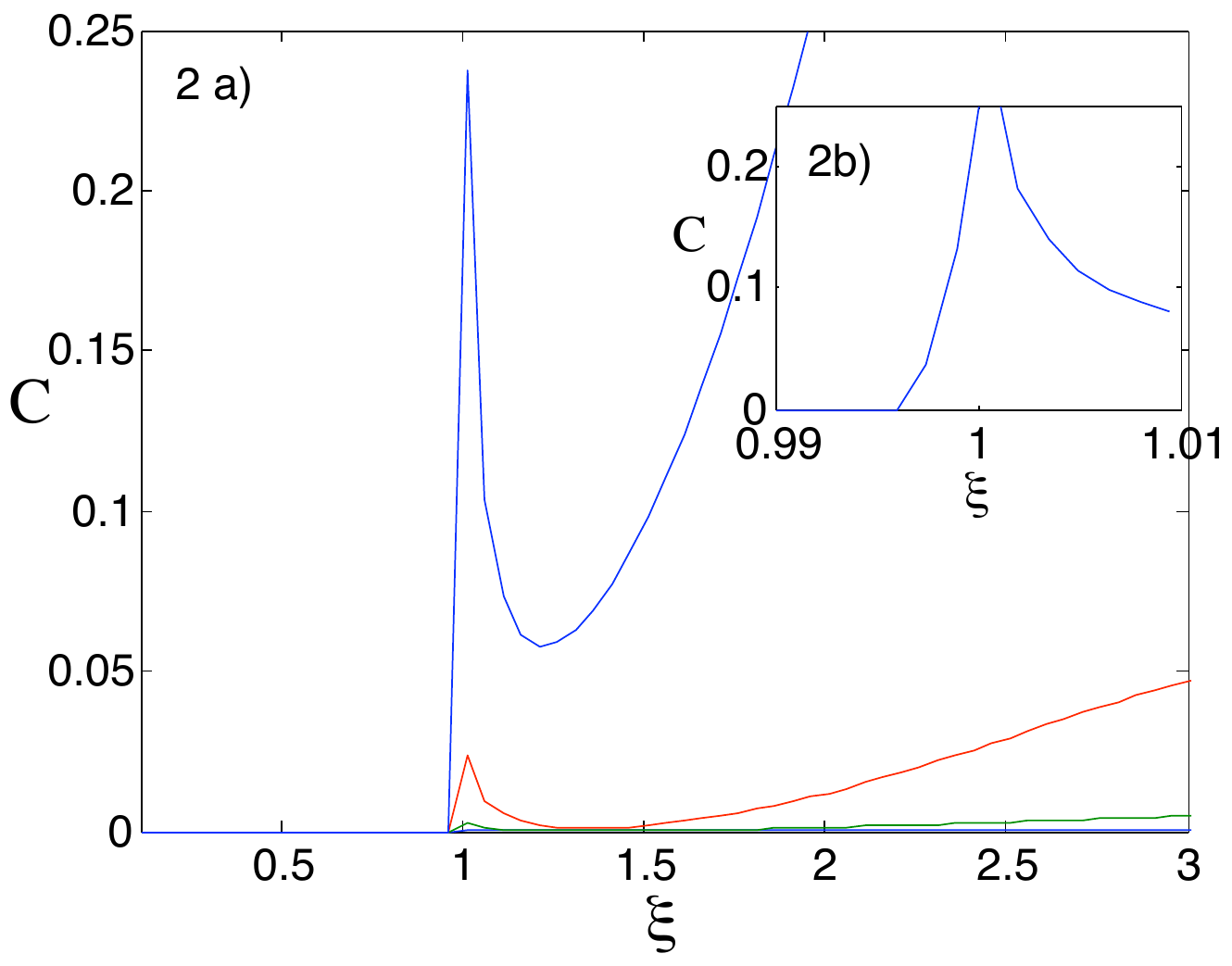}
\caption{ a) Concurrence vs. dimensionless separation $\xi$ for $r =
  \pi v / 4\Omega\sim \lambda/8$ and couplings $K= K_0$ , $10 K_0$ ,
  $100 K_0$ and $1000 K_0$ (bottom to top) b) Zoom around
  $\xi=1$ for the strongest coupling $K= 1000 K_0$. (Color online). }
\label{fig:concurrence}
\end{figure}

In Fig.~\ref{fig:concurrence} we plot the value of the concurrence for two qubits which are separated a distance $r=\lambda/8,$ using the couplings discussed before. Note how the entanglement jumps discontinuously to a measurable value right inside the light cone $(\xi > 1),$ signaling the arrival of photons. Furthermore, even a certain amount of entanglement appears outside the light cone, before photons could be exchanged. This is best seen for the largest couplings, as Fig.~\ref{fig:concurrence}b illustrates.

\begin{figure}[h!]
\includegraphics[width=\linewidth]{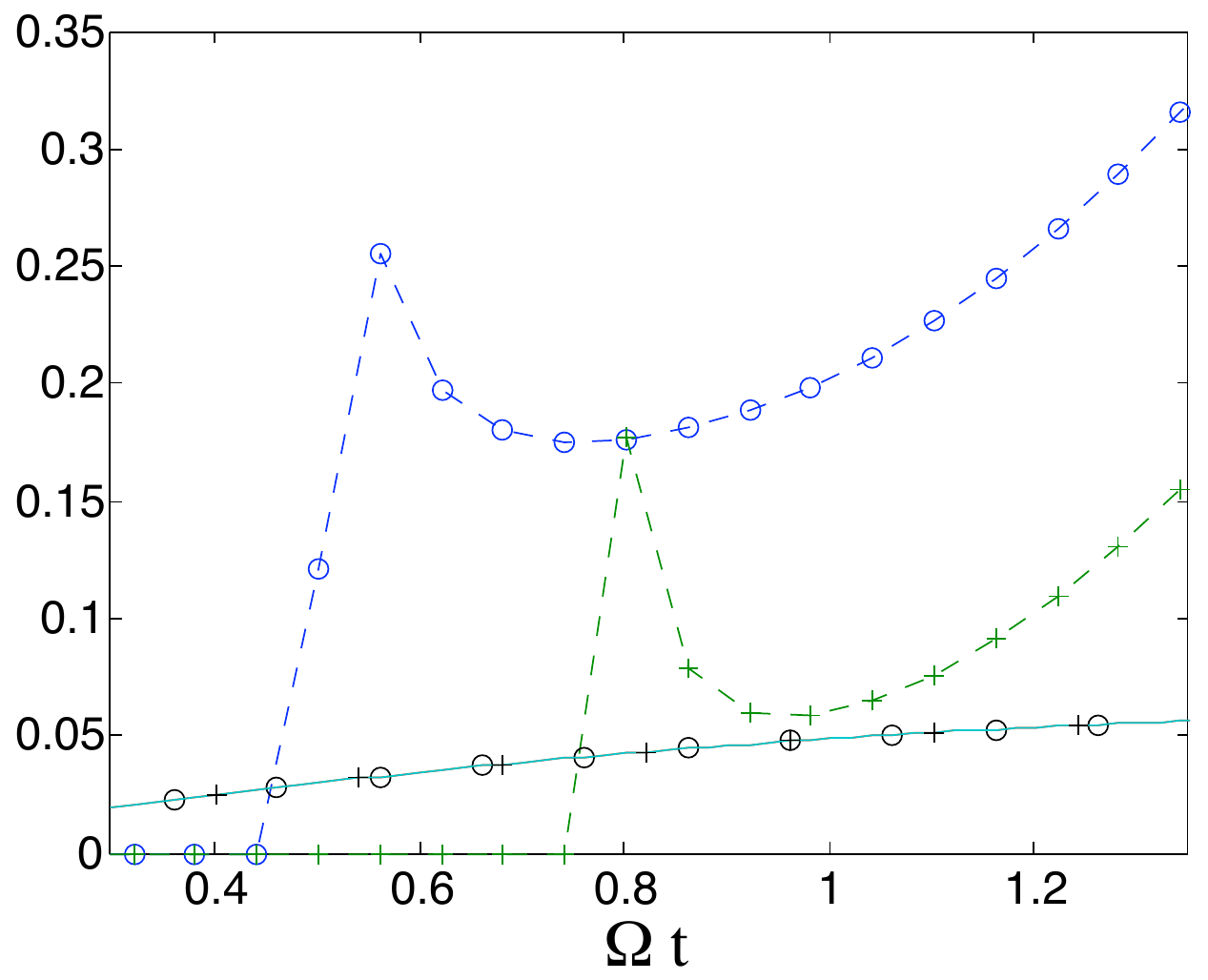}
\caption{Concurrence (dash) and probability of excitation of atom B (solid) vs. dimensionless time, $\Omega t.$ Qubits are separated by $r=\lambda/12$ (circles) and $\lambda/8$ (crosses) and have a coupling strength $K= 1000 K_0.$ Note that, following microcausality, the excitation probabilities do not depend on the separation $r$ outside the light cone (Color online).}
\label{fig:concurrence-exp}
\end{figure}

The dynamics looks even more exciting when we go back to lab time and space.  Fig.~\ref{fig:concurrence-exp} shows the concurrence and the excitation probability of qubit B, 
\begin{equation}
p_B=|V_B|^2/c+ {\cal O}(d^4),\label{excitation} 
\end{equation}
for two different separations, $r=\lambda/12$ and $r=\lambda/8.$ The probability of excitation appears as independent of the qubit separation. This is exactly the case for the lowest order considered here, which only accounts for B self-interaction, and at all orders in perturbation theory ~\cite{powerthiru} outside the light cone of this setup (region I in Fig.~\ref{fig:setup}a). This is in full agreement with microcausality. However, as can be seen in  Fig.~\ref{fig:concurrence-exp}, what was a tiny concurrence jumps to a sizable value when crossing the light cone $\Omega\, t = 2\pi/12$ and $\Omega\, t = 2\pi/8$. In other words, from the experimental point of view, it is the entanglement between the qubits and not the excitation probability $p_B$ what best signals the presence of a light cone and a finite propagation speed.

\subsection{Experimental implementation}

In order to study the dynamics of quantum correlations between the two superconducting qubits, one has to perform a partial or full tomography of their state. In the first and simpler case, performing measurements in different basis should be enough to gather an entanglement quantifier, such as a Bell inequality violation or, as studied in this paper, the concurrence. This has to be repeated many times, not only to gather sufficient statistics, but also to resolve different of instants of time before and after the light-cone boundary. This may seem a daunting task, but thanks to the speed at which quantum circuits operate and their fast repetition rate, it will be as demanding as recent experiments realizing a controlled-NOT gate~\cite{mooij07} or full two-qubit tomography~\cite{bialczak09}.

The actual experimental challenge, though, arises from the need to perform quantum measurements of the qubit state and ensuring that this state is not altered by the ongoing dynamics. One possibility is to perform very fast measurements of the qubits, which means faster than $1/\Omega.$ The typical response of measurement apparatus, which in the case of SQUIDs is around a few nanoseconds, sets an upper limit on the qubit and photon frequencies of a few hundreds of megahertzs, though we expect this to be improved in the near future.

Another more reliable approach is to connect and disconnect the coupling between the qubit and the transmission line. In this manner, we could prepare, entangle, and finally measure the qubits without interference or decay processes. If we work with flux qubits, a simple approach is to apply a very large magnetic flux on both qubits, taking the qubit away from its symmetry point. From a mathematical point of view, this amounts to adding a large contribution $E \sigma^x_{A,B}$ to the Hamiltonian. If done very quickly, the field projects the qubit on the same basis on which the coupling operates, eliminating the possibility of spontaneous emission. One would still need to combine the switching of this flux with short pulses that rotate the qubit basis in order to perform a complete set of measurements. The last and most elegant possibility is to effectively switch off all couplings between the qubit and the surrounding field. This can be achieved using a direct coupling between the qubit and the transmission line, with an scheme that incorporates an intermediate loop~[Fig.~\ref{fig:setup}c]. As Peropadre et al.  show in a recent work~\cite{peropadre10}, the result is a coupling that can be rotated and completely deactivated in a time of about 0.1 ns, that is the time needed to inject flux through the loop. The advantage is that, contrary to the case of a large external flux, the influence of the line is completely suppressed and makes it possible to easily rotate the qubits to perform all needed measurements.

\subsection{Conclusions}

Summing up, in this work we have proposed a circuit-QED experiment to study the dynamics of entanglement between two qubits that interact by exchanging traveling photons. Our work focuses on the existence of a finite propagation speed, the appearance of a light cone, the notion of microcausality and the possibility of achieving entanglement both by means of the correlated fluctuations of the vacuum and by photon exchange. The resulting predictions have a wide interest that goes beyond the assessment of microcausality in the QED of quantum circuits, demonstrating that the open transmission line is a useful mediator of entanglement, much like cavities and zero-dimensional resonators. Furthermore, the experiment we propose is also among the simplest ones that can probe the effective QFT for waveguides, both asserting the existence of propagating single photons and probing the dispersion relation at the single-photon level.  Finally, we have shown that entanglement via traveling photons works better for stronger qubit-line couplings, making it one of the first potential applications of the ultrastrong coupling regime \cite{niemczyk10, forn-diaz10}.

\chapter{Causality in matter-radiation interactions}
\begin{quote}
``She was particularly exasperated by the behavior of subatomic particles. She wanted the universe to behave sensibly'' (Martin Amis, \textit{The pregnant widow})
\end{quote}
Throughout this thesis we have analyzed several features of nonlocality in Quantum Mechanics, like the generation and destruction of entanglement between qubits outside their mutual light cone. We have remarked many times that these effects do not represent a violation of causality of any kind. In section 5. 1 of this chapter we will show explicitly that this is indeed the case, in the context of the so-called Fermi problem and the long-lasting controversy associated to it. Although the results are valid also for the 3-D interaction of real atoms and the electromagnetic field, we choose the theoretical framework of 1-D circuit QED in order to make a realistic experimental proposal. Since this proposal involves measurements of a qubit excitation probability at short times in a regime at which non-RWA contributions are relevant, in section 5.2 we deal with the effect of these contributions in the interpretation of the readouts.

\section{The Fermi problem with artificial atoms in circuit QED}
\subsection{Introduction}
Information cannot travel faster than light. But in quantum theory, as we have seen in chapters 3 and 4, correlations may be established between spacelike separated events. We remark again that these facts are not contradictory, since correlations need to be assisted with classical communication in order to transmit information.

 The two physical phenomena above arise in a natural fashion in the following situation,  which is the so-called Fermi problem \cite{fermi}, originally proposed by Fermi to check causality at a microscopic level. At $t=0$ a two-level neutral atom $A$ is in its excited state and a two-level neutral atom $B$ in its ground state, with no photons present. If $A$ and $B$ are separated by a distance $r$ and $v$ is the speed of light, can $A$ excite $B$ at times $t<r/v$?  Fermi 's answer was negative but his argument had a mathematical flaw. When a proper analysis is carried on, fundamental quantum theory questions arise due to the interplay between causal signaling and quantum non-local phenomena. 
 
These issues led to a controversy \cite{hegerfeldtfer,yngvasonfer,yngvasontipo3,powerthiru} on the causal behavior of the excitation probability of qubit B, whose conclusions were never put to experimental test. A notorious claim on causality problems in Fermi's two-atom system was given in \cite{hegerfeldtfer}. The reply of \cite{yngvasonfer} was in the abstract language of algebraic field theory and the proof of strict causality of \cite{powerthiru} is perturbative,  although given to all orders in perturbation theory.  The Fermi problem is usually regarded just as a gedanken experiment, and remains untested, essentially because interactions between real atoms cannot be switched on and off.

In this section we give a complete description of the problem in a physical framework in which predictions can be verified. This framework will be circuit QED  which  can be regarded as a 1-D version of Quantum Electrodynamics (QED) with two-level (artificial) atoms, a testbed which makes it possible to control the interaction and tune the physical parameters. We complete previous descriptions made of the problem and explain how there are no real causality issues for Fermi's two-atom system. We give an explicit non-perturbative proof of strict causality in these setups, showing that the probability of excitation of qubit $B$ is completely independent of qubit $A$ for times $t<r/v$ and for arbitrary initial states. As a matter of fact, this comes as a manifestation of the nonsignaling character of the quantum theory~\cite{gisin}. We also show how this is compatible with the existence of nonlocal correlations at times $0<t<r/v,$ a fact pointed out in various theoretical proposals to entangle qubits at arbitrarily short times~\cite{reznik,franson} (and chapters 3 and 4 of this Thesis). More precisely, we give a non-perturbative proof of the fact that the probability of $B$ being excited and $A$ in the ground state is finite and $r$-dependent  at any time, even for $t<r/v$. We provide also a physical and intuitive explanation of why the conclusions in \cite{hegerfeldtfer}, even if mathematically sound, do not apply to the causality problem. At the end of the Section we discuss the time dependence predicted in our model for the various excitation probabilities and suggest a feasible experimental test of causality using superconducting circuits.
\subsection{There are no causality problem in Fermi's two-qubit system}
 In what follows  we focus
on a practical setup of circuit-QED, with two qubits, $A$ and $B,$
interacting via a quantum field. The qubits have two stationary states
$\ket{e}$ and $\ket{g}$ separated by an energy $\hbar\Omega$ and
interact with a one-dimensional field, $V(x),$ (\ref{field}) which propagates
along a one-dimensional microwave guide that connects them.
This field has a continuum of Fock operators $[a_k,a^{\dag}_{k'}]=\delta(k-k'),$ and a linear spectrum, $\omega_k =
v|k|$, where $v$ is the propagation velocity of the field. The normalization and the speed of photons, $v$ (\ref{propagationvelocity}) depend on the microscopic details such as the capacitance and inductance per unit length, $c$ and $l.$ We will assume qubits that are much smaller than the relevant wavelengths, $\lambda=v/\Omega,$ and are well separated. Under these conditions the Hamiltonian, $H = H_0 + H_I,$  splits into a free part (\ref{free}) for the qubits and the field and a point-like interaction between them (\ref{interaction}).

The original formulation of the Fermi problem begins with an initial state
\begin{equation}
 \ket{in} = \ket{e_A\,g_B\,0} \label{51d}
\end{equation}
in which only qubit $A$ has been excited, while $B$ and the field remain in their ground and vacuum states, respectively. The total probability of excitation of qubit $J$ is the expectation value of the projector onto the excited state 
\begin{equation}
\mathcal{P}^e_J=\ket{e_J}\bra{e_J}. \label{51e}
\end{equation}
In the Heisenberg picture
\begin{equation}
P_{eJ}=\bra{in}\mathcal{P}^e_J (t)\ket{in},\quad J\in\{A,B\}.
\label{eq:probability}
\end{equation}
We will prove that for $vt<r$ the probability $P_{eB}$ is \textit{completely independent of the state of qubit $A$ for all initial states.} In the Heisenberg picture this amounts to showing that there appears no observable of $A$ in the projector $\mathcal{P}^e_B(t)$ for $vt<r.$ Our proof begins by solving formally the Heisenberg equations for $\mathcal{P}^e_J$
\begin{equation}
\mathcal{P}^{e}_J (t)- \mathcal{P}^{e}_J (0)=- \frac{d_J}{\hbar} \int_0^t dt'\sigma^y_J (t') V(x_J,t').
\label{eq:evolution}
\end{equation}
Integrating the Heisenberg equations of the operators $a_k$ and $a_k^{\dagger}$ and inserting them in Eq.~(\ref{a}), the total field evaluated at $x$ in Heisenberg picture is decomposed
\begin{equation}
V(x,t)= V_0(x,t)+V_A(x,t)+V_B(x, t)
\label{eq:fielddecomposed} 
\end{equation}
into the homogenous part of the field
\begin{equation}
V_0(x,t)=i\,\int_{-\infty}^{\infty}dk\,\sqrt{N\omega_k}\,e^{i(kx-\omega t)}a_k +\mathrm{H.c.}
\label{eq:fielddecomposed1} 
\end{equation}
and the back-action of $A$ and $B$ onto the field
\begin{eqnarray}
&&V_J(x,t)=\frac{-id_J\,N}{\hbar}\times
\label{eq:fieldgeneral}\\
{}&&\quad\times \int_0^t \sigma_J^x (t')
\int_{-\infty}^{\infty}\omega_ke^{ik(x-x_J)-i\omega_k\,(t-t')}dkdt'
+\mathrm{H.c.}\nonumber
\end{eqnarray}
Eqs.~(\ref{eq:evolution}) translates into a similar decomposition for the  probabilty $\mathcal{P}^e_B$ with three terms
\begin{equation}
\mathcal{P}^e_B (t)= \mathcal{P}^{e}_{B0} (t)+ \mathcal{P}^{e}_{BB} (t)+ \mathcal{P}^{e}_{BA} (t)
\label{eq:causality}
\end{equation}
which are proportional to $V_0, V_B$ and $V_A,$ respectively.  The only explicit dependence on $A$ may come from $\mathcal{P}^e_{BA}$ through $V_A(x_B,t).$ Manipulating Eq.~(\ref{eq:fieldgeneral}) gives
\begin{eqnarray}
V_A (x_B,t)=\frac{-2\pi d_A N}{\hbar} \frac{d}{dr}\left[\sigma_A^x\left(t-\frac{r}{v}\right)\,\theta\left(t-\frac{r}{v}\right)\right]\label{eq:field2}
\end{eqnarray}
where the Heaviside function $\theta$ shows that strictly $\mathcal{P}^e_{BA}(x_B,t)=0$ for $vt<r,$ and no such dependence is possible. We still have to analyze a possible implicit dependence on A through $\mathcal{P}^e_{BB},$ whose expression is
\begin{eqnarray}
\mathcal{P}^e_{BB} (t)&=&\frac{id_B^2N}{\hbar^2}\int_0^t dt'\int_0^{t'}dt''\sigma_B^y (t')\sigma_B^x (t'')\nonumber \\ & &\int_{-\infty}^{\infty}dk\,\omega_k e^{-i\,\omega_k(t'-t'')}+ \mathrm{H.c.} \label{eq:implicit}
\end{eqnarray}
The only implicit dependence could come through the evolution of  $\sigma_B^{x,y} (t),$ but again this is not the case. Since 
\begin{equation}
[\sigma_B^x,H_I]=0, \label{51f}
\end{equation}
the evolution of $\sigma_B^x$ does not involve the field in any way, and for $\sigma_B^y (t)$ we have that 
\begin{equation}
\dot{\sigma}_B^y (t)=\Omega\,\sigma_B^x(t)/2\, - \frac{d_B}{\hbar} V(x_B,t)\sigma_B^z (t) \label{51g}
\end{equation}
so using again Eq. (\ref{eq:fielddecomposed}) and Eq. (\ref{eq:field2}) we see that  the A-dependent part of $P^e_{BB}$ is 0 for $vt<r$.  Thus $\mathcal{P}^e_B$ may be finite but is completely independent of qubit $A$ for $vt<r$, as we wanted to show.

So far, we have demonstrated that although $P^e_B (t)$ is non-zero for $vt<r,$ the only non-zero contribution is $\mathcal{P}^e_{B0},$ which is not sensitive to the qubit $A$ and thus cannot be used to transmit information between the qubits. Now we will show that this result is compatible with the existence of correlations for $vt<r$. For instance, we consider the probability of finding qubit $B$ excited and qubit $A$ on the ground state $P_{eB,gA}$, which is:
\begin{equation}  
P_{eB,gA}=\bra{in}\mathcal{P}^e_B (t)\mathcal{P}^g_A (t)\ket{in},
\label{eq:correlation}
\end{equation}
where 
\begin{equation}
\mathcal{P}^g_A=\mathbf{1}-\mathcal{P}^e_A. \label{51h}
\end{equation}
Using Eqs.~(\ref{eq:evolution}), (\ref{eq:fielddecomposed}) and (\ref{eq:causality}), we find a term in this probability which is proportional to $\mathcal{P}^{e}_{BB} \, \mathcal{P}^{g}_{AA}$ and thus to  $V_B(x_B,t)\,V_A(x_A,t)$. From Eq.~(\ref{eq:fieldgeneral}) we obtain: 
\begin{equation}
V_J(x_J,t)=\frac{2\pi d_JN}{\hbar v}\frac{d}{dt}\{\sigma_J^x(t) \theta(t)\} \label{51i}
\end{equation}
Therefore, we conclude that in  (\ref{eq:correlation}) there is an unavoidable dependence on $A$ at any $t>0,$ but this is not a causality violation because correlations alone cannot transmit information.

At this point it remains a single question:  How can the A-dependent part of  $P_{eB}$ be zero while the one of  $P_{eB,gA}$  is nonzero for $vt<r$? To better understand it we need less formal results that rely on perturbative expansions, but we would like to remark here that the conclusions above are valid to all orders in perturbation theory.
\begin{figure}[h!]
\centering
\includegraphics[width=\linewidth]{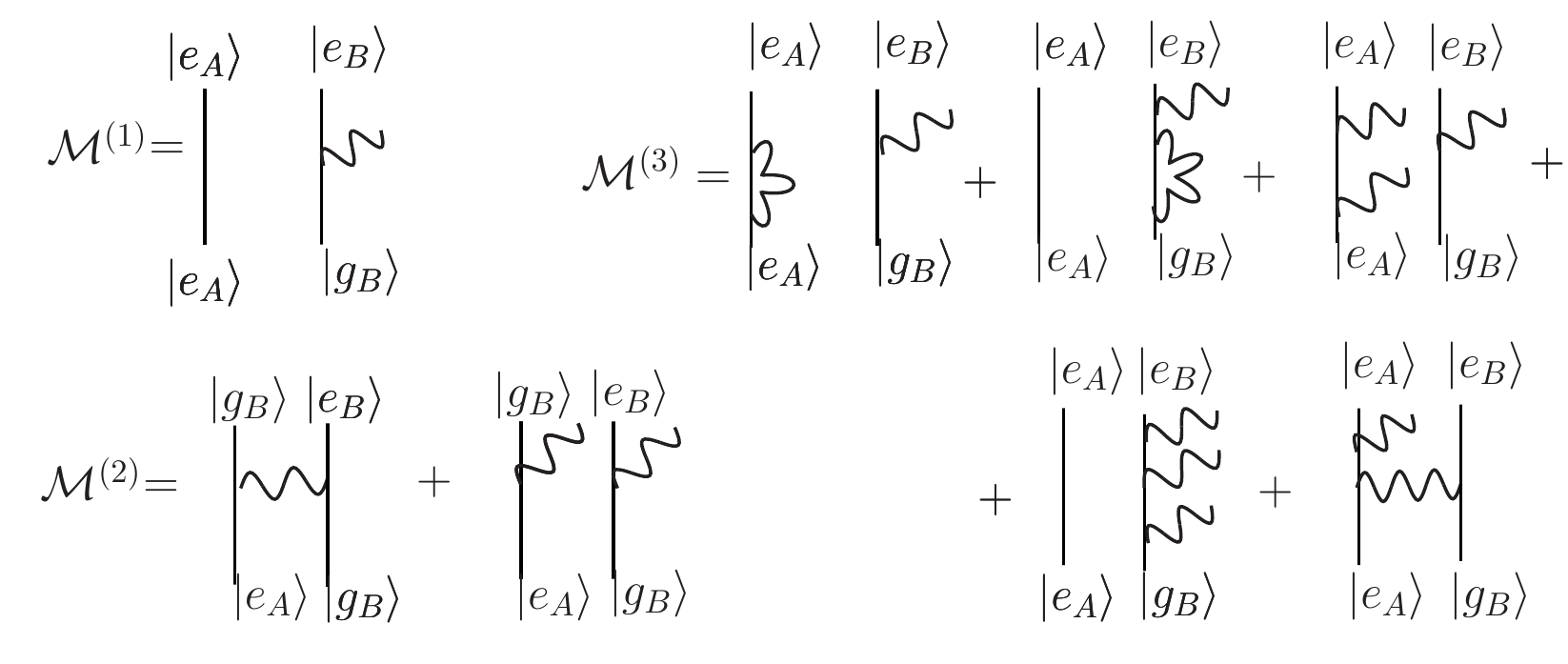}
\caption{The terms of order $d_J, d_J^2$ and $d_J^3$ contributing to the amplitude for exciting qubit $B$. } \label{Fig.31}
\end{figure}
To obtain the total probability of excitation of qubit $B$ $P_{eB}$ to a given order in perturbation theory, one has to expand to a certain order the operators appearing in Eqs.~(\ref{eq:evolution}), (\ref{eq:fielddecomposed}), (\ref{eq:fieldgeneral}). The different terms in the expansion can be related to the probabilities of the different physical processes involved. Fig.~\ref{Fig.31} shows the diagrams of the different amplitudes contributing to $P_{eB}$ up to the fourth order in $d_J.$ The lowest order amplitude contributing to a final excited $B$ qubit is of order $d_J,$ which means that terms up to order $d_J^3$ have to be considered. The only terms leading to this final state will be 
\begin{equation}
\mathcal{M}^{(1)} = V_B,\;\, \mathcal{M}^{(2)} = X + U_A\,V_B,\,\;  \mathcal{M}^{(3)}= A' V_B +U_AV_AV_B+ V_BU_BV_B+\mathcal{\delta M}^{(3)} \label{51j}
\end{equation}
where $U_{J}$ ($V_J$) represent the amplitude for single photon emission at qubit $J$ when the qubit is initially in the ground (excited) state, $x$ is the amplitude for photon exchange, $A_J$  are the  radiative corrections of qubit $J$, and finally  $\mathcal{\delta M}^{(3)}$ is the amplitude for photon exchange accompanied by a single photon emission at qubit $A$. Notice that some of these processes are only possible beyond the rotating wave approximation, which breaks down for strongly coupled circuit-QED setups~\cite{bourassa09} as the ones considered later. Keeping only terms up to fourth order, we have for the probability to get B excited at a time $t$
\begin{eqnarray}
P_{eB}(t) &=& |\mathcal{M}^{(1)}|^2 +|\mathcal{M}^{(2)}|^2 + 2\, Re\{\mathcal{M}^{(1)^*} \mathcal{M}^{(2)}\} \nonumber\\ 
&+&2\, Re\{\mathcal{M}^{(1)^*} \mathcal{M}^{(3)}\}+ \mathcal{O}(d^5)\label{51s}
\end{eqnarray}
The final states in $\mathcal{M}^{(1)}$ are orthogonal to those in $\mathcal{M}^{(2)}$ and to the three photon terms in $\mathcal{M}^{(3)}$. Hence, their interference vanishes. Besides, we are only interested in the $A$-dependent part of the probability, so we can  remove the $r$-independent terms left in (\ref{51s}), marking the remaining contributions with a superscript ${}^{(r)}$
\begin{equation}
\label{eq:perturbation}
P_{eB}^{(r)}(t) = |\mathcal{M}^{(2)}|^{2^{(r)}} + 2 Re\{\mathcal{M}^{(1)^*}\mathcal{\delta M}^{(3)}\} + \mathcal{O}(d^5).
\end{equation}
The first term actually gives $P_{eB,gA}$ up to the fourth order, it is positive and $A$-dependent at all times, as shown in Fig.~\ref{fig32}a. The second term is not a projector onto any physical state, but an interference term which has the effect of canceling out exactly the first term for $vt<r$ but not for $vt>r$ (cf. Fig.~\ref{fig32}b). In a nutshell, interference seems to be the physical mechanism that operates at all orders in perturbation theory to give the causal behavior of the total probability of excitation that we had previously shown. (See Appendix C for more details on the computation of $\mathcal{\delta M}^{(3)}$.)

These perturbative results cast new light on the controversy on the Fermi problem and help us understand \textit{why} our results do not contradict those of Hegerfeldt~\cite{hegerfeldtfer}. Hegerfeldt proved mathematically that the expectation value of an operator consisting of a sum of projectors cannot be zero for all the times $vt<r,$ unless it is zero at any time. Indeed, the expectation value of $\mathcal{P}^e_B (t)$ cannot be zero for all $vt<r,$ for it always contains the contribution $\mathcal{P}^e_{B0}$ from Eq.~(\ref{eq:fielddecomposed1}).  However, as we showed non-perturbatively, the actual relevant question for causality is whether the expectation value of $\mathcal{P}^{e}_{BA}(t)$ vanishes for $vt<r$ or not, since only this part of the probability is sensitive to qubit $A$ and could be used to transmit information. Besides, according to our above perturbative results to fourth order, the $r$-dependent part of the probability, that is the expectation value of  $\mathcal{P}^{e}_{BA}(t)$, is not a mere sum of projectors, but also contains interfering terms. Thus, Hegerfeldt's result does not apply  and  $\mathcal{P}^{e}_{BA}(t)$ can be zero for $vt<r$ as is actually the case. Both results are in accord with a general fact of Relativistic Quantum Field Theory: two global states can not be distinguished locally with the aid of a local projector annihilating one of the states, since the local observable algebras are Type III von Neumann algebras (See \cite{yngvasonfer,yngvasontipo3} for a discussion).
\begin{figure}[h!]
\begin{center}
\includegraphics[width=\linewidth]{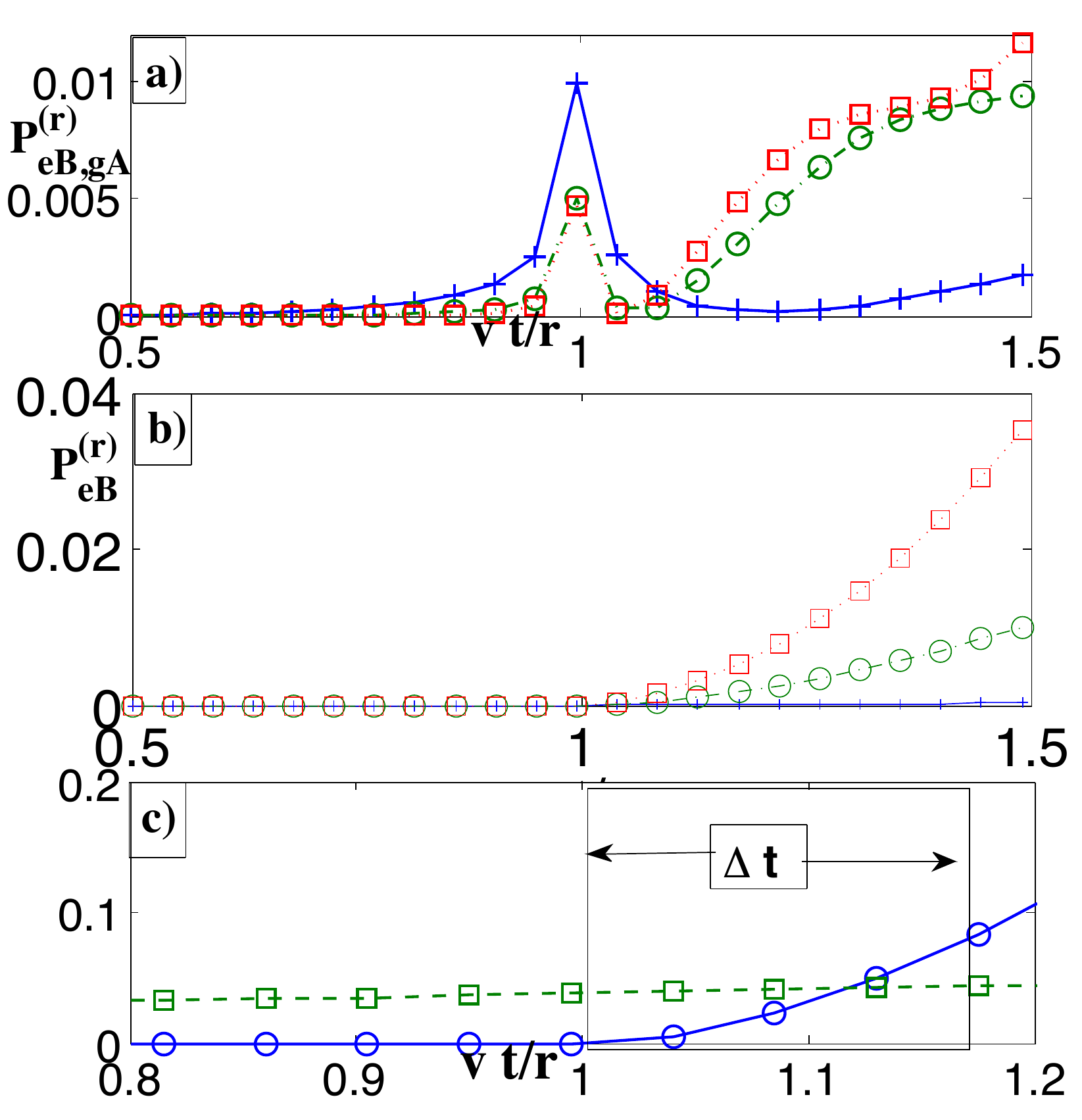}
\end{center}
\caption{(a) $P^{(r)}_{eB,gA}$ and (b) $P^{(r)}_{eB}$ versus $v t/r$ for $\Omega r/v= \frac{\pi}{2}$ (blue, crosses), $\pi$ (red, squares), and $2\pi$ (green, circles) with $K_{A,B}=0.0225$. For $vt<r$ the qubits are spacelike separated, but there are correlations between them and figure (b) shows the expected causal behavior. (c) $P^{(r)}_{eB}$  (blue, circles) and  $|\mathcal{M}^{(1)}|^2$ (green, squares) vs.  $v t/r$ for $K_A=0.20$, $K_B=0.04$ and a separation of one wavelength $r = 2\pi v/\Omega_{A,B}.$ With this data and $\Omega/2\pi\simeq1$GHz we have $\Delta\,t\simeq1$ns. (Color online) } \label{fig32}
\end{figure}
\subsection{Experimental proposal}
 We will now suggest an experiment to check the causal behavior of  $P_{eB}.$ For this we need to control the interaction time at will to access the regions at both sides of $t=r/v.$ This, which is highly unrealistic with real atoms, becomes feasible in circuit-QED. While the ideas are valid for both inductive and capacitive couplings, we will focus on using a pair of three-junction flux qubits \cite{mooij,reviewflux}. Each of the qubits is governed by the Hamiltonian 
 \begin{equation}
 H_{0J}= \frac{1}{2}\epsilon_J \sigma^z_J+ \frac{1}{2}\Delta_J\sigma^x_J.\label{landauzener1}
 \end{equation}
 The energy $\epsilon_J=2I_p\delta\Phi_{xJ},$ is approximately linear in the external magnetic flux, $\delta\Phi_{xJ},$ measured from the degeneracy point, and we assume that the gap $\Delta_J$ is fixed. The result is a qubit energy difference 
 \begin{equation}
 \Omega_J(\delta\Phi_{xJ})=\sqrt{(2\,I_p\,\delta\Phi_{xJ})^2+\Delta_J^2}.\label{landauzener2}
\end{equation}
The coupling between the qubit and the microwave photons is ruled by the dimensionless ratio
\begin{equation} \label{eq:couplingstrength}
K_{J}=\frac{4d_{J}^2N}{\hbar^2 v}= 2\left(g/\Omega_J\right)^2.
\end{equation}
Here 
\begin{equation}
g_J=\frac{d_J\sqrt{N\Omega}}{\hbar} \label{coupling} 
\end{equation}
is the coupling strength between a qubit and the cavity that would be obtained by cutting the transmission line to be perfectly resonant with the qubit transition. These numbers enter the qubit excitation probability computed before (\ref{eq:perturbation}) through the product 
\begin{equation}
P^{(r)}_{eB}(t) \propto K_AK_B. \label{probabilitik}
\end{equation}
Since 
\begin{equation}
K_J\propto 1/\Omega_J, \label{kajotaomegaj}
\end{equation}
we may use the external fluxes to move from a weakly coupled regime with no qubit excitations, $\Omega_J \ll g,$ to the maximum coupling strength, $\Omega_J\simeq \Delta_J~(\delta\Phi_{xJ}=0).$

Let us first discuss how to prepare the initial state (\ref{51d}) of the Fermi problem. We assume that the system starts in a ground state of the form $\ket{g_A\,g_B\,0}$. This is achieved cooling with a large negative value of $\delta\Phi_{xJ}$ on both qubits, which ensures a small value of $g/\Omega_J$ and $K_J.$ We estimate that couplings  $g/\Omega_J<0.15$ and $\Omega_J \sim 1.5$GHz lower the probability of finding photons in the initial state below $5\times10^{-3},$ both for vacuum and thermal excitations. Both magnetic fluxes are then raised up linearly in time, 
\begin{equation}
\delta\Phi_{xJ}=\alpha_J t, \label{externalflux}
\end{equation}
to prepare the qubits. Using a Landau-Zener analysis~\cite{zener} of the process we conclude that an adiabatic ramp   
\begin{equation}
\alpha_B\ll \pi\Delta_B^2 / 4 \hbar I_p \label{adiabaticramp}
\end{equation}
of qubit $B$ followed by a diabatic ramp \cite{johanssoncasimir1, johanssoncasimir2} 
\begin{equation}
\alpha_A\gg \Delta_A^2/\hbar2I_p \label{diabaticramp}
 \end{equation}
 of qubit A, leads to the desired  state $\ket{e_A\,g_B\,0}$ with a fidelity that can be close to 1, depending only of $ \alpha_A$, $\alpha_B$ as derived from the Landau-Zener formula:
\begin{equation}
P_{J}=e^{-\frac{2\pi}{\hbar}\, \frac{(\frac{\Delta_J}{2})^2}{2\,I_p\,\alpha_J}}.\label{landauzener2}
\end{equation}
 Note that the minimum gap $\Delta_B$ has to be large enough to ensure that the qubit-line coupling of B remains weak and the qubit does not ``dress" the field with photons.

Once we have the initial state, both magnetic fluxes must take a constant value during the desired interaction time.  After that, measurements of the probability of excitation of qubit B can be performed with a pulsed DC-SQUID scheme \cite{measurementqubits1, measurementqubits2}. The timescale of the ``jump''   of the probability around $t=r/v$ for qubit frequencies in the range of GHz and a separation of one wavelength $r=2\pi v/\Omega$ is 
\begin{equation}
\Delta t\simeq \,1ns \label{jump}
\end{equation}
 [Fig.~\ref{fig32}c].  Although the total measurement of the SQUID may take a few $\mu$s, the crucial part is the activation pulse ($\sim 15$ns) in which the SQUID approaches its critical current and may switch depending on the qubit state. During this activation period the SQUID and the qubit are very strongly coupled $(g\sim\mathrm{GHz}),$~\cite{measurementqubits1} and the qubit is effectively frozen. The time resolution of the measurement is thus determined by the ramp time of the activation pulse, which may be below nanoseconds. Among the sources of noise that are expected, the short duration of the experiment, well below $T_1$ and $T_2$ of usual qubits, makes the ambient noise and decoherence pretty much irrelevant. Thermal excitations of the qubits and the line may be strongly suppressed by using larger frequencies ($>1.5$GHz). The most challenging aspect is the low accuracy of SQUID measurements, which are stochastic, have moderate visibilities~\cite{measurementqubits1} and will demand a large and careful statistics.

On the technical side, it is important to choose carefully the coupling regimes. If we wish to compare with perturbation theory, we need $K_J \ll 1.$ However, at the same time the product $K_AK_B$ must take sizable values for $P_{eB}\propto K_AK_B$ to be large. And we need to discriminate the causal signal from the $r$-independent background of the probability of excitation, whose main contribution is $|\mathcal{M}^{(1)}|^2\propto K_B$. Thus, a good strategy would be to work with $K_A>K_B.$  In Fig. \ref{fig32}c we show that it is possible to achieve a regime in which the perturbative approximations are still valid and the $r$-dependent part of $P_{eB}$ is comparable to 
$|\mathcal{M}^{(1)}|^2$ in the spacetime region of interest $v t\simeq r$. 
\subsection{Conclusions}
In this section, we have considered a system of two superconducting qubits coupled to a transmission line, which can be suitably described in the framework of 1-D QED with two-level (artificial) atoms. Starting from an initial state with qubit A excited, qubit B in the ground state and no photons, we have illustrated the causal character of the model  showing that the probability of excitation of qubit B is completely independent of qubit A when $vt<r$. We have also shown that this is in agreement with the existence of nonlocal correlations and we have used perturbative computations to see the physical mechanism underlying the causal behavior. Finally, we have suggested an experiment feasible with current technology that would solve the controversy on the Fermi problem.
 \section{Quantum fluctuations and short-time quantum detection}
\subsection{Introduction}
 A counterintuitive direct consequence of the breakdown of the RWA is that a detector in its ground state interacting with the vacuum of the field has a certain probability of getting excited and emitting a photon. There is  however not a widespread consensus on the physical reality of this effect. Introducing counterrotating terms is interpreted by some to be a problem as the processes described by those terms seem virtual.  It seems difficult to accept that a detector in a ground state in the vacuum could get excited.  As a matter of fact, there have been attempts of suggesting effective detector models by imposing this phenomenon to be impossible \cite{piazzacosta}. 
We should however recall here that these peculiar effects should not be that discomforting. They are linked to the fact that the initial state considered has not  definite energy, since the state ``detector and field in their ground states" is not an eigenstate of the full Hamiltonian beyond RWA.   Indeed, in section 4.1 we have devised an experimental protocol to detect the trace of this ground-state qubit self-excitations in a simplified-setup with a qubit in a cavity.


In this section we will keep using the theoretical framework applicable to circuit QED, without imposing any additional constraint. We will study the following setup:  a source $A$ initially excited, a detector $B$ initially in the ground state and both interacting with the electromagnetic field in its vacuum state.  If the detector clicks at a given time, does it mean that the source is now in the ground state? This problem amounts to compute the probability of decay of the source, conditioned to the excitation of the detector. We will show that, unlike Glauber's RWA detector in which this conditioned probability would be equal to 1 at any time, this circuit QED  detector only achieves this value at long times due to the impact of non-RWA effects, like the ground-state qubit self-excitations described above. We will see how these theoretical results have to be taken into account for the interpretation of the readouts of real ongoing experiments.
\subsection{What does a detector's click mean?}
More precisely, let us consider an initial moment $t=0$ where $A$ is excited, $B$  is in its ground state and there are no excitations in the transmission line, that is, the initial state would be Eq. \ref{51d}.
After a certain time $t$, if  we measure qubit $B$  and it results excited, that would na\"{i}vely lead us to think $A$ has decayed and produced a photon which  has then later been absorbed by $B$. We intend to proof otherwise by quantifying what information about the state of $A$  can be extracted by knowing qubit $B$ state after a certain time $t$. For that we will compute  the probability $P_{gA/eB} (t)$ of $A$ to have decayed at a certain instant $t$, conditioned we have measured $B$ excited at that same moment: 
\begin{equation}
P_{gA/eB} (t)=\frac{P_{eB,gA}}{P_{eB}},
\label{eq:conditionedprobability}
\end{equation}
 $P_{eB,gA}$ being the probability of $A$ being in the ground state and $B$ excited and $P_{eB}$ the total probability of excitation of $B$, as they have been defined in the previous section 5.1. Within RWA $P_{gA/eB} = 1$ at any time. But beyond RWA and up to the fourth order in perturbation theory, we have seen in the previous section that $P_{eB}$ is given by Eq. \ref{51s} and that
 \begin{equation}
 P_{eB,gA}=|\mathcal{M}^{(2)}|^{2} 
  \label{52a}
 \end{equation}
 \begin{figure}[h!]
\begin{center}
\includegraphics[width=\textwidth]{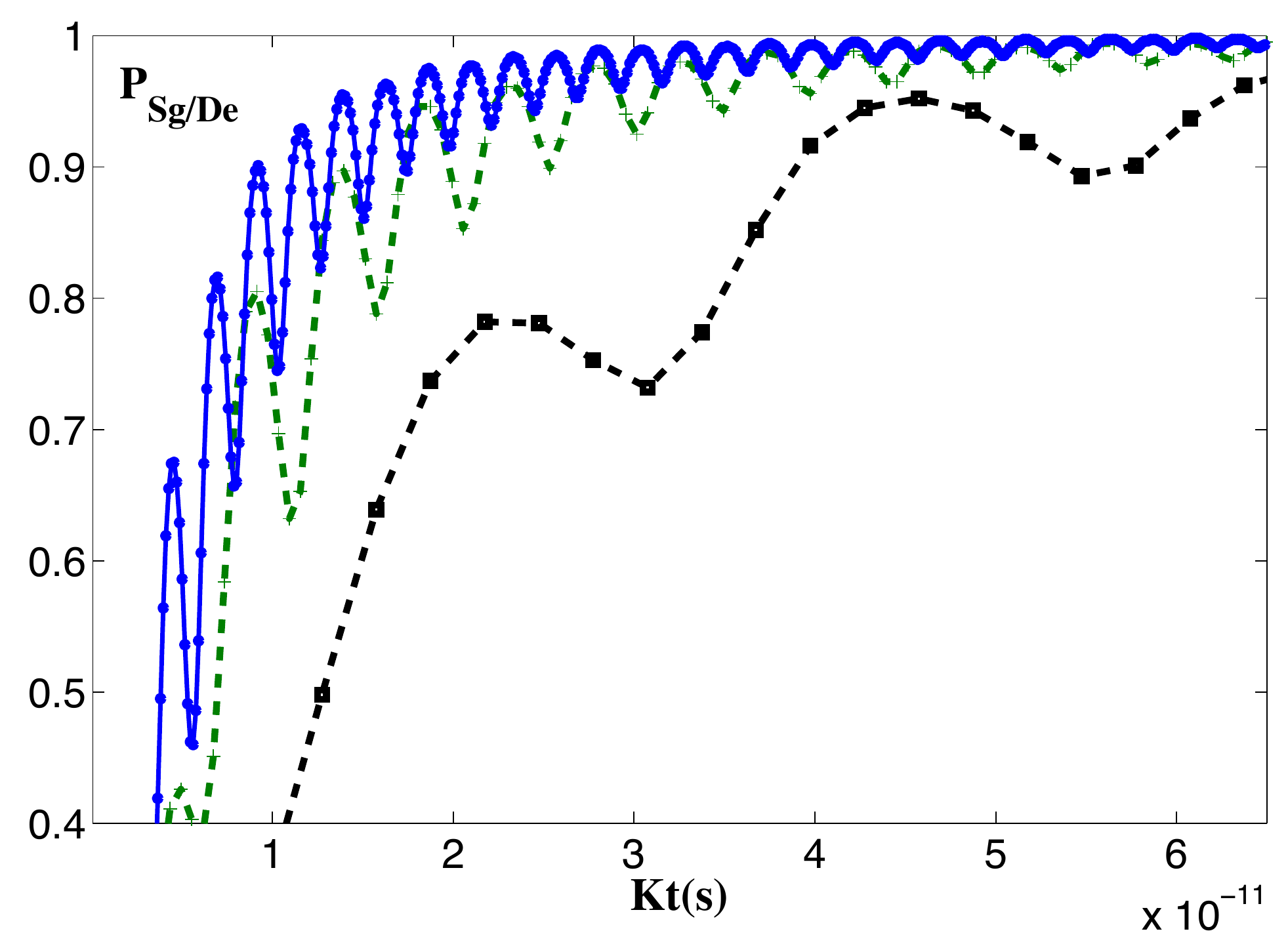}
\end{center}
\caption{$\mathcal{P}_{gA/eB} (t)$ (\ref{eq:conditionedprobability}) in front of $Kt$ for three different values of the  coupling strength of $K=\,K_A= K_B=7.5\cdot 10^{-3}$ (solid, blue, circles), $1.5\, 10^{-2}$(dashed, green, crosses), $7.5 \cdot 10^{-2}$  (dashed, black, squares). In the three cases  $2\pi\frac{r}{\lambda}= 1$ and  $\Omega/(2\pi) = 1\, GHz$ ($\Omega=\Omega_A=\Omega_B$).  } 
\label{fig:521}
\end{figure}
\begin{figure}[h!]
\begin{center}
\includegraphics[width=\textwidth]{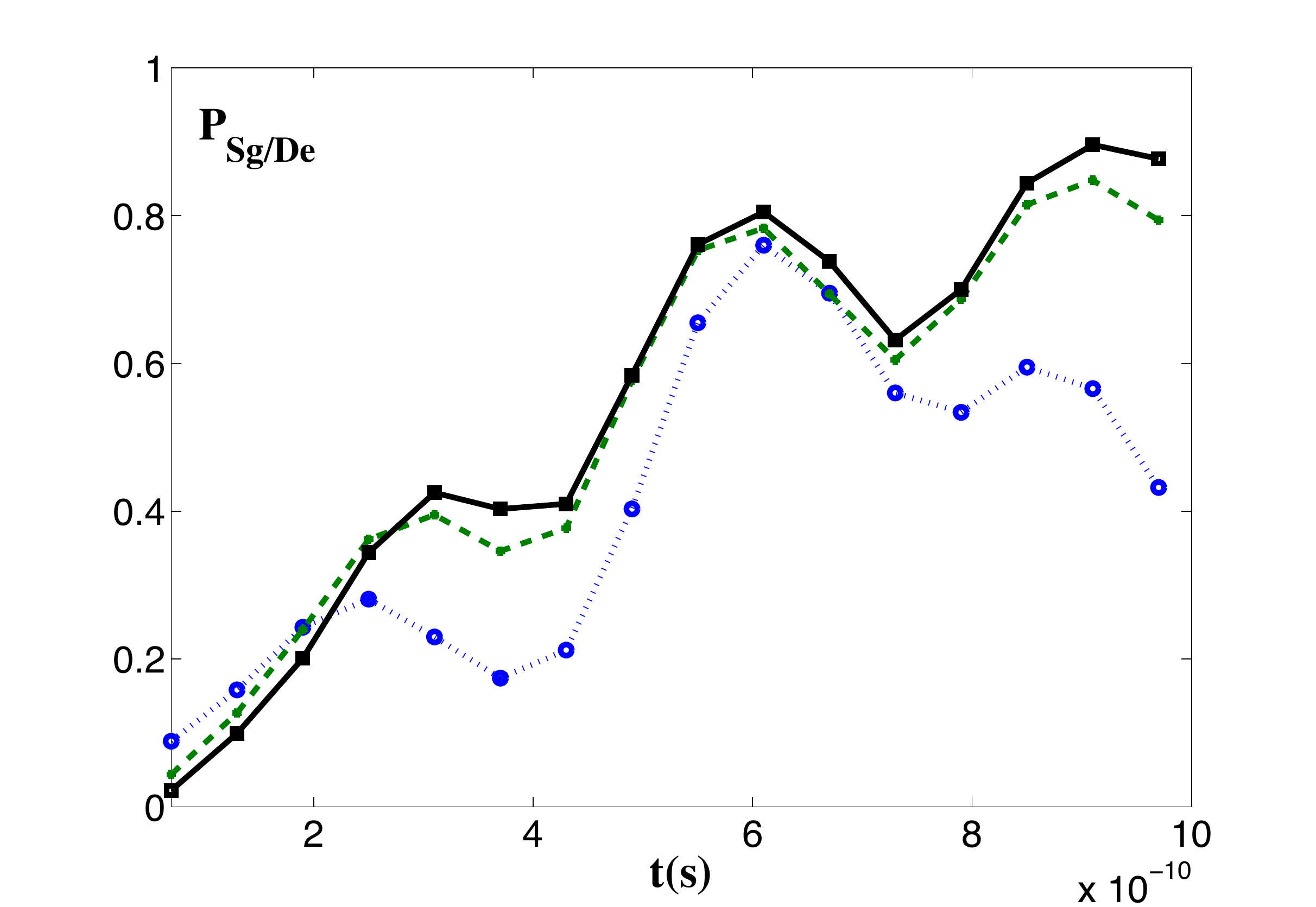}
\end{center}
\caption{$\mathcal{P}_{gA/eB} (t)$ (\ref{eq:conditionedprobability}) in front of $t$ in $s$ for three different values of the distance $2\pi\frac{r}{\lambda}=0.5$ (dotted, blue, circles), $0.75$ (dashed, green, crosses), $1$ (solid, black, squares). In the three cases the  coupling strength $K=\,K_A= K_B=1.5\cdot 10^{-2}$  and $\Omega/(2\pi)=1\, GHz$ ($\Omega=\Omega_A=\Omega_B$).} 
\label{fig:522}
\end{figure}

The effect of the non-RWA contributions to the evolution of  $\mathcal{P}_{gA/eB} (t)$ can be seen in Figs.\ref{fig:521} and \ref{fig:522}, where the consequences of changing the coupling and the distance between qubits are considered. The first thing we notice in Fig.\ref{fig:521} is that for short times the information provided by the detector is not very much related to the state of the source, that is, self-excitations and other non-RWA phenomena dominate over the photon exchange between source and detector. For the cases considered, only at interaction times $t \gtrsim 1\,ns	\simeq  1/\Omega$  the conditioned probability converges to the RWA prediction, that is, the excitation of the detector is a reliable way to detect the decay of the source.  Since the non-RWA contributions are more relevant for large couplings and short distances, the convergence is faster as the distance grows and the couplings diminish, as can be seen in Figs.\ref{fig:521} and \ref{fig:522}. 
It is convenient to remark here that the ripple frequency we see for instance in Fig. \ref{fig:521}  comes from higher harmonics of the qubit frequency $\Omega\ (= \Omega_A=\Omega_B$  in our case) and so does not represent the emergence of a new time-scale. It can be thought as a process similar to that of a Rabi oscillation, where the qubits would be absorbing in cycles (in a self-deexcitation fashion) the photons previously emitted in self-excitations.


The above theoretical results could have an impact in real experiments of circuit QED. In particular, a typical setup to measure the internal state of a flux qubit coupled to a transmission line consists of a SQUID surrounding the qubit. Although the total measurement process could take up to tens of nanoseconds, most of the time the coupling SQUID-qubit is much stronger than $K$, \cite{measurementqubits1} and the dynamics qubit-transmission line is effectively frozen. Thus this dynamics is only important during the activation of the SQUID, a process that may be in the nanosecond regime. For those measurement times, as we have proved, self-excitation effects cannot be disregarded and should manifest themselves. 

Besides, it should, in principle, be possible to prepare experiments in the near future to test our predictions directly. We here intend to give just a rough sketch. Such experiments would involve the preparation of the system at $t=0$ in the initial state of Eq. \ref{51d}, the switching of the interaction for a certain time $t$  (in the line of previous proposals, as in section 5.1)   and then the SQUID-measurement of both qubits $A$ and $B$. By repeating the experiment several times, the result frequencies should theoretically match our probability predictions.
\chapter{Quantum simulations of relativistic dynamics. Majorana physics} 
\begin{quote}
``But is this really the poet'' I asked. ``There are two brothers, I know; and both have attained reputation in letters. The Minister I believe has written learnedly on the Differential Calculus. He is a mathematician, and no poet."
`You are mistaken; I know him well; he is both. As poet and mathematician, he would reason well; as mere mathematician, he could not have reasoned at all, and thus would have been at the mercy of the Prefect."
(Edgar Allan Poe, \textit{The purloined letter})
\end{quote}
\begin{quote}
``Il mare mi ha rifiutato e ritorner{\'o} domani all'albergo Bologna, viaggiando forse con questo stesso foglio.'' (Ettore Majorana, last letter to Prof. Carelli. English free translation: ``The sea rejected me and I'll be back tomorrow at Hotel Bologna traveling perhaps with this sheet.'')
\end{quote}
In chapters 4 and 5 circuit QED  was considered as a quantum simulation of matter-radiation interaction. In this chapter we will deal with different quantum simulations, in particular quantum simulations of Relativistic Quantum Mechanical systems. To this end, we will introduce in section 6.1 a new kind of pseudo-Hamiltonian, referred to as ``Majorana Hamiltonian''. An example of this class is the object appearing in the Majorana equation, that is, the relativistic equation of a fermion with a Majorana mass term instead of the standard mass term of the Dirac equation. We will propose a trapped ion simulation of the 1-D version of this equation in section 6.2. Finally, in section 6.3 we will show a way to extend the quantum simulations of free relativistic quantum-mechanical systems to systems under the action of potentials, both for single-particle and bipartite systems, by simulating the potentials with free hamiltonians. 

Note: Throughout this chapter we will use natural units ($\hbar=c=1$)
\section{On Majorana Hamiltonians}
\subsection{Introduction}
The existence of spin $1/2$  fermions that are their own antiparticles, a possibility opened up by Majorana~\cite{donettore}, is under close scrutiny in terms of neutrino properties, theoretical schemes beyond the Standard Model or even in solid state systems~\cite{wilczek}. These particles could have a mass with properties crucially different from those of the standard Dirac mass of charged fermions. Their very equation --called Majorana Equation-- was introduced by Jehle~\cite{majoranaeq1} and by Case~\cite{majoranaeq3}, who also  made a thorough study of their behaviour under space-time transformations, quantized the field and also analyzed their possible (weak) interactions. Primers on these topics from a modern perspective can be found in Refs.~\cite{aste:2008dc,Pal:2010dc} and~\cite{zee}.

 On studying the quantum mechanics of a fermion obeying the Majorana equation we have identified a structure of much more general applicability. In this section, we present this structure as a very general theoretical framework. We introduce a new kind of generalized Hamiltonians with different mathematical features of the textbook Hamiltonians: they are neither linear nor antilinear, and thus the definition of hermiticity is questionable. We will refer to them as Majorana Hamiltonians, since the Majorana equation is a good example, although their scope is more general. Our main result is that these objects enjoy the physically meaningful property of inducing a temporal evolution which conserves the norm, as expected from total probability conservation. However, this result comes with a bag of surprises: particles with a Majorana Hamiltonian do not have stationary states, probability amplitudes are thus not conserved and an initial global phase is observable in the evolution of the expectation values of physical magnitudes. Moreover, even the notion of density matrix is no longer adequate for a description of this new Majorana dynamics.

Alternatively, we will also show that all the above physics can be equivalently analyzed with a standard Hamiltonian in a real Hilbert space of higher dimensionality, which is actually the way in which a Majorana Hamiltonian would be implemented in the lab, as we will show in section 6.2. As we shall see,  this complementary focus sheds light on the shocking features of the Majorana physics commented above.

\subsection{Majorana Hamiltonians.}%
In what follows, we will present the general formalism, that will be illustrated in the text with some cases of the Majorana equation. We will refer to the following object as a Majorana Hamiltonian:
\begin{equation}\label{eq:hamilt}
 \opH_\eta=\opA+i\eta\opB\opK\,,
\end{equation}
where $\eta$ is a complex number such that 
\begin{equation}
|\eta|=1, \label{phase}
\end{equation}
$\opA$ and  $\opB$ are linear Hermitian operators and the novelty is in the complex conjugation operator $\opK$, which obeys the following properties:  
\begin{equation}
\hat{K}[\alpha\psi]=\alpha^* \psi^*\,, \hat{K}^2=\hat{1}, \label{complex}
\end{equation}
and, for all $\psi$ and $\phi$ in the Hilbert space,  
\begin{equation}
(\phi,\opK\psi)=(\psi,\opK\phi). \label{selfadjoint}
\end{equation}
The general antilinear product operator $\opB\opK$ results antiunitary if $\opB$ is unitary~\cite{wigner}. We restrict this last operator by  the condition 
\begin{equation}
\{\opK,\opB\}=0\,,\label{condition}
\end{equation}
 i.e. $\opK$ and $\opB$ anticommute. We can express this by saying that $\opB$ is an imaginary operator (with respect to $\opK$). Thus, $\opH_\eta$ is neither linear nor antilinear, but it is well defined on (a domain of) the Hilbert space. 

A very important result is that if the evolution in a Hilbert space is governed by a Majorana operator, that is by the generalized Schr\"odinger equation
\begin{equation}
  \label{eq:schroedinger}
  i\partial_t\psi=\opH_\eta \psi = (\opA+i\eta\opB\opK) \psi,
\end{equation}
then the norm is still conserved,
\begin{equation}
 \partial_t\left(\psi(t),\psi(t)\right)=0.
\end{equation}
and we can still think of $|\psi|^2$ as a probability distribution. The proof is simple but subtle. First of all, the term with $\opA$ leads to norm conservation due to the standard hermiticity argument. Next, one can readily check that 
\begin{equation}
(\psi,\opB\opK\psi)= -(\psi,\opB\opK\psi)=0\,.  \label{conservationproof}
\end{equation}
This last result follows from the Hermiticity of $\opB,$ the fact that $\opK$ and $\opB$ anticommute, and the relation $(\phi,\opK\psi)=(\psi,\opK^{\dag}\phi)$. It is important to remark that while the norm is preserved, amplitudes are not.
 
The formalism introduced so far appears to respect all the postulates of quantum mechanics, but there is one usual convention that will have to be dropped: the equivalence of state vectors under global phases. This follows from the fact that the constant $\eta$ in Eq.~(\ref{eq:hamilt}) can be traded for a global phase change. In other words, if $\psi(t)$ is a solution of the generalized Schr\"odinger equation~(\ref{eq:schroedinger}) associated to the phase $\eta,$ then 
\begin{equation}
\psi_1(t)=\eta^{-1/2}\psi(t) \label{phase1}
\end{equation}
will evolve according to a different operator, 
\begin{equation}
\opH_1=\opA+i\opB\opK, \label{majoranahamiltonian1}
\end{equation}
and it will behave, in general, differently.

This inequivalent evolution of states initially related through a global phase entails that the standard description of mixed states with projector density matrices is not suitable.  A mixed state in the present context will be a collection of couples whose elements are unitary vectors $\psi_i$ and corresponding probabilities $p_i$, with evolution given by 
\begin{equation}
\rho_M(t)=\{p_i,\psi_i(t)\}_{i=1}^N \label{mixedmajorana}
\end{equation}
with each $\psi_i(t)$ a solution of Eq. (\ref{eq:schroedinger}). 
This gives rise to the following interesting point: consider $\psi(0)$ and $\phi(0)$  such that 
\begin{equation}
\phi(0)=\exp(i\varphi)\psi(0). \label{initialglobalphase}
\end{equation}
Then there exists a family of mixed states $\left\{\left(p, \psi(0)\right),\left(1-p, \phi(0)\right)\right\}$ which result in identical measurements at time $0$ but are nonetheless generically inequivalent under evolution. In other words, density matrices can be the adequate tool for the description or computation of static quantities; but not for dynamic ones. We will illustrate all the above in the Examples section 6.1.4.

\subsection{Hamiltonization.}%
We shall now provide an alternative description of the previous results in terms of a doubled (real) Hilbert space.   We introduce the objects
\begin{equation}
 P_+ = \tfrac{1}{2}(1+\opK),\quad P_-=\tfrac{1}{2}(1-\opK), \label{quasiprojectors}
\end{equation}
which are not projectors due to the antilinearity of $\opK$ though satisfy the useful relations 
\begin{equation}
\hat{P}_{\pm}^2=\hat{P}_{\pm}. \label{quasiprojectorsproperty}
\end{equation}
Using these operators we reconstruct the isomorphism
\begin{equation}
  \label{eq:isomorphism}
  \psi\in\Hcal \to\Psi=
  \begin{pmatrix}\hat{P}_+\psi\\ -i\,\hat{P}_-\psi\end{pmatrix}=
  \begin{pmatrix}\re{\psi}\\ \im{\psi}\end{pmatrix}\in\HHcal,
\end{equation}
between the complex Hilbert space $\Hcal$ and the direct sum $\Hcal^{(2)}$ of two real Hilbert spaces, $\hat{P}_+\Hcal$,  $-i\,\hat{P}_- \Hcal$. The norm is conserved under the isomorphism, naturally, but it is important to notice that the inner products are not preserved, namely, 
\begin{equation}
\left(\Phi,\Psi\right)=\mathrm{Re}\left[\left(\phi,\psi\right)\right]. \label{innerproduct}
\end{equation}
An alternative way of writing the isomorphism is as \cite{maccague-mosca}
\begin{equation}
\Psi = \re{\psi}\otimes|0\rangle + \im{\psi}\otimes|1\rangle\,, \label{encodedform}
\end{equation}
and demanding that the coefficients be real. Note that the reverse mapping, from the doubled real Hilbert space to the initial one is achieved explicitly as
\begin{equation}
 \psi=V\Psi=\begin{pmatrix}\hat{1}_\Hcal& i\hat{1}_\Hcal\end{pmatrix}\Psi=\re{\psi}+i\im{\psi}. \label{reversemapping}
\end{equation}
In order to compute how the previous isomorphism acts on operators, we have to introduce the real and imaginary parts of an arbitrary operator,
\begin{equation}
  \label{eq:3}
  \opO = \re{\opO}+i\,\im{\opO},\;\left\{
    \begin{array}{lll}
      \re{\opO} &=& \tfrac{1}{2}(\opA + \opK \opA \opK)\\
      \im{\opO} &=& \tfrac{i}{2}(\opK \opA \opK - \opA)
    \end{array}
    \right..
\end{equation}
Both the real and imaginary parts of a linear operator are also linear, but while the real part is Hermitian, $\im{\opO}$ is anti-Hermitian. Using these components the Hermitian operator $\opO$ acting on $\Hcal$ is mapped onto an operator
\begin{equation}
  \opO^{(2)} = \begin{pmatrix}
    \re{\opO} &-\im{\opO} \\
    \im{\opO} & \re{\opO}
  \end{pmatrix} = \re{\opO}\otimes\mathbf{1}_2-i\,\im{\opO}\otimes\sigma_2, \label{eq:observabledoubled}
\end{equation}
acting on $\HHcal$ that has all components real with respect to the induced complex conjugation.

Using these tools we can find the evolution equation for the mapped vector $\Psi\in\HHcal,$ up from the Majorana Hamiltonian~(\ref{eq:schroedinger}). Taking for simplicity the operator without phase, $\opH_1,$ we obtain the Schr\"odinger-like equation
\begin{equation}
\label{eq:schrodouble}
  i\partial_t\Psi=\begin{pmatrix}i\,\im{\opA}&\opB+i\,\re{\opA}\\ \opB-i\,\re{\opA}&i\,\im{\opA}\end{pmatrix}\Psi=\opH^{(2)}\Psi.
\end{equation}
This equation deserves a detailed explanation. First of all, the operator $\opH^{(2)}$ is designed to act on the doubled complex Hilbert space $\Hcal\oplus\Hcal,$ which contains our real Hilbert space, $\HHcal,$ as a subset.  Second, note the operator $i\opH^{(2)}$ is real and thus if $\Psi(0)$ is initially real, then it remains so throughout the evolution. In other words, if $\Psi(0)\in\HHcal\subset\Hcal\oplus\Hcal,$ it remains in the real sector at all times, preserving our isomorphism. Finally, while the doubled Hamiltonian $\opH^{(2)}$ is Hermitian and its eigenvalues are real, this only happens \emph{when this operator is defined on the complex doubled Hilbert space $\Hcal\oplus\Hcal$}. As a consequence of this and the fact that $\opH^{(2)}$ is purely imaginary, the  eigenvectors corresponding to nonzero eigenvalues are necessarily nonreal, so they live outside $\HHcal$. It follows that $\opH_1$ does not have any eigenvectors with nonzero eigenvalues, and the evolution under its generalized Schr\"odinger equation will of necessity be either constant or nonstationary. In other words, the Majorana Hamiltonians and Eq.~(\ref{eq:schroedinger}) have no stationary states.

Now we can understand the role of global phases in Majorana physics in a different way. Two states in the original Hilbert space $\Hcal$ related by a global phase are mapped to states which are no longer related by a global phase in $\HHcal$, but by a rotation
\begin{equation}
\begin{array}{ccc}
  \phi(0) &=& \exp{(i\varphi)}\psi(0)\\
  \updownarrow &=& \updownarrow \\
  \Phi(0) &=& \exp{(-i\sigma_2\varphi)}\Psi(0).  
\end{array}
\end{equation}
Thus clearly $\Phi(t)$ and $\Psi(t)$ evolve in a non-equivalent way, as do $\phi(t)=V\,\Phi(t)$ and $\psi(t)=V\,\Psi(t)$. 

\subsection{Examples}%
We now apply our formalism to a series of cases whose first examples were given in  the 1+1 Majorana equation~\cite{majoranaeq1,majoranaeq3}: 
\begin{equation}
i\,\partial_t\,\psi=\alpha\,p\,\psi\pm i\,m\,\sigma_2\,K\psi, \label{majoranaeqgeneral}
\end{equation}
where $\alpha$ is one of the Pauli matrices $\sigma_i$ ($i=1,\,2,\,3$), depending on the chosen representation of the Clifford algebra (see Appendix D). This equation is derived by replacing the $\psi$ in the mass term of the usual Dirac equation by its charge conjugate $\psi^C$, as we will see with more detail in the next section and in the Appendix D. For simplicity and without loss of generality, we will use now the example with $p=0$ and $m=1$:
\begin{equation}
\label{eq:example}
i\partial_t\psi=i\sigma_2\psi^*. 
\end{equation}

Let us see how the previous statements are realized in  (\ref{eq:example}). One can readily compute the solution, which is given by
\begin{equation}
 \label{eq:examplesol}
\psi(t)=\cos t\,\psi(0)+\sin t\,\sigma_2\psi(0)^*\,.
\end{equation}
Norm conservation is an immediate consequence of the antisymmetry of $\sigma^2$. On the other hand, amplitudes are not conserved: consider 
\begin{equation}
\psi(0)=\begin{pmatrix}1\\0\end{pmatrix} \, ,\phi(0)=\begin{pmatrix}0\\1\end{pmatrix}.\label{initialstates}
\end{equation}
 It is then the case that 
 \begin{equation}
 \left(\phi(t),\psi(t)\right)=i\sin 2 t\,. \label{amplitudes}
 \end{equation}
 The family of transformations given by (\ref{eq:examplesol}) is not a family of linear or antilinear isometries~\cite{wigner1959};  we could define an evolution operator 
 \begin{equation}
 \hat{U}(t)=\cos t\,\hat{1}+\sin t\,\sigma_2\opK\,, \label{evolutionoperator}
 \end{equation}
which is not unitary.

Let us, as before, study the case  (\ref{eq:example}). Consider two initial vectors,
\begin{equation}
\psi(0)=\begin{pmatrix}1\\0\end{pmatrix}\,, \phi(0)=\exp(i\varphi)\psi(0). \label{initialglobalphaseex}
\end{equation}
We then obtain 
\begin{equation}
\psi(t)= \begin{pmatrix}\cos t\\ i\sin t \end{pmatrix};\; \phi(t) = \begin{pmatrix}e^{i\varphi}\cos t\\ ie^{-i\varphi}\sin t \end{pmatrix}. \label{initialglobalphaseevolution}
\end{equation}
Furthermore, this phase is observable through the evolution in time of a measurable magnitude. Consider for instance the expectation value of the Hermitian operator $\sigma_2$ as a function of time: 
\begin{equation}
\langle\sigma_2\rangle(t)=\sin(2t) \cos(2\varphi). \label{observingglobalphases}
\end{equation}
\begin{figure}[h!]
  \centering
  \includegraphics[width=\linewidth]{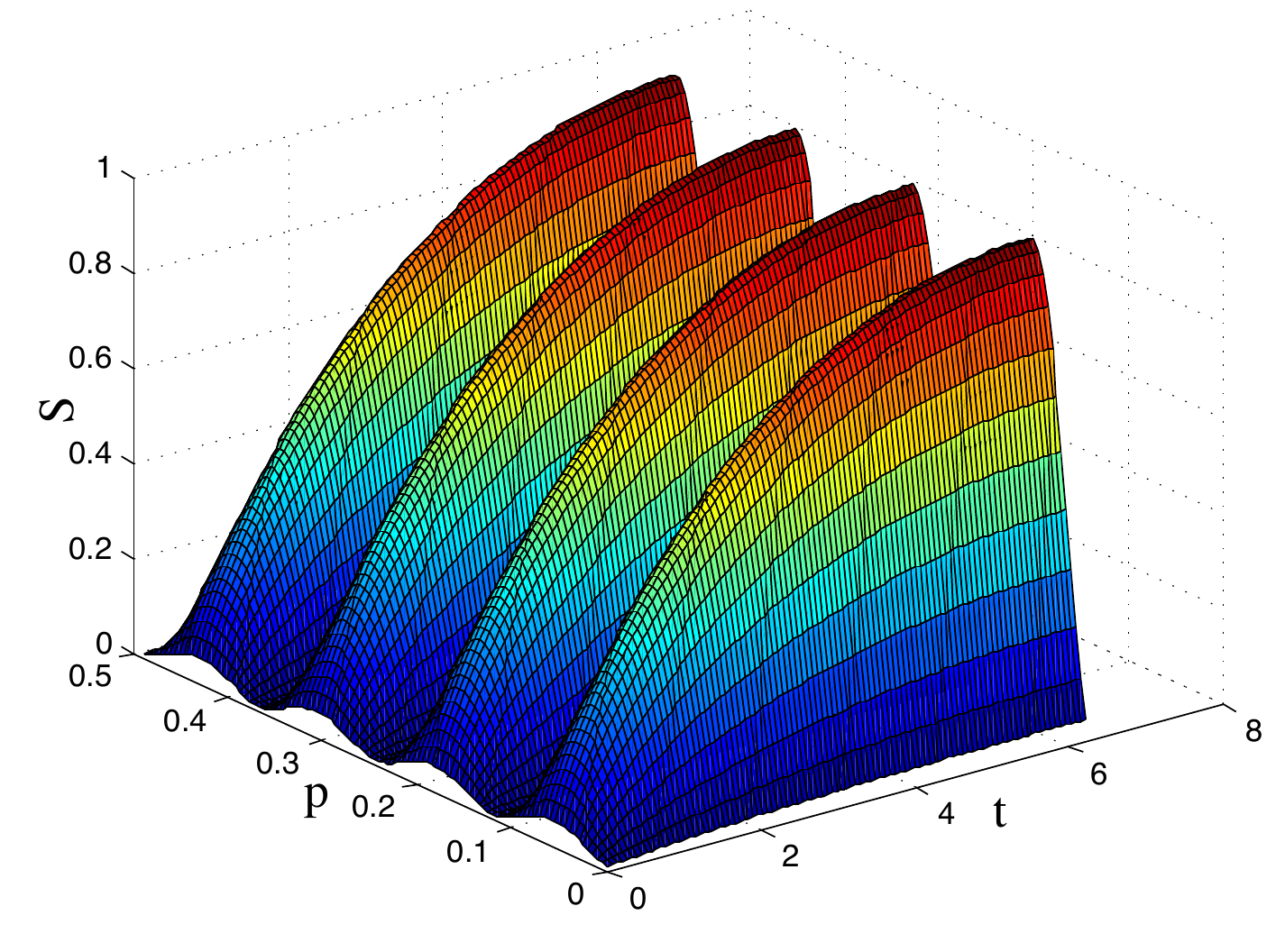}
  \caption{Von Neumann entropy of the state $\rho(t)$ Eq. (\ref{eq:rho}) for $\varphi=\pi/2$ in front of $p$ and $t$. In the standard quantum mechanical picture, this would be a pure state for any value of $p$ and $t$ and so entropy would be zero. }
  \label{fig:fig33}
\end{figure}

We now consider the mixed state defined by $\psi(t)$ with probability $p$ and $\phi(t)$ with probability $1-p$. If $\varrho(t)$ is the density matrix that would correspond to  $\psi(t)$, then the density matrix $\rho(t)$, corresponding to the mixed state at time $t$, is
\begin{equation}
\rho(t)=\varrho(t)+(1-p)\sin\varphi\sin(2t)\left(\cos\varphi\sigma_1-\sin\varphi\sigma_2\right)\,\label{eq:rho}.
\end{equation}
The von Neumann entropy of this mixed state is depicted in Fig. \ref{fig:fig33} as a function of time and $p$ for the case of $\varphi=\pi/2$. In the standard quantum mechanical picture, the initial state would be considered as a pure state, and unitary evolution would conserve entropy. Indeed, if the state were the reduced state of a composite system this would mean that the global state is non-entangled. All the above suggests that the standard quantum information theory, as it is, does not fit to Majorana physics.

As a general example of the hamiltonization we can consider Eq. (\ref{eq:hamilt}) with $\eta=1$, 
\begin{equation}
\opB=m\,\sigma_2\,,\opA=p_x\sigma_1+p_y\sigma_2, \label{1+12+1}
\end{equation}
which accommodates both the 1+1 and 2+1 Majorana equation. Thus, using Eq. (\ref{eq:schrodouble}), we obtain the following Hamiltonian in the doubled real Hilbert space: 
\begin{equation}
\opH^{(2)}=\begin{pmatrix}p_x\sigma_1&i\,p_y\sigma_2+m\,\sigma_2\\ -i\,p_y\sigma_2+m\,\sigma_2&p_x\sigma_1\end{pmatrix},\label{hamiltoniandoubled}
\end{equation}
 which is the Hamiltonian of a Dirac equation with 
 \begin{equation}
 \beta=\sigma_1\otimes\sigma_2,\, \alpha^1=\mathbf{1}_{2\times2}\otimes\sigma_1,\, \alpha^2=-\sigma_2\otimes\sigma_2. \label{alphabetasdoubled} 
 \end{equation}
 It is straightforward to check that 
\begin{equation}
 \gamma^0=\beta,\, \gamma^1=\beta\alpha^1=-i\,\sigma_1\otimes\sigma_3,\, \gamma^2=\beta\alpha^2=-i\,\sigma_3\otimes\mathbf{1}_{2\times2} \label{gammasdoubled}
 \end{equation}
 form a  Clifford algebra, and also $\gamma^0$, $\gamma^1$. Thus Majorana equation in 2+1 and 1+1 dimensions is mapped to a 3+1 Dirac equation with $p_z=0$ and $p_y=p_z=0$ respectively. This result is more general: with this techniques  the 3+1 Majorana equation is mapped to a 7+1 Dirac equation with four components of the momentum set to 0.

As another example, let us consider the Hilbert space of qubits, where a general operator $\opA=a_\mu\sigma_\mu$ with $\sigma_\mu=\left(\hat{1},\vec{\sigma}\right)$  and real $a_\mu$ will be decomposed as $\re{\opA} = a_0\hat{1}+a_1\sigma_1+a_3\sigma_3,$ $\im{\opA} = -i a_2\sigma_2.$  Without losing generality, we can choose $\opB=\sigma_2.$ The scale of $\opB$ would be fixed by the timescale. Thus we can explicitly write 
\begin{equation}
\opH^{(2)}=\begin{pmatrix} a_2\sigma_2 & \sigma_2+i\re{\opA} \\ \sigma_2 - i\,\re{\opA} & a_2\sigma_2 \end{pmatrix}\,, \label{hamiltoniandoubled2}
\end{equation}
whose eigenvalues appear in pairs, 
\begin{eqnarray}
\lambda_1=|\mathbf{a}|+\sqrt{1+a_0^2},\, \lambda_2=-\lambda_1,\nonumber\\ \lambda_3=|\mathbf{a}|-\sqrt{1+a_0^2} \,,\lambda_4=-\lambda_3.\label{eigenvaluesdoubled}
\end{eqnarray}

\subsection{Conclusions and outlook.}%
Putting all together, we have introduced a new kind of generalized Hamiltonians Eq. (\ref{eq:hamilt}).  We refer to them as Majorana Hamiltonians since the Majorana equation is a good example, but the framework is far more general. These objects are neither linear nor antilinear and thus there is not a clear definition of hermiticity, but they induce a physical dynamics in which the norm is conserved. This new physics has very interesting features: for instance, initial global phases are relevant since they can be observed in the dynamics of expectation values of physical magnitudes. As a consequence, the notion of density matrix is not adequate for a dynamical description of the system, and has to be replaced. Alternatively, all this can be seen within the usual Quantum Mechanics framework, since a Majorana Hamiltonian can always be mapped to a standard Hamiltonian in a real Hilbert space of double dimensionality. A global phase transformation in the original Hilbert space is not mapped to a global phase transformation in the new one, which explains the differences in the evolution. This mapping between Hilbert spaces is actually the tool to implement Majorana Hamiltonians and also antiunitary operations like time reversal, charge conjugation or partial transpose in the lab, as we will see in the next section. A deeper look into the physics of real Hilbert spaces will imply also a better understanding of the new Majorana physics, for instance the quantum information theory of  Majorana systems, since it has been noted that  quantum information properties in real Hilbert spaces are different from the standard ones~\cite{woottersreal},~\cite{batleetalreal}. 
\section{Quantum simulation of Majorana equation and unphysical operations}
\subsection{Introduction}
The Majorana equation \cite{majoranaeq1}, \cite{majoranaeq3} is a relativistic wave equation for fermions where the mass term contains the charge conjugate of the complex spinor, $\psi_{c}$,
\begin{equation}
\label{majorana}
i  \slpar \psi = m  \psi_{c} .
\end{equation}
Here, $\slpar = \gamma^{\mu} \partial_{\mu}$ and $\gamma_{\mu}$ are the Dirac matrices~\cite{thaller}, while the non-Hamiltonian character stems from the simultaneous presence of  $\psi$ and $\psi_{c}$. The significance of the Majorana equation rests on the fact that it can be derived from first principles in a similar fashion as the Dirac equation~\cite{zee}. Both wave equations are Lorentz invariant but the former preserves helicity and does not enjoy stationary solutions. The Majorana equation is considered a possible model~\cite{aste:2008dc} for describing exotic particles in supersymmetric theories --photinos and gluinos--, or in grand unified theories, as is the case of neutrinos. Indeed, the discussion of whether neutrinos are Dirac or Majorana particles still remains open~\cite{wilczek}.  Relativistic quantum models can be simulated, but they can also emerge as a natural description of certain systems, as happens with the 2+1 Dirac equation in graphene systems. Nevertheless, note that despite the similar naming, this work is neither related to the Majorana fermions (modes) in many-body systems, nor to the Majorana fermions (spinors) in the Dirac equation~\cite{aste:2008dc}.

In order to simulate physics described by the Majorana equation, we have to solve a fundamental problem: the physical implementation of antilinear and antiunitary operations in a quantum simulator. In this section, we apply a mapping introduced in the previous section by which complex conjugation, an unphysical operation, becomes a unitary operation acting on an enlarged Hilbert space. The mapping works in arbitrary dimensions and can be immediately applied on advanced quantum simulation platforms. As a key application, we show how to simulate the Majorana equation in 1+1 dimensions and other unphysical operations --time reversal and complex conjugation-- using only two trapped ions. This is completed with a recipe for measuring relevant observables and a roadmap towards more general scenarios, including the combination of Majorana and Dirac physics. Finally, we discuss further scopes of quantum simulations in the context of fundamental and relativistic quantum physics.
\subsection{Quantum simulation of antiunitary operations}
\begin{figure}[h!] 
\hspace*{-0.4cm}
\includegraphics[width=0.95\linewidth]{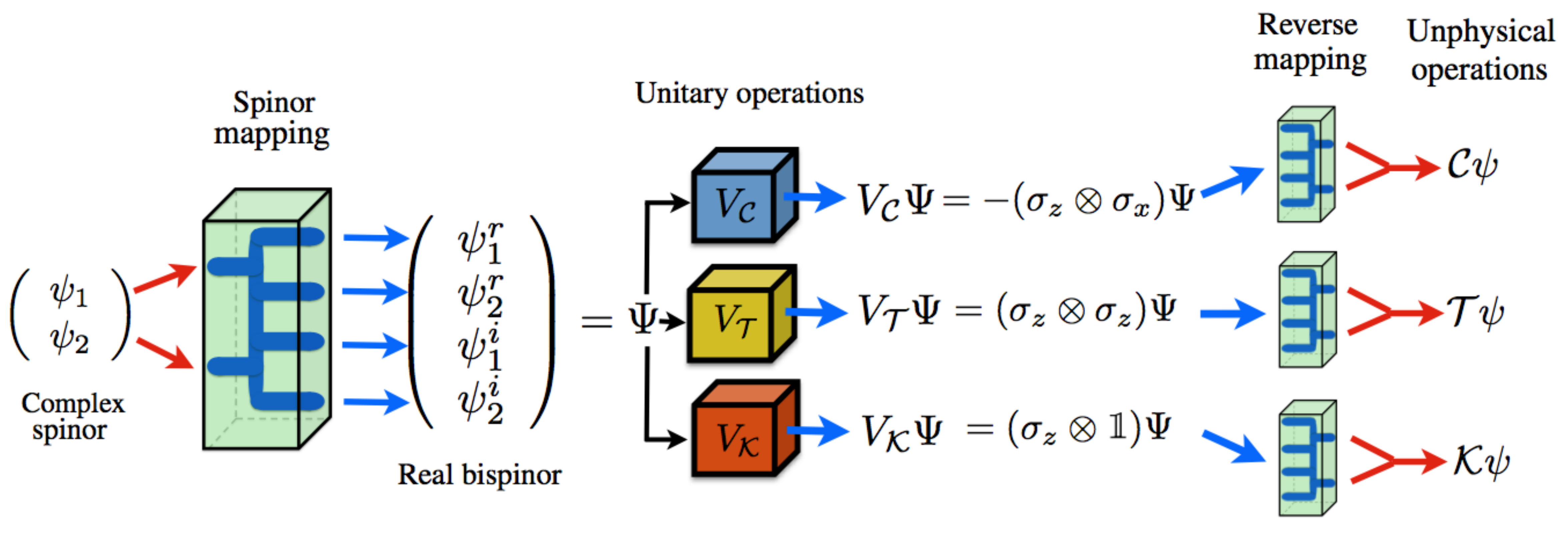}
  \caption{Diagram showing the different steps involved in the quantum simulation of unphysical operations in 1+1 dimensions.}
  \label{fig:34}
\end{figure}
\subsection{Simulating antiunitary operations}
There are three discrete symmetries~\cite{Streater} which are central to quantum mechanics and our understanding of particles, fields and their interactions: parity, $\mathcal{P}$, time reversal, $\mathcal{T}$, and charge conjugation, $\mathcal{C}$. None of these operations can be carried out in the real world: $\mathcal{P}$  involves a global change of the whole physical space, while $\mathcal{C}$ and $\mathcal{T}$ are antiunitaries. However, there is no apparent restriction for implementing them in a physical system that simulates quantum mechanics. We will focus on the study of antiunitary operations, which can be decomposed into a product of a unitary, ${\mathcal{U}}_{\mathcal{C}}$ (for charge conjugation) or ${\mathcal{U}}_{\mathcal{T}}$ (for time reversal), and complex conjugation, $\mathcal{K}\psi=\psi^*$. We consider the mapping \ref{eq:isomorphism} of the quantum states of an $n$-dimensional complex Hilbert space, $\mathbb{C}_n$, onto an real Hilbert space, $\mathbb{ R}_{2n}$.
This mapping can be physically implemented by means of an auxiliary two-level system, such that $\mathbb{R}_{2n} \in \mathcal{H}_{2} \otimes \mathcal{H}_{n}$. In this manner, the complex conjugation of the simulated state becomes a local unitary $V_{\mathcal K}$ acting solely on the ancillary space,
\begin{equation}
 \mathcal{K}\psi = \psi^*\to V_{\mathcal K} \Psi = (\sigma_z\otimes\mathbf{1})\Psi,\label{eq:complex1}
\end{equation}
and thus physically implementable. Furthermore, unitaries and observables can be also mapped onto the real space, (\ref{eq:observabledoubled}) preserving unitarity and Hermiticity. In addition to complex conjugation, unitaries and Hermitian operators, the proposed simulator also accomodates  the antiunitary operations 
\begin{equation}
\mathcal{C} = {\mathcal U}_{\mathcal{C}}\mathcal{K} \label{eq:chargeconjugation1}
\end{equation}
and 
\begin{equation}
\mathcal{T} = {\mathcal U}_{\mathcal{T}}\mathcal{K}. \label{eq:timereversal1}
\end{equation}
To this end, we have to choose a particular representation (see Appendix D) that fixes the unitaries ${\mathcal U}_{\mathcal{C}}$ and ${\mathcal U}_{\mathcal{T}}$, as will be shown below.

At this point, we possess the basic tools to simulate the Majorana equation  in the enlarged space. The expression for the charge conjugate spinor is given by
\begin{equation}
  \psi_c = \eta\mathcal{C}\gamma^0\mathcal{K} \psi , \label{eq:chargeconjugation2}
\end{equation}
with $\mathcal{C}$ a unitary matrix satisfying 
\begin{equation}
\mathcal{C}^{-1}\gamma^\mu\mathcal{C}=-\left(\gamma^\mu\right)^T. \label{eq:chargeconjugationproperty}
\end{equation}
We illustrate now with the case of 1+1 dimensions (Appendix D). Here, a suitable representation of charge conjugation is 
\begin{equation}
\psi_c = i \sigma_y\sigma_z \psi^*, \label{eq:chargeconjugation3}
\end{equation}
that is ${\eta\mathcal{C}} =  i \sigma_y$, and the Majorana equation reads
\begin{equation}
  \label{1+1}
  i\partial_{t}\psi = \sigma_xp_{x}\psi - im\sigma_{y} \psi^{*},
\end{equation}
where $p_{x}= -i\hbar\partial_{x}$ is the momentum operator. Note that Eq.~(\ref{1+1}) is not  Hamiltonian, $i\hbar\partial_t \psi = H \psi$, as is the case of Schr\"odinger and Dirac equations. This is due to the presence of a complex conjugate operation in the right-hand side of Eq.~(\ref{1+1}), which is not a linear Hermitian operator. Through our mapping~(\ref{eq:isomorphism}),
\begin{equation}
\left(
\begin{array}{c}
\psi_1 \\ \psi_2
\end{array}
\right)
\in \mathbb{C}_2 \to \Psi = \left(
\begin{array}{c}
\psi_1^r \\ \psi_2^r \\ \psi_1^i \\ \psi_2^i
\end{array}
  \right) \in \mathbb{R}_{4} ,
\label{map 1+1}
\end{equation}
the Majorana equation for a complex spinor becomes a 3+1 Dirac equation with dimensional reduction, $p_y,p_z=0$, and a four-component real bispinor
\begin{equation}
\label{eq:majorspin31}
i\hbar \partial_{t}\Psi =
\left[(\mathbf{1} \otimes \sigma_x) p_x  - m\sigma_x\otimes\sigma_y\right] \Psi .
\end{equation}
Note that, here, the dynamics preserves the reality of the bispinor $\Psi$ and, in general, cannot be reduced to a single 1+1 Dirac particle. The result of Eq.~(\ref{eq:majorspin31}) is even more general and the complex-to-real map in arbitrary dimensions transforms always a Majorana equation into a higher dimensional Dirac equation. Since Eq.~(\ref{eq:majorspin31}) is a Hamiltonian wave equation, it can be simulated in a conventional quantum system while suitably encoding the Majorana dynamics.

The mapping of wavefunctions into larger spinors may allow us not only to implement Majorana equations in the lab, but also to explore exotic symmetries and unphysical operations, otherwise impossible in nature. From Eqs.~(\ref{eq:complex1}), (\ref{eq:chargeconjugation3}), and (\ref{map 1+1}), for the 1+1 dimensional case, we can deduce that charge conjugation is implemented in the enlarged space via the unitary operation $V_{\mathcal C}$
\begin{equation}
\psi_c  = {\mathcal C} \psi = {\mathcal U_C K} \psi \to V_{\mathcal C} \Psi = -(\sigma_z\otimes \sigma_x)\Psi. \label{eq:chargeconjugationdoubled}
\end{equation}
We can do something similar with time reversal, defined as the change $t\to(-t).$ In this case, we expect~\cite{zee} (Appendix D)
\begin{equation}
  i\partial_\tau\psi'(\tau) = H\psi'(\tau) ,
\end{equation}
where the time variable $\tau = -t$ and the modified spinor 
\begin{equation}
\psi'(\tau)=\mathcal{T}\psi(t). 
\end{equation}
In order to preserve scalar products and distances, the time reversal operator must be an anti-unitary operator and thus decomposable as the product 
\begin{equation}
{\cal T} = {\mathcal U}_{\mathcal T}\mathcal{K}. 
\end{equation}
In $1+1$ dimensions, imposing that the Hamiltonian be invariant under time reversal, 
\begin{equation}
H'=\mathcal{T}^{-1}H\mathcal{T}, 
\end{equation}
implies that the unitary satisfies 
\begin{equation}
{\mathcal U}_{\mathcal T}^{-1}(i\sigma_x\partial_x) {\mathcal U}_{\mathcal T} =-i\sigma_x\partial_x,
\end{equation}
 with a possible choice being 
 \begin{equation}
 {\mathcal U}_{\mathcal T} =\sigma_z. 
 \end{equation}
In other words, in the enlarged simulation space
\begin{equation}
 \mathcal{T}\psi = {\mathcal U}_{\mathcal{T}}\mathcal{K}\psi \to V_{\mathcal T} \Psi = (\sigma_z\otimes\sigma_z)\Psi.
\end{equation}
See Fig.~\ref{fig:34} for a scheme of the simulated symmetries.

\begin{figure}[t]
  \centering
  \resizebox{\linewidth}{!}{\includegraphics{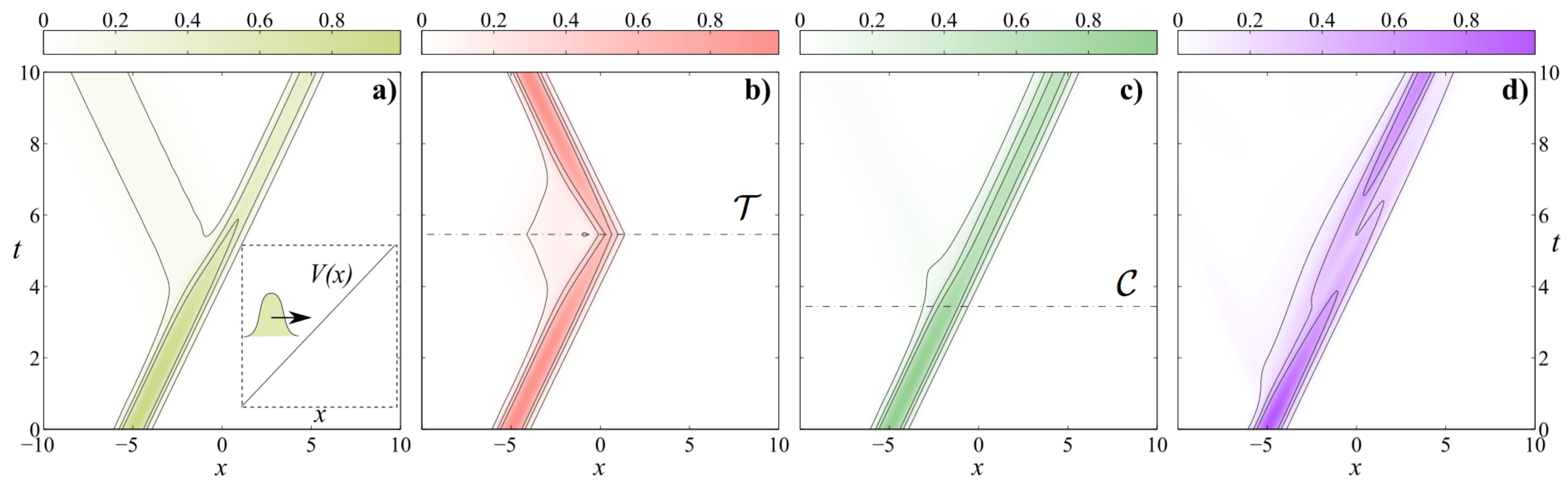}}
  \caption{Scattering of a fermion against a linearly growing potential (inset). (a) Ordinary Klein process: a fraction of a Dirac fermion turns into an antiparticle, entering the potential. (b) At an instant of time we apply the time reversal operator $\mathcal{T}$ causing the particle to retrace its own trajectory. (c) Similar to (b) but now we apply charge conjugation, converting the particle in its antiparticle. (d) Scattering of a Majorana fermion, which propagates through the potential. Parameters are $m=0.5$ and $V(x)=x,$ in dimensionless units.}
  \label{fig:scattering}
\end{figure}

Equation~(\ref{eq:majorspin31}), a Dirac equation in 3+1 dimensions with dimensional reduction $p_y,p_z=0$, can also be interpreted as a recipe for the quantum simulation of the Majorana equation in the laboratory. In a recent experiment, the dynamics of a free Dirac particle was simulated using a single trapped ion~\cite{naturekike}, a quantum platform that has proved instrumental for quantum information implementations. Unfortunately, Eq.~(\ref{eq:majorspin31}) has a more complex structure and a different setup is required, which is outlined in the Appendix E. Moreover, the encoded Majorana dynamics requires a systematic decoding via a suitable reverse mapping of observables, as we show below. In short, the real bispinor $\Psi\in\mathbb{R}_{4}$ can be encoded in the internal state of two ions, while the position and momentum of the Majorana particle are mapped onto the quadratures of a collective motional mode, e. g. the center-of-mass mode, of the ions~\cite{lucas, naturekike}. The Hamiltonian of Eq.~(\ref{eq:majorspin31}) can be implemented term-by-term, in principle, in the trapped-ion system by a number of lasers coupling the motional and internal states of the ions. However, our proposal is valid for a general quantum simulator and we do not discard its implementation in other quantum platforms.

A relevant feature of the Majorana equation in 3+1 dimensions is the conservation of helicity.  A reminiscent of the latter in $1+1$ dimensions is the observable called hereafter as {\it pseudo-helicity} 
\begin{equation}
\Sigma=\sigma_xp_x. 
\end{equation}
This quantity is conserved in the 1+1 Majorana dynamics of Eq.~(\ref{1+1}) but not in the 1+1 Dirac equation. We will use this observable to illustrate measurements on the Majorana wavefunction.  The mapping for operators can be simplified if we are only interested in expectation values. Reconstructing the complex spinor 
\begin{equation}
\psi = M \Psi
\end{equation}
 with matrix
\begin{equation}
M = \left(\begin{array}{lccr} \mathbf{1} &i \mathbf{1} \end{array}\right)
\end{equation}
associated with Eqs.~(\ref{map 1+1}) and (\ref{eq:majorspin31}), we can write the following equivalence
\begin{equation}
  \label{rule}
  \langle O \rangle_\psi = \langle\psi| O |\psi\rangle =
  \langle \Psi| M^{\dag} O \,  M |\Psi\rangle =: \langle \tilde O\rangle_\Psi.
\end{equation}
According to this, in order to measure the pseudo-helicity $\Sigma$, we have to measure
\begin{equation}\label{eq:pshelicity}
\tilde\Sigma = M^{\dag} \sigma_x p_x \  M =(\mathbf{1} \otimes \sigma_x - \sigma_{y} \otimes \sigma_{x})\otimes p_x
\end{equation}
in the enlarged simulation space. In an ion trap implementation, the first term of this observable, $(\mathbf{1} \otimes \sigma_x )\otimes p_x$, is measurable with recently developed techniques~\cite{naturekike}. The second term is a three-operator correlation, $(\sigma_{y} \otimes \sigma_{x})  \otimes p_x$, and will require a specific design with measurements involving short-interaction times~\cite{Lougovski06}, as explained in the Methods section.

We want to emphasize that the previous mappings and the implementation of discrete symmetries are not only valid for Majorana equations, but also for Dirac spinors. Equally interesting is the possibility of combining both Dirac and Majorana mass terms in the same equation~\cite{aste:2008dc},
\begin{equation}
  i  \slpar \psi = m_M  \psi_{c} + m_D  \psi,
\end{equation}
which still requires only two ions for a 1+1 quantum simulation. It also becomes feasible to have CP violating phases in the Dirac mass term, $m_D\exp(i\theta\gamma^5)$. Furthermore, we could study the dynamics of coupled Majorana neutrinos with a term $\bar{M}\psi_c$, where $\bar M$ is now a matrix and $\psi =\psi(x_1,x_2)$ is the combination of two such particles, simulated with three ions and two vibrational modes.

So far, we have presented a complete toolbox of unphysical operations, $\mathcal{C}$, $\mathcal{T}$, and $\mathcal{K}$, that are available in the proposed quantum simulator. We can combine all these tools to study dynamical properties of the transformed wavefunctions. To exemplify the kind of experiments that become available, we have studied the scattering of wavepackets against a linearly growing potential, 
\begin{equation}
V(x)=\alpha x.\label{eq:simpotential} 
\end{equation}
It is known that repulsive potentials are partially penetrated by Dirac particles~\cite{thaller}, an effect called the Klein paradox~\cite{klein,  jorge1, jorge2}. This is shown in Fig.~\ref{fig:scattering}a, where a Dirac particle splits into a fraction of a particle, that bounces back, and a large antiparticle component that penetrates the barrier. This numerical experiment has been combined with the discrete symmetries and the Majorana equation. In Fig.~\ref{fig:scattering}b we show a Dirac wavepacket that suffers the time reversal operation some time after entering the barrier: all momenta are reversed and the wavepacket is refocused, tracing back exactly its original trajectory. In Fig.~\ref{fig:scattering}c we repeat the same procedure but using charge conjugation. This operation changes the sign of the charge turning a repulsive electric potential into an attractive one, which can be easily penetrated by the antiparticle. In our last example, Fig.~\ref{fig:scattering}d, we show the scattering of a Majorana particle. While the evolution is not so smooth ---there are no plane wave solutions in the Majorana equation---, we can still identify a wavepacket penetrating the barrier, showing a counter-intuitive insensitivity to the presence of the barrier, that will be explained in the next section 6.3.
\subsection{Conclusions}
In summary, we have introduced a general method to implement the quantum simulation of unphysical operations and the non-Hamiltonian Majorana equation in a Hamiltonian system. To this end, we have designed a suitable mapping that enlarges the simulation space by means of an ancillary system to allow for complex conjugation, charge conjugation, and time reversal. We have exemplified the implementation of the 1+1 dimensional case in the context of trapped-ion physics. The flexibility of this protocol allows to explore a novel front of quantum simulations, that of unphysical operations and exotic quantum relativistic processes that go beyond ordinary Schr\"odinger and Dirac quantum mechanics.

\section{Quantum simulation of relativistic potentials without potentials}
\subsection{Introduction}
Last years had witnessed an increasing interest in simulating dynamics coming from the Relativistic Quantum Mechanics realm with physical systems of Quantum Optics, such as trapped ions. Striking theoretical predictions like \textit{Zitterbewegung}  and Klein paradox \cite{klein} has been observed in these simulations. In particular, the proposal of simulation with one trapped ion \cite{lucas}  of  the single particle free 1-D Dirac equation has been successfully implemented in the lab \cite{naturekike}. This is also the case for the single particle 1-D  Dirac equation with some external potential \cite{jorge1} whose simulation involves two ions \cite{jorge2}. In the last section we have proposed that the free Majorana equation and unphysical operations like complex conjugation, charge conjugation and time reversal  can also be simulated with two trapped ions. Besides, two-body Dirac equations have been the subject of recent research \cite{bermudez}.

In this section we will show that the same setups built up for the simulations of the free single particle Dirac and Majorana equations can also be used for simulations of these equations with the addition of some external potential, for a large class of potentials. This is based in the following idea, which is the main result of this paper: some states which are solutions of the  Dirac or Majorana equation with the potential can be related through a phase transformation  with a solution of the free corresponding equation. So, if we want to simulate the dynamics of a given state under certain potential the transformation tells us which is the state whose dynamics under the free equation does the job. In other words, the potential is codified in the phase involved in the transformation. The trick does not work for any potential, but only for potentials belonging to a certain class. In some cases, it works for potentials in the Majorana equation but not for the same potentials in the Dirac equation, which can be used to analyze the different character of Majorana and Dirac dynamics. In general, the transformation does not leave the probability density unchanged but this happens in some particular cases, showing us an additional amazing feature: under certain potentials the particle behaves as a free particle. We extend our results also to two-body equations. 

Although our method works in principle in any dimension and any representation, we will focus in this work in the 1-D case and the particular representations commonly employed in the experiments. 
\subsection{One particle systems} 
Let us come to explain our results with more detail. We  will consider  the following 1-D Dirac equation in natural units:
\begin{equation}
i\dot{\psi}=-i\,\sigma_x \psi '+(\sigma_z\,m+V(x))\,\psi \label{eq:dirac}
\end{equation}
where $\dot{}$, $'$ denote time and space partial derivatives respectively. In the same representation, Majorana equation looks like:
\begin{equation}
i\dot{\psi}=-i\sigma_x\, \psi'-i\sigma_y\,m\psi^*+V(x)\psi \label{eq:majorana}
\end{equation}
In 1-D a general potential can be written down as \cite{jorge1}:
\begin{equation}
V= f_1(x)+ f_2(x)\,\sigma_z+f_3(x)\,\sigma_y+f_4(x)\,\sigma_x \label{eq:generalpotential}
\end{equation}

We start from  the Eq.(\ref{eq:majorana}) with the potential in Eq.(\ref{eq:generalpotential}). We will     analyze under what conditions a  transformation of the form
\begin{equation}
\psi=\prod_j e^{-i\Theta_j(x)}\phi\label{eq:generaltransformation}
\end{equation}
where 
\begin{equation}
\prod_j e^{-i\Theta_j(x)}=e^{-i\,F_1(x)\sigma_x}e^{-i\,F_2(x)\sigma_y}e^{-i\,F_3(x)\sigma_z}e^{-i\,F_4(x)}\label{eq:phase}
\end{equation}
 can convert  Eq. (\ref{eq:majorana}) in the corresponding free Majorana equation for $\phi$. (Please recall that $e^A\,e^B\ne e^{A+B}$ unless $[A,B]=0$, which in general is not the case here.) To this end we first notice that after applying Eq.(\ref{eq:generaltransformation}) the LHS of Eq.(\ref{eq:majorana}) becomes:
\begin{equation}
\prod_je^{-i\Theta_j(x)}i\dot{\phi} \label{eq:lhs}
\end{equation}
 while due to the anticommutation properties of the Pauli matrices the first term of the RHS transforms as:
 \begin{equation}
\prod_j e^{-i\Delta_j(x)}(-i\,\sigma_x \phi)' + (\prod_j e^{-i\Delta_j(x)})' (-i\sigma_x\phi) \label{eq:rhs1}
\end{equation}
with 
\begin{equation}
\prod_j e^{-i\Delta_j(x)}=e^{-i\,F_1(x)\sigma_x}e^{i\,F_2(x)\sigma_y} e^{i\,F_3(x)\sigma_z}e^{-i\,F_4(x)}.\label{eq:phase2}
\end{equation}
 If we choose the $F's$ such that:
 \begin{eqnarray}
F_1 '(x)&=& f_1(x), \, F_2'(x)= -i\,f_2 (x),\,\nonumber\\  F_3'(x)&=&i\,f_3(x) \, ,F_4'(x)=f_4(x).\label{eq:generalproperty2}
\end{eqnarray}
the second term of Eq.(\ref{eq:rhs1}) cancels out the one coming from the third term of Eq.(\ref{eq:majorana}), that is, the potential is removed. Finally, if $F_1$ is real and $F_2$, $F_3$, $F_4$ imaginary, we have:
\begin{equation}
\sigma_y\,(\prod_j e^{-i\Theta_j(x)})^*=(\prod_j e^{-i\Delta_j(x)})\sigma_y. \label{eq:generalproperty1}
\end{equation}
Putting all together, we have that applying Eq.(\ref{eq:generaltransformation}) in Eq.(\ref{eq:majorana}) we obtain the following equation for $\phi$, provided that $F_1$ is real  and the other $F$'s imaginary, and that they are related with the $f$'s in the potential through Eq.(\ref{eq:generalproperty2}):
\begin{equation}
\prod_j e^{-i\Theta_j(x)} i\dot{\phi}=\prod_j e^{-i\Delta_j(x)}(i\,\sigma_x \phi '-i\sigma_y\,m\phi^*).\label{eq:result}
\end{equation}
Then, if $F_2(x)=F_3(x)=0$, $\phi$ verifies a free Majorana equation. That is, if we want to simulate  the dynamics of a particle in the state $\psi(x,t)$ under the 1-D  Majorana equation with a potential of the form $V(x)= f_1(x)+f_4 (x) \sigma_x$, we only have to study the state $\phi(x,t)$ which is related to it through $\psi(x,t)=e^{-i\,F_1(x)\sigma_x}e^{-i\,F_4(x)}\sigma_x\,\phi(x,t)$. Since, $F_1$ is real and $F_4$ imaginary, we have the following relationship between the probability densities:
\begin{equation}
|\psi(x,t)|^2=e^{-2\,i\,F_4(x)}\,|\phi(x,t)|^2.\label{eq:probabilities}
\end{equation}
So the probability density observed for $\phi$ can be easily related with the simulated probability density for $\psi$. Also, analogous relations can be derived, for instance for expectation values of observables. Of particular interest is the case in which $F_4(x)=0$ (then $f_4(x)=0$) and the probability density is the same for $\psi$ and $\phi$. Therefore, the probability density of a state under the potential is always the same as the one of a free state. We shall illustrate this with an example below. The method does not work in general for $F_2(x)\neq0$, $F_3(x)\neq0$ unless approximatively in the regions of space in which $F_2(x)\simeq0$,  $F_3(x)\simeq0$, which does not necessarily entails $f_2(x)\simeq0$, $f_3(x)\simeq0$.
\begin{figure}[h!]
  \centering
  \includegraphics[width=0.7\linewidth]{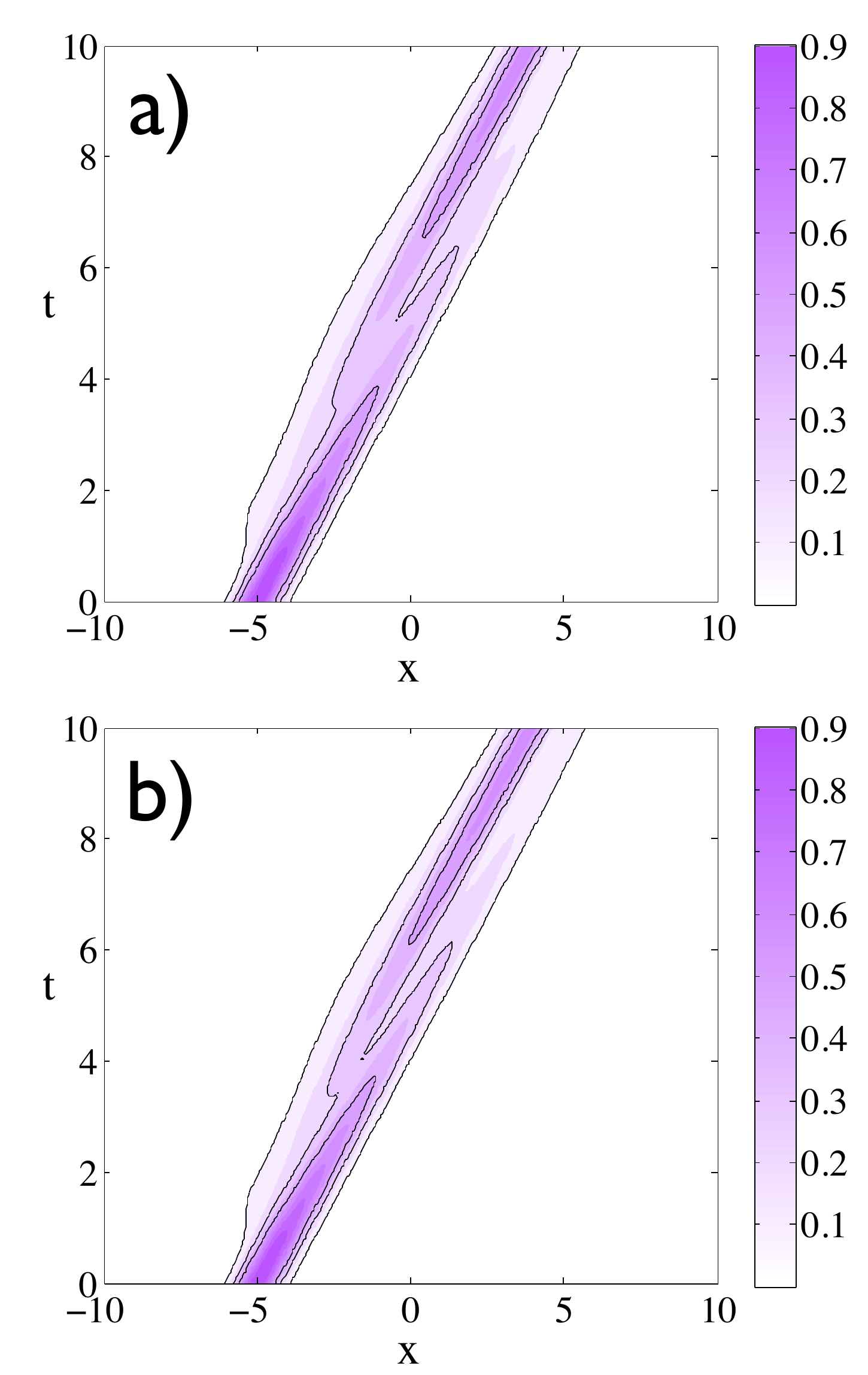}
  \caption{a) Evolution of a fermionic  wavepacket under Eq.(\ref{eq:majorana}) with the potential of Eq.(\ref{eq:potential}). b) Evolution of the corresponding state with the transformation in Eq.(\ref{eq:change}) under a free Majorana equation. In both cases: $m=0.5$ and $g=1$.}
  \label{fig:fig36}
\end{figure}

We now analyze the case of the 1-D Dirac equation Eq.(\ref{eq:dirac}). Using the same techniques as in the Majorana case, if we consider a potential with $f_1(x)=f_3(x)=0$:
\begin{equation}
V(x)=f_2(x)\,\sigma_z+f_4(x)\,\sigma_x, \label{eq:generalpotentialdirac}
\end{equation}
the transformation 
\begin{equation}
\psi=e^{-i\,F_2(x)\sigma_y}e^{-i\,F_4(x)}\phi\label{eq:generaltransformationdirac}
\end{equation}
transforms Eq.(\ref{eq:dirac}) into:
\begin{eqnarray}
e^{-i\,F_2(x)\sigma_y}e^{-i\,F_4(x)}i\dot{\phi} =e^{i\,F_2(x)\sigma_y}e^{-i\,F_4(x)}\nonumber\\(-i\sigma_x \phi '\,+\sigma_z\,m\phi), \label{eq:resultdirac}
\end{eqnarray}
and similar conclusions as in the Majorana case are reached for this class of potentials. But the absence of the complex conjugation in the mass term, prevents the possibility of extract a phase in the RHS if we include $F_1$ or $F_3$, so potentials with $f_1$ and/or $f_3$ cannot be simulated, not even approximatively, with free Dirac Hamiltonians.
\subsection{Examples}
Now  we will illustrate all the above with some examples.  First, we will consider the 1-D Dirac  and Majorana equations with a linear potential 
\begin{equation}
V(x)=g\,x, \label{eq:potential} 
\end{equation}
and the transformation:
 \begin{equation}
\psi(x,t)=e^{\frac{-i\,g\,x^2\,\sigma_x}{2}}\phi(x,t).\label{eq:change}
\end{equation}
\begin{figure}[h!]
  \centering
  \includegraphics[width=0.7\linewidth]{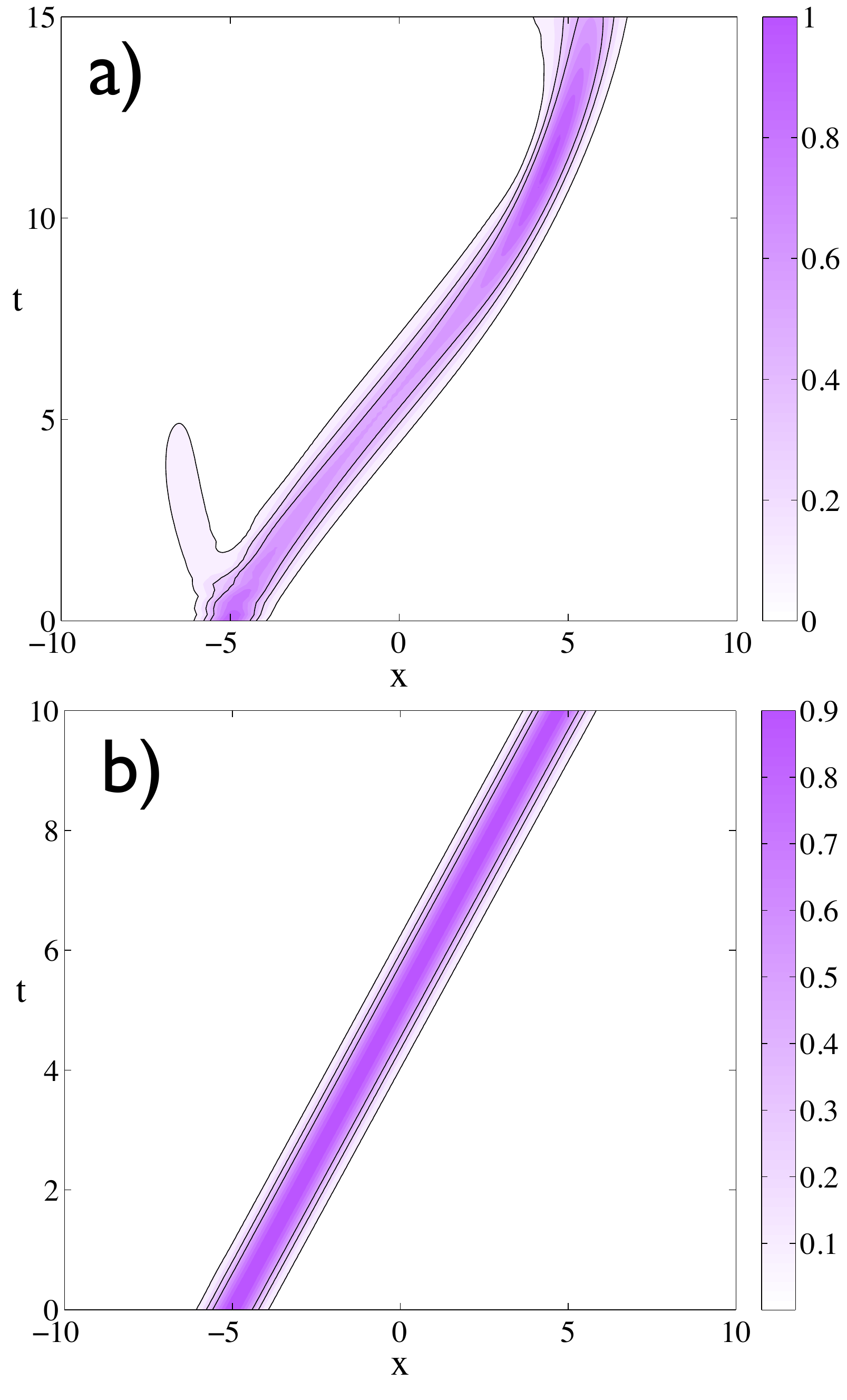}
  \caption{a) Evolution of a fermionic  wavepacket under Eq.(\ref{eq:dirac}) with the potential of Eq.(\ref{eq:potential2}). b)  Evolution of the corresponding state with the transformation in Eq.(\ref{eq:change}) under a free Dirac equation. In both cases: $m=0.5$ and $g=-1$.}
  \label{fig:fig37}
\end{figure}
In that case $\phi$ satisfies the free Majorana equation and  $|\psi|^2=|\phi|^2$, so any solution of the Majorana equation with potential has the same probability density of a solution of the free Majorana equation. The same transformation does not work in the Dirac case, shedding light to the different behavior of Majorana and Dirac particles against such potentials, as can be seen in Fig. \ref{fig:fig36}. Notice that the Majorana equation does not have stationary solutions and that an initial global phase is not conserved as a global phase during the evolution, as commented in section 6.1. Thus Eq.(\ref{eq:change}) cannot be written in terms of the evolution of some $\psi(0)$ and $\phi(0)$. In other words, a solution $\phi$ of the equation with potential is always equivalent to some solution $\psi$ of the free equation, but in general this  $\psi$  is a different state at each instant of time, not the evolution of one state  $\psi(0)$. 
In Fig.\ref{fig:fig37} we show that our method works as a good approximation in a certain region of space in the case of the Dirac equation with a potential:
\begin{equation}
V(x)=g\,x\,\sigma_z, \label{eq:potential2} 
\end{equation}
and the corresponding transformation:
 \begin{equation}
\psi(x,t)=e^{\frac{-\,g\,x^2\,\sigma_y}{2}}\phi(x,t).\label{eq:change2}
\end{equation}

\subsection{Bipartite systems}

All the above can be extended to two-particle systems. For instance, it has been shown that a Lorentz-invariant two-body Dirac equation with an oscillator-like mutual interaction can be written in the center of mass reference frame \cite{moshinski}. In 1-D and the particular representation we are using, we have:
\begin{equation}
i\dot{\psi}=-\frac{i}{\sqrt2}\,(\alpha_{1}-\alpha_{2}) (\psi '+m\omega x\beta_{12}\psi)+(\beta_1+\beta_2)m\,\psi \label{eq:dirac-osc2}
\end{equation}
with $\alpha_1=\sigma_x\otimes\mathbf{1}$, $\alpha_2=\mathbf{1}\otimes\sigma_x$, $\beta_1=\sigma_z\otimes\mathbf{1}$, $\beta_2\mathbf{1}\otimes\sigma_z$, $\beta_{12}=\sigma_y\otimes\sigma_y$, $x=(x_1-x_2)/\sqrt2$. 
Thus the corresponding two-body Majorana oscillator equation is :
\begin{equation}
i\dot{\psi}=-\frac{i}{\sqrt2}\,(\alpha_{1}-\alpha_{2}) (\psi '+m\omega x\beta_{12}\psi)-i(\hat{\beta}_1+\hat{\beta_2})m\,\psi^* \label{eq:majorana-osc2}
\end{equation}
with $\hat{\beta_1}=\sigma_y\otimes\mathbf{1}$ and $\hat{\beta_2}=\mathbf{1}\otimes\sigma_y$. With the techniques explained above, we find that the transformation 
 \begin{equation}
\psi(x,t)=e^{\frac{-\,m\,\omega x^2\,\beta_{12}}{2}}\phi(x,t).\label{eq:change3}
\end{equation}
transforms Eq.(\ref{eq:majorana-osc2}) into a quasi-free two-body Majorana equation:
\begin{equation}
e^{\frac{-\,m\,\omega x^2\,\beta_{12}}{2}}i\dot{\phi}=e^{\frac{\,m\,\omega x^2\,\beta_{12}}{2}}(-\frac{i}{\sqrt2}\,(\alpha_{1}-\alpha_{2}) \phi '+(\hat{\beta}_1+\hat{\beta_2})m\,\phi^*)\label{eq:majorana-free2}
\end{equation}
Thus, in that case the situation would be similar to the one-particle example of Fig.\ref{fig:fig37}: Eq.(\ref{eq:majorana-osc2}) can be simulated with good approximation with a two-body free Majorana equation in a large region of space. Interestingly, the trick does not work for the two-body Dirac oscillator.

\subsection{Conclusions}
We have shown that Majorana and Dirac potentials can be simulated with free Hamiltonians, since there is a mapping between free states and states under the action of the potential. The method works better in the Majorana case, in which there are potentials that cannot be simulated with a free Dirac Hamiltonian, as has been illustrated with a particular case. This example also exhibits the peculiarity of conservation of probability density between free and non-free states, illuminating the issue of the different behavior of Majorana and Dirac particles against such a potential. In other cases the method works only as a good approximation in certain regions of space. We have extended our results to two-particle interacting systems.

\appendix
\chapter{}
In this appendix, we will give details on the computations of the quantities of interest in Chapter 3. We first start with
$A$ and $X$ in (\ref{d}). Both are a sum of second order transition amplitudes, which can be written as:
\begin{eqnarray}
(\frac {-i}{\hbar})^2\sum_{k}\langle\,f|\,H_I\,|k\rangle\langle\,k|H_I|\,i\rangle\,\int^t_0 dt_1\int^{t_1}_0 dt_2 
e^{i(E_f-E_k)t_2/\hbar}e^{i(E_k-E_i)t_1/\hbar} \label{Ao}
\end{eqnarray}
being $E_f$, $E_k$ and $E_i$ the energies of the final $|\,f\rangle$, intermediate $|\,k\rangle$ and initial $|\,i\rangle$
states of the system, respectively. The sum over $k$ is a sum over all the possible intermediate states of the system which in
the case of fixed two-level atoms reduces to a sum over all the momenta and polarizations of the emitted photon. The time integrations in (\ref{Ao}) are
just
\begin{equation}
-\hbar^2(\frac{e^{i(E_f-E_i)t/\hbar}-1}{(E_f-E_k)(E_f-E_i)}-\frac{e^{i(E_k-E_i)t/\hbar}-1}{(E_f-E_k)(E_k-E_i)}) \label{Ap}
\end{equation}
The second term in (\ref{Ap}) is usually neglected, but give rise to a very different short time behavior \cite{thiru} ($\propto t^4$, not $\propto
t^2$). Therefore, it is of interest for our purposes.

In order to obtain $X$ we have to sum over the amplitudes for single photon emission at atom $A$ ($B$) followed by absorption at
atom $B$ ($A$). The case where a photon is emitted and absorbed by the same atom corresponds to $A$, that we will consider below. Using the mode expansion for the electric field:
\begin{eqnarray}
\mathbf{E}(\mathbf{x})=i\sqrt{\frac{\hbar\, c}{2\varepsilon_0\,(2\pi)^3}}\sum_{\lambda}\int d^3k
\sqrt{k}(e^{i\mathbf{k}\,\mathbf{x}}\mathbf{\epsilon}(\mathbf{k},\lambda)\,a_{k\lambda}-
e^{-i\mathbf{k}\,\mathbf{x}}\mathbf{\epsilon}^*\,( \mathbf{k},\lambda)\,a^{\dag}_{k\lambda}), \label{Aq}
\end{eqnarray}
(with $[\,a_{k\lambda},\,a^{\dag}_{k'\lambda'}\,]=\delta^3(\mathbf{k}-\mathbf{k'})\,\delta_{\lambda\,\lambda'}$)  taking into
account (\ref{Ao}) and (\ref{Ap}), recalling that
\begin{equation}
\sum_{\lambda}\mathbf{\epsilon}_i^*\,(\mathbf{k},\lambda)\,\mathbf{\epsilon}_j\,(\mathbf{k},\lambda)=\delta_{ij}-\hat{k_i}\hat{k_j},
\end{equation}
and using the tabulated integrals that we list at the end of the appendix, a somewhat tedious although straightforward
computation leads to:
\begin{equation}
X= \frac{\alpha\,d^i d^j}{\pi\,e^2}(-\mathbf{\nabla}^2\delta_{ij}+\nabla_i\nabla_j)\,I \label{Ar}
\end{equation}
$\alpha$ being the fine structure constant and 
\begin{equation}
I=I_+ + I_- , 
\end{equation}
which, in terms of the dimensionless parameters $z=\Omega\,r/c$ and $x=r/c\,t$ are
\begin{eqnarray}
I_{\pm}={\frac{1\pm\frac{1}{x}}{2}}\{e^{iz}[Ei(-iz)-Ei(-iz(1\pm1/x))]+e^{-iz}[Ei(iz)-Ei(iz(1\pm1/x))]\}
\label{s}
\end{eqnarray}
for $x>1$, $I$ having the extra term $i\pi(1-1/x)e^{-i\,z}$ for $x<1$. We use the conventions of \cite{bateman}. As noted in
\cite{thiru}, the non-zero contributions for $x>1$ come from the second term of (\ref{Ap}). We display here the results of the
derivatives in (\ref{Ar}) only for the particular case where the dipoles are parallel along the $z$ axis
($\mathbf{d}_A=\mathbf{d}_B=\mathbf{d}=d\,\mathbf{u}_z$) and the atoms are along the $x$ axis, corresponding to the physical
situation considered previously in, for instance, \cite{franson} and in this paper. Actually, $|\,E\,\rangle$ is a triply
degenerate state $|\,E\,,m\rangle$ with $m=0,\pm1$ and our scheme holds for a transition with $\Delta m=0$ \cite{milonniII}.
Another independent possibility would be to consider transitions with $\Delta m=\pm1$ that corresponds to
$\mathbf{d}=d\,(\mathbf{u}_x \pm i \mathbf{u}_y)/\sqrt{2}$ \cite{milonniII}. We find that:
\begin{eqnarray}
X&=-\frac{\alpha|\mathbf{d}|^2}{2\pi \,x\,r^2\,e^2}\{4 x (-1+\frac{(-2+x^2) \cos{\frac{z}{x}}}{-1+x^2})+e^{i z} [-2\,x\,z^2 Ei(-i z)\nonumber\\
&+\left(2+z \left(-2 i+(-1+x) z\right)\right)Ei(-\frac{i (-1+x) z}{x})\nonumber\\&+\left(-2+z (2 i+z+x z)\right) Ei(-\frac{i
(1+x) z}{x})]\nonumber\\&+2e^{-iz} \left(Ei(\frac{i (-1+x) z}{x})-Ei(\frac{i (1+x) z}{x})\right)+ \nonumber\\&z\,e^{-iz}
[-2\,x\, z Ei(i z)+\left(2 i+(-1+x) z\right) \nonumber\\&Ei(\frac{i (-1+x) z}{x})+(-2 i+z+x z) Ei(\frac{i (1+x)
z}{x})]\}\label{At}
\end{eqnarray}
for $x>1$, with the additional term $i\alpha\,e^{-iz}d^2\left(2+z\left(2i+(-1+x)z\right)\right)/(r^2\,x)$ for $x<1$. 

Now we come to $A$, which is the sum of the radiative corrections of atoms $A$ and $B$. As can be seen in the main text, $A$ appears in our results only as a higher order correction to $X$. Therefore, instead of finding an exact expression for it, we are mainly concerned with removing the divergencies. We followed the standard treatment (see,
for instance, \cite{cohentannoudji}) which is valid for the times $\Omega\,t>1$ we are considering. From (\ref{Ao}), it is
possible to arrive at:
\begin{equation}
\frac{-2\,i\,\alpha|\,d\,|^2\,t}{3\,\pi\,e^2\,c^2}\,\lim_{\epsilon\rightarrow0_+}\int_0^{\infty}d\omega\,\omega^3\,
({1\over{\Omega-\,\omega+i\,\epsilon}}-{1\over{\Omega+\,\omega-i\,\epsilon}}) \label{Au}.
\end{equation}
Now, using in (\ref{Au}) the identities
\begin{equation}
 {\omega^3\over\Omega\pm\, \omega}=\pm (\omega^2-\mp\,\Omega\,\omega+\Omega^2-{\Omega^3\over\Omega\pm\,\omega}), \label{Av}
\end{equation}
the first term of (\ref{Av}) cancels out the contribution of the Hamiltonian self-interaction terms Eq. (\ref{eq:2p}), the
second is the state-independent contribution that can be absorbed in the definition of the zero of energy \cite{cohentannoudji},
the third cancels the counterterm coming from the mass renormalization \cite{cohentannoudji} and finally the last term has
logarithmic divergences and a cut-off, related with the fact that we are in the electric dipole representation could be imposed
at $t_{min}={a_0\over c}=1.76\cdot 10^{-19}\,s$. Please notice that the times relevant in our computations are of the order of
$t\cong {10\over\Omega}\approx 4\cdot 10^{-15}\,s$. Therefore:
\begin{equation}
A=\frac{2\,i\,\alpha\,|\mathbf{d}|^2\,z^3}{3\,\pi\,L^2\,e^2\,x}\,\ln{(|\frac{1-\frac{z_{max}}{z}}{1+\frac{z_{max}}{z}}|})],\label{w}
\end{equation}
with $z_{max}/z=c\,/(\Omega\,a_0)$.

Another quantity of interest is what we called $l$ in the main text and it is given by (\ref{k}) and (\ref{31l}). Performing the
integration in (\ref{31l}), we obtained $M(z,x)=M_+ (z,x)+ M_- (z,x)$, where:
\begin{eqnarray}
M_{\pm}(z,x) &=& \frac{e^{i\frac{z}{x}}}{4\pi^2\,z}\{\sin{(z(1\pm\frac{1}{x}))}[ci(z)-ci(z(1\pm\frac{1}{x}))]-\nonumber\\
&{}&\cos{(z(1\pm\frac{1}{x}))}[si(z)-si(z(1\pm\frac{1}{x}))]\}\label{Ax}
\end{eqnarray}
The derivatives in (\ref{k}) were performed in the same particular situation as in $X$.

$|U|^2=\langle0|\mathcal{S}^+_A \mathcal{S}^-_A|0\rangle,\,$ and $|V|^2=\langle0|\mathcal{S}^-_B\mathcal{S}^+_B|0\rangle,\,$ are
just the two terms that contribute to $A$ without the time ordering and therefore their divergencies are removed by
the application of (\ref{Au}). Taking this into account, we obtain:
\begin{eqnarray}
&|U|^2&=\frac{2\alpha|d|^2 z^2}{3\pi\,e^2\,L^2} (-2+\pi \frac{z}{x}+2 \cos(\frac{z}{x})+2
(\frac{z}{x})si(\frac{z}{x}))\nonumber\\
&|V|^2&=\frac{2\alpha|d|^2 z^2}{3\pi\,e^2\,L^2} (2+\pi \frac{z}{x}-2 \cos(\frac{z}{x})-2 (\frac{z}{x})si(\frac{z}{x}))\ \ \ \ \
\ \ \ \ \label{Ay}
\end{eqnarray}
$F$ and $G$ can be written in terms of previously computed quantities, taking into account that:
\begin{eqnarray}
F&=&\theta(t_1-t_2)\,(\,u_A\,(t_1)v'_A(t_2)\,+\,v'_A\,(t_1)u_A\,(t_2)\nonumber\\
&+&\,u_B\,(t_1)v'_B\,(t_2)+\,u'_B\,(t_1)v_B(t_2)\,)\nonumber\\
G&=&u_A\,v'_B\,+\,u'_A\,v_B, \label{Az}
\end{eqnarray}
being $V_A\,=\,\langle\,1\,|\, \mathcal{S}^+_A\,|\,0\,\rangle,\,$ and $U_B\,=\,\langle\,1\,|\, \mathcal{S}^-_B\,|\,0\,\rangle$.
The primes are introduced to label two different single photons. Therefore, in the computation of $|\,F\,|^2$,
$|\,G\,|^2$ and $F\,G^*$ we will only need, besides $|\,U\,|^2$, $|\,V\,|^2$ and $l$, the following:
\begin{eqnarray}
V_A\,U_A^* &=&V_B\,U_B^*=\frac{2\,\alpha|\,d\,|^2\,z^2}{3\,\pi\,L^2}\,e^{i\,\frac{z}{x}}\,\sin{\frac{z}{x}} \nonumber\\
V_A\,V_B^*&=& \frac{\alpha\,d^i d^j}{\pi}(-\mathbf{\nabla}^2\delta_{ij}+\nabla_i\nabla_j)\,I' \label{Azz}
\end{eqnarray}
being $I'=I'_+ + I'_- $, with:
\begin{eqnarray}
I'_{\pm}&=&{\frac{1\pm\frac{1}{x}}{2}}\{e^{i\,z}\,Ei(-iz(1\pm\frac{1}{x}))+e^{-i\,z}\,Ei(iz(1\pm\frac{1}{x}))\}\nonumber\\
&-&e^{\mp\,i\,z}\,Ei(\pm\,i\,z)\} \label{Azzz}
\end{eqnarray}
and $U_A\,U_B^*=V_A\,V_B^*$ when $x>1$, with the additional term $-2\,\pi\sin{z}\,(1-1/x)$ when $x<1$. Again the derivatives
were performed as in $X$ and $l$.

The following integrals are useful to obtain the results of this appendix \cite{bateman}:
\begin{eqnarray}
\int_0^\infty\,d\omega\,\frac{e^{\pm\,i\,\omega\,\gamma}}{\omega\,+\,\beta}&=&-e^{\mp\,i\,\gamma\,\beta}\,Ei(\pm\,i\,\gamma\,\beta)\label{Azzzz}\\
\int_0^\infty\,d\omega\,\frac{e^{\pm\,i\,\,\omega\,\gamma}}{\omega\,-\,\beta}&=&-e^{\pm\,i\,\gamma\,\beta}\,(Ei(\mp\,i\,\gamma\,\beta)\,\mp\,i\,\pi),
\nonumber
\end{eqnarray}
with $a>0$, $arg\, \beta\leq\pi$.

\chapter{}

In this appendix we will give further details on the computations of the relevant magnitudes $|X|$, $|U_A|^2$, $|V_B|^2$ which are necessary to compute the concurrence Eq.(\ref{eq:31s}) in section 4.2. With Eqs. (\ref{42f})- (\ref{exchange}) and Eq. (\ref{commutation}):
\begin{eqnarray} 
X=\frac{d^2\,N\,v}{\hbar^2}\int_{-\infty}^{\infty} dk |k|\,(e^{ikr}I_{t1}+e^{-ikr}I_{t2}) \label{exchange1}
\end{eqnarray}
with 
\begin{eqnarray}
I_{t1,2}=\int^{t}_0 dt_{2,1}\int^{t_{2,1}}_0dt_{1,2}e^{i\Omega\,(t_2-t_1)}e^{-iv|k|(t_{2,1}-t_{1,2})}.\label{exchange2}
\end{eqnarray}
Notice that the term with $I_{t2}$ gives the non-RWA probability amplitude associated to a single photon emission of qubit B followed by an absorption of qubit A. Performing the time integrations, inserting them in Eq. (\ref{exchange1}) and after some algebra, $X$ can be given as a combination of integrals of the form:
\begin{eqnarray}
\int_0^\infty\,d\,k\,\frac{\cos(\,k\,\gamma)}{k\,+\,\beta}&=&-\sin(\gamma\beta) si(\gamma\beta) -\cos(\gamma\beta) ci(\gamma\beta)\nonumber\\
\int_0^\infty\,d\,k\,\frac{\cos(\,k\,\gamma)}{k\,-\beta}&=&-\sin(\gamma\beta) si(\gamma\beta) -\cos(\gamma\beta) ci(\gamma\beta)\nonumber\\&-&\pi\sin(\gamma\beta)\label{integrals}\\
\int_0^\infty\,d\,k\,\frac{\sin(\,k\,\gamma)}{k\,+\,\beta}&=&\sin(\gamma\beta) Ci(\gamma\beta) -\cos(\gamma\beta) si(\gamma\beta)\nonumber\\
\int_0^\infty\,d\,k\,\frac{\sin(\,k\,\gamma)}{k\,-\,\beta}&=&-\sin(\gamma\beta) Ci(\gamma\beta) +\cos(\gamma\beta) si(\gamma\beta)\nonumber\\&+&\pi\cos(\gamma\beta)\nonumber
\end{eqnarray}
with $\gamma,\beta>0$ and the conventions in \cite{bateman} for the $si$ and $Ci$. Putting all together we find:
\begin{eqnarray}
&&X=\frac{K}{2}(i\pi  \rho\xi \sin(\rho)-e^{ \frac{i\rho\xi}{2}} ((1+i \tau_{-})(C(\tau_{-})-S(\tau_{-})\nonumber\\&&-\pi\sin(\tau_{-})\Theta(1-\xi) )+(1-i\tau_{+})(-C(\tau_{+})-S(\tau_{+})\nonumber\\&&-\pi\sin(\tau_{+}))+)(-\tau_{-}+i)(-SC(\tau_{-})+CS(\tau_{-})+\pi\nonumber\\&&\cos(\tau_{-}))+(-\tau_{+}-i)(-SC(\tau_{+})+CS(\tau_{+})+\pi \cos(\tau_{+}))\nonumber\\&&-2)-e^{\frac{-i \rho\xi}{2}} ((1-i \tau_{-})(-C(\tau_{-})-S(\tau_{-})-\pi\sin(\tau_{-})\nonumber\\&&\Theta(\xi-1))+(1+i\tau_{+})(-C(\tau_{+})-S(\tau_{+}))+(\tau_{-}+i)\nonumber\\&&(SC(\tau_{-})-CS(\tau_{-}))+(\tau_{-}-i)(SC(\tau_{+})-CS(\tau_{+})+\pi\nonumber\\&&\cos(\tau_{+}))-2) - 2 - 2C(\rho)-2 S(\rho)-\rho(-2SC(\rho) \nonumber\\
&&+2 CS(\rho))\label{exchange3}
\end{eqnarray}
where $\xi$ has been defined in Eq. (\ref{chi}), 
\begin{equation}
\rho=\Omega\,r/v \label{dimensionlessdistance} 
\end{equation}
is a dimensionless distance, 
\begin{eqnarray}
\tau_{-}&=&\rho(1-\xi)=\rho-\Omega\,t, \nonumber\\
\tau_{+}&=&\rho (1+\xi)=\rho+\Omega\,t, \label{taumastaumenos}
\end{eqnarray}
and we define 
\begin{eqnarray}
C(x)=\cos(x)\,Ci(x),S(x)=\sin(x)\,si(x), \nonumber\\
CS(x)=\cos(x)\,si(x),  SC(x)=\sin (x)\,Ci(x). \label{defsvarias}
\end{eqnarray}
Notice the dependence with the spacetime region through the factors with the Heaviside function $\Theta$.

Now we come to the emission probabilities $|U_A|^2$, $|V_B|^2$, which are given by
\begin{equation}
|U_A|^2  \!\! =  \bra{0}\mathcal{S}^+_A\mathcal{S}^-_A\ket{0},
|V_B|^2 = \bra{0}\mathcal{S}^-_B\mathcal{S}^+_B\ket{0}
\end{equation}
Following similar techniques we find that 
\begin{equation}
|U_A|^2=f_{+}(\Omega\,t), |V_B|^2=f_{-}(\Omega\,t), \label{1demissions}
\end{equation}
where:
\begin{equation}
f_{\pm}=\frac{K}{2} (\pi\,\Omega\,t\pm2 (\cos(\Omega\,t)+ \Omega\,t \,Si(\Omega\,t)-1))
\end{equation}
where $Si$ must not be mistaken by $si$, $Si=si\,+\pi/2$ as usual.

Finally, notice that $A$ (Eq.(\ref{42e})) is a sum of two terms like $|U_A|^2$ and $|V_B|^2$ with the time ordering operator $T$ and that $U_A*V_B$ (Eq.(\ref{42j})) is similar to $X$ without $T$. The treatment of the divergencies has followed the lines sketched in Appendix A.

\chapter{}
In this appendix the task is to compute the term $\mathcal{\delta M}^{(3)}$ introduced in section 5.1. This term is included in the whole third order amplitude
\begin{eqnarray}
\label{eq:1}
\mathcal{M}^{(3)} = (\frac{1}{i\hbar})^3 \sum_{\alpha \beta}
\langle f |V| \alpha \rangle \langle \alpha |V| \beta \rangle \langle \beta |V| i \rangle \times \\ \nonumber
\times \int_0^t dt_1 e^{iw_{f\alpha}t_1}
\int_0^{t_1} dt_2 e^{iw_{\alpha\beta}t_2}
\int_0^{t_2} dt_3 e^{iw_{\beta i}t_3} 
\end{eqnarray}
where $\alpha,\beta$ stand for arbitrary sets of quantum numbers and the sum includes integrals over momenta and polarization sums. We start computing the time integrals:

\begin{eqnarray}
\label{eq:2}
&&\int_0^t dt_1 e^{iw_{f\alpha}t_1}
\int_0^{t_1} dt_2 e^{iw_{\alpha\beta}t_2}
\int_0^{t_2} dt_3 e^{iw_{\beta i}t_3} =\nonumber\\
&&\frac{1}{(iw_{\beta i})(iw_{\alpha i})}
[\frac{e^{iw_{f i}t}-1}{iw_{f i}}-
\frac{e^{iw_{f \alpha}t}-1}{iw_{f \alpha}}]-\nonumber \\
&& \frac{1}{(iw_{\beta i})(iw_{\alpha \beta})}
[\frac{e^{iw_{f \beta}t}-1}{iw_{f \beta}}-
\frac{e^{iw_{f \alpha}t}-1}{iw_{f \alpha}}] 
\end{eqnarray}
The last two lines in Eq.(\ref{eq:2}) contains all the time dependence along with that in the final and exchanged photon frequencies. Among all the terms that can interfere with $V_B$, we select only those that depend on the interatomic distance $r $ whose contribution $\mathcal{\delta M}^{(3)}$ we will now compute. We first consider the three cases where the exchange ``goes from A to B",
\begin{figure}[h]
\begin{center}
\includegraphics[width=\textwidth]{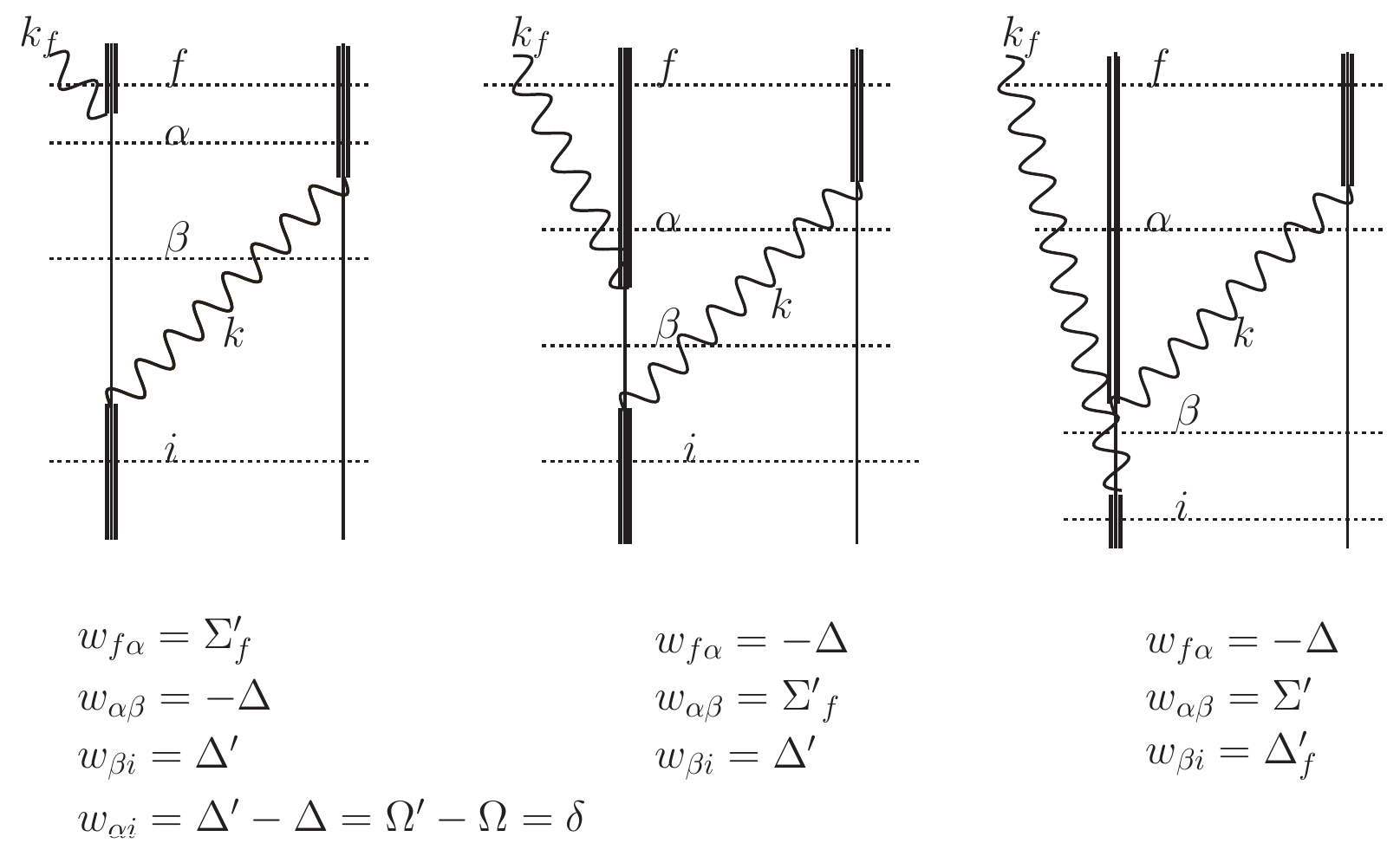}
\end{center}
\caption{The three diagrams contributing to $\mathcal{\delta M}^{(3)}$ in which qubit A emits } \label{Fig.38}
\end{figure}
\noindent where we display the frequencies that enter in Eq.(\ref{eq:1}) and Eq.(\ref{eq:2}) in terms of 
\begin{eqnarray}
\Sigma_f = w_f + \Omega, \Delta_f = w_f - \Omega, \nonumber\\
\Sigma = w + \Omega and \Delta = w - \Omega, 
\end{eqnarray}
where $w_f,\,w$ are the final and exchanged photon frequencies, and $\Omega = (E_e - E_g)/\hbar$ is two level frequency gap. The primed objects that will appear in the equations are associated to the energy $E'_g$ of the intermediate atomic ground states; we will take $E'_g-E_g \rightarrow 0$ at the end of the calculations.

The contribution to $\mathcal{M}^{(3)}$ coming from Fig.2 can be written as
\begin{eqnarray}
\label{eq:3}
&&-\frac{1}{\hbar^3} V^\dag_{e g«} (\mathbf{x}_A)
\bcontraction{}{V}{_{e g}(\mathbf{x}_B)}{ V}
V_{e g} (\mathbf{x}_B) V^\dag_{\,g' e} (\mathbf{x}_A)\{\frac{1}{\Delta' (\Sigma_f-\Sigma'_f)}(\mathcal{S}(\Sigma_f,t)\nonumber\\
&&-\mathcal{S}(\Sigma'_f,t))
+\frac{1}{\Delta' \Delta}(\mathcal{S}(\Sigma'_f-\Delta,t)-\mathcal{S}(\Sigma'_f,t))\}-\nonumber\\ &&\frac{1}{\hbar^3} \bcontraction{}{V}{_{e g}(\mathbf{x}_B)V^\dag_{e g«} (\mathbf{x}_A)}{ V}
V_{e g} (\mathbf{x}_B) V^\dag_{e g«} (\mathbf{x}_A) V^\dag_{\,g' e} (\mathbf{x}_A)\{\frac{1}{\Delta' (\Sigma_f+\Delta)}(\mathcal{S}(\Sigma_f,t)-\nonumber\\ &&\mathcal{S}(-\Delta,t))-
\frac{1}{\Delta' \Sigma'_f}(\mathcal{S}(\Sigma'_f-\Delta,t)-\mathcal{S}(-\Delta,t))\}-\nonumber\\ 
&&\frac{1}{\hbar^3} \bcontraction{}{V}{_{e g} (\mathbf{x}_B)}{ V}
V_{e g} (\mathbf{x}_B) V^\dag_{e g«} (\mathbf{x}_A) V^\dag_{\,g' e} (\mathbf{x}_A)\{
\frac{1}{\Delta'_f (\Sigma_f+\Delta)}(\mathcal{S}(\Sigma_f,t)-\nonumber\\ &&\mathcal{S}(-\Delta,t))-
\frac{1}{\Delta'_f \Sigma'}(\mathcal{S}(\Sigma'-\Delta,t)-\mathcal{S}(-\Delta,t))\}
\end{eqnarray}
In the above formula $\bcontraction{}{P}{(x)\cdots }{ Q} P(x)\cdots Q(y)$ stands for the contraction of the operators $P$ and $Q$, namely
\begin{equation}
\label{eq:4}
\bcontraction{}{V}{_{fi}(\mathbf{x}\cdots) }{ V} V_{fi}(\mathbf{x})\cdots V^\dagger_{f'i'}(\mathbf{y})=
\int dk\, N d_{fi} \cdots d_{f'i'}
e^{ik(x-y)}
\end{equation}
and the function $\mathcal{S}$ is defined as
\begin{equation}
S(z,t) = \frac{e^{izt}-1}{z}
\label{p}
\end{equation}
Finally, we recall some issues about the contractions. First, the frequency that appears in Eq.(\ref{eq:4}) and in the definitions for $\Delta$ and $\Sigma$ is $w = v\, k$. Second, we can pull out the lone operator $V^\dagger$ from the contractions. Third, we will consider real dipole moments, so that $V_{e g}=V_{ge}$, etc. This allows to factorize from equation (\ref{eq:3}) a common factor $-(1/\hbar^3)V^\dagger(x_A)
\bcontraction{}{V}{(x_B)}{V}
V(x_B)V^\dagger(x_A)I_A^B(w,t)$,
where $I_A^B(w,t)$ contains all the terms within braces.

\begin{figure}[h]
\begin{center}
\includegraphics[width=\textwidth]{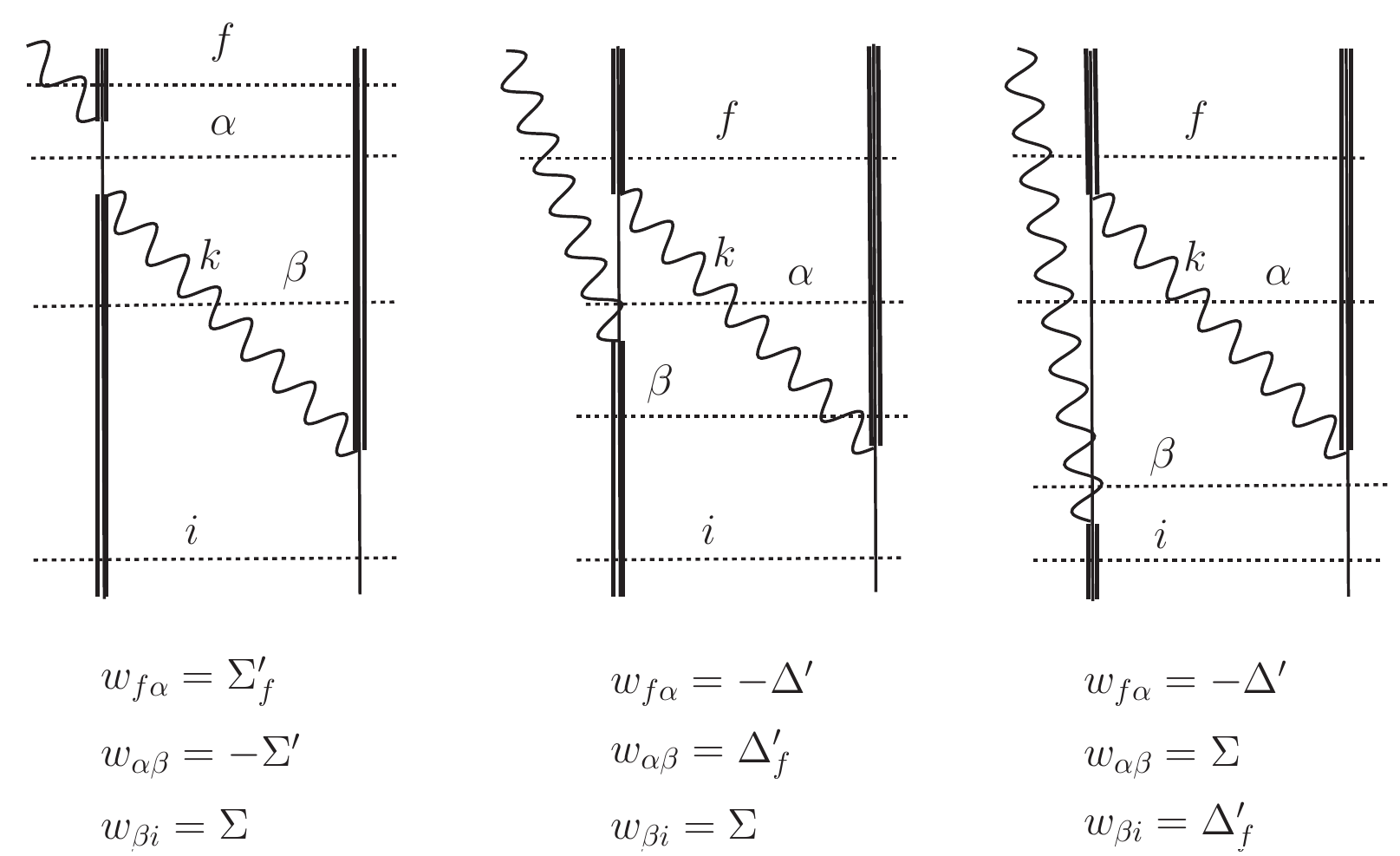}
\end{center}
\caption{The three diagrams contributing to $\mathcal{\delta M}^{(3)}$ in which qubit A absorbs } \label{Fig.39}
\end{figure}
The cases where the exchange ``goes from B to A" are depicted in Fig.3 below. From arguments similar to those of the previous case, we can factorize out the $V$ terms obtaining
$-(1/\hbar^3)V^\dagger(x_A)
\bcontraction{}{V}{(x_A)}{V}
V(x_A)V^\dagger(x_B)I^A_B(w,t)$, where
\begin{eqnarray}
\label{eq:5}
I^A_B(w,t) &&= \frac{1}{\Sigma (\Sigma_f-\Sigma'_f)}(\mathcal{S}(\Sigma_f,t)-\mathcal{S}(\Sigma'_f,t))
+\nonumber\\ &&\frac{1}{\Sigma \Sigma'}(\mathcal{S}(\Sigma'_f-\Sigma,t)-\mathcal{S}(\Sigma'_f,t))
+\frac{1}{\Sigma (\Sigma+\Delta'_f)}\nonumber\\ &&(\mathcal{S}(\Sigma_f,t)-\mathcal{S}(-\Delta',t))-
\frac{1}{\Sigma \Delta'_f}(\mathcal{S}(\Sigma'_f-\Delta',t)-\nonumber\\ &&\mathcal{S}(-\Delta',t))
+\frac{1}{\Delta'_f (\Sigma+\Delta'_f)}(\mathcal{S}(\Sigma_f,t)\mathcal{S}(-\Delta',t))-\nonumber\\ &&-
\frac{1}{\Delta'_f \Sigma}(\mathcal{S}(\Sigma-\Delta',t)-\mathcal{S}(-\Delta',t))
\end{eqnarray}
Finally, the contribution of third order to the amplitude we are interested in is given by
\begin{eqnarray}
\label{eq:6}
\mathcal{\delta M }^{(3)}&& = -(1/\hbar^3)\{
V^\dagger(x_A)
\bcontraction{}{V}{(x_B)}{V}
V(x_B)V^\dagger(x_A)I_A^B(w,t)+\nonumber \\ 
&&V^\dagger(x_A)
\bcontraction{}{V}{(x_A)}{V}
V(x_A)V^\dagger(x_B)I^A_B(w,t)\}
\end{eqnarray}

\chapter{Majorana equation in 1+1 dimensions}
We start from the Dirac equation in 1+1 dimensions and in covariant form \cite{thaller}:
\begin{equation}
-i\gamma^0\,\partial_0\psi-i\gamma^1\,\partial_1\psi+m\psi=0 \label{dirac}
\end{equation}
with $\hbar=c=1$ and the gamma matrices $\gamma^{\mu}$, $\mu=0,1$ verifying the anticommutation rules of a Clifford algebra 
\begin{equation}
 \{\gamma^{\mu},\gamma^{\nu}\}=2\,g^{\mu\nu} , \label{cliffordluyk}
 \end{equation}
 $g^{\mu\nu}$ being the 1+1 Minkowski metric. Multiplying by $\gamma^0$ and with the usual notation 
 \begin{equation}
 \beta=\gamma^0, \alpha=\gamma^0\gamma^1, \label{alfasbetasgammas}
 \end{equation}
 we can write it in the Schr\"odinger-like form:
\begin{equation}
i\partial_t\psi=\alpha\,p\psi+m\beta\psi\label{dirac2}
\end{equation}
If we want $\alpha$ and $\beta$ to be  involutions ($\alpha^2=\beta^2=1$) they must be Pauli matrices. In  table ~\ref{tab:1} we spell the six possible cases.
\begin{table}[h] 
\[ \begin{array}{|c|c|c|c|c|}\hline
\beta & \gamma^1 & \alpha & H & Cases \nonumber\\ 
\hline
&i \sigma_y & - \sigma_z&  -\sigma_z p - i m\sigma_y K &\mbox{I.a}\nonumber\\ \cline{2-4}
\mbox{\raisebox{1.5ex}[0pt]{$ \sigma_{x}$}}& i \sigma_z & \sigma_y& \sigma_y p - i m \sigma_y K &\mbox{I.b}\nonumber\\ \hline
 & i \sigma_x &   \sigma_z&  \sigma_z p +i m \sigma_y K&\mbox{II.a}\nonumber\\ \cline{2-4}
 \mbox{\raisebox{1.5ex}[0pt]{$ \sigma_{y}$}}& i \sigma_z & -\sigma_x& -\sigma_x p + i m \sigma_y K &\mbox{II.b}\nonumber\\ \hline
& i \sigma_x &  - \sigma_y&  -\sigma_y p -i m \sigma_y K &\mbox{III.a}\nonumber\\ \cline{2-4}
\mbox{\raisebox{1.5ex}[0pt]{$ \sigma_{z}$}}& i \sigma_y & \sigma_x& \sigma_x p - i m \sigma_y K&\mbox{III.b}\nonumber\\ \hline 
\end{array}\]
\caption{The six possible representations of the 1+1 Clifford algebra, with the corresponding Majorana Hamiltonian associated.}\label{tab:1}
\end{table}

The Majorana equation is obtained by replacing the $\Psi$ in the mass term of Eq. (\ref{dirac}) or Eq. (\ref{dirac2}) by the charge conjugate $\Psi^C$. Then:
\begin{equation}
i\partial_t\psi=\alpha\,p\psi+m\beta\psi^C\label{majorana}
\end{equation}
The charge conjugation is defined in the following way \cite{zee}: 
\begin{equation}
\psi^C=\eta\tilde{C}\psi^* \label{conjugating} 
\end{equation}
where the matrix $\tilde{C}$ must verify 
\begin{equation}
\tilde{C}\gamma_{\mu}^* \tilde{C}^{-1}=-\gamma_{\mu}\label{tildec}
\end{equation}
and $\eta$ is an arbitrary phase.
Now, defining 
\begin{equation}
\tilde{C}=C\,\beta \label{tildecyc}
\end{equation}
and using that
 \begin{equation}
 \beta\gamma_{\mu}^*\beta=\gamma_{\mu}^T \label{stepbetas}
 \end{equation}
 if $\beta$ is real or purely imaginary, we have:
\begin{equation}
C\gamma_{\mu}^T C=-\gamma_{\mu} \label{stepc}
\end{equation}
and then $C$ must commute with $\sigma_y$ and anticonmute with $\sigma_{x,z}$, so 
\begin{equation}
C=\sigma_y \label{crepind}
\end{equation}
\textit{for all representations}, and 
\begin{equation}
\tilde{C}=\sigma_y\,\beta. \label{tildecend}
\end{equation}

Therefore, the Majorana equation can be written in the following form:
\begin{equation}
i\partial_t\psi=\alpha\,p\psi+\eta m\tilde{\beta}\psi^*\label{majorana2}
\end{equation}
where $\tilde{\beta}$ is given by 
\begin{equation}
\tilde{\beta}=\beta\tilde{C}=\pm \sigma_y \label{tiedebetaend}
\end{equation}
 and the plus sign corresponds to the representations in which $\beta=\sigma_y$ (Cases II in  table \ref{tab:1}) and the minus to the ones with $\beta=\sigma_{x,z}$ (Cases I and III). Finally we define the antiunitary involution $K$ \cite{wigner}, given by 
\begin{equation}
\psi^*=K\,\psi \label{ka}
\end{equation}
and set the global phase to $\eta=i$ by convenience. Thus the Majorana equation is 
\begin{equation}
i\partial_t\psi=\alpha\,p\psi\pm i m\sigma_y\,K\psi\label{majorana3}
\end{equation}
where now the plus corresponds to II.  In chapter 6 we choose the particular representation III. b and then
\begin{equation}
i\partial_t\psi=\sigma_x\,p\psi- i m\sigma_y\,K\psi\label{majorana3-b}
\end{equation}
This last equation can alternatively be thought of as the dimensional reduction of one of the equations in which the 3+1 Majorana equation in Weyl representation can be decomposed.

Eq.(\ref{majorana3}) is a Schr\"odinger-like equation in which 
\begin{equation}
H= \alpha p \pm i m \sigma_y K \label{hache}
\end{equation}
plays the role of a Hamiltonian. 
Hitherto we have considered the ``free'' Majorana equation, but we could add a potential $V$ to Eq. (\ref{majorana3}), obtaining:
\begin{equation}
H'= V + \alpha p \pm i m\sigma_y K \label{hacheprime}
\end{equation}

\chapter{Implementation of the quantum simulation of the Majorana equation in trapped ions}

We can simulate Eq.~(\ref{eq:majorspin31}) using two trapped ions that are subject to two dynamical terms, coupling both to the internal states and motion of the ions. The kinetic part, $cp_x (\mathbf{1}\otimes \sigma_x)$, is created with a laser tuned to both the blue and the red motional sideband of an electronic transition~\cite{lucas, jorge1}, and focussed on ion $2$. The spin-spin interaction term, $\sigma_x\otimes \sigma_y$, is derived from detuned red and blue sideband excitations acting on each ion in addition to the previous one~\cite{Molmer99,Roos08}. The Hamiltonian describing this situation reads
\begin{eqnarray}
\label{H}
H&=&\hbar\frac{\omega_0}{2}\sigma_1^z+\hbar\frac{\omega_0}{2}\sigma_2^z+\hbar\nu a^\dag a +\hbar\nu_r b^\dag b \nonumber \\
&+& \hbar\Omega           \left[ (e^{i(qz_1 -\omega_1t+\phi_1)}+e^{i(qz_1 - \omega_{1}'t+\phi'_1)})  \sigma_1^+  + {\rm H. c.} \right]\nonumber\\
&+& \hbar\Omega           \left[ (e^{i(qz_2 - \omega_2t+\phi_2)}+e^{i(qz_2 - \omega_2't+\phi'_2)})  \sigma_2^+  + {\rm H. c.} \right]\nonumber\\
&+& \hbar{\tilde\Omega} \left[ (e^{i(qz_2-\omega t +\phi_{}  )} + e^{i(qz_2-\omega' t +\phi'_{} )})  \sigma_2^+ + {\rm H.c.}\right] . \nonumber \\
\end{eqnarray}
Here $z_{1,2} = Z \pm \frac{z}{2}$ are the ion positions, measured from the center of mass, $Z,$ and relative coordinate, $z$. The phases of the lasers $\phi_i$ for $i=1,2$, ($\phi$,  $\phi'$), are controlled to perform the interaction term (kinetic term). The frequencies of the center of mass and stretch mode are given by $\nu$ and $\nu_r=\sqrt{3}\nu$, while $a^{\dag}, \ a$, $b^{\dag}$, and $b$, are the corresponding creation and annihilation operators. Finally, $\Omega$ and ${\tilde \Omega}$ are the Rabi frequencies of the lasers within the bounds of applicability of the rotating-wave approximation. With an appropriate choice of parameters
\begin{equation}
\label{frecuencias}
\begin{array}{ccc}
\omega_1&=&\omega_0 + \nu_r -\delta\\
\omega_1'&=&\omega_0-\nu_r + \delta\\
\omega_2&=&\omega_0-\nu_r + \delta\\
\omega_2'&=&\omega_0 + \nu_r -\delta ,\\
\end{array} \
\begin{array}{lcc}
\omega&=&\omega_0-\nu\\
\omega'&=&\omega_0+\nu\\
\phi&=&\pi\\
\phi'&=&0,\\
\end{array} \
\begin{array}{ccl}
\phi_1&=&\pi / 2\\
\phi'_1&=&\pi /2\\
\phi_2&=&0\\
\phi'_2&=&0,
\end{array}
\end{equation}
the Hamitonian~(\ref{H}) in the interaction picture reads
\begin{eqnarray}
  H&=& \hbar\eta_r\Omega(\sigma_x\otimes\mathbf{1} - \mathbf{1}\otimes\sigma_y ) (b^{\dag}e^{i\delta t} + be^{-i\delta t}) , \nonumber
 \\
  &&+\hbar\eta{\tilde\Omega}(\mathbf{1} \otimes \sigma_x)i(a^{\dag} - a)
  \end{eqnarray}
where $\eta\equiv \eta_r3^{1/4}\equiv\sqrt{{\hbar}/{4 m' \nu}} \ll 1$ is the Lamb-Dicke parameter and $m'$ the ion mass. In the limit of large detuning, we have
\begin{equation}
  \label{eq:dispersive}
  \delta\gg \eta_r\Omega\sqrt{\langle b^\dag b \rangle}, \eta{\tilde\Omega}|\langle {a^\dag -a} \rangle | .
\end{equation}
We recover Eq.~(\ref{eq:majorspin31}) with the momentum operator $p_x=i\hbar(a^{\dag}{-}a)/2\Delta$ and the equivalences
\begin{equation}
\label{eq:equivalence}
c = 2\eta\Delta\tilde{\Omega}, \ \ \ mc^2=\frac{2\hbar\eta_r^2\Omega^2}{\delta}.
\end{equation}

\noindent Here, $\Delta=\sqrt{\frac{\hbar}{4 m' \nu}}$ is the size of the harmonic oscillator ground state. Note that it is possible to explore all velocity regimes. Introducing the ratio $\gamma = |mc^2 / \langle c p_x \rangle |$,
\begin{equation}
  \gamma = \frac{2 (\eta_r\Omega/\delta)^2}{|\langle i( a^\dagger - a)\rangle|(\eta\tilde\Omega/\delta)},
\end{equation}
it is possible to tune the numerator and denominator independently so as to preserve the dispersive regime, while exploring simultaneously the range from $\gamma \simeq 0$ (ultrarelativistic limit) to $\gamma \to\infty$ (nonrelativistic limit).

We could also consider other implementations that do not require synchronization of laser phases for different beams. Consider, for example, the Hamiltonian
\begin{equation}\label{eq:Hequivalent}
  H = c\,\mathbf{1}\otimes(p_x \sigma_z) - mc^2 \sigma_y\otimes\sigma_y,
\end{equation}

\noindent which is equivalent to that of equation~(\ref{eq:majorspin31}) up to local unitary rotations. Using a detuning of $\pm \nu/2$ for the blue and red sideband, respectively, in the laser focussed onto ion 2, leads to an interaction of the form $8\Delta\eta\tilde{\Omega}^2\mathbf{1}\otimes(p_x \sigma_z)/\nu$ and the equivalence $c = 8\Delta\eta\tilde{\Omega}^2/\nu$. The spin-spin interaction in the second term can be implemented by an additional global bichromatic light field acting on the red and blue sidebands of the stretch mode, similar to the case above. In the present case, however, the ions experience laser light with the same phase $\phi_1=\phi_2=\phi'_1=\phi'_2=0$ and a single laser beam can be used. 

\vspace{0.5cm}
{\bf Measurement of the pseudo-helicity}\\

In experiments using trapped ions, the only observable that can be directly measured by fluorescence detection for each ion is $\sigma_z$, and additional laser pulses are needed to map other observables onto it. The application of a state-dependent displacement operation on ion 2, $U_2=\exp(-ik(\mathbf{1}\otimes\sigma_y)\otimes p_x/2)$, generated by a resonant blue and red sideband, followed by a measurement of $\mathbf{1}\otimes\sigma_z$ is equivalent to measuring the observable
\begin{eqnarray}
A(k) & = & U_2^\dagger(\mathbf{1}\otimes\sigma_z)U_2 \nonumber \\
& = & \cos(k\,p_x)(\mathbf{1}\otimes\sigma_z)+\sin(k\,p_x)(\mathbf{1}\otimes\sigma_x).
\end{eqnarray}

Here, $k$ is proportional to the probe time $t_{probe}$~\cite{naturekike}. In order to measure the first term in Eq.~(\ref{eq:pshelicity}), we note that $\left.\frac{d}{dk}\langle A(k)\rangle\right\vert_{k=0}\propto\langle(\mathbf{1}\otimes\sigma_x)\otimes p_x\rangle$. Therefore, this term can be measured by applying a short probe pulse to the ions and measuring the initial slope of the observable $A(k)$~\cite{Lougovski06,naturekike}.

To measure the second term in Eq.~(\ref{eq:pshelicity}), we have to apply a different state-dependent displacement operation to ion 1: $U_1=\exp(-ik(\sigma_x\otimes\mathbf{1})\otimes p_x/2)$, and measure the spin correlation $\sigma_z\otimes\sigma_x$, which requires an additional $\pi/2$ pulse on ion 2. Again taking the initial slope in this observable, we have
\begin{equation}
\label{correlation}
\frac{\partial \langle \sigma_z \otimes \sigma_x \rangle}{\partial k}\bigg|_{k=0} =  2 \langle (\sigma_y \otimes \sigma_x)\otimes p_x\rangle.
\end{equation}
\addcontentsline{toc}{chapter}{Conclusions}
\chapter*{Conclusions}
Throughout this Thesis, the specific results have been summarized at the end of each section.  Now we will outline the most important results and conclusions.
\section*{Dynamics of entanglement}
We have characterized the dynamics of the generation and destruction of entanglement between qubits mediated by a quantum field, focusing in the relationship with the spacetime region at which the qubits are placed. More precisely, we have considered a pair of two-level atoms -real or artificial, that is, superconducting qubits-  A and B separated by a fixed distance $r$ interacting through a quantum electromagnetic field. 
\begin{itemize}
\item (Sections 3.1, 3.2 and 4.2) If the qubits are initially in a separable state, two different spacetime regions emerge from our computations. If $v\,t<r$ - $t$ being the time of the interaction and $v$ the propagation velocity of the field quanta- entanglement can be generated between the atoms due to the non-locality of the Feynman propagator - or in other language, ``vacuum entanglement'' -  which give rise to correlations at any time. In the weak coupling regime, which is the case with real atoms and the electromagnetic field, we show that these correlations are only classical, but entanglement -even maximal- can be generated through a measurement of the state of the field. For stronger couplings, as is the case in 1-D circuit QED, a small amount of entanglement shows up even in the absence of a measurement. If $v\,t>r$, a very different behavior of entanglement appears due to photon exchange, making apparent that entanglement generation may be a good signature of single photon propagation, a fact of particular significance in the framework of circuit QED. In all the cases the point $v\,t =r$ plays the role of a frontier between two different spacetime regions regarding the behavior of entanglement dynamics. We have devised a detailed circuit QED proposal of an experiment to test all the above physics.
\item (Section 3.3) If the qubits are initially entangled, we have characterized the remarkable phenomenon that a full Entanglement Sudden Death-Sudden Birth cycle can take place for $v\,t<r$. In this case, these entanglement destructions and revivals are related with the revivals and destructions of the atom-field entanglement, that is to say, there is sort of an entanglement flux between the different pairs of the tripartite system atoms-field. That this can happen at times at which photon exchange is not allowed represent an additional insight on the nature of the generation and destruction of quantum correlations.
\end{itemize}
\section*{Non-RWA effects}
We have seen that the reasons for going beyond the ubiquitous and celebrated Rotating Wave approximation are twofold. On the fundamental side, it is necessary to include all the non-RWA terms for a complete theoretical short-time analysis of the matter-radiation interaction or to properly deal with causality questions, although in these systems these effects are beyond experimental reach. On a more applied viewpoint, circuit QED in the ultrastrong coupling provides a framework in  which the non-RWA effects are available to experiment. We have explored a striking consequence of working beyond RWA, the fact that a qubit in the ground state can get excited and emit a photon, even if the field is also in the vacuum state. 
\begin{itemize}
\item (Section 4.1) In the simplified model of a circuit QED analog of a cavity QED system, we have introduced the ``slow quantum anti-Zeno effect'', providing a full experimental proposal to detect a qubit in its excited state with certainty after a few measurements starting from the ground state of the whole system. The protocol is based in the certain probability of excitation already present in the system for strong enough couplings and in the non-equilibrium dynamics after a single measurement. This feasible and realistic experiment would shed light onto the physical reality of a phenomenon commonly considered as ``virtual''.
\item (Section 5.2) In the case of the full multimode Hamiltonian of circuit QED, we have analyzed the implications of the mentioned ground state qubit self excitations and in general the non-RWA dynamics in the interpretation of the measurements of the probability of excitation. In particular, we have seen that only for long interaction times we can say with certainty that a qubit detector's click entails the decay of an excited source. For short -although experimentally accessible in circuit QED- times, non-RWA contributions dominate in the dynamics of the probability of excitation conditioned to the decay of the source, meaning that in this regime a click is more likely linked with a self-excitation.
\end{itemize}
\section*{Causality}
We have revisited the old question -Fermi problem- on causality in matter-radiation interactions, which we can rephrase as follows: ``Is the probability of excitation of the atom B causal?'' To properly deal with this issue, we have first to pose a different question: ``What does causality mean?'' As we have seen, quantum correlations may show up at superluminal rates, thus the probability of a given nonlocal state such as one atom in the excited state and the other in the ground state becomes non-zero at $v\,t<r$. Also we have to take into account the mentioned fact that an atom can get self-excited at any time. But all these is compatible with causality since cannot be used to transmit information faster than light. The relevant question is actually ``Is the probability of excitation of atom B independent of  atom A at times $t<r/v$?''
\begin{itemize}
\item (Section 5.1) We have shown that the answer to the latter question is affirmative, by providing a complete non-perturbative proof of the independence of the probability of excitation of atom B with regard to atom A.  
\item (Section 5.1) We have exploited again the experimental amenability of circuit QED to propose a realistic experiment to check the causal behavior of the probability of excitation, coming true a hitherto considered as ``gedanken'' experiment. The experiment, together with the theoretical clarification given above, would close a long-lasting controversy on the Fermi problem, confirming the causal behavior of Nature at this level and shedding light on the meaning and implications of the notion of causality.
\end{itemize} 
\section*{Quantum simulations}
Although circuit QED can be understood as a closed 1D universe and its physics is interesting by itself, it can also be understood as a quantum simulator for matter-radiation interactions. Besides, we have also introduced some results on quantum simulations of relativistic quantum mechanics with trapped ions.
\begin{itemize}
\item (Sections 6.1 and 6.2) We have introduced a new family of generalized hamiltonians, `Majorana Hamiltonians" which do not follow the standard textbook definition of hermiticity but we have shown that induce a norm-conserving dynamics and can always be simulated with regular Hamiltonians. In particular, we have proposed a trapped ion quantum simulation of one conspicuous member of the family, namely the pseudo-hamiltonian appearing in the Majorana equation- the Relativistic Quantum Mechanics equation of a fermion with a Majorana mass term-. We have explored astounding properties of the Majorana Hamiltonians, as the possibility of measure initial global phases in the evolution of physical observables.  Thus, this experiment would be useful to explore fundamental questions being at the edges of Quantum Mechanics and Quantum Field Theory. But the scope is more general and includes possible simulations of operations like time reversal, charge conjugation or partial transpose in the lab.
\item (Section 6.3) We have introduced a method to use the current simulations of free relativistic Hamiltonians to simulate also Hamiltonians including potentials, opening the way to more complex simulations of interacting many-body relativistic equations, in which the results of the first parts of this Thesis could be explored within different frameworks.
\end{itemize}

\end{document}